\documentclass[12pt,oneside,letterpaper]{article}
\usepackage{caption}
\usepackage{tabu}
\usepackage{boldline,multirow}
\usepackage{multirow,array}
\usepackage{arydshln}
\usepackage{makecell}
\usepackage{longtable}
\usepackage{amssymb}
\usepackage{amsmath}
\numberwithin{equation}{section}
\usepackage[dvips]{graphicx}
\usepackage[dvipsnames]{xcolor}
\usepackage{setspace}
\usepackage{amsfonts}
\usepackage{fancyhdr}
\usepackage{xcolor}
\usepackage{graphicx}
\usepackage{rotating}
\usepackage{comment}
\usepackage{color}
\usepackage{cite}
\usepackage{braket}
\usepackage{amssymb,amsmath,amsthm,mathrsfs,latexsym,amsxtra,graphicx,appendix,epstopdf,feynmf,subfig,setspace,fix-cm,color,cite}
\usepackage[nottoc,numbib]{tocbibind}
\usepackage{pgfplots}
\usepgfplotslibrary{fillbetween}
\pgfplotsset{compat=newest}
\usepackage{mathtools}
\usepackage{tensor}
\usepackage{siunitx}
\usepackage{tikz}
\usetikzlibrary{trees}
\usetikzlibrary{decorations.pathmorphing}
\usetikzlibrary{decorations.markings}
\usetikzlibrary{decorations.markings}
\usepackage{float}
\usepackage{rotating}
\usepackage{simplewick}
\usepackage{amssymb}
\usepackage{amsmath}
\usepackage{setspace}
\usepackage{amsfonts}
\usepackage{fancyhdr}
\usepackage{xcolor}
\usepackage{graphicx}
\usepackage{rotating}
\usepackage{comment}
\usepackage{color}
\usepackage{cite}
\usepackage{braket}
\usepackage[colorlinks=true, linkcolor  = blue,urlcolor=blue,citecolor=violet]{hyperref}
\usepackage{simpler-wick}

\definecolor{darkgreen}{rgb}{0,0.5,0}
\definecolor{darkblue}{rgb}{0,0,0.6}
\definecolor{purple}{rgb}{0.4,0.15,0.21}
\definecolor{black}{rgb}{.2,.2,.2}

\DeclareSymbolFont{myletters}{OML}{ztmcm}{m}{it}
\DeclareMathSymbol{\uplambda}{\mathord}{myletters}{"15}

\begin{document} 

\tikzset{middlearrow/.style={
        decoration={markings,
            mark= at position 0.5 with {\arrow{#1}} ,
        },
        postaction={decorate}
    }
}

\newcommand{\Vast}{\bBigg@{5}}
\newtheorem*{theorem}{Theorem}
\newtheorem{conjecture}{Conjecture}[section]
\newtheorem{corollary}{Corollary}
\newtheorem{lemma}{Lemma}
\newtheorem*{remark}{Remark}
\newtheorem{claim}{Claim}
\newtheorem{prop}{Proposition}
\theoremstyle{definition}
\newtheorem{definition}{Definition}[section]
\renewcommand\qedsymbol{$\blacksquare$}

\newcommand{\be}{\begin{equation}}
\newcommand{\ee}{\end{equation}}
\newcommand{\bea}{\begin{eqnarray}}
\newcommand{\eea}{\end{eqnarray}}
\newcommand{\ba}{\begin{align}}
\newcommand{\ea}{\end{align}}
\newcommand{\bb}{\mathbb}
\newcommand{\mrm}{\mathrm}
\newcommand{\scr}{\mathscr}
\newcommand{\p}{\partial}
\newcommand{\dd}{\mathrm{d}}
\newcommand{\tr}{\mathrm{tr}}
\newcommand{\RR}{\mathrm{R}}
\newcommand{\FF}{\mathrm{F}}
\newcommand{\DD}{\mathrm{D}}
\newcommand{\ff}{\mathrm{f}}
\newcommand{\rr}{\mathrm{r}}
\newcommand{\Tr}{\mathrm{Tr}}
\def\e{{\rm e}}
\def\bz{{\bar z}}
\def\bw{{\bar w}}
\def\p{{$|\Phi\rangle$}}
\def\pp{{$|\Phi^\prime\rangle$}}
\def\cO{{\mathcal O}}
\def\cH{\mathcal{H}}
\def\cF{\mathcal{F}}
\def\cL{\mathcal{L}}
\def\Xint#1{\mathchoice
   {\XXint\displaystyle\textstyle{#1}}%
   {\XXint\textstyle\scriptstyle{#1}}%
   {\XXint\scriptstyle\scriptscriptstyle{#1}}%
   {\XXint\scriptscriptstyle\scriptscriptstyle{#1}}%
   \!\int}
\def\XXint#1#2#3{{\setbox0=\hbox{$#1{#2#3}{\int}$}
     \vcenter{\hbox{$#2#3$}}\kern-.5\wd0}}
\def\ddashint{\Xint=}
\def\dashint{\Xint-}

\newcommand{\CC}{\mathbb{C}} 
\newcommand{\ZZ}{\mathbb{Z}} 
\newcommand{\II}{\mathrm{I}} 
\newcommand{\NN}{\mathbb{N}} 
\newcommand{\QQ}{\mathrm{Q}}
\newcommand{\xx}{\mathrm{x}}
\newcommand{\Pp}{\mathrm{p}}

\newcommand{\T}{R}

\newcommand{\lcm}{\mathrm{lcm}}

\newcommand{\ie}{{\it i.e.~}}
\def\eg{{\it e.g.~}}

\onehalfspacing

\begin{center}

~
\vskip5mm

{{\Huge {\textsc{Notes on Matrix Models}}}
}
\vskip2mm
{{ {\textsc{(Matrix Musings)}}}
}

\vskip10mm

{{Dionysios Anninos$^{1}$ and Beatrix M\"uhlmann$^{2}$}}\\ 

\end{center}
\vskip4mm
\begin{center}
{\footnotesize{$^{1}$ Department of Mathematics, King's College London, Strand, London WC2R 2LS, UK \newline\newline
$^{2}$ Institute for Theoretical Physics and $\Delta$ Institute for Theoretical Physics, University of Amsterdam, Science Park 904, 1098 XH Amsterdam, The Netherlands\\
}}
\end{center}
\begin{center}
{\textsf{\footnotesize{dionysios.anninos@kcl.ac.uk \, \& \,  b.muhlmann@uva.nl}} } 
\end{center}
\vskip5mm

\vspace{4mm}

\begin{abstract} 
In these notes we explore a variety of models comprising a large number of constituents. An emphasis is placed on integrals over large Hermitian matrices, as well as quantum mechanical models whose degrees of freedom are organised in a matrix-like fashion. We discuss the relation of matrix models to two-dimensional quantum gravity coupled to conformal matter. We provide a brief overview of a variety of more general quantum mechanical matrix models and their putative worldsheet interpretation, as well as other recent developments on large $N$ systems. We end with a discussion of open questions and future directions. 
\newline\newline

 \end{abstract}

\pagebreak
\pagestyle{plain}

\setcounter{tocdepth}{2}
{}
\vfill
\tableofcontents


\section{Introduction$^{\color{magenta} A, B}$}

The Universe is replete with spacetime. Although there is no known complete theory of spacetime at the quantum level, various ideas in modern theoretical physics suggest the radical notion that the classical description of spacetime must somehow be replaced by a microscopic theory whose basic variables are no longer built from a metric field. For general spacetimes, particularly those dominated by a vanishing or positive cosmological constant, little is known about this microscopic theory. Nevertheless, there are important hints that this theory must contain a large number $N$ of building blocks which are strongly interacting amongst each other. In these notes\footnote{The content is an extended version of lectures delivered at the theoretical physics Solvay Doctoral School in December of 2017.} we explore several systems composed of a large number $N$ of constituents and discuss a variety of computational techniques used to solve them in the large $N$ limit. A significant focus will be placed on large $N$ systems whose components are organised in a matrix-like fashion. We will focus on both ordinary integrals over matrices as well as quantum mechanical models whose degrees of freedom and interactions are organised in a matrix-like structure. 

The theory of random matrices has a rich history in physics and mathematics. It is not our aim to review this history, but we would like to mention a few examples. Perhaps the earliest application, dating back to the 1950s, was Wigner's proposal \cite{Bohr:1936zz, wigner, wigner2} that the distribution of adjacent energy level spacings for the spectra of heavy nuclei is well approximated by those of a random real symmetric matrix drawn from a Gaussian ensemble. This remarkable hypothesis constitutes a conceptual leap -- a single theory is approximated by an ensemble of theories. This is different to performing averages over an ensemble of configurations within a single theory.\footnote{A method that is conceptually similar to Wigner's hypothesis is the quenched disorder approximation used in certain condensed matter and statistical systems such as spin glasses \cite{spinglass}. In this context, certain couplings in the theory, rather than the entire Hamiltonian, are chosen to be random.} Wigner surmised that the distribution of level spacings $s$ of a random $N\times N$ real symmetric matrix whose elements are independently drawn from a Gaussian distribution obeys the following universal form in the large $N$ limit
\begin{equation}\label{levelspacing}
p(s) = \frac{\pi s}{2} e^{-\pi s^2/4}~. \nonumber
\end{equation} 
The above formula has been successfully compared to numerous spectra of heavy nuclei such as $^{238}$U and $^{166}$Er. The precise form of $p(s)$ was obtained by Mehta and Gaudin  \cite{mehta}, and is indeed described to good approximation by the above expression. The class of ensembles was broadened to random Hermitian and random symplectic matrices by Dyson \cite{Dyson}. The choice of ensemble is tied to what global symmetries are present in the underlying system. An important characteristic feature of the above distribution is the suppression of $p(s)$ at small values of $s$. More generally, level repulsion in the eigenvalue spacing is a common litmus of complex systems. 


Wigner's hypothesis is that certain properties of systems governed by sufficiently `complex' Hamiltonians may be well approximated by a random ensemble. There is no reason this hypothesis should be restricted to the spectra of heavy nuclei. Indeed, another context where the ideas of random matrices have been applied is the theory of quantum chaos. At the classical level, a dynamical system can be said to be non-integrable if the number of conserved quantities is less than half the dimension of its phase space. A characteristic feature is exponential sensitivity to variations in the initial conditions, encoded in quantities such as the Lyapunov coefficients, the Kolmogorov-Sinai entropy, and the study of Poincar{\'e} sections. Any analogous framework for quantum systems, whose wavefunction obeys a linear equation, is significantly more involved. Nevertheless, it has been observed in numerous systems that the spectra of quantum Hamiltonians stemming from the quantisation of classically chaotic theories also exhibit a level spacing of the type in $p(s)$. This has been postulated by Bohigas-Giannoni-Schmit \cite{Bohigas:1983er} as a general principle. Although exceptions to the rule have been observed, the BGS postulate is an intriguing tenet.

Matrix integrals, and hence the theory of random matrices, are connected to another class of physical systems. Consider a quantum theory whose degrees of freedom transform in the adjoint representation of an $SU(N)$ symmetry. Such theories include, but are in no sense limited to, Yang-Mills theory with an $SU(N)$ gauge group. If the theories admit a perturbative diagrammatic expansion, we can calculate terms using a path integral over the matrix degrees of freedom. The crucial observation due to `t Hooft \cite{tHooft:1973alw} is that at large $N$ the Feynman diagrams of such theories can be organised in terms of a genus expansion of triangulated Riemann surfaces $\Sigma_h$. The lattice points of $\Sigma_h$ are the interaction vertices in the Feynman graph. This perturbative expansion resembles the perturbative expansion of a string theory. If it so happens that for certain values of the coupling the number of vertices in $\Sigma_h$ diverges, one might imagine associating a string theory to the large $N$ limit of a quantum mechanical theory of matrices. Upon quantisation, the spectrum of a string often includes a massless spin-two particle. In such circumstances, the low energy description of the string theory is given by general relativity coupled to matter \cite{Callan:1985ia}. As such, assuming the hypothesis that large $N$ matrix models are associated to string theories is indeed valid, one encounters a powerful avenue to explore theories of quantum gravity from an entirely different perspective. There is overwhelming evidence that this hypothesis is indeed correct for certain matrix models. Perhaps the most well known example \cite{Maldacena:1997re} is maximally supersymmetric $SU(N)$ Yang-Mills theory in four-dimensions which is captured by the type IIB superstring on AdS$_5\times S^5$ -- an instance of the AdS/CFT correspondence. 

Finally, we note that the relation between the diagrammatic expansion of large $N$ matrices and triangulated Riemann surfaces also finds interesting roots in the realm of two-dimensional geometry and quantum gravity. Early and remarkable numerical work on random triangulation motivated by Polyakov's path integral formulation of string theory \cite{Polyakov:1981rd} as well as the ideas of Regge-calculus \cite{Regge:1961px} has been performed in \cite{Ambjorn:1985az, David:1985nj, Kazakov:1985ea}. From a more mathematical perspective, Kontsevich has shown that Witten's conjecture \cite{Witten:1990hr} on the intersection theory of Riemann surfaces is captured by a particular matrix integral resembling the Airy integral \cite{Kontsevich:1992ti}. The beautiful results obtained by  Maryam Mirzakhani for volumes of certain hyperbolic moduli spaces (Weil-Petersson volumes) \cite{Mirzakhani:2006fta} fit nicely into the topological recursion of Eynard and Orantin \cite{Eynard:2007kz}, itself having connections to matrix integrals. 

%

It is natural to suspect that the theory of large $N$ random matrices as well as the scope of Wigner's hypothesis -- particularly in the context of string theory and the theory of black holes -- will continue to surprise. This is an important motivation for our choice of content. There are many excellent and detailed reviews   \cite{mehtabook, Klebanov:1991qa, Ginsparg:1993is, DiFrancesco:1993cyw,Polchinski:1994mb,Taylor:2001vb,tHooft:2002ufq, Alexandrov:2003ut, Nakayama:2004vk,Marino:2004eq,Martinec:2004td,Marino:2012zq,Eynard:2015aea, PolchinskiBook } on the subjects we discuss, many of which we have drawn enormous inspiration from. Nevertheless, it seems appropriate to bring together several of the subjects that are often presented in a somewhat disjoint fashion, and to provide a discussion on the relation to more recent developments. We develop our presentation in order of simplicity starting with integrals over vector-like variables, followed by integrals over a single as well as multiple matrices, quantum mechanical matrix theories, and finally the relation to two-dimensional quantum gravity and Liouville theory. We provide a brief overview of the broader scope of the ideas explored and end with a speculative overview. Details of some calculations and further results can be found in the various appendices. Finally, the bibliography is partitioned in terms of the various themes and ideas discussed. The relevant pieces of the bibliography for a given section are encoded in the superscripts at the end of the section titles. 

\section{Large $N$ vector integrals}

In this section we provide a brief discussion of the saddle point approximation in its simplest form. We subsequently apply it to a class of integrals over a large number $N$ of variables organised in a vector-like structure.

\subsection{Saddle point approximation}

We begin by considering the following family of integrals
\begin{equation}
\mathcal{I}_{N} = \int_{\mathbb{R}} \dd x \, e^{-N f(x)} \label{eq:saddle1D}~,
\end{equation}
where $N$ is a positive integer which will eventually be taken to be large, and $f(x)$ is a real valued function. In the limit where $N$ becomes large, the exponential causes the integrand to peak sharply at the minima of the function $f(x)$, while all other values are suppressed. Of all extrema, the integral $\mathcal{I}_N$ will be dominated by the one which minimises $f(x)$ as $N$ becomes large, and we denote this by $x_e$. Expanding $f(x)$ in a Taylor series around $x_e$
\begin{equation}
f(x)= f(x_e) + \frac{f''(x_e)}{2!} \, \delta x^2 + \ldots~, \quad\quad \delta x \equiv x-x_e~,
\end{equation}
one can approximate $\mathcal{I}_N$ as
\begin{equation}
\mathcal{I}_{N} \approx e^{-N f(x_e)}\int_{\mathbb{R}}\dd \delta x \, e^{-\frac{N}{2}f''(x_e)\delta x^2}\left(1-\frac{N}{3!}f^{(3)}(x_e) \, \delta x^3  + ...\right)~.
\end{equation}
Upon redefining $\delta x = \delta \xi/ \sqrt{N}$, we see that all higher powers of $\delta \xi$ are suppressed at large $N$, and we can compute the sub-leading correction to $\mathcal{I}_N$ in the large $N$ expansion:
\begin{equation}
\mathcal{I}_{N} \approx 
\sqrt{\frac{2\pi}{{N f''(x_e)}}} e^{-Nf(x_e)} \left(1+ {\mathcal{O}}(1/ N) \right)~.
\end{equation}
If our function $f(x)$ has multiple minima, we must scan for the global one since the others will give contributions that are exponentially suppressed. If there are degenerate minima, on which $f(x)$ takes the same value, we must include them all to get a good approximation. Finally, if there are an infinite number of minima care must be exercised in our approximation scheme. 
\newline\newline
\textbf{Example.}
As a simple example we consider the integral 
\begin{equation}
\mathcal{J}_N = \int_{0}^\infty \dd x \, x^N \, e^{-N x} = \int_{0}^\infty\dd x \, e^{-N\left(x-\log x \right)}~.
\end{equation}
This integral can be solved exactly by performing $N$ partial integrations yielding $(N-1)!/N^N$. Moreover, we note that taking $f(x) = {x}-\log x$ the integral is of the form $(\ref{eq:saddle1D})$. The extremum of $f(x)$ is given by $x_e= 1$, with $f(x_e)= 1$. Expanding $f(x)$ around $x_e$ in a Taylor series up to second order and inserting it into $(\ref{eq:saddle1D})$ we get
\begin{equation}
\lim_{N\to \infty} \mathcal{J}_N  \approx \sqrt{\frac{2\pi}{{N}}} \, e^{-N} \left(1 + \mathcal{O}(1/N) \right)~.
\end{equation}
The same expression can be obtained by expanding the exact result $\mathcal{J}_N = (N-1)!/N^N$ using the Stirling approximation for the factorial.

\begin{center}  {\it Complex saddles $\&$ contour deformations}  \end{center}

The above discussion captures the basic gist of the saddle point approximation in its simplest form. This dates back to Laplace's work in the eighteenth century.  It is worth emphasising, however, that in general it may be the case that some of the extrema do not lie on the original contour of integration. As a very simple example, we might consider $f_c(x) = x^2 + \log \left(x^2+1\right)$ with $x\in\mathbb{R}$. Though this gives rise to a perfectly well defined integral (\ref{eq:saddle1D}), the extrema lie at $x_\pm = \pm i \sqrt{2} $ and $x_0=0$. 
It is straightforward to check that only one of the three saddles, in this case the one lying on the original contour of integration, contributes. More generally, for a complex function $g(z)$ with $z\in \mathbb{C}$ integrated along some contour $\mathcal{C}$ in the complex plane, the saddle point approximation is implemented by identifying a novel contour $\tilde{\mathcal{C}}$ that crosses through some subset of the critical points of $g(z)$ in such a way that the imaginary part of $g(z)$ remains constant along $\tilde{\mathcal{C}}$. 
Though we will generally not require such a treatment in what follows, it is important to keep it in mind. 

\subsection{Vector integrals}
As a next step we consider the saddle point approximation for integrals with $N$ variables $x_I \in \mathbb{R}$ containing a vector index $I = 1,2,\ldots,N$. Consider the family of integrals
\begin{equation} \label{eq:VI}
\mathcal{V}_{N}= \int_{\mathbb{R}^N}\prod_{I=1}^N\dd x_{I}\,e^{-Nf(x_I x_I/N)}~,
\end{equation}
where we use the double index notation $x_I x_I \equiv \sum_{I=1}^N x_I x_I$ here and throughout. The integrals $\mathcal{V}_N$ are invariant under $O(N)$ rotations of the $x_I$. By changing to spherical coordinates 
\begin{equation}
x_I= \sqrt{N} R \, \Omega_I~, \quad\quad\quad \sum_{I=1}^N \Omega_I^2=1~,
\end{equation}
we reduce the integral to the one-dimensional case we already worked out in the previous section. Importantly, this change of variables produces a nontrivial Jacobian leading to the following expression 
\begin{equation}
\mathcal{V}_{N} =  \mathrm{vol} \, S^{N-1} \, N^{N/2} \, \int_0^{\infty}\dd R \, e^{(N-1)\log R  -Nf(R^2)}~.
\label{eq:vecsaddle}
\end{equation}
Happily, in the large $N$ limit the Jacobian of the coordinate change competes at the same order with the original integrand in $(\ref{eq:VI})$. Thus, we can employ the saddle point approximation in its simplest form to estimate $\mathcal{V}_N$ at large $N$. 
\newline\newline
\textbf{Example}. As a straightforward example we can approximate the volume of the $N$-sphere $S^{N}$. To this end, we take $f(x_I x_I)= x_I x_I$ in $(\ref{eq:VI})$ and compute the integral 
\be
\mathcal{V}_{N} = \int_{\mathbb{R}^N} \prod_{I=1}^N \dd x_I  \, e^{- x_I x_I } \label{eq:volsphere}
\ee
in two different ways. An explicit calculation using Gaussian integrals tells us that the value of $(\ref{eq:volsphere})$ gives $\mathcal{V}_{N} = \pi^{N/2}$. On the other hand, we can change to spherical coordinates and perform a saddle point approximation. Taking care of the Jacobian we rewrite $(\ref{eq:volsphere})$ as 
\be
\mathcal{V}_{N} = \mathrm{vol} \, S^{N-1} \, N^{N/2} \,  \int_0^\infty \dd R \, e^{-N F(R)}~,
\ee
with $F(R)\equiv R^2-(1-1/N)\log R$ which at large $N$ can be approximated as $F(R) \approx  R^2-\log R$. The dominant extremum of $F(R)$ is given by $R_e= {1}/{\sqrt{2}}$ and so we obtain for the leading term of the integral 
\be \label{eq:Ivgaussian}
\lim_{N \to \infty} \mathcal{V}_{N} =\mathrm{vol} \, S^{N-1} \sqrt{\frac{\pi}{8N}} \left(\frac{N}{2}\right)^{N/2}e^{-N/2}~.
\ee
Setting this equal to $(\ref{eq:volsphere})$ we find an expression for the volume of $S^N$ in the limit $N \gg 1$
\be
\mathrm{vol} \, S^N\approx \sqrt{N}\left(\frac{2\pi e}{N}\right)^{N/2}~. 
\ee
This agrees with the large $N$ limit of the volume of an $N$-sphere with known exact expression $\text{vol} \, S^{N-1} =  N \pi^{N/2} / \Gamma(N/2+1)$. 

\subsection{A perturbative expansion: the Cactus diagrams}

We would now like to consider the vector-like integrals (\ref{eq:VI}) from a slightly different perspective that is motivated by perturbative expansions often employed in the context of quantum field theory. We will do so by studying a specific example given by
\begin{align}
\mathcal{Z}_N(\alpha)= \int_{\mathbb{R}^N}\prod_{I=1}^N \dd x_I ~e^{-N\left(x_I x_I /N + {\alpha}(x_I x_I/N)^2\right)}~, \label{eq:quarticvec}
\end{align}
where $\alpha \in \mathbb{R}$ will be taken to be a small parameter. For $\mathcal{Z}_N(\alpha)$ to be well-defined we should further take $\alpha$ to be positive, but we shall see shortly that in the large $N$ limit, it may be sensible to allow $\alpha$ to also take small negative values. 
We will first consider the integral (\ref{eq:quarticvec}) in the small $\alpha$ limit, for which we can Taylor expand the exponential and calculate the correction terms using Wick contractions. To this end, it is convenient to introduce the `propagator' $\langle x_I x_J\rangle$ given by
\begin{align} \label{eq:propvi}
\langle x_I x_J\rangle= \mathcal{Z}^{-1}_N(0) \int_{\mathbb{R}^N} \prod_{K=1}^N \dd x_K \, x_I x_J \, e^{-x_I x_I} = \frac{1}{2} \, \delta_{IJ}~.
\end{align}
The corresponding graphical representation for the propagator and the quartic vertex in the integrand of (\ref{eq:propvi})  are dispayed in the figure below
\begin{figure}[H]
\begin{center}
\begin{tikzpicture}[scale=.9]
\draw (0,0) --(1.5,0)[line width=.4] node[pos=-.2,scale=.7]{$x_I$} node[pos=1.2,scale=.7]{$x_J$}node[pos=2,scale=.8]{$\sim \alpha\, N^{-1}~.$};
\draw (.75,.75) --(.75,-.75)[line width=.4] node[pos=-.2,scale=.7]{$x_I$} node[pos=1.2,scale=.7]{$x_J$};
\draw [fill] (1.5,0) circle [radius=0.03];
\draw [fill] (0,0) circle [radius=0.03];
\draw [fill] (.75,.75) circle [radius=0.03];
\draw [fill] (.75,-.75) circle [radius=0.03];
\draw (-5,0) --(-3.5,0)[line width=.4] node[pos=-.2,scale=.7]{$x_I$} node[pos=1.2,scale=.8]{$x_J$}node[pos=1.8,scale=.8]{$\sim 1$~  ,};
\draw [fill] (-3.5,0) circle [radius=0.03];
\draw [fill] (-5,0) circle [radius=0.03];
\end{tikzpicture}
\end{center}
\caption{Propagator and quartic vertex.}
\end{figure}
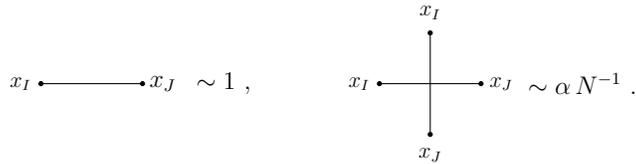
Using the propagator and the quartic vertex we can form closed `bubble' diagrams which compute perturbative contributions to  $\mathcal{Z}_N(\alpha)$. The set of connected bubble diagrams is generated by $\mathcal{F}_N(\alpha) \equiv -\log \left( \mathcal{Z}_N(\alpha)/\mathcal{Z}_N(0)\right)$. To leading order, we can approximate $\mathcal{Z}_N(\alpha)$ by
\begin{align}
\mathcal{Z}_N(\alpha) &=  \int_{\mathbb{R}^N}\prod_{K=1}^N \, \dd x_K  \, e^{-x_I x_I}\left(1 - \alpha \, N \left(\frac{x_I x_I}{N} \right)^2 + \cdots \right)~.
\end{align} 
One can evaluate the perturbative terms in the above integral by standard Gaussian integration. The first order correction in the small $\alpha$ expansion yields
\begin{equation}\label{pertZ}
\alpha \, N \, \int_{\mathbb{R}^N}\prod_{K=1}^N \, \dd x_K  \, e^{-x_I x_I} \left(\frac{x_I x_I}{N} \right)^2  = \frac{\alpha}{4N}\sum_{I,J}\left(\delta_{II}\delta_{JJ}+2\delta_{IJ}\delta_{IJ}\right) = \frac{\alpha}{4} ( N + {2} )~.
\end{equation}
Diagrammatically, the above integral corresponds to a bubble diagram where we have closed the propagators emanating from the quartic vertex. Thus, to linear order in $\alpha$ we have 
\begin{align} \label{eq:genconnectedVI}
 \lim_{\alpha\to0} {\mathcal{F}_N(\alpha)} =   \frac{\alpha}{4} \left(N+{2}\right) + \mathcal{O}(\alpha^2)~.
\end{align}
As we proceed with higher order corrections, we encounter a variety of diagrams. For any given power of $\alpha$ a certain class of diagrams, known as {\it cactus diagram}, will dominate at large $N$. In the figure below, we demonstrate several diagrams of which the first three belong to the cactus family
\begin{figure}[H]
\begin{center}
{
\begin{tikzpicture}[scale=.7]
\draw (0,0) circle [radius=0.5cm];
\draw [fill] (0.50,0) circle [radius=0.05];
\draw (1,0) circle [radius=0.5cm];
\node[scale=.7] at (0.5,-1)   {$\sim \alpha N~$};
\end{tikzpicture}
}
{
\begin{tikzpicture}[scale=.7]
\draw (0,0) circle [radius=0.5cm];
\draw [fill] (0.50,0) circle [radius=0.05];
\draw (1,0) circle [radius=0.5cm];
\draw [fill] (1.50,0) circle [radius=0.05];
\draw (2,0) circle [radius=0.5cm];
\node[scale=.7] at (1,-1)   {$\sim \alpha^2 N~$};
\end{tikzpicture}
}
{
\begin{tikzpicture}[scale=.7]
\draw (0,-1) circle [radius=0.5cm];
\draw [fill] (0,-.5) circle [radius=0.05];
\draw (0,0) circle [radius=0.5cm];
\draw [fill] (0,.5) circle [radius=0.05];
\draw (0,1) circle [radius=0.5cm];
\draw [fill] (.47,1.2) circle [radius=0.05];
\draw (.92,1.4) circle [radius=0.5cm];
\draw [fill] (-.47,1.2) circle [radius=0.05];
\draw (-.92,1.4) circle [radius=0.5cm];
\draw [fill] (1.2,1.8) circle [radius=0.05];
\draw (1.52,2.2) circle [radius=0.5cm];
\draw [fill] (-1.2,1.8) circle [radius=0.05];
\draw (-1.52,2.2) circle [radius=0.5cm];
\draw [fill] (2.,2.42) circle [radius=0.05];
\draw (2.38,2.7) circle [radius=0.5cm];
\draw [fill] (1.2,2.56) circle [radius=0.05];
\draw (.8,2.9) circle [radius=0.5cm];
\node[scale=.7] at (1.5,-1)   {$\sim \alpha^8 N~$};
\end{tikzpicture}
}
{
\begin{tikzpicture}[scale=.7]
\draw (0,0) circle [radius=0.5cm];
\draw (0,-0.5) to[out=180,in=180] (0,0.5);
\draw [fill] (0,.5) circle [radius=0.05];
\draw [fill] (0,-.5) circle [radius=0.05];
\draw (0,-0.5) to[out=15,in=5] (0,0.5);
\node[scale=.7] at (0,-1)   {$\sim \alpha^2~$};
\end{tikzpicture}
}
\end{center}
\caption{Some bubble diagrams. The first three are cactus diagrams.}
\end{figure}
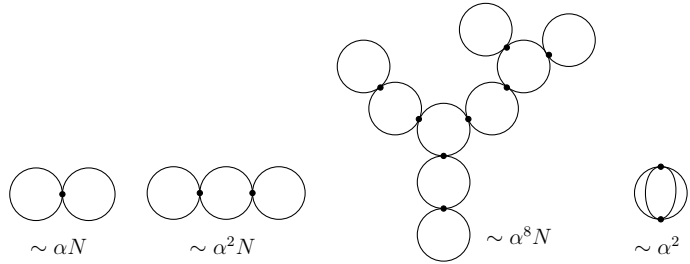

\subsection{Diagrams resummed}

The fact that at large $N$ only a subclass of diagrams survives leads to a dramatic simplification of the theory, and to the hope that a resummation can be performed. This resummation is precisely the saddle-point approximation of $\mathcal{Z}_N(\alpha)$. Let us perform the saddle point analysis to confirm the $\mathcal{O}({\alpha})$ correction we calculated perturbatively. To do so, let us go to spherical coordinates and perform the same steps as in $(\ref{eq:vecsaddle})$. We find
\begin{equation} \label{eq:partfcritb}
\mathcal{Z}_N(\alpha) = \mathrm{vol} \, S^{N-1}  \, N^{N/2} \,\int_{0}^\infty \dd R\, e^{-N F_\alpha(R)}~,
\end{equation}
where we have defined $F_\alpha(R) \equiv \left(R^2+\alpha R^4-(1-1/N) \log R\right)$. To leading order in the large $N$ limit, the extrema of $F_\alpha(R)$ 
are given by
\begin{equation} \label{eq:Rsqucritvec}
R^2_{\pm} = \frac{1}{4\alpha}\left(-1 \pm \sqrt{1+{4\alpha}}\right)~.
\end{equation}
Of the two saddles, only $R^2_{+}$ lies on the original contour of integration and has a well-defined limit as $\alpha$ tends to zero. The other saddle $R^2_{-}$ goes to infinity in the limit of vanishing $\alpha$ and hence cannot be relevant to compare to the perturbative analysis. 
Using the saddle point approximation, we are thus led to the leading $N$ approximation of $\mathcal{F}_N(\alpha)$:
\begin{equation}\label{ZlargeN}
\lim_{N\to\infty} \mathcal{F}_N(\alpha) =\frac{N}{8\alpha} \left(-1+ \sqrt{1+4 \alpha}- \alpha (2-4\log 2)-4 \alpha \log \left(\frac{-1+\sqrt{1+4 \alpha}}{\alpha}~ \right)\right) + \mathcal{O}(1)~. 
\end{equation}
The above result is valid to {\it all} orders in the small $\alpha$ expansion. Expanding (\ref{ZlargeN}) to leading order in $\alpha$ one can readily recover our perturbative correction (\ref{pertZ}). It is remarkable that the resummation of all the diagrams was encoded in the solution of the simple algebraic equation $\partial_R F_N(R) = 0$. Moreover (\ref{ZlargeN}) reveals that the expression gives a sensible result for $\text{Re}\, \alpha \ge -1/4$ which includes a negative interval. To continue beyond the critical value $\alpha_c =-1/4$ one has to specify a branch cut due to the non-analytic behaviour. That negative values of $\alpha$ might be sensible in the large $N$ limit, even though the finite $N$ integral is ill-defined for negative $\alpha$, comes about precisely because at large $N$ most of the contribution to the integral is localised near the critical value of $R_{+}$. Fortuitously, the integral (\ref{eq:partfcritb}) can be evaluated exactly for any value of $N$. One finds 
\begin{equation}\label{nonpertV}
\exp \left(-\mathcal{F}_N(\alpha) \right) = \left(\frac{N}{4\alpha }\right)^{\frac{N}{4}} U\left(\frac{N}{4},\frac{1}{2},\frac{N}{4 \alpha }\right)~,
\end{equation}
where $U(a,b,z)$ is Tricomi's confluent hypergeometric function. We have 
\begin{equation}
\lim_{N\rightarrow \infty}\lim_{\alpha\rightarrow 0} \left(\frac{N}{4\alpha }\right)^{\frac{N}{4}} U\left(\frac{N}{4},\frac{1}{2},\frac{N}{4 \alpha }\right)= 1-\frac{\alpha}{4}\left(N+2\right)+\mathcal{O}\left(\alpha^2\right)~.
\end{equation} 
This agrees again with the exponential of $(\ref{eq:genconnectedVI})$.
In relation to the critical value $R_+$, it is worth mentioning a quantity that will become of increasing relevance throughout our discussion. Recall that the generating function of connected bubble diagrams admits a large $N$ expansion of the form
\begin{equation} \label{eq:expansionVI}
\mathcal{F}_N(\alpha) = N \mathcal{F}^{(0)}(\alpha) + \mathcal{F}^{(1)}(\alpha) + \frac{1}{N} \, \mathcal{F}^{(2)}(\alpha) + \ldots~.
\end{equation}
Moreover, each of the functions $\mathcal{F}^{(k)}(\alpha)$ themselves admits a Taylor expansion near $\alpha=0$. For instance
\begin{equation}
\mathcal{F}^{(0)}(\alpha) = \sum_{n=0}^\infty f^{(0)}_n \alpha^n~.
\end{equation}
The power of $\alpha$ counts the number of bubbles in a given cactus diagram. Consequently, we can define the moments of the `bubble number' as
\begin{equation}\label{bubbleno}
\langle n^L \rangle = \partial^{(L)}_{\log \alpha} \, \log \mathcal{F}^{(0)}(\alpha)~.
\end{equation}
Near the critical value $\alpha_c$, and for $L \ge 2$, the above moments will diverge. For instance $\langle n^2 \rangle \sim (\alpha-\alpha_c)^{-1/2}$. This suggests that in the limit $\alpha \to \alpha_c$ the contribution of cacti with arbitrarily large numbers of bubbles becomes increasingly important. Moreover, the behaviour near $\alpha_c$ is somewhat reminiscent of critical phenomena.

\begin{center}{\it Order of limits $\&$ analytic continuation}\end{center}

Expanding (\ref{nonpertV}) at large $N$ gives us expressions for all the $\mathcal{F}^{(n)}(\alpha)$. We can moreover study (\ref{nonpertV}) for finite values of $N$. For positive integers $N$, the function (\ref{nonpertV}) can be expressed either in terms of Bessel functions (for $N$ odd) or the complementary error function (for $N$ even). At finite $N$ the integral (\ref{eq:partfcritb}) is defined for $\textrm{Re} \, \alpha > 0$, though it can be analytically continued to certain regions of complex $\alpha$-plane depending on whether $N$ is even or odd. On the other hand, the infinite $N$ approximation (\ref{ZlargeN}) can be extended to $\textrm{Re} \, \alpha > -1/4$, after which we encounter a branch cut. We see that the analytic structure of $\mathcal{F}_N(\alpha)$ across the complex $\alpha$-plane can depend on whether we first take the $N\to \infty$ limit and then analytically continue $\alpha$ or vice versa. In other words, the infinite $N$ limit can qualitatively affect the analytic structure of a family of functions labelled by $N \in \mathbb{Z}$. We will not explore this aspect of the large $N$ limit and the accompanying theoretical tools in any detail. We felt it important, nevertheless, to give a flavour of what they entail.

\section{Large $N$ integrals over a single matrix I$^{\color{magenta} C}$}\label{singlematrix}
Having introduced the saddle point approximation and its application to vector integrals, we will now proceed to discuss matrix integrals, whose variables are organised in the form of an $N\times N$ Hermitian matrix $M_{IJ}$ with $I,J=1,2,\ldots,N$. As briefly mentioned in the introduction, such integrals play a role in a remarkably rich set of physical systems ranging from nuclear theory to quantum chaos and string theory. The class of integrals we will study is of the following form
\begin{equation}
\mathcal{M}_N =  \int_{\mathbb{R}^{N^2}} [\dd M] \, e^{-N \, \Tr \, V(M)} \label{eq:1MM}~,
\end{equation}
where $V(M)$ is a matrix valued function of $M$, and the trace is only taken at the end. The measure factor is given by 
\begin{equation}
[\dd M ] \equiv \prod_{J}\dd M_{JJ}\prod_{I <  J}\dd \mathrm{Re} M_{IJ} \,\dd \mathrm{Im} M_{IJ} \label{eq:matrixpartitionfunction}~.
\end{equation}
The integrals are invariant under a conjugation of $M$ by an $N\times N$ unitary matrix $U \in U(N)$. 

\subsection{Eigenvalue distribution}\label{eigdist}



Every Hermitian matrix can be diagonalised using a unitary matrix $\mathcal{U}$ as $M = \mathcal{U} D_M \mathcal{U}^\dagger$ with $D_M \equiv \mathrm{diag}(\lambda_1,\lambda_2,\ldots ,\lambda_N)$ a real diagonal matrix. Consequently, the exponent of the integrand in $(\ref{eq:1MM})$ depends purely on the $N$ eigenvalues $\lambda_I \in \mathbb{R}$ of $M$. Thus, it is natural to consider the problem using a new set of variables given by an $N\times N$ unitary matrix $\mathcal{U} \in U(N)/U(1)^N$, and $N$ real variables $\lambda_I$, such that $M = \mathcal{U} D_M \mathcal{U}^\dag$. Notice that $\mathcal{U}$ resides in the coset space $U(N)/U(1)^N$, rather than the full $U(N)$, because those unitary matrices where each diagonal entry is multiplied by a pure phase are redundant. 
One must also compute the Jacobian associated to the change of variables $M = \mathcal{U} D_M \mathcal{U}^\dag$. To this end, it is useful to write $\mathcal{U}= \e^{i \mathcal{L}}$, with $\mathcal{L}$ an $N\times N$ Hermitian generator of $U(N)/U(1)^N$. The infinitesimal line element on the space of Hermitian matrices $M$ is
\begin{equation}\label{le}
\dd s^2 = \Tr \, \dd M \dd M^\dagger = \sum_{I}\dd \lambda_I^2+ 2\sum_{I<J}(\lambda_I- \lambda_J)^2|\dd \mathcal{L}_{IJ}|^2~.
\end{equation}
The Jabobian is the determinant of the above line element. It follows that 
\begin{equation}\label{dM}
[\dd M]= \frac{1}{\mathrm{vol}~ S_N} \prod_{I}\dd\lambda_I\prod_{I\neq J}|\lambda_I- \lambda_J|[\dd \mathcal{L}]~,
\end{equation}
where $[\dd \mathcal{L}]$ is the volume element of the coset space  $U(N)/ U(1)^N$ and we have further divided by the volume of the permutation group $S_N$ to avoid multiple counting of eigenvalue matrices $D_M$ with permuted elements.
\newline\newline
\textbf{Example}. As an example, let us consider the case $N=2$. We can parametrise the unitary matrix $\mathcal{U}$ as 
\begin{equation}\label{eq:2by2}
\mathcal{U} = \begin{pmatrix} 
      \sin \frac{\theta}{2} &  e^{i \varphi} \cos \frac{\theta}{2}  \\
      - e^{-i \varphi} \cos \frac{\theta}{2} & \sin \frac{\theta}{2} \\
   \end{pmatrix}~,
\end{equation}
where $\theta \in [0,\pi)$ and $\varphi \in (0,2\pi]$ are coordinates on the coset space $U(2)/U(1)^2$, whose geometry is the round two-sphere. Using that the relation $\mathcal{U}=\e^{i\mathcal{L}}$ implies
\begin{equation}
\dd\mathcal{L}= -i \;\mathcal{U}^{-1}\dd \mathcal{U}~.
\end{equation}
The line element (\ref{le}) can be written out explicitly:
\begin{equation}\label{eq:ds2by2}
\dd s^2 = \dd\lambda_1^2 + \dd\lambda_2^2 + \frac{\left( \lambda_1 -\lambda_2 \right)^2}{2} \left( \dd\theta^2 + \sin^2\theta \dd\varphi^2 \right)~.
\end{equation}
Defining $r = (\lambda_1-\lambda_2)/\sqrt{2}$ and $z = (\lambda_1 + \lambda_2 )/\sqrt{2}$ the above becomes 
\begin{equation}
 {\dd s^2} =  {\dd z^2} + {\dd r^2} + r^2 \left( \dd\theta^2 + \sin^2\theta \dd\varphi^2 \right)~.
\end{equation}
We observe that the above metric is indeed the flat metric on $\mathbb{R}^4$ in cylindrical coordinates
up to the fact that $r\in \mathbb{R}$ as opposed to $r \ge 0$.  However, the map $r \to -r$ is in fact a permutation of the eigenvalues $\lambda_1 \leftrightarrow \lambda_2$. This is a double counting in our configuration space, since conjugation with respect to $\mathcal{U}$ in (\ref{eq:2by2}) with $\theta=0$ exchanges $\lambda_1$ and $\lambda_2$.
Thus, after dividing by the volume of the two-dimensional permutation group the volume element becomes
\begin{equation}
[\dd M ] = r^2\dd r \, \dd z \, \dd\Omega_2~, \quad\quad r~>0~,
\end{equation}
where $\dd\Omega_2$ is the standard volume element on the unit sphere $S^2$. The right hand side is indeed the flat metric on $\mathbb{R}^4$ expressed in cylindrical coordinates.

\begin{center}{\it The Vandermonde contribution}\end{center}

The product appearing in the volume element (\ref{dM}) is the determinant of the Vandermonde matrix
\begin{equation}\label{eq:Vandermonde}
\textbf{V}_N \equiv
   \begin{pmatrix} 
      1 & \lambda_1& \lambda_1^2 &  \cdots & \lambda_1^{N-1} \\
      1 & \lambda_2 & \lambda_2^2 &\hdots & \hdots  \\
      \vdots &\vdots&\vdots &\ddots& \\
      1& \lambda_N &\lambda_N^2 &\hdots & \lambda_N^{N-1}\\
   \end{pmatrix}~.
\end{equation}
Since it will appear several times throughout our discussion, we introduce the following notation for the determinant of the Vandermonde matrix
\begin{equation}\label{eq:vandermondedet}
\Delta_N(\lambda) \equiv \prod_{I < J} \left( \lambda_I - \lambda_J \right)~.
\end{equation}
We can bring our original integral ($\ref{eq:1MM}$) to the following form
\begin{equation}\label{eq:partitionfunctioneigenvalues}
\mathcal{M}_N
=\mathrm{vol}\,\frac{U(N)}{U(1)^N\times S_{N}}  \, \int_{\mathbb{R}^N}\prod_{I=1}^N \dd\lambda_{I} \, e^{-N^2 S[\lambda ]}~,
\end{equation}
where we have defined the multivariable function 
\begin{equation}
S[\lambda] \equiv \frac{1}{N}\sum_{I=1}^N V(\lambda_I)  - \frac{1}{N^2}\sum_{J\neq I}\log|\lambda_I- \lambda_J|~. \label{eq:effaction}
\end{equation}
As for the vector integral case, the change of coordinates adds a nontrivial Jacobian that will compete with the original integrand. Moreover, the Jacobian causes an effective `repulsive' pressure against the eigenvalues all lying on top of each other. The repulsive effect is the matrix analogue of the logarithmic term encountered in (\ref{eq:vecsaddle}) for the vector integrals. 

Before proceeding to solve $(\ref{eq:1MM})$ in the large $N$ limit, let us summarise what we have achieved so far. By a unitary transformation of the Hermitian matrix $M$ we reduced the matrix integral from a theory with $N^2$ independent degrees of freedom to a theory described by the $N$ real eigenvalues of the Hemitian matrix $M$. The Jacobian arising from this change of coordinates gives rise to the square of the Vandermonde determinant $\Delta_N(\lambda)^2$. This contribution competes with $V(\lambda)$ such that the eigenvalues are distributed around the mininum of $V(\lambda)$.

\subsection{Saddle point approximation $\&$ the resolvent}
 
In order to solve the matrix integrals at large $N$, we will once again employ the saddle point approximation. The saddle point equations for $S[\lambda]$ in (\ref{eq:effaction}) are given by 
\begin{equation}
{V'(\lambda_I)}  = \frac{2}{N}\sum_{J\neq I}\left({\lambda_I - \lambda_J}\right)^{-1}~.  \label{eq:SP}
\end{equation}
To solve these equations, it is convenient to introduce the following normalised eigenvalue distribution
\begin{equation}
\rho(\lambda)=\frac{1}{N}\sum_{I=1}^N\delta(\lambda- \lambda_I)~, \quad\quad\quad \int_{-a}^a \dd \lambda\,\rho(\lambda)=1~.  \label{eq:CH}
\end{equation}
In the large $N$ limit, and under the assumption that the range $[-a,a]$ is compact and does not grow with $N$, we can approximate $\rho(\lambda)$ by a continuous, non-negative function. For simplicity, we further assume that the range of $\rho(\lambda)$ is symmetric about the origin. This is not at great cost since we can always shift all the $\lambda$ by a constant to accommodate this. We can then rewrite $(\ref{eq:SP})$ as the following integral equation
\begin{equation}\label{matrixsaddle}
V'(\lambda)= 2\, \dashint_{-a}^a \dd \mu \, \frac{\rho(\mu)}{\lambda - \mu}~,
\end{equation}
where it is understood that we are taking the principal value for the integral. Our goal is to find the eigenvalue distribution $\rho(\lambda)$ which solves the above equation. To do so, it is instructive to introduce the resolvent \cite{Brezin:1977sv}
\begin{equation}\label{eq:defres}
\RR_N(z)\equiv \frac{1}{N}\text{Tr} \left( z \, \mathbb{I}_N-M \right)^{-1} = \frac{1}{N}\sum_{I=1}^N\frac{1}{z-\lambda_I}~, \quad\quad z\in \mathbb{C}/\{\lambda_I\}~.
\end{equation}
Sending $N\rightarrow \infty$ the sum can be replaced by an integral where each eigenvalue is weighted by its average density
\begin{equation} \label{RlargeN}
\lim_{N\rightarrow \infty}\RR_N(z)\equiv \RR(z)= \int_{-a}^{a}\dd \mu \,\frac{\rho(\mu)}{z-\mu}~.
\end{equation}
By evaluating the resolvent close to the real axis  $z= x\pm i\epsilon$ we find
\begin{equation}
\RR(x+i\epsilon)= \int_{-a}^a \dd \mu \,\frac{\rho(\mu)}{x-\mu + i\epsilon}=  \int_{-a}^a\dd \mu \,\frac{\rho(\mu)(x-\mu)}{(x-\mu)^2 +\epsilon^2} - i\int_{-a}^a \dd \mu\,\frac{\rho(\mu)\epsilon}{(x-\mu)^2+\epsilon^2}~, \label{eq:resolvent}
\end{equation}
where in the limit $\epsilon\rightarrow 0$ the first integral is a principal value integral, while we evaluate the second integral using delta function identities. Finally, we arrive at 
\begin{align}
\RR(x\pm i\epsilon)= \dashint_{-a}^a \dd \mu\, \frac{\rho(\mu)}{(x-\mu)} \mp i\pi \rho(x)~,
\end{align}
where we have employed the Sokhotski-Plemelj theorem. Using $(\ref{eq:SP})$ and $(\ref{eq:CH})$
\be
\RR(x\pm i \epsilon)= \frac{1}{2}\,V'(x) \mp i\pi \rho(x)~.
\ee
Using (\ref{RlargeN}) and the above expressions we obtain the following properties the resolvent must satisfy: 
\begin{align}
\mathrm{res}_a~:\quad &\lim_{z\rightarrow \infty}\RR(z)= \frac{1}{z}~, \label{norm} \\
\mathrm{res}_b~:\quad &\rho(x)= \frac{1}{2\pi i}\left(\RR(x-i\epsilon) - \RR(x+i\epsilon)\right),  \quad x\in\mathrm{supp}(\rho)~, \label{evaldensity}\\ 
\mathrm{res}_c~:\quad &V'(x)= \RR(x+i\epsilon) + \RR(x-i\epsilon)~, \quad x\in \mathrm{supp}(\rho)~. \label{jump}
\end{align}
The condition $\mathrm{res}_b$ specifies the jump of the resolvent across the branch cut $x \in [-a,a]$. Elsewhere, the resolvent is analytic. Finally, $\mathrm{res}_c$ fixes the real part of the resolvent to the derivative of the potential. Thus, the class of problems we are trying to solve is a Riemann-Hilbert type problem. 

In the large $N$ limit the exponent of our original integral ($\ref{eq:partitionfunctioneigenvalues}$) takes the form
\begin{equation}
S[\rho(\lambda)]= \int_{-a}^{a}\dd \lambda\, \rho(\lambda)V(\lambda) - \int_{-a}^{a} \dd \lambda\,\rho(\lambda)\,\int_{-a}^{a}  \dd \mu \,\rho(\mu)\log|\lambda - \mu|~.  \label{eq:Seffcont}
\end{equation}
Using the saddle point equations, we can further simplify
\begin{equation}
S[\rho_{\mathrm{ext}}(\lambda)]= \frac{1}{2}\int_{-a}^{a} \dd \lambda\,\rho_{\mathrm{ext}}(\lambda) V(\lambda) - \int_{-a}^{a} \dd\lambda\,\rho_{\mathrm{ext}}(\lambda)\,\log |\lambda|~,
\end{equation}
where, for simplicity,  we have further assumed that $V(0)=0$ and $\rho_{\mathrm{ext}}(\lambda)$ is that solution of (\ref{matrixsaddle}) which minimises $S[\rho(\lambda)]$ in (\ref{eq:Seffcont}). Thus, our original integral ($\ref{eq:partitionfunctioneigenvalues}$) is approximated by the expression
\begin{equation}
\mathcal{M}_N \approx \mathrm{vol} \, U(N)  \, e^{-N^2S[\rho_{\mathrm{ext}}(\lambda)]}~,
\end{equation}
where we have dropped the volume of $U(1)^N \times S_N$ since it does not compete with terms that grow exponentially in $N^2$.
\newline\newline
\textbf{Gaussian example}. We now consider the simplest concrete example. Let us take $V(M)= \frac{1}{2}M^2$ such that
\begin{equation}
\mathcal{M}_N =\int_{\mathbb{R}^{N^2}} [\dd M] \, e^{-\frac{N}{2}\Tr \, M^2}= \left(\frac{2\pi}{N}\right)^{N^{2}/2}~,  \label{eq:Gaussianmatrix}
\end{equation}
where we obtained the exact result using Gaussian integrals. 
We will now approximate (\ref{eq:Gaussianmatrix}) using the saddle point approximation. As was already outlined, this involves finding the resolvent R$(z)$. Let us consider the following ansatz
\begin{equation} \label{eq:resolventevenpot}
\RR(z)= \frac{1}{2}V'(z) - P(z)\sqrt{z^2-a^2}~,
\end{equation}
where $P(z)$ is a polynomial in $z$. The degree of $P(z)$ must be chosen such that it cancels all positive powers of R$(z)$ to guarantee $\mathrm{res}_a$. For $V(z)=z^2/2$ we find that $P(z)$ must be constant. Moreover, imposing the remaining conditions $\text{res}_{b,c}$ on the resolvent one obtains
\begin{equation} \label{eq:Wigner}
\RR(z)= \frac{1}{2}z - \frac{1}{2}\sqrt{z^2-4} \quad\quad  \text{and} \quad \quad  \rho_{\mathrm{ext}}(x)= \frac{1}{2\pi}\sqrt{4 - x^2}~, \quad x \in [-2,2]~.
\end{equation}
The eigenvalue distribution $(\ref{eq:Wigner})$ is known as {\it Wigner's semicircle law} and appears in a wide range of physical and mathematical examples.
Notice that it is connected and has compact support. Using the eigenvalue distribution we obtain the large $N$ approximation
\begin{equation}
\mathcal{M}_N \approx \mathrm{vol} \, U(N)  \, e^{-\frac{3}{4}N^2}~.
\end{equation}
Setting this expression equal to the exact result in $(\ref{eq:Gaussianmatrix})$, we can provide an approximation for the volume of the unitary group in the limit of large rank
\begin{equation}
\lim_{N\to\infty} \log \, \mathrm{vol} \, U(N) \approx \frac{N^2}{2} \left( \log \frac{2\pi}{N} + \frac{3}{2}\right) + \mathcal{O}(N)~.
\end{equation}
We can readily verify that the above expression agrees with the large $N$ approximation of the exact expression
\begin{equation}\label{eq:volUN}
\mathrm{vol} \, U(N)= \frac{(2\pi)^{N(N+1)/2}}{G(N+1)}~, 
\end{equation}
where $G(N+1)$ is Barnes $G$-function and we recall that at large $N$
\begin{equation}
\log G(N+1) =-\frac{3 N^2}{4}+\frac{N^2}{2}  \log N+\frac{N}{2}  \log 2 \pi -\frac{1}{12}\log N+ \ldots ~.
\end{equation} 
\textbf{Quartic example.} Having done the Gaussian case, we now move on to a slightly more involved example. We will consider the integral
\begin{equation} \label{eq:resolventex}
\mathcal{Z}_N(\alpha) = \int_{\mathbb{R}^{N^2}} [\dd M] \, e^{-N \, \Tr V_\alpha(M)}~,\quad V_\alpha(M)= \frac{1}{2}M^2 + \alpha M^4~.
\end{equation}
Using the ansatz $(\ref{eq:resolventevenpot})$ and following the approach delineated for the Gaussian case, we can write down the resolvent with $P(z)$ a quadratic polynomial
\begin{equation}
\RR(z) = \frac{1}{2}z + 2\alpha z^3 - (p_1+p_2z + p_3 z^2)\sqrt{z^2-a^2}~. \label{eq:Ansatz}
\end{equation}
The parameters $p_1$, $p_2$ and $p_3$ are fixed by $\mathrm{res}_a$.
\begin{equation}
p_1= \frac{1}{2} + \alpha \, a^2~,\quad p_2=0~,\quad p_3=2\alpha~.
\end{equation}
The parameter $a$ solves the equation
\begin{equation}
3\alpha  \,a^4 + a^2-4=0 \quad \Rightarrow \quad a_{\pm}^2= -\frac{1}{6\alpha}\pm \frac{1}{6\alpha}\sqrt{1+48\alpha}~. \label{eq:d}
\end{equation}
We are interested in the solution that is continuously connected to the Gaussian solution as $\alpha \to 0$, which is $a_+$. Notice that the saddle point solution (\ref{eq:d}) exhibits non-analytic behaviour as $\alpha$ approaches $\alpha_c= -{1}/{48}$, somewhat reminiscent of what we saw in our vector integral example (\ref{eq:Rsqucritvec}). If we wish to continue past $\alpha =\alpha_c$, the argument of the square root becomes negative and we have to specify a branch cut. Our expression for the resolvent and eigenvalue density are given by 
\begin{eqnarray} \label{eq:densityM4}
\RR(z)&=&  \frac{1}{2}z + 2\alpha z^3 - \left(\frac{1}{2}+ \alpha\, a_+^2+ 2\alpha z^2\right)\sqrt{z^2-a_+^2}~,\\
\rho_{\mathrm{ext}}(\lambda)&=& \frac{1}{\pi}\left(\frac{1}{2}+ \alpha \,a_+^2 + 2\alpha \lambda^2\right)\sqrt{a_+^2-\lambda^2}~,
\end{eqnarray}
with $a_+$ given by (\ref{eq:d}).
\begin{figure}[H]
\begin{center}
 \includegraphics[scale=.32]{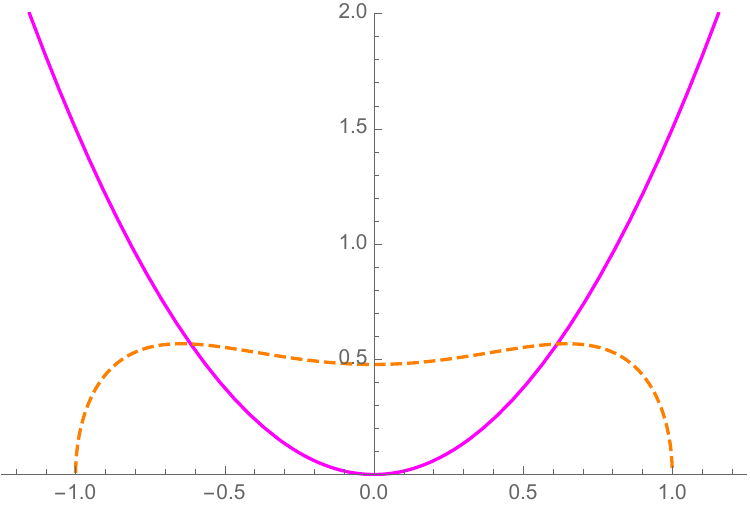} \quad\qquad \includegraphics[scale=.32]{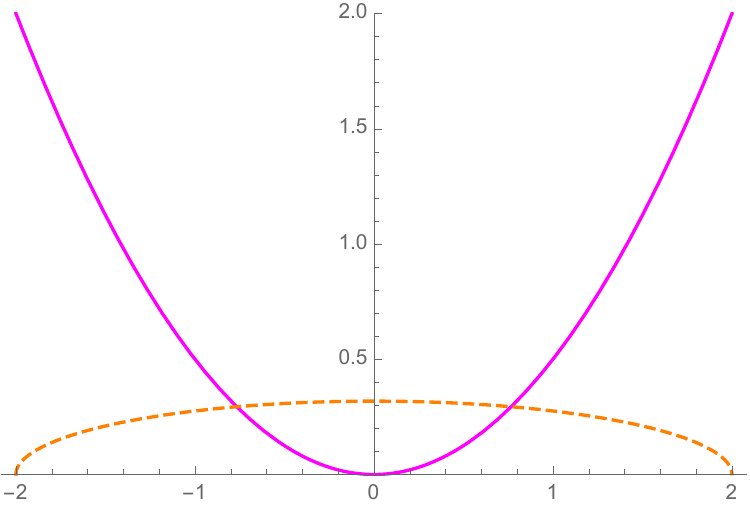} \quad\qquad \includegraphics[scale=.32]{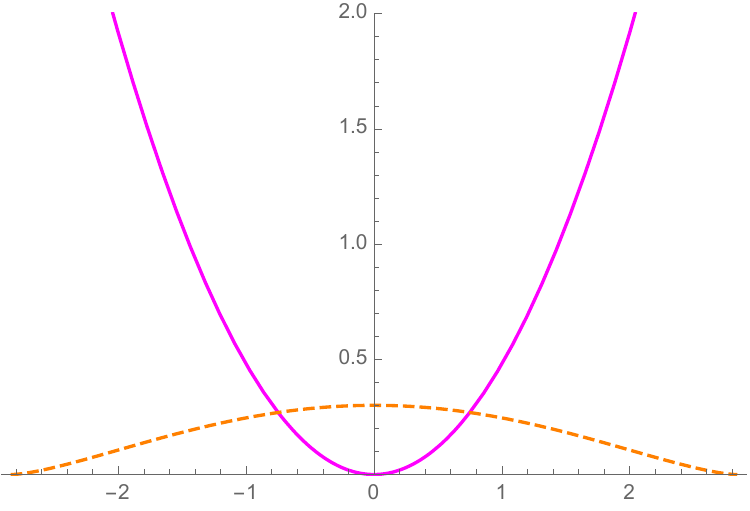}
 \caption{Quartic polynomial $V_\alpha(\lambda)$ (magenta) and corresponding eigenvalue distribution $\rho_{\mathrm{ext}}(\lambda)$ (orange, dashed) for $\alpha=1$ (left), $\alpha=0$ (center), and $\alpha=\alpha_c$ (right).}
 \end{center}
\end{figure}
To approximate our original integral $(\ref{eq:resolventex})$ at large $N$, we must evaluate
\begin{equation}
S[\rho_{\mathrm{ext}}(\lambda)]= \int_{0}^{a_+}\dd\lambda \,\rho_{\mathrm{ext}}(\lambda)\left(\frac{1}{2}\lambda^2 + \alpha\lambda^4\right)- 2 \int_0^{a_+}\dd\lambda \, \rho_{\mathrm{ext}}(\lambda)\log{|\lambda|}
\end{equation}
on the saddle point solution $\rho_{\mathrm{ext}}(\lambda)$. In the above expression we also used that our integrands are symmetric around the origin to adjust the boundaries of the integrals. 
Evaluating this integral at the saddle $\rho_{\mathrm{ext}}(\lambda)$ we find
\begin{equation}\label{onshellalpha}
S[\rho_{\mathrm{ext}}(\lambda)]- S[\rho_{\mathrm{ext}}(\lambda)]_{\alpha=0}=  -\frac{3}{8}-\frac{1}{384} \left(a_+^2-40\right) a_+^2+\log 
\frac{2}{a_+}~,
\end{equation}
where we used $(\ref{eq:d})$ to write the above as a function of $a_+$. Substituting $a_+$, a small $\alpha$ expansion reveals the following \cite{Brezin:1977sv, Bessis:1979is, Bessis:1980ss}
\begin{equation}\label{eq:expF0}
 \lim_{N\rightarrow \infty}  \frac{\mathcal{F}_N(\alpha)}{N^2} =   2 \alpha-18 \alpha^2+288 \alpha^3-6048 \alpha^4+\frac{746496}{5} \alpha^5+\mathcal{O}\left(\alpha^6\right)~,
\end{equation}
where we have defined $\mathcal{F}_N(\alpha) \equiv -\log \mathcal{Z}_N(\alpha)/\mathcal{Z}_N(0)$, analogously to the vector integral case. More generally, and somewhat similar to the vector integral case $(\ref{eq:expansionVI})$, the matrix integrals allow for an expansion in large $N^2$
\begin{equation}\label{eq:Nsquaredexp}
\mathcal{F}_N(\alpha) = N^2 \mathcal{F}^{(0)}(\alpha) + \mathcal{F}^{(1)}(\alpha) +\frac{1}{N^2}\mathcal{F}^{(2)}(\alpha) + \cdots~,
\end{equation}
where each of the $\mathcal{F}^{(n)}(\alpha) $ itself admits a power series expansion around $\alpha=0$. 
A careful examination of (\ref{onshellalpha}) and (\ref{eq:expF0}) reveals \cite{Brezin:1977sv}
\begin{equation}
\mathcal{F}^{(0)}(\alpha)=-\sum_{n=1}^{\infty}(-12\alpha)^{n}\frac{(2n-1)!}{n!(n+2)!} \,=2\alpha \,_3F_2\left(1,1,\frac{3}{2};2,4;-48\alpha\right)~. \label{eq:gzero}
\end{equation}
We can expand the above near the critical value $\alpha_c = -1/48$ to find the leading non-analytic behaviour (which we denote with a subscript n.a.) 
\begin{equation}\label{leadingna}
\lim_{\alpha\rightarrow \alpha_c} \partial^{(3)}_\alpha \, \mathcal{F}^{(0)}_{\mathrm{n.a.}}(\alpha) = 4608\sqrt{3} \left({\alpha - \alpha_c}\right)^{-1/2} + \ldots~.
\end{equation}
We will analyse the terms $\mathcal{F}^{(h)}(\alpha)$ in (\ref{eq:Nsquaredexp}) in the next section. In particular we will provide detailed evidence that 
%
near the critical value $\alpha_c$ the leading non-analytic behaviour goes as
\begin{equation}\label{gengenus}
\lim_{\alpha\to\alpha_c}\mathcal{F}_{\mathrm{n.a.}}^{(h)}(\alpha)  \sim (\alpha-\alpha_c)^{5\chi_h/4}~, \quad\quad h \in \mathbb{N}~,
\end{equation}
where a logarithmic behaviour is understood for $h=1$ and $\chi_h = 2-2h$. 

\subsection{General polynomial $\&$ multicritical models}
The planar solution for the eigenvalue distribution stemming from a general $V(\lambda)$ can be written down rather concisely. We discuss here the case for which the eigenvalue distribution is connected and has compact support on a single real interval $\lambda \in [b,a]$. In that case the resolvent is given by \cite{Migdal:1984gj}
\begin{equation}\label{eq:generalResolvent}
\text{R}(z) = \frac{1}{2} \oint_{\mathcal{C}} \frac{\dd u}{2\pi i} \frac{V'(u)}{z-u} \sqrt{\frac{(z-a)(z-b)}{(u-a)(u-b)}}~,
\end{equation}
where the contour $\mathcal{C}$ goes around the branch cut $z \in [b,a]$. The end points $a$ and $b$ follow from the conditions (\ref{norm}) and (\ref{jump}) imposed on the resolvent. Given $\text{R}(z)$, we can extract the eigenvalue density from (\ref{evaldensity}). 

It is interesting to note that there exist a variety of polynomials which upon tuning the coefficients give rise to non-analytic behaviour different from (\ref{gengenus}). These matrix integrals  are known as multicritical matrix integrals and were originally introduced in \cite{Kazakov:1989bc}. For $V(M)$ an even polynomial of degree $2m$ we denote the matrix integral as an $m^{\mathrm{th}}$ multicritical model. For instance, using the above techniques, the following polynomial \cite{Kazakov:1989bc, Ambjorn:2016lkl}
\begin{equation}\label{eq:multcritmodel}
V(M) = \frac{1}{\gamma} \left( \frac{1}{2}M^2 -\frac{1}{3}M^4+\frac{4}{45}M^6\right)~,\quad \gamma= \frac{1}{12}-\frac{1}{12}(1-a^2)^3~,
\end{equation}
with the eigenvalues of $M$ distributed in the range [$-a,a$] gives rise to a non-analytic behaviour of the form 
\begin{equation}\label{nonanaMultcrit}
\lim_{\gamma\rightarrow \gamma_c}\partial_{\gamma}^{(3)}\mathcal{F}_{\text{n.a.}}^{(0)}(\gamma) \sim (\gamma-\gamma_c)^{-2/3}~,
\end{equation}
with $\gamma_c=1/12$. Note that we have determined $\gamma$ in (\ref{eq:multcritmodel}) by combining (\ref{eq:generalResolvent}) with $\mathrm{res}_a$ (\ref{norm}).  We note that the branch cut of the density is still determined by a square root:
\begin{equation}
\rho_{\mathrm{ext}}(\lambda)= \frac{1}{\gamma\,\pi}\left(\frac{1}{2}-\frac{1}{3}a^2+\frac{1}{10} a^4+\frac{2}{15}(a^2-5)\lambda^2+\frac{4}{15}\lambda^4\right)\sqrt{a^2-\lambda^2}~,\quad \lambda \in [-a,a]\,.
\end{equation}
\begin{figure}[H]\label{figurem3}
\begin{center}
 \includegraphics[scale=.33]{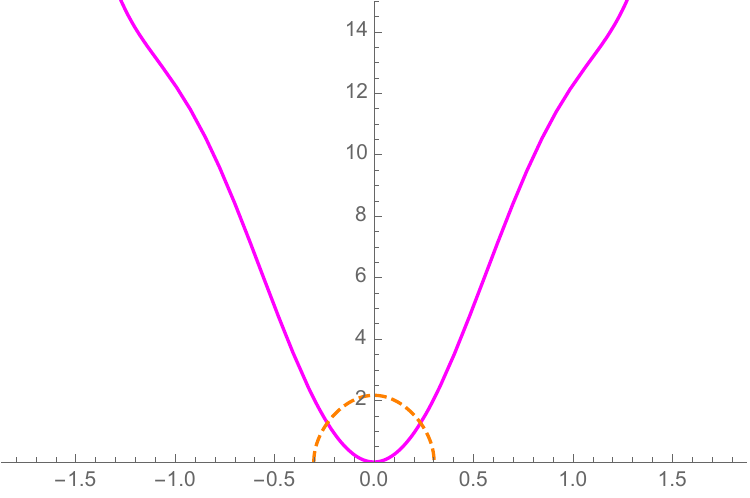} \qquad \qquad \includegraphics[scale=.33]{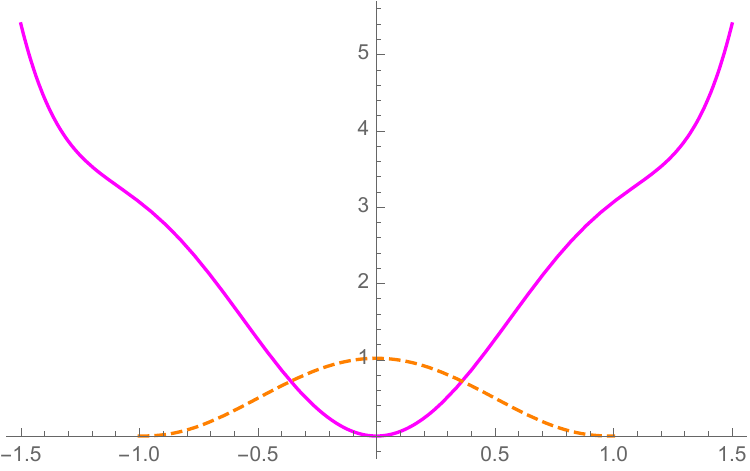}
 \caption{Multicritical polynomial $V(\lambda)$ for $m= 3$  (magenta) and corresponding eigenvalue distribution $\rho_{\mathrm{ext}}(\lambda)$ (orange, dashed) for $\gamma= 1/48$ (left) and $\gamma= \gamma_c$ (right).}
 \end{center}
\end{figure}
\subsection{Large $N$ factorisation $\&$ loop equations}\label{loopsec}

A natural set of integrals living in the same class as (\ref{eq:1MM}) are the following
\begin{equation}\label{eq:correlators}
\mathcal{M}^{(k_1,k_2,\ldots ,k_n)}_N \equiv \frac{1}{\mathcal{M}_N} \,  \int_{\mathbb{R}^{N^2}} [\dd M] e^{-N \text{Tr} V(M)} \prod_{i=1}^n \frac{1}{N} \text{Tr} \, M^{k_i} = \frac{1}{N^n} \Big\langle   \prod_{i=1}^n \text{Tr} M^{k_i} \Big\rangle ~.
\end{equation}
These integrals also preserve the $U(N)$ symmetry. To leading order in the large $N$ limit we can use the eigenvalue density to estimate the $\mathcal{M}^{(k_1,k_2,\ldots ,k_n)}_N$. For example
\begin{equation}
\mathcal{M}^{(k_1,k_2,\ldots, k_n)}_N = \prod_{i=1}^n \int_{-a}^a \dd \lambda \, \rho_{\mathrm{ext}}(\lambda) \, \lambda^{k_i} + \mathcal{O}(1/N^2)~.
\end{equation}
So long as neither $n$ nor $k_i$ scale with $N$ we see that to leading order in the large $N$ limit we cannot distinguish $\mathcal{M}^{(k_1,\ldots, k_n)}_N$ from $\mathcal{M}_N^{(k_1)}\mathcal{M}_N^{(k_2)}\ldots\mathcal{M}^{(k_n)}_N$.  This is a characteristic property of large $N$ systems known as large $N$ factorisation \cite{Yaffe:1981vf}. From this perspective, $\text{Tr} M^k/N$ can be viewed as a collection of weakly correlated quantities, the correlation strength going as some inverse power of $N$. The phenomenon of large $N$ factorisation thus provides us with a novel type of perturbative expansion for a certain class of matrix functions. 

As an example, let us use large $N$ factorisation to recover the equations governing the resolvent at large $N$. The starting point is the observation that the integral $\mathcal{M}_N$ should be left unchanged under a change in variables $M \to F(M)$ where $F$ is a matrix-valued function. For instance, we can consider the family of functions $F_\ell(M) = M + g \, e^{\ell M}$ where $g$ is taken to be parametrically small and $\ell$ is a real number. Invariance of $\mathcal{M}_N$ under these transformations imposes a set of interesting constraints \cite{Migdal:1984gj}. For parametrically small $g$ one finds 
\begin{equation}\label{loopeqns}
\frac{\ell}{N^2} \sum_{k=0}^\infty \frac{\ell^k}{k!} \sum_{j=0}^k \left\langle \, \text{Tr} \, M^j \text{Tr} \, M^{k-j} \right\rangle  =  \hat{G}_\ell^{(V)} \, \frac{1}{N} \left\langle \, \text{Tr} \, e^{\ell M} \, \right\rangle~.
\end{equation} 
The left hand side stems from the variation of the measure, whereas the right hand side stems from the variation of the exponent. $\hat{G}_\ell^{(V)}$ is a differential operator acting on functions of $\ell$. If $V(M)$ admits a power series,  $V(M) = \sum_{n=0}^\infty \alpha_n M^n$, 
we have 
\begin{equation}
\hat{G}_\ell^{(V)} = \sum_{n=1}^\infty  {n\, \alpha_n} \,\partial_\ell^{(n-1)}~.
\end{equation}
To leading order in the large $N$ expansion, we can use large $N$ factorisation to break the left hand side of (\ref{loopeqns}) into a sum over a product of traces.  The leading large $N$ expression is then given by
\begin{equation}\label{eq:leadingNloop}
\int_0^\ell \dd u\, \left\langle W_u \right\rangle \left\langle W_{\ell-u} \right\rangle = \hat{G}_\ell^{(V)}   \left\langle W_\ell \right\rangle~.
\end{equation}
The above equation is a type of equation known as a loop equation  \cite{Makeenko:1979pb, Polyakov:1980ca, Wadia:SD1981,Migdal:1984gj, Ambjorn:1990wg}, and the object 
\begin{equation}\label{eq:defloopop}
W_\ell \equiv  \frac{1}{N}\text{Tr} \, e^{\ell M}
\end{equation}
is known as a macroscopic loop operator. The resolvent $\text{R}_N(z)$ defined in (\ref{eq:defres}) can be obtained from the loop operator by the following Laplace transform
\begin{equation}\label{eq:laplacetrafoR}
\text{R}_N(z)  =   \int_0^\infty \dd\ell \, e^{-\ell z} \, W_\ell~\quad\quad \implies \quad\quad \text{R}(z) = \lim_{N\to\infty} \int_0^\infty \dd\ell \, e^{-\ell z} \left\langle W_\ell \right\rangle~.
\end{equation}
We can take the Laplace transform of (\ref{eq:leadingNloop}) with respect to $\ell$ and find 
\begin{equation}
\text{R}(z)^2 =V'(z)\RR(z)- \mathcal{P}(z)~,\quad\quad \mathcal{P}(z)= -\int_0^{\infty}\dd \ell \, \hat{G}_\ell^{(V)}\left\langle W_\ell \, e^{-\ell z} \right\rangle~.
\end{equation}
Solving the quadratic equation leads to 
\begin{equation}\label{eq:resolventloop}
 \RR(z)= \frac{1}{2}\, V'(z)- \frac{1}{2}\sqrt{V'(z)^2-4 \mathcal{P}(z)}~,
\end{equation}
where we choose the negative root to satisfy $\mathrm{res}_a$ (\ref{norm}). Equation (\ref{eq:resolventloop}) constitutes a derivation of (\ref{eq:generalResolvent}). 
\newline\newline
{\textbf{Examples.}} We present the loop operators for the three polynomials discussed in the beginning of this section. 
For the Gaussian case we can use the density in (\ref{eq:Wigner}) to calculate the loop operator
\begin{equation}\label{WLG}
\lim_{N \to \infty}  \left\langle W_\ell \right\rangle = \frac{1}{\ell}\, I_1(2\ell)~,
\end{equation}
where $I_n(z)$ is the modified Bessel function of the first kind. The differential operator is $\hat{G}_\ell^{(V)} = \partial_\ell$ and it confirms (\ref{eq:leadingNloop}). In the large $z$ limit we obtain 
\begin{align}
\mathcal{P}(z)= - \int_0^\infty \dd \ell \, \partial_\ell\left(\frac{1}{\ell}\, I_1(2\ell)\, e^{-\ell z}\right)= 1~,
\end{align}
such that (\ref{eq:resolventloop}) agrees perfectly with the resolvent in (\ref{eq:Wigner}). 
For the quartic polynomial we find  $\hat{G}_\ell^{(V_\alpha)}= \partial_\ell+4\alpha\partial_\ell^3$ and using the density in (\ref{eq:densityM4}) we obtain
\begin{equation}
\lim_{N \to \infty}  \left\langle W_\ell \right\rangle = \frac{a_+(1+6 \,a_+^2 \alpha)}{2\ell}\,I_1(a_+\,\ell)-  \frac{6\, a_+^2 \alpha}{\ell^2}\,I_2(a_+\,\ell) ~,
\end{equation}
where $a_+$ has been defined in (\ref{eq:d}). Combining $W_\ell$ with $\hat{G}_\ell^{(V_\alpha)}$ we easily confirm (\ref{eq:leadingNloop}) and combining it with (\ref{eq:resolventloop}) we confirm the polynomial in (\ref{eq:densityM4}). 
Finally, for the multicritical polynomial introduced in (\ref{eq:multcritmodel}) we obtain 
\begin{equation}\label{eq:loopmult}
\lim_{N \to \infty}  \left\langle W_\ell\right\rangle = \frac{a}{2\gamma\ell^4}\left(8a^2\ell +(1-a^2)^2\ell^3\right)I_1(a\ell)- \frac{2a^2}{\gamma\ell^4}\left(8-(1-a^2)\ell^2\right)I_2(a\ell)~,
\end{equation}
and the differential operator $\hat{G}_\ell^{(V)}$ is given by 
\begin{equation}
\hat{G}_\ell^{(V)}= \frac{1}{\gamma}\left(\partial_\ell-\frac{4}{3}\,\partial_\ell^3+\frac{4}{45}\,\partial_\ell^5\right)~,
\end{equation}
with $\gamma$ defined in (\ref{eq:multcritmodel}). Using (\ref{eq:loopmult}), it is straightforward to verify (\ref{eq:leadingNloop}).

\subsection{A perturbative expansion: the Riemann surfaces}\label{Riemannsurfaces}

We now proceed to study how the perturbative expansion of a matrix type integral is organised in the large $N$ limit. We will uncover a remarkable connection, originally observed by 't Hooft \cite{tHooft:1973alw}, to Riemann surfaces.

Let us focus on matrix integrals with the following structure
\begin{equation} \label{eq:sMI}
\mathcal{M}_N = \int_{\mathbb{R}^{N^2}} [\dd M] \, e^{-N \, \Tr \, V(M)}~,
\end{equation}
where $V(M)$ is an arbitrary polynomial in $M$ containing parameters $\alpha$ which will be taken to be small. Though the lessons will  be general, for the sake of concreteness we will consider again a purely quartic example $V_\alpha(M)$ (\ref{eq:resolventex}).
Since we will be interested in a perturbative expansion in small $\alpha$, it is convenient to write down the propagator
\begin{equation}\label{eq:Mprop}
\langle M_{IK}M_{JL}\rangle =\mathcal{Z}^{-1}_N(0) \, \int_{\mathbb{R}^{N^2}} [\dd M] \, e^{-\frac{N}{2}\Tr \, M^2}M_{IJ}M_{KL} =\frac{1}{N}\delta_{IL}\delta_{KJ}~.
\end{equation}
For small coupling $\alpha$ we expand the exponential ($\ref{eq:resolventex}$) leading to 
\begin{equation}
\mathcal{Z}_N(\alpha)  = \int_{\mathbb{R}^{N^2}} [\dd M] \, e^{-\frac{N}{2}\Tr \,M^2}\sum_{k \in \mathbb{N}} (-1)^k\frac{\left(N \alpha \right)^k}{k!} \, (\Tr M^4 )^k~.
\end{equation}
The graphical representation of the propagator and the quartic vertex is given in the figure below: 
\begin{figure}[H]
\begin{center}
\begin{tikzpicture}[scale=.8]
\draw[middlearrow={<},line width=0.15mm] (.4,1) --(1,1)node[pos=3,scale=.7]{L};
\draw[middlearrow={>},line width=0.15mm]  (.4,.6) --(1,.6)node[pos=3,scale=.7]{K};
\draw[middlearrow={>},line width=0.15mm]  (1,.6) --(1,0)node[pos=1.3,scale=.7]{J};
\draw[middlearrow={<},line width=0.15mm]  (1.4,.6) --(1.4,0)node[pos=1.3,scale=.7]{K};
\draw[middlearrow={>},line width=0.15mm]  (1.4,.6) --(2.,.6)node[pos=-2,scale=.7]{J};
\draw[middlearrow={<},line width=0.15mm]  (1.4,1) --(2.,1)node[pos=-2,scale=.7]{I};
\draw[middlearrow={>},line width=0.15mm]  (1.4,1) --(1.4,1.6)node[pos=1.4,scale=.7]{L};
\draw[middlearrow={<},line width=0.15mm]  (1,1) --(1,1.6)node[pos=1.4,scale=.7]{I};
\draw[middlearrow={<},line width=0.15mm] (-3.5,1) --(-5,1)node[pos= -.2, scale=.7]{$I$}node[pos= 1.2,scale=.7]{$I$};
\node[scale=.8] at (3.5,.8)   {$\sim \alpha\, N~.$};
\draw[middlearrow={>},line width=0.15mm]  (-3.5,.6) --(-5,.6)node[pos= -.2,scale=.7]{$J$}node[pos= 1.2,scale=.7]{$J$};
\node[scale=.8] at (-2, .8)   {$\sim  {N^{-1}}$ ~, };
\end{tikzpicture}
\end{center}
\caption{Propagator and quartic vertex.}
\label{fig:diagramsValpha}
\end{figure}
Using standard Gaussian integration, we can compute the various terms in the small $\alpha$ expansion. For example, to linear order in $\alpha$ the large $N$ integral $(\ref{eq:resolventex})$ becomes
\begin{equation}
\mathcal{F}_N(\alpha) \equiv -\log \frac{\mathcal{Z}_N(\alpha)}{\mathcal{Z}_N(0)} = (2N^2 +1)\alpha + \mathcal{O}(\alpha^2)~.
\end{equation}
Each of the terms in the $\alpha$ expansion has a diagrammatic representation. We give some examples in the figure below. Due to their double line representation, these diagrams are often called ribbon diagrams.
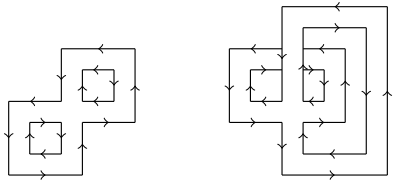
\begin{figure}[H]
\begin{center}
\begin{tikzpicture}[scale=.7]
\draw[middlearrow={<},line width=0.15mm] (0,1) --(1,1);
\draw[middlearrow={>},line width=0.15mm]  (0.4,.6) --(1,.6);
\draw[middlearrow={>},line width=0.15mm]  (1,.6) --(1,0);
\draw[middlearrow={<},line width=0.15mm]  (1.4,.6) --(1.4,-.4);
\draw[middlearrow={>},line width=0.15mm]  (1.4,.6) --(2.4,.6);
\draw[middlearrow={<},line width=0.15mm]  (1.4,1) --(2.,1);
\draw[middlearrow={>},line width=0.15mm]  (1.4,1) --(1.4,1.6);
\draw[middlearrow={<},line width=0.15mm]  (1,1) --(1,2);
\draw[middlearrow={<},line width=0.15mm]  (1.4,1.6) --(2,1.6);
\draw[middlearrow={>},line width=0.15mm]  (2,1.6) --(2,1);
\draw[middlearrow={>},line width=0.15mm]  (2.4,.6) --(2.4,2);
\draw[middlearrow={>},line width=0.15mm]  (2.4,2) --(1,2);
\draw[middlearrow={>},line width=0.15mm]  (0,1) --(0,-.4);
\draw[middlearrow={<},line width=0.15mm]  (.4,.6) --(.4,0);
\draw[middlearrow={<},line width=0.15mm]  (.4,0) --(1,0);
\draw[middlearrow={>},line width=0.15mm]  (0,-.4) --(1.4,-.4);
\end{tikzpicture}\quad\quad \quad
\begin{tikzpicture}[scale=.7]
\draw[middlearrow={<},line width=0.15mm] (.4,1) --(1,1);
\draw[middlearrow={>},line width=0.15mm] (.4,1) --(.4,1.6);
\draw[middlearrow={>},line width=0.15mm] (.4,1.6) --(1,1.6);
\draw[middlearrow={>},line width=0.15mm] (1.4,1.6) --(1.8,1.6);
\draw[middlearrow={>},line width=0.15mm] (1.8,1.6) --(1.8,1);
\draw[middlearrow={>},line width=0.15mm]  (0,.6) --(1,.6);
\draw[middlearrow={<},line width=0.15mm]  (0,.6) --(0,2);
\draw[middlearrow={<},line width=0.15mm]  (0,2) --(1,2);
\draw[middlearrow={<},line width=0.15mm]  (1.4,2) --(2.2,2);
\draw[middlearrow={<},line width=0.15mm]  (2.2,2) --(2.2,.6);
\draw[middlearrow={>},line width=0.15mm]  (1,.6) --(1,-.4);
\draw[middlearrow={>},line width=0.15mm]  (1,-.4) --(3,-.4);
\draw[middlearrow={>},line width=0.15mm]  (3,-.4) --(3,2.8);
\draw[middlearrow={>},line width=0.15mm]  (3,2.8) --(1,2.8);
\draw[middlearrow={<},line width=0.15mm]  (1.4,.6) --(1.4,0);
\draw[middlearrow={<},line width=0.15mm]  (1.4,0) --(2.6,0);
\draw[middlearrow={<},line width=0.15mm]  (2.6,0) --(2.6,2.4);
\draw[middlearrow={<},line width=0.15mm]  (2.6,2.4) --(1.4,2.4);
\draw[middlearrow={>},line width=0.15mm]  (1.4,.6) --(2.2,.6);
\draw[middlearrow={<},line width=0.15mm]  (1.4,1) --(1.8,1);
\draw[middlearrow={>},line width=0.15mm]  (1.4,1) --(1.4,2.4);
\draw[middlearrow={<},line width=0.15mm]  (1,1) --(1,2.8);
\end{tikzpicture}
\caption{Figure of a planar (left) and a non-planar (right) diagram.}
\end{center}
\end{figure}
From the small $\alpha$ expansion it is clear that the diagrams contributing to the perturbative expansion of a matrix integral can scale with different powers of $N$. In $(\ref{eq:Mprop})$ we observed that the propagator scales with inverse power of $N$. Moreover, each loop contributes a factor of $N$, and each vertex adds a factor $\alpha N$. This implies that a diagram with $L$ loops, $V$ vertices, and $P$ propagators scales as $\alpha^VN^{V+L-P}$.  
Making the identification \cite{tHooft:1973alw}
\begin{equation}\label{eq:discreteEuler}
\mathrm{loops}\; L \;\hat{=}\;\mathrm{faces}\; F~, \quad \mathrm{propagators}\; P\; \hat{=}\;\mathrm{edges}\; E~, \quad \mathrm{vertices}\; V\; \hat{=}\; \mathrm{vertices}\; V~,
\end{equation}
one can identify the power of $N$ associated to a particular ribbon diagram with the Euler characteristic $\chi_h$ of the two-dimensional compact surface $\Sigma_h$ of genus $h$ it can be drawn on. Since $\chi_h = 2-2h$, at least pictorially, there seems to be a natural genus expansion in the large $N$ limit for the perturbative expansion of our matrix integral $\mathcal{Z}_N(\alpha)$. We leave it to the reader to verify that this conclusion will not be affected by allowing $V_\alpha(M)$ to be an arbitrary polynomial, so long as the couplings are scaled appropriately as we take $N$ large. 
\newline\newline
To recapitulate: the large $N$ organisation of our perturbative expansion can be expressed as
\begin{equation} \label{eq:connecteddiagrams}
\mathcal{F}_N(\alpha) = \sum_{h=0}^\infty e^{\chi_h\log N} \, \mathcal{F}^{(h)}(\alpha)~,
\end{equation}
where $h$ labels the genus of the Riemann surface $\Sigma_h$, $\chi_h$ its Euler characteristic, and each $ \mathcal{F}_{h}(\alpha)$ is a sum of connected diagrams that can be drawn on a surface of genus $h$
\begin{equation}\label{Foriginal}
\mathcal{F}^{(h)}(\alpha) = \sum_{V=0}^\infty f_V^{(h)} \, \alpha^V~.
\end{equation}
The leading contribution in the large $N$ expansion is known as the {\it planar} contribution, and sub-leading contributions in the large $N$ expansion are known as non-planar contributions. 
We argued that there exists a natural map between fat graphs and the polygonisation of $\Sigma_h$. For the quartic theory, the surfaces of the graph dual to the polygonisation of $\Sigma_h$ are squares. 
\begin{figure}[H]
\begin{center}
{\begin{tikzpicture}[scale=1.6]
\draw[middlearrow={<},line width=0.15mm] (0,1) --(1.7,1);
\draw[middlearrow={>},line width=0.15mm]  (0,.8) --(1.7,.8);
\draw[middlearrow={<},line width=0.15mm]  (1.7,1) --(1.7,2.2);
\draw[middlearrow={>},line width=0.15mm]  (1.9,1) --(1.9,2.2);
\draw[middlearrow={<},line width=0.15mm]  (1.9,1) --(2.7,1);
\draw[middlearrow={>},line width=0.15mm]  (1.9,.8) --(2.7,.8);
\draw[middlearrow={>},line width=0.15mm]  (1.9,0) --(1.9,.8);
\draw[middlearrow={<},line width=0.15mm]  (1.7,0) --(1.7,.8);
\draw[middlearrow={<},line width=0.15mm]  (2.7,1) --(2.7,2.2);
\draw[middlearrow={>},line width=0.15mm]  (2.9,1) --(2.9,2.2);
\draw[middlearrow={>},line width=0.15mm]  (2.9,0) --(2.9,.8);
\draw[middlearrow={<},line width=0.15mm]  (2.7,0) --(2.7,.8);
\draw[middlearrow={>},line width=0.15mm]  (1.9,2.2) --(2.7,2.2);
\draw[middlearrow={<},line width=0.15mm]  (1.9,0) --(2.7,0);
\draw[middlearrow={>},line width=0.15mm]  (0,2.2) --(1.7,2.2);
\draw[middlearrow={>},line width=0.15mm]  (0,1) --(0,2.2);
\draw[middlearrow={<},line width=0.15mm]  (0,0) --(1.7,0);
\draw[middlearrow={<},line width=0.15mm]  (0,2.4) --(1.7,2.4);
\draw[middlearrow={<},line width=0.15mm]  (1.9,2.4) --(2.7,2.4);
\draw [middlearrow={<},line width=0.15mm] (1.7,2.4) --(1.7,3);
\draw [middlearrow={>},line width=0.15mm] (1.9,2.4) --(1.9,3);
\draw[middlearrow={>},line width=0.15mm]  (0,2.4) --(0,3);
\draw[middlearrow={>},line width=0.15mm]  (0,3) --(1.7,3);
\draw[middlearrow={<},line width=0.15mm]  (0,3.2) --(1.7,3.2);
\draw[middlearrow={>},line width=0.15mm]  (0,3.2) --(0,3.7);
\draw[middlearrow={<},line width=0.15mm]  (-.2,0) --(-.2,.8);
\draw[middlearrow={<},line width=0.15mm]  (-.2,.8) --(-1.5,.8);
\draw[middlearrow={>},line width=0.15mm]  (-.2,0) --(-1.5,0);
\draw[middlearrow={<},line width=0.15mm]  (-.2,-.2) --(-1.5,-.2);
\draw[middlearrow={>},line width=0.15mm]  (-.2,-.2) --(-.2,-1);
\draw[middlearrow={<},line width=0.15mm]  (0,-.2) --(0,-1);
\draw[middlearrow={>},line width=0.15mm]  (0,0) --(0,.8);
\draw[middlearrow={>},line width=0.15mm]  (-1.5,0) --(-1.5,.8);
\draw[middlearrow={<},line width=0.15mm]  (-1.5,1) --(-.2,1);
\draw[middlearrow={<},line width=0.15mm]  (-.2,1) --(-.2,2.2);
\draw[middlearrow={<},line width=0.15mm]  (-.2,2.2) --(-1.5,2.2);
\draw[middlearrow={<},line width=0.15mm]  (-1.5,2.4) --(-.2,2.4);
\draw[middlearrow={<},line width=0.15mm]  (-.2,2.4) --(-.2,3);
\draw[middlearrow={<},line width=0.15mm]  (-.2,3) --(-1.5,3);
\draw[middlearrow={<},line width=0.15mm]  (-1.5,3) --(-1.5,2.4);
\draw[middlearrow={>},line width=0.15mm]  (0,-.2) --(1.7,-.2);
\draw[middlearrow={<},line width=0.15mm]  (-1.5,-.2) --(-1.5,-1.);
\draw[middlearrow={<},line width=0.15mm]  (1.9,-.2) --(1.9,-1);
\draw[middlearrow={>},line width=0.15mm]  (-1.5,1) --(-1.5,2.2);
\draw[middlearrow={<},line width=0.15mm]  (1.7,-.2) --(1.7,-1);
\draw[middlearrow={>},line width=0.15mm]  (1.9,-.2) --(2.7,-.2);
\draw[middlearrow={>},line width=0.15mm]  (2.7,-.2) --(2.7,-.8);
\draw[middlearrow={>},line width=0.15mm]  (2.9,-.2) --(3.7,-.2);
\draw[middlearrow={<},line width=0.15mm]  (2.9,-.2) --(2.9,-.8);
\draw[middlearrow={<},line width=0.15mm]  (2.9,0) --(3.9,0);
\draw[middlearrow={>},line width=0.15mm]  (2.9,.8) --(3.9,.8);
\draw[middlearrow={<},line width=0.15mm]  (2.9,1) --(3.6,1);
\draw[middlearrow={>},line width=0.15mm]  (2.9,2.2) --(3.5,2.2);
\draw[middlearrow={<},line width=0.15mm]  (2.9,2.4) --(3.7,2.4);
\draw[middlearrow={>},line width=0.15mm]  (2.9,2.4) --(2.9,3);
\draw[middlearrow={<},line width=0.15mm]  (2.7,2.4) --(2.7,3);
\draw [middlearrow={>},line width=0.15mm] (1.9,3) --(2.7,3);
\draw[middlearrow={<},line width=0.15mm]  (1.9,3.2) --(2.7,3.2);
\draw [middlearrow={>},line width=0.15mm] (1.9,3.2) --(1.9,3.8);
\draw[middlearrow={<},line width=0.15mm]  (1.7,3.2) --(1.7,3.6);
\draw[middlearrow={<},line width=0.15mm]  (2.7,3.2) --(2.7,3.7);
\draw[middlearrow={>},line width=0.15mm]  (2.9,3.2) --(2.9,3.7);
\draw[middlearrow={<},line width=0.15mm]  (2.9,3.2) --(3.5,3.2);
\draw[middlearrow={>},line width=0.15mm]  (2.9,3) --(3.5,3);
\draw[middlearrow={<},line width=0.15mm]  (-.2,3.2) --(-.2,3.8);
\draw[middlearrow={<},line width=0.15mm]  (-1.5,3.2) --(-.2,3.2);
\draw[middlearrow={>},line width=0.15mm]  (-1.5,3.2) --(-1.5,3.9);
\draw[middlearrow={<},line width=0.15mm]  (-1.7,3.2) --(-1.7,3.9);
\draw[middlearrow={>},line width=0.15mm]  (-1.7,3.2) --(-2.5,3.2);
\draw [middlearrow={<},line width=0.15mm] (-1.7,3) --(-2.7,3);
\draw[middlearrow={>},line width=0.15mm]  (-1.7,2.4) --(-2.8,2.4);
\draw[middlearrow={<},line width=0.15mm]  (-1.7,2.2) --(-2.7,2.2);
\draw[middlearrow={>},line width=0.15mm]  (-1.7,1) --(-2.7,1);
\draw[middlearrow={>},line width=0.15mm]  (-1.7,0) --(-2.7,0);
\draw[middlearrow={<},line width=0.15mm]  (-1.7,-.2) --(-2.7,-.2);
\draw [middlearrow={<},line width=0.15mm] (-1.7,.8) --(-2.8,.8);
\draw[middlearrow={<},line width=0.15mm]  (-1.7,-.8) --(-1.7,-.2);
\draw[middlearrow={<},line width=0.15mm]  (-1.7,0) --(-1.7,.8);
\draw[middlearrow={<},line width=0.15mm]  (-1.7,1) --(-1.7,2.2);
\draw [middlearrow={<},line width=0.15mm] (-1.7,2.4) --(-1.7,3);
\draw[line width=0.25mm, magenta] (.5,.3) --(.5,1.3);
\draw[line width=0.25mm, magenta] (.5,1.3) --(2.2,1.8);
\draw[line width=0.25mm, magenta] (2.2,1.8) --(2.1,.4);
\draw[line width=0.25mm, magenta] (2.1,.4) --(.5,.3);
\draw[line width=0.25mm, magenta] (2.1,.4) --(3.3,.3);
\draw[line width=0.25mm, magenta] (3.3,.3) --(3.1,1.3);
\draw[line width=0.25mm, magenta] (3.3,.3) --(3.1,-.6);
\draw[line width=0.25mm, magenta] (3.1,-.6) --(2.2,-.7);
\draw[line width=0.25mm, magenta] (2.2,-.7) --(2.1,.4);
\draw[line width=0.25mm, magenta] (2.2,-.7) --(2.2,-.9);
\draw[line width=0.25mm, magenta] (2.2,-.7) --(.9,-.6);
\draw[line width=0.25mm, magenta] (3.1,-.6) --(3.5,-.9);
\draw[line width=0.25mm, magenta] (.9,-.6) --(.5,.3);
\draw[line width=0.25mm, magenta] (.9,-.6) --(-.7,-.9);
\draw[line width=0.25mm, magenta] (.9,-.6) --(.9,-.9);
\draw[line width=0.25mm, magenta] (-.7,-.9) --(-.9,.4);
\draw[line width=0.25mm, magenta] (-.7,-.9) --(-.7,-1.2);
\draw[line width=0.25mm, magenta] (-.7,-.9) --(-2.3,-.6);
\draw[line width=0.25mm, magenta] (-2.3,-.6) --(-2.1,.3);
\draw[line width=0.25mm, magenta] (-2.3,-.6) --(-2.5,-.9);
\draw[line width=0.25mm, magenta] (-2.1,.3) --(-.9,.4);
\draw[line width=0.25mm, magenta] (-2.1,.3) --(-2.6,.4);
\draw[line width=0.25mm, magenta] (-2.1,.3) --(-2.2,1.8);
\draw[line width=0.25mm, magenta] (-.9,.4) --(-.8,1.6);
\draw[line width=0.25mm, magenta] (-.8,1.6) --(-.9,.4);
\draw[line width=0.25mm, magenta] (-.8,1.6) --(-2.2,1.8);
\draw[line width=0.25mm, magenta] (-.8,1.6) --(.5,1.3);
\draw[line width=0.25mm, magenta] (.5,.3) --(-.9,.4);
\draw[line width=0.25mm, magenta] (.5,1.3) --(.9,2.6);
\draw[line width=0.25mm, magenta] (.9,2.6) --(-1.2,2.6);
\draw[line width=0.25mm, magenta] (-1.2,2.6) --(-.8,1.6);
\draw[line width=0.25mm, magenta] (-1.2,2.6) --(-2.3,2.5);
\draw[line width=0.25mm, magenta] (-1.2,2.6) --(-1.2,3.6);
\draw[line width=0.25mm, magenta] (-2.3,2.5) --(-2.2,1.8);
\draw[line width=0.25mm, magenta] (-2.3,2.5) --(-2.6,2.6);
\draw[line width=0.25mm, magenta] (-2.2,1.8) --(-2.7,1.5);
\draw[line width=0.25mm, magenta] (-2.3,2.5) --(-2.2,3.4);
\draw[line width=0.25mm, magenta] (-2.2,3.4) --(-1.2,3.6);
\draw[line width=0.25mm, magenta] (-2.2,3.4) --(-2.4,3.6);
\draw[line width=0.25mm, magenta] (-1.2,3.6) --(-1,3.8);
\draw[line width=0.25mm, magenta] (-1.2,3.6) --(1,3.4);
\draw[line width=0.25mm, magenta] (1,3.4) --(.9,3.8);
\draw[line width=0.25mm, magenta] (1,3.4) --(.9,2.6);
\draw[line width=0.25mm, magenta] (.9,2.6) --(2.4,2.7);
\draw[line width=0.25mm, magenta] (2.4,2.7) --(2.3,3.5);
\draw[line width=0.25mm, magenta] (2.3,3.5) --(1,3.4);
\draw[line width=0.25mm, magenta] (2.3,3.5) --(2.4,3.8);
\draw[line width=0.25mm, magenta] (2.3,3.5) --(3.4,3.4);
\draw[line width=0.25mm, magenta] (2.4,2.7) --(3.2,2.6);
\draw[line width=0.25mm, magenta] (2.4,2.7) --(2.2,1.8);
\draw[line width=0.25mm, magenta] (2.2,1.8) --(3.1,1.3);
\draw[line width=0.25mm, magenta] (3.3,.3) --(3.6,.5);
\draw[line width=0.25mm, magenta] (3.1,1.3) --(3.2,2.6);
\draw[line width=0.25mm, magenta] (3.1,1.3) --(3.5,1.4);
\draw[line width=0.25mm, magenta] (3.2,2.6) --(3.6,2.5);
\draw[line width=0.25mm, magenta] (3.2,2.6) --(3.4,3.4);
\draw[line width=0.25mm, magenta] (3.4,3.4) --(3.6,3.8);
\end{tikzpicture}}
\end{center}
\caption{Polygonisation: the black lines are the ribbon diagrams, whereas the magenta lines correspond to the dual lattice. }
\end{figure}
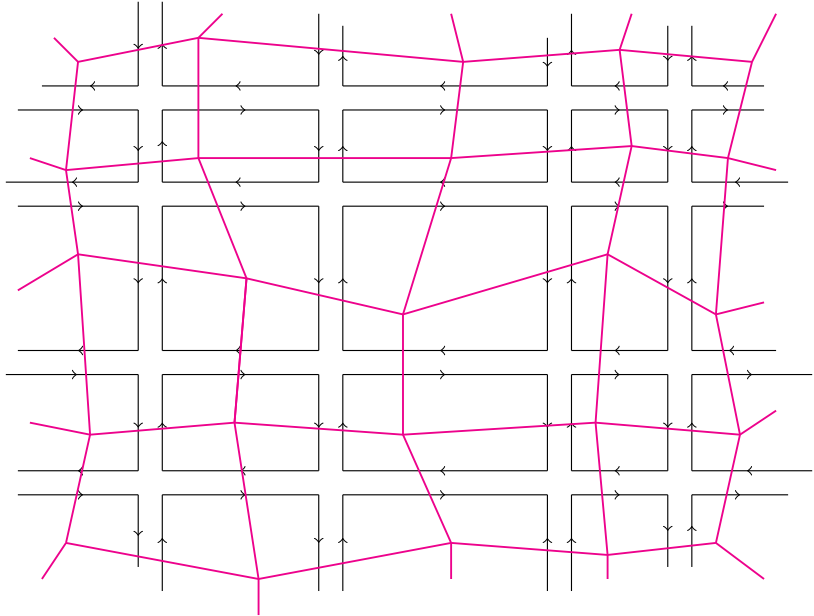
\begin{center}{{\it Riemann surfaces with boundaries}}\end{center}
In addition to ribbon diagrams corresponding to discretised Riemann surfaces without boundary, we can also examine diagrams living on Riemann surfaces with boundaries.
We start by coupling $N$ dimensional complex Grassmann valued vectors $\bar{\xi}^{(s)}$ and $\xi^{(s)}$ to the matrix integral (\ref{eq:1MM}) 
\begin{align}\label{eq:matrixplusvectors}
 \mathcal{K}_N(\alpha; z)\equiv \frac{1}{\mathcal{M}_N(\alpha)}\int_{\mathbb{R}^{N^2}} [\dd M]\,\int \prod_{s=1}^{N_f}\prod_{I=1}^N\dd\bar{\xi}^{(s)}_I\dd\xi^{(s)}_I\,e^{-N \Tr\, V_\alpha(M)- \bar{\xi}^{(s)}(z\,\mathbb{I}_N-M)\xi^{(s)}}~, \quad z\in \mathbb{C}~.
\end{align}
The index $s$ can be viewed as a flavour index ranging over $N_f$ `flavours'. For concreteness, although not necessary, we choose the quartic polynomial  (\ref{eq:resolventex}). The ribbon diagrams in figure \ref{fig:diagramsValpha} get enhanced by graphs with a single line. 
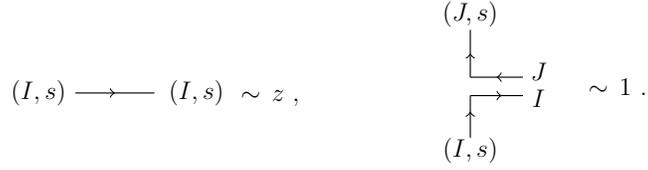
\begin{figure}[H]
\begin{center}
{\begin{tikzpicture}[scale=.7]
\draw[middlearrow={>},line width=0.15mm] (0,-1) --(-1,-1);
\draw[middlearrow={<},line width=0.15mm] (0,-1.35) --(-1,-1.35);
\draw[middlearrow={>},line width=0.15mm] (-1,-1) --(-1,-.1);
\draw[middlearrow={<},line width=0.15mm] (-1,-1.35) --(-1,-2.15);
\draw[middlearrow={>},line width=0.15mm] (-8.5,-1.3) --(-7,-1.3);
\node[scale=.8] at (-6.2, -1.3)   {$(I,s) $};
\node[scale=.8] at (-9.2, -1.3)   {$(I,s) $};
\node[scale=.8] at (-4.8, -1.4)   {$\sim \,z~, $};
\node[scale=.8] at (1.8, -1.175)   {$\sim \,1~.$};
\node[scale=.8] at (.3, -1.4)   {$I$};
\node[scale=.8] at (.3, -.9)   {$J$};
\node[scale=.8] at (-1., .2)   {$(J,s)$};
\node[scale=.8] at (-1., -2.4)   {$(I,s)$};
\end{tikzpicture}}
\end{center}
\caption{Propagator and vertex.}
\label{fig:vMv}
\end{figure}
Integrating out $\bar{\xi}^{(s)}$ and $\xi^{(s)}$ in (\ref{eq:matrixplusvectors}) we obtain
\begin{equation}\label{eq:integratedoutK}
\mathcal{K}_N(\alpha; z)=\frac{1}{\mathcal{M}_N(\alpha)} \int_{\mathbb{R}^{N^2}} [\dd M]\e^{-N\, \Tr V_\alpha(M)+N_f\,\mathcal{W}(z)}= 1+ \;\big\langle N_f\;\mathcal{W}(z)\big\rangle +\frac{1}{2!}\big\langle N_f^2\;\mathcal{W}(z)^2\big\rangle \;+ \cdots~,
\end{equation}
where we defined 
\begin{equation}\label{eq:macroscopicloop}
\mathcal{W}(z)\equiv \Tr\log(z\,\mathbb{I}_N-M)~.
\end{equation}
It is worth noting that $\partial_z  \mathcal{W}(z) $ is equal to the resolvent introduced in (\ref{eq:defres}). Further defining $\mathcal{B}_N(\alpha; z)\equiv \log\mathcal{K}_N(\alpha; z)$ we obtain the large $N$ expansion
\begin{equation}\label{eq:BNNf}
\mathcal{B}_N(\alpha; z)= \sum_{h=0}^\infty\sum_{ b=1}^\infty\,e^{\chi_{h,b}\log N}N_f^b\,\mathcal{B}^{(h,b)}(\alpha; z)~,\quad \chi_{h,b}\equiv 2-2h-b~.
\end{equation}
Here, $\mathcal{B}_N(\alpha; z)$ encodes the sum of connected diagrams corresponding to discretised Riemann surfaces with $h$ holes and $b$ boundaries. In the figure below we display an example of a ribbon diagram and its dual lattice for $h=0$ and $b=1$.
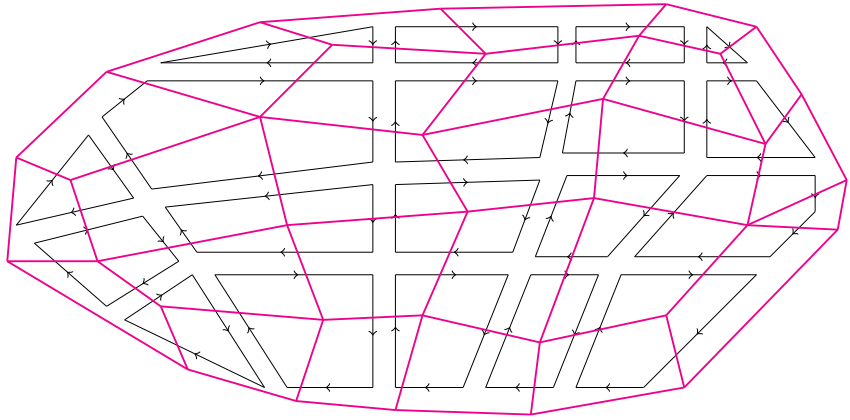
\begin{figure}[H]
\begin{center}
{\begin{tikzpicture}[scale=1.2]
\draw[middlearrow={<},line width=0.1mm] (-1.2,0) --(-.25,0);
\draw[middlearrow={<},line width=0.1mm] (0,0) --(.75,0);
\draw[middlearrow={<},line width=0.1mm] (1,0) --(1.75,0);
\draw[middlearrow={<},line width=0.1mm] (2,0) --(2.75,0);
\draw[middlearrow={<},line width=0.1mm] (2.75,0) --(4,1.25);
\draw[middlearrow={<},line width=0.1mm] (4,1.25) --(2.5,1.25);
\draw[middlearrow={<},line width=0.1mm] (2.5,1.25) --(2,0);
\draw[middlearrow={<},line width=0.1mm] (1.75,0) --(2.25,1.25);
\draw[middlearrow={<},line width=0.1mm] (2.25,1.25) --(1.5,1.25);
\draw[middlearrow={<},line width=0.1mm] (1.5,1.25)--(1,0);
\draw[middlearrow={<},line width=0.1mm] (.75,0)--(1.25,1.25);
\draw[middlearrow={<},line width=0.1mm] (1.25,1.25)--(0,1.25);
\draw[middlearrow={<},line width=0.1mm] (0,1.25)--(0,0);
\draw[middlearrow={<},line width=0.1mm] (-.25,0) --(-.25,1.25);
\draw[middlearrow={<},line width=0.1mm] (-.25,1.25) --(-2,1.25);
\draw[middlearrow={<},line width=0.1mm] (-2,1.25) --(-1.2,0);
\draw[middlearrow={>},line width=0.1mm] (-2.25,1.25) --(-1.45,0);
\draw[middlearrow={>},line width=0.1mm] (-1.45,0) --(-3,.75);
\draw[middlearrow={<},line width=0.1mm] (-2.25,1.25) --(-3,.75);
\draw[middlearrow={>},line width=0.1mm] (-2.4,1.4) --(-3.2,.9);
\draw[middlearrow={<},line width=0.1mm] (-4,1.6) --(-3.2,.9);
\draw[middlearrow={>},line width=0.1mm] (-4,1.6) --(-2.8,1.9);
\draw[middlearrow={<},line width=0.1mm] (-2.4,1.4) --(-2.8,1.9);
\draw[middlearrow={>},line width=0.1mm] (-2.2,1.5) --(-2.55,2);
\draw[middlearrow={<},line width=0.1mm] (-2.2,1.5) --(-.25,1.5);
\draw[middlearrow={<},line width=0.1mm] (-.25,1.5) --(-.25,2.25);
\draw[middlearrow={<},line width=0.1mm](-2.55,2)--(-.25,2.25);
\draw[middlearrow={<},line width=0.1mm] (-4.2,1.8) --(-2.9,2.1);
\draw[middlearrow={>},line width=0.1mm] (-4.2,1.8) --(-3.4,2.8);
\draw[middlearrow={<},line width=0.1mm] (-2.9,2.1) --(-3.4,2.8);
\draw[middlearrow={>},line width=0.1mm] (-2.7,2.2) --(-3.25,3);
\draw[middlearrow={<},line width=0.1mm] (-2.7,2.2) --(-.25,2.5);
\draw[middlearrow={>},line width=0.1mm] (-.25,3.4) --(-.25,2.5);
\draw[middlearrow={<},line width=0.1mm] (-.25,3.4) --(-2.75,3.4);
\draw[middlearrow={>},line width=0.1mm] (-3.25,3) --(-2.75,3.4);
\draw[middlearrow={>},line width=0.1mm] (-.25,3.6) --(-2.6,3.6);
\draw[middlearrow={<},line width=0.1mm] (-.25,4) --(-2.6,3.6);
\draw[middlearrow={>},line width=0.1mm] (-.25,4) --(-.25,3.6);
\draw[middlearrow={<},line width=0.1mm] (0,4) --(-0,3.6);
\draw[middlearrow={>},line width=0.1mm] (0,4) --(1.8,4);
\draw[middlearrow={<},line width=0.1mm] (0,3.6) --(1.8,3.6);
\draw[middlearrow={>},line width=0.1mm] (1.8,4) --(1.8,3.6);
\draw[middlearrow={>},line width=0.1mm] (0,2.5) --(-0,3.4);
\draw[middlearrow={<},line width=0.1mm] (0,2.25) --(-0,1.5);
\draw[middlearrow={>},line width=0.1mm] (1.3,1.5) --(-0,1.5);
\draw[middlearrow={<},line width=0.1mm] (1.3,1.5) --(1.6,2.3);
\draw[middlearrow={>},line width=0.1mm] (0,2.25) --(1.6,2.3);
\draw[middlearrow={<},line width=0.1mm] (0,2.5) --(1.6,2.55);
\draw[middlearrow={>},line width=0.1mm] (0,3.4) --(1.8,3.4);
\draw[middlearrow={<},line width=0.1mm](1.6,2.55) --(1.8,3.4);
\draw[middlearrow={<},line width=0.1mm] (2.,4) --(2.,3.6);
\draw[middlearrow={>},line width=0.1mm] (3.2,4) --(3.2,3.6);
\draw[middlearrow={<},line width=0.1mm] (2,3.6) --(3.2,3.6);
\draw[middlearrow={>},line width=0.1mm] (2,4) --(3.2,4);
\draw[middlearrow={>},line width=0.1mm] (2,3.4) --(3.2,3.4);
\draw[middlearrow={<},line width=0.1mm] (2,3.4) --(1.85,2.6);
\draw[middlearrow={>},line width=0.1mm] (3.2,2.6) --(1.85,2.6);
\draw[middlearrow={<},line width=0.1mm] (3.2,2.6) --(3.2,3.4);
\draw[middlearrow={>},line width=0.1mm] (3.15,2.35) --(2.35,1.45);
\draw[middlearrow={<},line width=0.1mm] (1.55,1.45) --(2.35,1.45);
\draw[middlearrow={>},line width=0.1mm] (1.55,1.45) --(1.9,2.35);
\draw[middlearrow={<},line width=0.1mm] (3.15,2.35) --(1.9,2.35);
\draw[middlearrow={<},line width=0.1mm] (2.65,1.45) --(4.15,1.45);
\draw[middlearrow={>},line width=0.1mm] (4.65,1.95) --(4.15,1.45);
\draw[middlearrow={<},line width=0.1mm] (4.65,1.95) --(4.65,2.35);
\draw[middlearrow={>},line width=0.1mm] (3.45,2.35) --(4.65,2.35);
\draw[middlearrow={<},line width=0.1mm] (3.45,2.35) --(2.65,1.45);
\draw[middlearrow={<},line width=0.1mm] (3.45,2.55) --(4.65,2.55);
\draw[middlearrow={>},line width=0.1mm] (3.45,2.55) --(3.45,3.4);
\draw[middlearrow={>},line width=0.1mm] (3.45,3.4) --(4,3.4);
\draw[middlearrow={<},line width=0.1mm] (4.65,2.55) --(4,3.4);
\draw[middlearrow={<},line width=0.1mm] (3.45,3.6) --(3.9,3.6);
\draw[middlearrow={>},line width=0.1mm] (3.45,3.6) --(3.45,4);
\draw[middlearrow={<},line width=0.1mm] (3.9,3.6) --(3.45,4);
\draw[line width=0.25mm,magenta] (0,-.25) --(1.5,-.3);
\draw[line width=0.25mm,magenta] (1.5,-.3) --(1.6,.5);
\draw[line width=0.25mm,magenta] (.3,.8) --(0,-.25);
\draw[line width=0.25mm,magenta] (1.5,-.3) --(3.2,0);
\draw[line width=0.25mm,magenta] (3.2,-0) --(4.9,1.75);
\draw[line width=0.25mm,magenta] (3,.8) --(3.2,0);
\draw[line width=0.25mm,magenta] (4.9,1.75) --(5,2.3);
\draw[line width=0.25mm,magenta] (5,2.3) --(3.9,1.8);
\draw[line width=0.25mm,magenta] (4.5,3.25) --(5,2.3);
\draw[line width=0.25mm,magenta] (4.5,3.25) --(4,4);
\draw[line width=0.25mm,magenta] (4,4) --(3.6,3.7);
\draw[line width=0.25mm,magenta] (3,4.25) --(4,4);
\draw[line width=0.25mm,magenta] (2.7,3.9) --(3,4.25);
\draw[line width=0.25mm,magenta] (3,4.25) --(.5,4.2);
\draw[line width=0.25mm,magenta] (.5,4.2) --(1.,3.7);
\draw[line width=0.25mm,magenta] (-1.5,4.05) --(.5,4.2);
\draw[line width=0.25mm,magenta] (-1.5,4.05) --(-3.2,3.5);
\draw[line width=0.25mm,magenta] (-.7,3.8) --(-1.5,4.05);
\draw[line width=0.25mm,magenta] (-3.2,3.5) --(-1.5,3.);
\draw[line width=0.25mm,magenta] (-4.2,2.55) --(-3.2,3.5);
\draw[line width=0.25mm,magenta] (-3.6,2.3) --(-4.2,2.55);
\draw[line width=0.25mm,magenta] (-4.2,2.55) --(-4.3,1.4);
\draw[line width=0.25mm,magenta] (-4.3,1.4) --(-3.3,1.4);
\draw[line width=0.25mm,magenta] (-4.3,1.4) --(-2.3,.2);
\draw[line width=0.25mm,magenta] (-1.1,-.15) --(-2.3,.2);
\draw[line width=0.25mm,magenta] (-1.1,-.15) --(0,-.25);
\draw[line width=0.25mm,magenta] (-1.1,-.15) --(-.8,.75);
\draw[line width=0.25mm,magenta] (-2.6,.9) --(-.8,.75);
\draw[line width=0.25mm,magenta] (-2.6,.9) --(-2.3,.2);
\draw[line width=0.25mm,magenta] (-2.6,.9) --(-3.3,1.4);
\draw[line width=0.25mm,magenta] (-1.2,1.8) --(-3.3,1.4);
\draw[line width=0.25mm,magenta] (-3.6,2.3) --(-3.3,1.4);
\draw[line width=0.25mm,magenta] (-3.6,2.3) --(-1.5,3.);
\draw[line width=0.25mm,magenta] (-1.2,1.8) --(-1.5,3.);
\draw[line width=0.25mm,magenta] (-1.2,1.8) --(-.8,.75);
\draw[line width=0.25mm,magenta] (-.7,3.8) --(-1.5,3.);
\draw[line width=0.25mm,magenta] (-.7,3.8) --(1.,3.7);
\draw[line width=0.25mm,magenta] (2.7,3.9) --(1.,3.7);
\draw[line width=0.25mm,magenta] (.3,2.8) --(1.,3.7);
\draw[line width=0.25mm,magenta] (2.7,3.9) --(3.6,3.7);
\draw[line width=0.25mm,magenta] (4.1,2.7) --(3.6,3.7);
\draw[line width=0.25mm,magenta] (4.1,2.7) --(3.9,1.8);
\draw[line width=0.25mm,magenta] (2.2,2.1) --(3.9,1.8);
\draw[line width=0.25mm,magenta] (3,.8) --(3.9,1.8);
\draw[line width=0.25mm,magenta] (3,.8) --(1.6,.5);
\draw[line width=0.25mm,magenta] (2.2,2.1) --(1.6,.5);
\draw[line width=0.25mm,magenta] (2.2,2.1) --(.8,1.95);
\draw[line width=0.25mm,magenta] (.3,.8) --(1.6,.5);
\draw[line width=0.25mm,magenta] (.3,.8) --(-.8,.75);
\draw[line width=0.25mm,magenta] (.3,.8) --(.8,1.95);
\draw[line width=0.25mm,magenta] (.3,2.8) --(.8,1.95);
\draw[line width=0.25mm,magenta] (-1.2,1.8) --(.8,1.95);
\draw[line width=0.25mm,magenta] (.3,2.8) --(-1.5,3.);
\draw[line width=0.25mm,magenta] (.3,2.8) --(2.3,3.2);
\draw[line width=0.25mm,magenta] (2.2,2.1) --(2.3,3.2);
\draw[line width=0.25mm,magenta] (4.5,3.25) --(4.1,2.7);
\draw[line width=0.25mm,magenta] (4.9,1.75) --(3.9,1.8);
\draw[line width=0.25mm,magenta] (2.7,3.9) --(2.3,3.2);
\draw[line width=0.25mm,magenta] (4.1,2.7) --(2.3,3.2);
\end{tikzpicture}}
\end{center}
\caption{Polygonisation with boundary: the black lines are the ribbon diagrams, whereas the magenta lines correspond to the dual lattice.}
\end{figure}
Each term $\mathcal{B}^{(h,b)}(\alpha;z)$ in (\ref{eq:BNNf}) is itself a sum of connected diagrams with fixed $h$ and $b$:
\begin{equation}
\mathcal{B}^{(h,b)}(\alpha;z)=\sum_{V=0}^\infty \sum_{B=1}^\infty f_{V,B}^{(h,b)}\alpha^V z^B~.
\end{equation}
The parameter $z$ is the boundary analog of $\alpha$, while $N_f$ is the boundary analog of $N$. Keeping track of the powers of $N_f$ in (\ref{eq:BNNf}) allows one to distinguish Riemann surfaces which would be indistinguishable purely from their power of $N$. 
\begin{figure}[H]
\begin{center}
\begin{tikzpicture}[rotate= 180,scale=1.4]
\node[scale=.8] at (-3.7, -.1)   {$\sim N_f \,N^{-1}~,$};
\draw[] (-1.1,-0.5) arc (270:90:.2 and 0.40);
\draw[] (-1.1,-0.5) arc (-90:90:.2 and .40);
\draw[] (-1.8605,-0.3656) to[out=35,in=190] (-1.1,-0.5);
\draw[] (-1.8605,0.1656) to[out=-30,in=190] (-1.1,0.3);
\draw[] (-1.8605,0.1656) arc (45:315:.6 and 0.37);
\draw[rounded corners=8pt] (-2.5,-.1)--(-2.3,-.25)--(-2.1,-.1);
\draw[rounded corners=7pt] (-2,-.15)--(-2.3,.05)--(-2.6,-.15);
\end{tikzpicture}
\quad\quad \quad 
\begin{tikzpicture}[rotate= 180,scale=1.4]
\node[scale=.8] at (-1.2, -.2)   {$\sim N_f^3 \,N^{-1}~.$};
  \draw (0,0) ellipse (.26 and .1);
    \draw (-.44,-.9) ellipse (.26 and .12);
    \draw (.44,-.9) ellipse (.26 and .12);
   \draw (-.18,-.9) to[out=80,in=80] (.18,-.9);   
    \draw (-.26,0) to[out=-90,in=90] (-.7,-.9);
  \draw (.26,0) to[out=-90,in=90] (.7,-.9);
\end{tikzpicture}
\end{center}
\caption{The parameter $N_f$ allows one to distinguish otherwise indistinguishable surfaces.}
\end{figure}
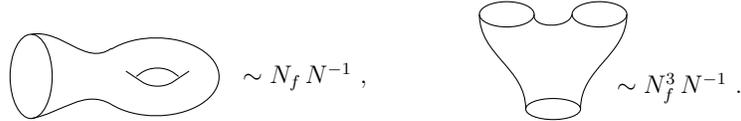

\begin{center}{{\it Continuum limit and double scaling}}\end{center}

In analogy to the bubble number we defined in (\ref{bubbleno}) for cactus diagrams, we can define the `vertex number' for the polygonisation of $\Sigma_h$ as
\begin{align} \label{eq:continuumlimit}
\langle n_h \rangle =  \partial_{\log \alpha} \log \mathcal{F}^{(h)}(\alpha)~.
\end{align}
For instance, given (\ref{eq:gzero}) we can compute $\langle n_0 \rangle$ near $\alpha_c$ to find $\langle n_0 \rangle \sim (\alpha-\alpha_c)^{-1}$. More generally, granting (\ref{gengenus}), we find $\langle n_h \rangle  \sim (\alpha-\alpha_c)^{-1}$ for all $h$. Consequently, as $\alpha$ approaches $\alpha_c$ the number of vertices diverges suggesting the possibility for a continuum limit of the discretised surfaces $\Sigma_h$. 

There is another limit of interest we can consider, known as the {\it double scaling limit}. The limit consists of simultaneously taking $N$ large as well as $\alpha$ to it's critical value $\alpha_c$, while keeping a particular combination $\alpha$ and $N$ fixed. Recall (\ref{gengenus}), namely that the non-analytic part of  $\mathcal{F}^{(h)}(\alpha)$ near $\alpha_c$ scales as
\begin{equation}\label{eq:naF}
\lim_{\alpha\to\alpha_c} \mathcal{F}_{\mathrm{n.a.}}^{(h)}(\alpha)= f_h (\alpha- \alpha_c)^{5\chi_h/4}~, \quad\quad h \in \mathbb{N}~,
\end{equation}
with $f_h$ being some proportionality constant and $h=1$ is understood to be logarithmic. It follows that the large $N$ expansion near $\alpha=\alpha_c$ is approximately
\begin{equation}\label{critgen}
\mathcal{F}_N(\alpha) \approx N^2 \mathcal{F}^{(0)}(\alpha_c) + \sum_{h=0}^\infty f_h \, e^{\chi_h \log N} (\alpha- \alpha_c)^{5\chi_h/4}~.
\end{equation}
The above expression suggests the introduction of a new parameter
\begin{equation}\label{eq:definitionkappa}
\kappa^{-1}\equiv N (\alpha- \alpha_c)^{5/4}~.
\end{equation}
The double scaling limit takes $N\rightarrow \infty$ and $\alpha\rightarrow \alpha_c$ while keeping $\kappa$ fixed. In this way, we see how the continuum limit and the large $N$ limit work harmoniously in producing a new function of potential interest 
\begin{equation}\label{twodimM}
{\mathcal{F}}(\kappa) \equiv  \lim_{N\to\infty,\,\, \kappa \,\, \text{fixed}}  \left( \mathcal{F}_N(\alpha) - N^2 \mathcal{F}^{(0)}(\alpha) \right) =  \sum_{h=0}^\infty f_h \, \kappa^{-\chi_h}~.
\end{equation}
In the next subsection, by introducing a new method for dealing with the matrix integrals, we will produce a non-linear differential equation whose solution encodes ${\mathcal{F}}(\kappa)$. 

\begin{center} *** \end{center}
An interesting system which naturally exhibits a genus expansion is the perturbative expansion of worldsheet string theory. At this stage there is no immediate reason for the two systems to be related, but one may speculate so. We shall see in section \ref{LiouvilleSec} that such speculations are materialised in a concrete and elegant sense.


\section{Large $N$ integrals over a single matrix II$^{\color{magenta}C,D}$}\label{singlematrixii}

In this section we introduce a different technique to solve matrix integrals. This technique will allow us to calculate contributions of the matrix integral beyond the planar approximation \cite{Bessis:1980ss}.

\subsection{Orthogonal polynomials}

Two polynomials are said to be orthogonal with respect to a weight function $w(x)$ if they satisfy
\begin{equation}
\mathrm{ortho}_a:\quad \int \dd x \, w(x) \, p_n(x)p_m(x)= h_n\delta_{m,n} \label{eq:ortho}
\end{equation}
for some $h_n$.  A familiar example is given by the Hermite polynomials 
\begin{equation}\label{eq:hermit}
H_n(x)=(-1)^n e^{x^2}\frac{\dd^n}{\dd x^n} e^{-x^2},\; n=1,2,\ldots
\end{equation}
which satisfy 
\begin{equation}
\int_{\mathbb{R}}\dd x \,   e^{-x^2} \, H_n(x)H_m(x) = 2^n \, \sqrt{\pi} \, n! \, \delta_{mn}~.
\end{equation}
In addition to $(\ref{eq:ortho})$, orthogonal polynomials satisfy the {\it{three-term recurrence relation}}
\begin{align}\nonumber
\mathrm{ortho}_b: \quad  &x\, p_n(x)= A_n\, p_n(x) + S_n\, p_{n+1}(x) + R_n\, p_{n-1}(x)\quad\quad \text{for} \quad\quad n>0~,\\ \label{eq:3termsrec}
& x\, p_0(x)= A_0\, p_0(x)+ S_0\, p_1(x)~,
\end{align}
where $A_n$, $S_n$, and $R_n$ are some real constants. 
The Hermite polynomials ($\ref{eq:hermit}$) satisfy 
\begin{equation}
x\,H_n = \frac{1}{2}\,H_{n+1}+n\,H_{n-1}~,
\end{equation}
and thus for $n\geq 0$ we have $A_n=0$, $S_n={1}/{2}$, and $R_n= n$. Other examples of orthogonal polynomials include the Legendre polynomials, Laguerre polynomials, and Chebychev polynomials. 

In what follows we will focus on monic polynomials
\begin{equation}\label{eq:monicp}
P_n(\lambda)\equiv \lambda^{n} + \sum_{j=0}^{n-1} a_j\lambda^j~, \quad\quad n=0,\ldots, N-1~.
\end{equation}
It is worth mentioning that when the measure in ($\ref{eq:ortho}$) is even under $\lambda \to -\lambda$, the monic polynomials $P_n(\lambda)$ transform as $P_n(-\lambda)= (-1)^n P_n(\lambda)$.
This implies that only even powers of $\lambda$ appear in $P_{2n}(\lambda)$ whereas only odd powers of $\lambda$  appear in $P_{2n-1}(\lambda)$. Moreover, invariance under $\lambda \to -\lambda$ implies that the coefficient $A_n$ in the three-term recurrence relation vanishes, while being monic implies $S_n=1$. 

\begin{center}{\it Orthogonal polynomials for matrix integrals}\end{center}

We will now use the properties of monic orthogonal polynomials to solve matrix integrals. In order to do so, we observe that the Vandermonde matrix can be expressed in the following way
\begin{equation}
\mathbf{V}_N = 
\begin{pmatrix}
P_0(\lambda_1) & P_1(\lambda_1)& \hdots & P_{N-1}(\lambda_1)\\ 
P_0(\lambda_2) & \hdots & \hdots & \hdots \\
\vdots & \hdots & \hdots & \hdots \\ 
P_0(\lambda_N) & P_1(\lambda_N)& \hdots & P_{N-1}(\lambda_N)
\end{pmatrix}~.
\end{equation}
The above expression is related to $(\ref{eq:Vandermonde})$ by a similarity transformation. The determinant $\Delta_N(\lambda)$ depends only on the leading degree of each polynomial.
Using the Leibniz formula for determinants, we can recast $\Delta_N(\lambda)$ as
\begin{equation}
\Delta_N(\lambda) = \sum_{\sigma\in S_N}\mathrm{sgn}(\sigma)\prod_{I=1}^N P_{\sigma(I)-1}(\lambda_I)\;.
\end{equation}
Consequently, we can re-express the matrix integral $(\ref{eq:1MM})$ as
\begin{equation} \label{eq:partfortho2}
\text{vol} \, \frac{U(1)^N \times S_N}{U(N)} \, \mathcal{M}_N=  \sum_{\sigma,\tau\in S_N}\mathrm{sgn}(\sigma)\,\mathrm{sgn}(\tau)\prod_{I=1}^N\int_{\mathbb{R}} \dd \mu(\lambda_I)P_{\sigma(I)-1}(\lambda_I)P_{\tau(I)-1}(\lambda_I)  =N!\prod_{I=0}^{N-1}h_{I}~,
\end{equation} 
where for notational convenience we have defined $\dd\mu(\lambda)\equiv \dd\lambda \, e^{-NV(\lambda)}$.
Using the three-term recurrence relation we can relate $h_n$ to $R_n$ in $(\ref{eq:3termsrec})$ and rewrite $(\ref{eq:partfortho2})$ entirely in terms of  $R_n$, which are themselves determined by the weight function $w(\lambda)= e^{-N V(\lambda)}$. It will also prove convenient to express $h_n$ in the following way
\begin{equation}
h_n= \int_{\mathbb{R}} \dd\mu(\lambda)  \lambda P_{n-1}(\lambda) P_{n}(\lambda)~.\label{eq:hn}
\end{equation}
An additional relation that will be useful is
\begin{equation}
n h_n= \int_{\mathbb{R}}\dd\mu(\lambda) P_n'(\lambda)\, \lambda P_n(\lambda)= NR_n \int_{\mathbb{R}}\dd\mu(\lambda)V'(\lambda)P_{n}(\lambda)P_{n-1}(\lambda)\;.\label{eq:recursionRn}
\end{equation}
For example, when combined with $(\ref{eq:hn})$ the above relation informs us that for a Gaussian potential $V(\lambda)= \lambda^2/2$ we have $h_0 = \sqrt{2\pi/N}$ and $R_n=n/N$. Hence the orthogonal polynomials are the Hermite polynomials for $V(\lambda)= \lambda^2/2$. From the above expressions, we can recursively calculate the sub-leading coefficients $a_J$ in (\ref{eq:monicp}). Applying ortho$_b$ to $(\ref{eq:hn})$ leads to the recursion $h_n=R_n \,h_{n-1}$,
which implies that $(\ref{eq:partfortho2})$ can be rewritten as
\begin{equation}
\mathcal{M}_N = \text{vol} \, \frac{U(N)}{U(1)^N \times S_N} \times  N! \, h_0^N R_1^{N-1}R_2^{N-2}\cdots R_{N-1}\;.\label{eq:hnRn}
\end{equation}
As a simple check of the above expression we note that for $\alpha=0$ 
\begin{equation}
\mathcal{M}_N =  \text{vol} \, \frac{U(N)}{U(1)^N \times S_N} \times N! \left( \frac{2\pi}{N} \right)^{N/2} \prod_{n=1}^{N-1} \left( \frac{n}{N} \right)^{N-n} = \left(\frac{2\pi}{N} \right)^{N^2/2}~,
\end{equation}
where we have used (\ref{eq:volUN}), $\text{vol} \, S_N = N!$, and $\text{vol} \, U(1) = 2\pi$.

Finally, we obtain a relation between the quantities defined and $\mathcal{F}_N(\alpha)$:
\begin{align}\label{eq:FEinRn}
\frac{1}{N^2}\mathcal{F}_N(\alpha) = -\frac{1}{N}\log \frac{h_0(\alpha)}{h_0(0)} - \frac{1}{N}\sum_{n=1}^{N-1}\left(1- \frac{n}{N}\right)\log \frac{R_n(\alpha)}{R_n(0)} \;.
\end{align}
The argument of $h_n$ and $R_n$ indicates their dependence on the coefficient $\alpha$ parametrising a non-Gaussian piece of $V(\lambda)$. We now delve into a detailed example.

\subsection{Non-planar contributions}\label{nonplanar}

We are now ready to see how the orthogonal polynomials can be used to go beyond the planar approximation of the large $N$ expansion of $\mathcal{F}_N(\alpha)$. For concreteness we  take $V_\alpha(\lambda)= \lambda^2/2+ \alpha \lambda^4$. From $(\ref{eq:recursionRn})$ we find
\begin{equation} \label{eq:equationqI}
\frac{n}{N} = {R_n(\alpha)}\Big(1  + 4\alpha\left(R_{n+1}(\alpha) + R_n(\alpha) + R_{n-1}(\alpha)\right)\Big)\;,
\end{equation}
where the term multiplying $4\alpha$ follows from the quartic interaction term in the potential and the relation
\begin{equation}\label{eq:RnRnp1Rnm1}
\int_{\mathbb{R}}\dd\mu(\lambda)\lambda^3 P_{n-1}(\lambda)P_n(\lambda)
 = \big(R_{n+1}(\alpha)+ R_n(\alpha)+R_{n-1}(\alpha)\big)h_n(\alpha)\;,
\end{equation}
which is obtained using the three-term-recurrence relation (\ref{eq:3termsrec}). In appendix \ref{Rnstaircase} we provide a graphical procedure to obtain relations similar to (\ref{eq:RnRnp1Rnm1}) also for higher order potentials.
We can use equation $(\ref{eq:equationqI})$ to study the large $N$ limit. Let us define the variables $\varepsilon\equiv {1}/{N}$ and $x\equiv n\varepsilon$. In the large $N$ limit, $x$ is well approximated by a continuous parameter. In view of this, it is convenient to set $r(x,\alpha)\equiv R_n(\alpha)$. We note that $r(x,\alpha)$ is also a function of $N$, but we suppress this dependence for notational simplicity. 
We can rewrite $(\ref{eq:equationqI})$ as 
\begin{equation}\label{eq:repsilon}
x= r(x,\alpha) +4 \alpha r(x,\alpha)\left[r(x+\varepsilon,\alpha)+r(x,\alpha) +r(x-\varepsilon,\alpha)\right]~.
\end{equation}
It follows from ($\ref{eq:repsilon}$) that $r(x,\alpha)$ is symmetric under $\varepsilon\leftrightarrow -\varepsilon$ and we can expand it in even powers of $\varepsilon$
\begin{equation}\label{eq:expre}
r(x,\alpha)= r_{0}(x,\alpha)+\varepsilon^2 \, r_{2}(x,\alpha) +\varepsilon^4 \, r_{4}(x,\alpha)+\cdots~.
\end{equation}
Collecting terms with equal powers of $\varepsilon$ we obtain the expression: 
\begin{equation}
 r(x+\varepsilon,\alpha)+ r(x-\varepsilon,\alpha)=2\sum_{n=0}^{\infty}\varepsilon^{2n}\sum_{k+p=n}\frac{1}{(2p)!}\frac{\dd^{2p}}{\dd x^{2p}} \, r_{2k}(x,\alpha)~.
\end{equation}
Re-inserting this into $(\ref{eq:repsilon})$ and comparing powers of $\varepsilon$ we conclude
\begin{equation}\label{eq:recrn}
x\, \delta_{s,0}=r_{2s}(x,\alpha)+4\alpha\sum_{m+n=s}r_{2m}(x,\alpha)\left[r_{2n}(x,\alpha) +2\sum_{k+p=n}\frac{1}{(2p)!}\frac{\dd^{2p}}{\dd x^{2p}}\, r_{2k}(x,\alpha)   \right]~.
\end{equation}
For the cases $s=0$ and $s=1$ we readily find 
\begin{eqnarray} \label{eq:r0r2}
r_{0}(x,\alpha) &=& \frac{-1+\sqrt{1+48 \alpha x}}{24\alpha}~  \quad\quad \mathrm{and}\quad\quad
r_{2}(x,\alpha) = \frac{96 \alpha^2 r_{0}(x,\alpha)}{(1+48\alpha x)^2}~.
\end{eqnarray}
Our final ingredient will be the Euler-Maclaurin formula
\begin{equation}\label{eq:eulerMac}
\frac{1}{N}\sum_{n=1}^N f\left(\frac{n}{N}\right)=  \int_0^1\dd x f(x)+\frac{1}{2N}f(x)\big|_0^1 +\sum_{n=1}^{p-1} \frac{B_{2n}}{(2n)!}\frac{1}{N^{2n}}f(x)^{(2n-1)}\big|_0^1+\mathcal{R}_N~.
\end{equation}
In the above, $f(x)$ is a $2p$ times continuously differentiable function, $\mathcal{R}_N$ is a remainder term scaling as $\mathcal{O}(1/N^{2p+1})$, and the $B_{2n}$ denote the Bernoulli numbers. Applying the Euler-Maclaurin formula to
\begin{equation}\label{eq:EulerMclaurinf}
f\left(x\right)= \left(1- x\right)\log \frac{r(x,\alpha)}{x}~,
\end{equation}
and expanding $(\ref{eq:FEinRn})$ in inverse powers of $N$, we find
\begin{align}\label{eq:Fexp} 
\frac{1}{N^2}\mathcal{F}_N(\alpha) = &- \int_0^1\dd x(1-x)\log\frac{r(x,\alpha)}{x} -\frac{1}{N}\log \frac{h_0(\alpha)}{h_0(0)}+\frac{1}{2N}\lim_{x\rightarrow 0}\log\frac{r(x,\alpha)}{x}\cr
&-\frac{1}{12 N^2} \left((1-x)\log \frac{r(x,\alpha)}{x}\right)^{(1)}\Bigg|_0^1\end{align}
up to order $\mathcal{O}(1/N^4)$ corrections. To obtain $h_0(\alpha)$ we simply evaluate 
\begin{equation}\label{eq:h0}
h_0(\alpha) = \int_{\mathbb{R}}\dd \mu(\lambda) =\frac{e^{\frac{N}{32\alpha}}}{2\sqrt{2}}\,\sqrt{\frac{1}{\alpha}}\,K_{\frac{1}{4}}\left(\frac{N}{32\alpha}\right)= \sqrt{\frac{2\pi}{N}}\left(1-\frac{3}{N}\,{\alpha} +\frac{105}{2N^2}\,{\alpha^2} + \dots\right)~,
\end{equation}
where $K_n(x)$ is the modified Bessel function of the second kind.
Expanding all three terms in ($\ref{eq:expre}$) up to powers of order $\mathcal{O}(1/N^2)$ we find \cite{Bessis:1980ss}
\begin{align}\label{eq:F0F1}\nonumber
\frac{1}{N^2}\mathcal{F}_N(\alpha)=&-\int_0^1 \dd x\,(1-x)\,\log\frac{r_{0}(x,\alpha)}{x} \\ 
&-\frac{1}{N^2}\left[\int_0^1 \dd x\,(1-x)\,\frac{r_{2}(x,\alpha)}{r_{0}(x,\alpha)}+\frac{1}{12}\left[(1-x)\log \frac{r_{0}(x,\alpha)}{x}\right]^{(1)}\Bigg|_0^1 -3\,\alpha\right]~.
\end{align}
Using ($\ref{eq:r0r2}$) both integrals in ($\ref{eq:F0F1}$) can be evaluated analytically. The expression obtained for $\mathcal{F}^{(0)}(\alpha)$ agrees with the re-summed expression in $(\ref{eq:gzero})$ and near $\alpha_c=-1/48$ we recover
\begin{equation}
\lim_{\alpha\rightarrow \alpha_c}\partial_\alpha^{(3)}\,\mathcal{F}_{\mathrm{n.a.}}^{(0)}(\alpha)\sim (\alpha-\alpha_c)^{-1/2}+ \ldots ~.
\end{equation}
For $\mathcal{F}^{(1)}(\alpha)$ we obtain 
\begin{equation}\label{eq:sumF1}
\mathcal{F}^{(1)}(\alpha) = \frac{1}{24}\left(\log 4 +\log(1+48\alpha) -2\log\left(1+\sqrt{1+48\alpha}\right)\right)~.
\end{equation}
Interestingly, in the case of $\mathcal{F}^{(1)}(\alpha)$, the non-analyticity near $\alpha_c$ comes in the form of a logarithm. Finally, a small $\alpha$ expansion leads to
\begin{align}
\mathcal{F}^{(0)}(\alpha)&= 2\alpha-18\alpha^2+288\alpha^3-6048\alpha^4+\frac{746496}{5}\alpha^5-4105728\alpha^6+\mathcal{O}\left(\alpha^7\right)~,\\ 
\mathcal{F}^{(1)}(\alpha)&=\alpha-30\alpha^2+1056\alpha^3-40176\alpha^4+\frac{8004096}{5}\alpha^5-65774592\alpha^6+\mathcal{O}\left(\alpha^7\right)~.
\end{align}
In addition to $\mathcal{F}^{(0)}(\alpha)$ and $\mathcal{F}^{(1)}(\alpha)$, we present $\mathcal{F}^{(h)}(\alpha)$ for $h=2$, $3$, and $4$ in appendix \ref{npc}. Their non-analytic behaviour near $\alpha_c$ agrees with the general form (\ref{eq:naF}).

\subsection{Full genus expansion $\&$ non-perturbative effects}\label{painlevesec}

The orthogonal polynomial method allows us to systematically compute non-planar contributions. Near the critical coupling, however, one might expect that only a small piece of the detailed functions $\mathcal{F}^{(h)}(\alpha)$ should matter. To this end, let us revisit equation (\ref{eq:repsilon}). Recall that $\varepsilon$ is parameterically small at large $N$, admitting an expansion of the type (\ref{eq:expre}). We also note from our expressions (\ref{eq:r0r2}) and (\ref{higherR}) that if we also take $\alpha \to \alpha_c$, expressions are dominated by the region near $x = 1$. To render the expressions finite near $x=1$ we must keep $\kappa^{-1} = (\alpha-\alpha_c)^{5/4} N$ fixed as we take $N\to\infty$. This is the double scaling limit we encountered earlier. Thus, we are prompted to study the full equation (\ref{eq:repsilon}) in the double scaling limit. If we take 
\begin{equation}\label{eq:doublescalingvariables}
x = 1+ (\alpha- \alpha_c) z~, \quad r(x,\alpha) = r_0(1,\alpha_c) + (\alpha-\alpha_c)^{1/2} \delta r(z)~, \quad \varepsilon =  (\alpha-\alpha_c)^{5/4} \kappa~,
\end{equation}
we can readily show that in the limit $ \alpha \to \alpha_c$, and recalling that $\alpha_c=-1/48$, equation (\ref{eq:repsilon}) implies
\begin{equation}\label{diffeq}
\frac{1}{4}\delta r(z)^2 + \frac{\kappa^2}{6}\delta r''(z) + z = 0~.
\end{equation}
Thus, at least in the double scaling limit, the full genus expansion (\ref{twodimM}) is reduced to solving the above Painlev{\'e} I equation \cite{Gross:1989vs,Douglas:1989ve,David:1990ge, Brezin:1990rb}. This is a remarkable simplification of the original problem.  

It is worth mentioning some features of the Painlev{\'e} I equation. For instance, the Painlev{\'e} I equation remains unchanged under the rescaling
\begin{equation}\label{scalingi}
z \to \lambda z~, \quad\quad \delta r(z) \to \lambda^{1/2} \delta r(z)~, \quad\quad \kappa \to \lambda^{5/4}\kappa~.
\end{equation}
Given the above scaling symmetry, it only makes sense to build perturbative expansions out of a scale invariant quantity such as $\kappa (-z)^{-5/4}$ rather than, say, small $\kappa$. The first few terms of this expansion are given by
\begin{equation}\label{borel}
\delta r_\pm (z) = (-z)^{1/2} \left( \pm 2 + \frac{1}{12}\frac{\kappa^2}{(-z)^{5/2}} \mp \frac{49}{576}\frac{\kappa^4}{(-z)^{5}} + \frac{1225}{3456}\frac{\kappa^6}{(-z)^{15/2}}   +\sum_{n=4}^\infty a_{n}^{(\pm)} \frac{\kappa^{2n}}{(-z)^{5n/2}}\right)~. 
\end{equation}
Expanding further, one observes that the coefficients of the negative branch $\delta r_-(z)$ are all positive while those of the positive branch $\delta r_+(z)$ alternate. To specify a solution of the Painlev{\'e} I equation, we must pick a set of initial conditions for $\delta r(z)$ and $\delta r'(z)$ at some value of $z$. 
Upon fixing half of the initial data, most choices for the remaining data lead to the oscillatory branch $\delta r_+(z)$  \cite{Bender:2015bja}.
\begin{figure}[H]\label{Painlevesolutions}
\begin{center}
 \includegraphics[scale=.52]{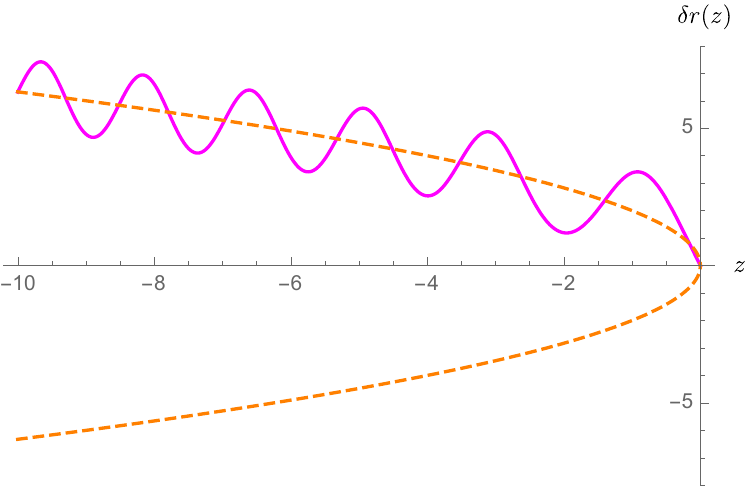} \quad\quad \qquad  \includegraphics[scale=.52]{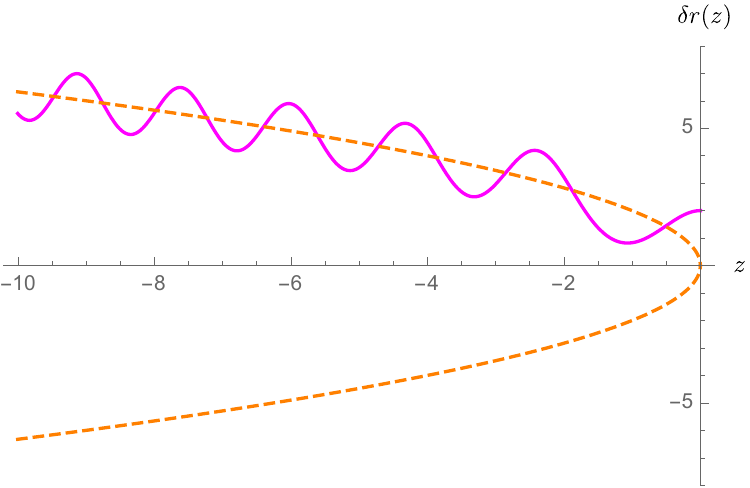}
\caption{Solutions of the Painlev{\'e} equation with $\kappa=1$. We take the initial conditions $\{\delta r(0),\delta r'(0)\}$= $\{0,-5\}$ (left) and $\{\delta r(0),\delta r'(0)\}$=$ \{2,0\}$ (right). The orange dashed lines are $\pm 2\sqrt{-z}$.}
 \end{center}
\end{figure}
\begin{figure}[H]\label{Painlevesolutionseigenvalues}
\begin{center}
 \includegraphics[scale=.52]{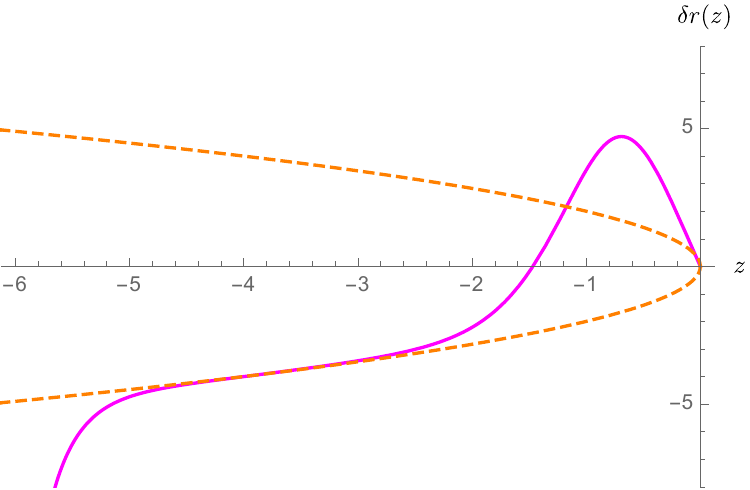} \quad\quad \qquad  \includegraphics[scale=.52]{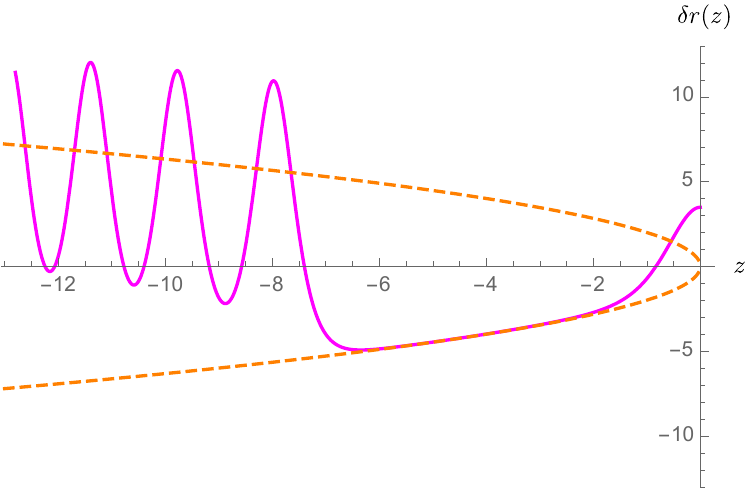}
 \caption{Solutions of the Painlev{\'e} equation with $\kappa=1$. We take the initial conditions $\{\delta r(0),\delta r'(0)\}$=  $\{0,-9.44761388485\}$ (left) and $\{\delta r(0),\delta r'(0)\}$=$\{3.482113278,0\}$ (right). The orange dashed lines are $\pm 2\sqrt{-z}$.}
 \end{center}
\end{figure}
Of the two branches (\ref{borel}), only the negative branch $\delta r_-(z)$ connects  to the double scaling limit stemming from the matrix integral. From now on we focus on this branch. The coefficients for $\delta r_-(z)$ satisfy the recursion relation
\begin{equation}\label{coefficients}
a_{n+1} =\frac{25n^2-1}{24}\,a_n +\frac{1}{4}\sum_{m=1}^n a_m  a_{n+1-m} ~, \quad\quad a_0 =-2~, \quad\quad n\geq 0~.
\end{equation}
For large $n$ the coefficients above grow as $\sim 5/(4\sqrt{6})\,\Gamma(2n-1/2)$ \cite{Ginsparg:1991ws}. This implies that the branch reproducing the perturbative expansion (\ref{borel}) is not Borel summable \cite{Gross:1989vs}. In fact, there is a family of solutions to (\ref{diffeq}) containing non-perturbative terms \cite{Shenker:1990uf, Ginsparg:1991ws} of the type 
\begin{equation}\label{epsilonPainleve}
\epsilon(z) = c\, \left(\frac{-z}{\kappa^{4/5}}\right)^{-1/8} \times e^{-\frac{4}{5\kappa}\sqrt{6}\,(-z)^{5/4}}~.
\end{equation} 
This family admits the same perturbative expansion (\ref{borel}) at large $(-z)^{5/4}/\kappa$. The parameter $c$ is an undetermined integration constant. The form (\ref{epsilonPainleve}) is obtained by considering a small deviation $\delta r(z)= \delta r_0(z)+\epsilon(z)$ and solving for $\epsilon(z)$ in a WKB approximation. To linear order in $\epsilon(z)$ we have 
\begin{equation}
\frac{1}{4}\delta r_0(z)^2 + \frac{\kappa^2}{6}\delta r_0''(z) + z = 0~, \quad \quad \frac{\kappa^2}{6}\epsilon''(z)+ \frac{1}{2}\delta r_0(z)\epsilon(z)=0~.
\end{equation}
For completeness, we mention that one can also construct a perturbative expansion for solutions near a double pole at $z=z_0$
\begin{equation}
\delta r(z) = -\frac{4}{(z-z_0)^2} + \frac{3z_0}{5}(z-z_0)^2+  (z-z_0)^3 + a_4 (z-z_0)^4 - \frac{3 z_0^2}{100}(z-z_0)^6 + \ldots~,
\end{equation}
where we have set $\kappa=1$. Both the location of the pole, $z_0$, as well as the value of $a_4$ are free parameters.

We now relate $\delta r_-(z)$ to the double scaling limit expression $\mathcal{F}(\kappa)$ introduced in (\ref{twodimM}).  To do so, we note that in the double scaling limit, the expression (\ref{eq:Fexp}) is dominated by 
\begin{equation}\label{eq:Fdscaling}
\mathcal{F}_N(\alpha) \approx - N^2\,\int_0^1\dd x (1-x)\log\frac{r(x,\alpha)}{x}+\cdots~.
\end{equation}
Combining the above expression with (\ref{eq:doublescalingvariables}) we obtain
\begin{multline}\label{Fkappa}
\mathcal{F}(\kappa) \equiv  \lim_{\text{d.s.l.}} \left( \mathcal{F}_N(\alpha) -  N^2  \mathcal{F}^{(0)}(\alpha) \right) \\ = \frac{12288\sqrt{3}}{5\kappa^2}  +  \frac{1}{2\kappa^2} \lim_{\varepsilon\to0^+} \int_{-(\kappa /\varepsilon)^{4/5}}^0 \dd z  z  \left(  \delta r_-(z) + 2(-z)^{1/2} \right)~.
\end{multline}
It follows from (\ref{eq:gzero}) that $\mathcal{F}^{(0)}(\alpha_c) = (7/24-\log{2}/2)$. We now recall the perturbative expansion (\ref{borel}) which is reliable for $(\kappa /\varepsilon)^{4/5} \ge -z \gg \kappa^{4/5}$. Since the first term in the expansion (\ref{borel}) is $\kappa$ independent and grows at large $z$ 
it is convenient to subtract it from $\delta r_-(z)$, as we have done in (\ref{Fkappa}). It is of interest to see how the perturbative expansion (\ref{borel}) is encoded in $\mathcal{F}(\kappa)$. We must recall that there is in fact no actual small $\kappa$ expansion since the scaling symmetry (\ref{scalingi}) always allows us to rescale $\kappa$. Instead, it is convenient to consider a slightly generalised integral given by
\begin{equation}\label{eq:Int_kappa}
{\mathcal{F}}(\kappa;\eta)  \equiv  \frac{1}{2\kappa^2} \lim_{\varepsilon \to 0^+}\int^{-\eta}_{-(\kappa /\varepsilon)^{4/5}} \dd z\, (z+\eta) \left( \delta r_-(z) + 2(-z)^{1/2} \right)~.
\end{equation}
We have introduced a cutoff at some small value $-\eta \gtrsim \kappa^{4/5}$ to avoid entering a region where the perturbative expansion (\ref{borel}) is invalidated. We note that $\partial_\eta^2 \mathcal{F}(\kappa;\eta)= -\left( \delta r_-(-\eta) + 2\eta^{1/2}\right)/2\kappa^2$.
Evaluating ${\mathcal{F}}(\kappa;\eta)$ in a small $\kappa/\eta^{5/4}$ expansion we find
\begin{equation}\label{eq:F_kappa}
\mathcal{F}(\kappa;\eta) = \frac{1}{24} \left(\log  \left(\eta \left( \frac{\varepsilon}{\kappa} \right)^{4/5}\right) +1 \right) - \frac{7 \kappa ^2}{1440\, \eta ^{5/2}}-\frac{245 \kappa ^4}{41472 \, \eta ^5}+ \ldots~. 
\end{equation}
We recognise the coefficient of the logarithm from our previous expression (\ref{eq:sumF1}). As a final remark we note that although the individual coefficients for genus $h\geq 2$ in (\ref{eq:F_kappa}) depend on the cutoff we choose in (\ref{eq:Int_kappa}), ratios of cutoff independent coefficients agree with the appropriate ratios upon calculating the subleading terms of (\ref{eq:Nsquaredexp}) explicitly using orthogonal polynomials. We provide some of these coefficients in (\ref{eq:f2f3f4}). 

\begin{center}\it{Double scaling limit for multicritical models}\end{center}

One can perform a similar analysis for the multicritical model introduced in (\ref{eq:multcritmodel}). Using the methods discussed in appendix \ref{Rnstaircase} it is straightforward to derive the equation
\begin{equation}\label{eq:planarmultcrit}
r_0(x,\gamma) -4 r_0(x,\gamma)^2 +\frac{16}{3}r_0(x,\gamma)^3=\gamma \, x~.
\end{equation}
The function $r_0(x,\gamma)$ encodes the planar contribution. Solving for $r_0(x)$ and selecting the branch relevant to the matrix integral, we obtain
\begin{equation}
r_0(x,\gamma)= \frac{1}{4}\left(1+(-1+12\gamma x)^{1/3}\right)~.
\end{equation}
From the above find that $\gamma_c= 1/12$, consistent with our previous discussion. We now define 
\begin{equation}
z \equiv N^{6/7}\left(1-\frac{\gamma}{\gamma_c} x\right) ,\quad\quad r(x,\gamma) \equiv r_0(1,\gamma_c)\left(1-N^{-2/7}\delta r(z)\right)~,\quad\quad \varepsilon  \equiv 1/N~.
\end{equation}
The double scaling limit is given by taking $N$ large while keeping $\vartheta^{-1}\equiv (\gamma-\gamma_c)^{7/6}N$ fixed. 
The equation governing $\delta r(z)$ in the double scaling limit reads
\begin{equation}\label{eq:painlevemulticrit}
\delta r(z)^3-\frac{1}{2}\delta r'(z)^2-\delta r(z)\delta r''(z)+ \frac{1}{10}\delta r^{(4)}(z)-z=0~.
\end{equation}
We can solve the above equation in a large $z$ expansion
\begin{equation}\label{pertmulticrit}
\delta r(z)= z^{1/3}\left(1+\sum_{n=1}^\infty b_k z^{-7n/3}\right)~,
\end{equation}
with the first few terms given by
\begin{multline}\label{largezII}
\delta r(z)= z^{1/3}\Big(1-\frac{1}{18}z^{-7/3}-\frac{7}{108}z^{-14/3}- \frac{4199}{17496}z^{-21/3}- \frac{409297}{262440}z^{-28/3} \cr - \frac{101108329}{9447840}z^{-35/3}
+ \frac{25947984239}{191318760}z^{-42/3}+ \ldots \Big)~.
\end{multline}
Expanding further the coefficients of the above expansion are found to be alternating. Contrary to the coefficients (\ref{coefficients}) of the Painlev{\'e} I equation, the coefficients (\ref{pertmulticrit}) of the perturbative expansion of (\ref{eq:painlevemulticrit}) are Borel summable \cite{Ginsparg:1991ws}. Related to this, $\mathcal{M}_N$ (\ref{eq:1MM}) for the polynomial $V(M)$ (\ref{eq:multcritmodel}) is well defined at finite $N$ for all positive $\gamma$. 

It is also interesting to assess whether the above series can admit non-perturbative corrections. Expanding about the $\delta r(z)$ solution given by the large $z$ expansion (\ref{largezII}), we find that to linear order the small deviation $\epsilon(z)$ must satisfy
\begin{equation}
\frac{1}{10}\epsilon^{(4)}(z)- z^{1/3}\epsilon''(z)-\frac{1}{3}z^{-2/3}\epsilon'(z)+3z^{2/3}\epsilon(z)=0~.
\end{equation}
Performing a WKB analysis we obtain $\epsilon(z)\approx c_\pm \, z^{-1/4}\,e^{-\frac{6}{7}\sqrt{5\pm i \sqrt{5}}\, z^{7/6}}$. It is interesting to note that the exponent is now complex. Interestingly, the polynomial governing the eigenvalues, $V(\lambda)$ given in (\ref{eq:multcritmodel}), has no maxima. Rather it has critical points at the complex values $\lambda_\pm^2 = \frac{1}{4}(5\pm i \sqrt{5})$.
Further multicritical models and their properties are presented in appendix \ref{m4}.

\subsection{Eigenvalues $\&$ instantons} 

As a final note, we discuss a connection between the non-perturbative corrections of the Painlev{\'e} I equation and the original matrix integral (\ref{eq:1MM}). To do so, we consider configurations for which one (or multiple) of the eigenvalues is sitting outside of the densely filled eigenvalue distribution. Such a configuration will give a subdominant contribution to the matrix integral whose size we will now compute.  

For simplicity, we focus on the case where a single eigenvalue $\lambda\equiv \lambda_N$ is separated from the remaining $N-1$ eigenvalues. The matrix integral
$\mathcal{M}_N$ can be written in the following form
\begin{equation} \label{MNinstanton}
\text{vol} \, \frac{U(1)^N \times S_N}{U(N)}  \mathcal{M}_N = \int_\mathbb{R} \dd \lambda  e^{- NV(\lambda)}  \int_{\mathbb{R}^{N-1}} \prod_{I=1}^{N-1} \dd \mu_I \Delta_{N-1}(\mu)^2 \, e^{-N\sum_{I=1}^{N-1}V(\mu_I)} \prod_{I=1}^{N-1}(\lambda-\mu_I)^2~,
\end{equation}
where the expectation value is defined in (\ref{eq:correlators}). 
We define the effective polynomial 
\begin{equation}\label{expansiondet}
V_{\mathrm{eff}}(\lambda)\equiv V(\lambda)-\frac{1}{N}\,\Big\langle \Tr \log\left(\lambda\, \mathbb{I}_{N-1}- M_{N-1}\right)^2\Big\rangle ~,
\end{equation}
which is the leading order correction of the polynomial $V(\lambda)$ in the large $N$ limit. Explicitly, assuming that $\mathcal{M}_{N}$ admits a single cut eigenvalue density $\rho(\lambda)$ symmetric about the origin, we find
\begin{equation}
V_{\mathrm{eff}}(\lambda) = V(\lambda)- \int_{-a}^a\dd \mu \, \rho(\mu)\log (\lambda-\mu)^2~,
\end{equation}
 to leading order in the large $N$ limit. It follows from (\ref{matrixsaddle}) that within the range $\lambda \in (-a,a)$, $V_{\mathrm{eff}}(\lambda)$ must be a constant $V_0$ given by 
\begin{equation}
V_0 \equiv V(0)- \int_{-a}^a\dd \mu\,\rho(\mu)\log \mu^2~.
\end{equation}
The effective polynomial ceases to be constant for $\lambda^2 > a^2$. To leading order in the large $N$ limit, the piece of the integral pertinent to the separated eigenvalue is given by
\begin{equation}\label{instantonint}
\mathcal{I}_N \equiv \int_{\mathbb{R}} \dd\lambda\, e^{-N V_{\text{eff}}(\lambda)}= 2 a \, e^{-N V_0}  \left(1 + \frac{1}{2a} \int_{\lambda	\notin [-a,a]}  \dd\lambda \,e^{-N \left(V_{\mathrm{eff}}(\lambda)-V_0 \right)}\right)~. 
\end{equation}
The critical points of $V_{\mathrm{eff}}(\lambda)$ for $\lambda^2 > a^2$ can be found by solving
\begin{equation}
{V'_{\mathrm{eff}}(\lambda)= \text{sign} \, \lambda \, \text{Re} \sqrt{V'(\lambda)^2-4 \mathcal{P}(\lambda)} = 0~,} 
\end{equation}
where $\mathcal{P}(\lambda)$ is defined in (\ref{eq:resolventloop}). 

At large $N$ we can apply the saddle point approximation to the integral in (\ref{instantonint}), which requires evaluating its exponent at its critical points.
We are interested in contributions that stem from a single eigenvalue sitting at a critical point outside the interval containing the dense set of remaining $N-1$ eigenvalues. These contributions can lead to corrections which are exponentially suppressed in $N$. Consequently they do not contribute to the perturbative $1/N$ expansion -- they are non-perturbative. In certain contexts, such suppressed configurations bear the name instantons. 
\newline\newline
{\textbf{Example.}} As a concrete example, we consider the quartic polynomial (\ref{eq:resolventex}). Taking $\alpha \in (\alpha_c,0)$, we find the following two maxima for the effective polynomial:
\begin{equation}
 \lambda_{\pm} =\pm \sqrt{- \frac{1}{6\alpha}- \frac{\sqrt{1+48\alpha}}{12\alpha}}~.
\end{equation}
For $\lambda \in (-a_+,a_+)$, with $a_+$ given in (\ref{eq:d}) and the eigenvalue density in (\ref{eq:densityM4}), we find
{
\begin{equation}
V_0 = -\frac{1}{48\alpha} +  \frac{\sqrt{1+48 \alpha}+24 \alpha  \left(1+\log 576 -2 \log  \frac{\sqrt{1+48 \alpha}-1}{\alpha }  \right)}{48 \alpha }~.
\end{equation} }
Evaluating $V_{\mathrm{eff}}(\lambda)$ at $\lambda = \lambda_{\pm}$ one finds
\begin{equation}\label{effectiveV}
V_{\mathrm{eff}}(\lambda_{\pm}) - V_0 =  {\sqrt{3}}\frac{\sqrt{2+\sqrt{y}}}{(1-y)}\,y^{1/4}-2\,\mathrm{Re}\left(\mathrm{arctanh}\frac{\sqrt{2+\sqrt{y}}}{\sqrt{3}y^{1/4}}\right)~,
\end{equation}
where we have defined $y\equiv 1-\alpha/\alpha_c$. 
A small $y$ expansion reveals 
\begin{equation}
V_{\mathrm{eff}}(\lambda_{\pm}) - V_0 =\frac{4}{5}\sqrt{6}\, y^{5/4}+  \frac{{2}}{7}\sqrt{\frac{3}{2}}\,y^{7/4}+\ldots~.
\end{equation}
Interestingly, in the double scaling limit, where $N$ is taken to infinity while keeping $ N y^{5/4}$ fixed, the leading term in the above expansion gives a contribution to (\ref{instantonint}) that remains finite \cite{Ginsparg:1991ws}. The correction is reminiscent of the non-perturbative correction (\ref{epsilonPainleve}) found in our discussion of the Painlev{\'e} I equation. 

\section{Large $N$ integrals over two matrices$^{\color{magenta} D,E}$}\label{section2MM}

In this section we will generalize the one matrix integral ($\ref{eq:1MM}$) to an integral involving two $N\times N$ Hermitian matrices $A$ and $B$  
\begin{equation}\label{eq:2MI}
\mathcal{T}_N(c)=\int_{\mathbb{R}^{2N^2}}[\dd A]\, [\dd B] \, e^{-N \, \Tr \,\left( V(A) + V(B) -  2 \, c \, A \, B \right) }~.
\end{equation}
$V(A)$ and $V(B)$ are two polynomials and $c\in \mathbb{R}$ . The measures $[\dd A]$ and $[\dd B]$ are as in (\ref{eq:matrixpartitionfunction}). Part of the motivation for studying such models is that they will form an interesting bridge between the quantum mechanical models explored in later sections and the matrix integrals explored so far. We note that the integral is invariant under a single $U(N)$ transformation that rotates both $A$ and $B$ concurrently. Given that we have two matrices and only a single $U(N)$ symmetry, the eigenvalue decomposition is slightly more involved.

\subsection{Eigenvalue decomposition}

Following an approach inspired by that pursued in section \ref{singlematrix}, we parameterise $A$ and $B$ in terms of their eigenvalues
\begin{equation}
A= \mathcal{U}_A D_A \, \mathcal{U}_A^\dagger~, \quad\quad\quad  D_A= \mathrm{diag}(a_1,\ldots ,a_N)~,
\end{equation} 
and similarly for $B$. Recalling the discussion in subsection $\ref{eigdist}$, the matrices $\mathcal{U}_A$ and $\mathcal{U}_B$ are general elements of $U(N)/U(1)^N$. Integrating over the angular variables in $(\ref{eq:2MI})$ leads to 
\begin{equation}\label{eq:2MIa}
\mathcal{T}_N(c)  =\mathrm{vol}\,\frac{U(N)}{U(1)^N\times S_N}\int_{\mathbb{R}^{2N}}\prod_{I=1}^N\dd a_I\,\dd b_I\,\Delta_N(a)^2 \Delta_N(b)^2 \, e^{-N \sum_{I=1}^N\left(V(a_I)+V(b_I)\right)}I(a,b,c)~,
\end{equation}
where $I(a,b,c)$ is known as the Harish-Chandra-Itzykson-Zuber (HCIZ) integral
\begin{equation}
I(a,b,c) \equiv \frac{1}{\mathrm{vol}\,S_N}\int [\dd \mathcal{L} ] \, e^{{2cN}\, \Tr\,  D_A  \mathcal{U}  D_B \mathcal{U}^{\dagger} }~, \quad\quad\quad ~\mathcal{U}\equiv \mathcal{U}_A^\dagger \mathcal{U}_B~.
\end{equation}
The integral $I(a,b,c)$ evaluates to \cite{ZinnJustin:2002pk}
\begin{align}\label{eq:IZHC}
I(a,b,c)= \mathrm{vol}\,\frac{U(N)}{U(1)^N\times S_N} \frac{G(N+1)}{\left({2cN}\right)^{N(N-1)/2}}\frac{\det\left(e^{{2cN}a_I b_J}\right)_{1\leq I,J\leq N}}{\Delta_N(a)\Delta_N(b)}~,
\end{align}
where $G(N+1)= \prod_{K=1}^{N-1}K!$ denotes Barnes $G$-function. Finally, using the Leibniz formula for the determinant we obtain for a measure $\dd \tilde{\mu}(a,b)$ completely anti-symmetric under $a_I\leftrightarrow a_J$ and $b_I\leftrightarrow b_J$
\begin{align}
&\int\dd \tilde{\mu}(a,b)\det\left(e^{{2cN}a_I b_J}\right)_{1\leq I,J\leq N}=\mathrm{vol}\,S_N \int \dd \tilde{\mu}(a,b)\,e^{{2cN}\sum_{I=1}^N a_I b_I}~.
\end{align}
We can thus write $(\ref{eq:2MIa})$ as
\begin{align}\label{eq:2Mf}
\mathcal{T}_N(c)&=\left(\mathrm{vol}\,\frac{U(N)}{U(1)^N}\right)^2\frac{G(N+1)}{\mathrm{vol}\,S_N}\int \prod^N_{I=1}\dd a_I\dd b_I\,\frac{\Delta_N(a)\Delta_N(b)}{(2cN)^{N(N-1)/2}}\,e^{-N\sum_{I} \left(V(a_I)+ V(b_I) - 2 \, c \, a_I b_I\right)}~,
\end{align}
The above expression resembles the single matrix expression (\ref{eq:partitionfunctioneigenvalues}). However, it has some differences such as the Vandermonde determinant for each collection of eigenvalues appearing with a single power. 
\newline\newline
\textbf{HCIZ integral for $N=2$.} As a simple example, we consider the case $N=2$.
We note that $[d\mathcal{L}]$ is the line element on the sphere
\begin{equation}
\dd s^2=\frac{1}{2}(\dd \theta^2+\sin\theta^2\dd\phi^2)~.
\end{equation}
The HCIZ integral for $N=2$ is thus equal to
\begin{eqnarray}
I(a,b,c) &=& \frac{1}{4}\int_0^{2\pi}\dd\phi\int_0^\pi\dd\theta\sin\theta\,e^{{4c}\,\left(\sin^2\frac{\theta}{2}(a_1b_1+a_2b_2)+\cos^2\frac{\theta}{2}(a_1b_2+a_2b_1)\right)} \cr
&=& \frac{\pi}{4c} \, \frac{1}{\Delta_2(a)\Delta_2(b)} \, \det
\begin{pmatrix}
e^{{4c}\,a_1b_1} & e^{{4c}\,a_1b_2}\\
e^{{4c}\,a_2b_1} &e^{{4c}\,a_2b_2}
\end{pmatrix}~,
\end{eqnarray}
confirming (\ref{eq:IZHC}) for $N=2$. 

\subsection{Orthogonal polynomials $\&$ the quartic polynomial}

Having reached the expression (\ref{eq:2Mf}), we can proceed to use the techniques developed in the previous to solve the integral at large $N$. One approach is to introduce an eigenvalue density for the $a_I$ and $b_I$ eigenvalues and study a generalisation of the loop equations introduced in section \ref{loopsec} for multi-matrix models \cite{Staudacher:1993xy, Lin:2020mme}. Here, we will consider the orthogonal polynomial approach. We take $V(a)$ and $V(b)$ to be the quartic polynomial previously considered in the single matrix case (\ref{eq:resolventex}). Thus, we would like to solve
\begin{equation}
\mathcal{T}_N(\alpha,c)= \mathcal{C}_N\int_{\mathbb{R}^{2N}} \prod_{I=1}^N\dd a_I\, \dd b_I\, \Delta_N(a)\Delta_N(b)\,e^{-N \sum_{I=1}^N W(a_I,b_I)}~,\label{eq:2matrixeigenvalueA}
\end{equation}
with
\begin{equation}\label{quarticW}
W(a,b)= \frac{1}{2} (a^2 + b ^2) + \alpha (a^4 + b^4) - 2 c \, a \, b~, \quad c^2\in (0,1/4)~.
\end{equation}
We have absorbed the volume pre-factors into an $N$-dependent constant $\mathcal{C}_N$. It is worth reiterating that the Vandermonde for each set of eigenvalues only appears linearly in $(\ref{eq:2matrixeigenvalueA})$.

At this stage we introduce a set of monic polynomials $Q_n(x)$, with $x\in \mathbb{R}$, subject to the condition $Q_n(-x)=(-1)^n Q_n(x)$. A priori we should have introduced two different sets of monic polynomials for the eigenvalues of $A$ and $B$ respectively. However due to the symmetry of $W(a,b)$ under interchanging these eigenvalues we can 
choose the same set of monic polynomials. 
Orthogonality of the $Q_n(x)$ is expressed as
\begin{equation}\label{eq:2ortho}
\int_{\mathbb{R}^2} \dd\mu(a,b)\,Q_n(a)Q_m(b) = k_n(\alpha,c)\delta_{mn}~, ~\quad\quad \dd\mu(a,b)\equiv \dd a \,\dd b \, e^{-N W(a,b)}~,
\end{equation}
and the three-term recurrence relation is given by
\begin{equation}
x\, Q_n (x) = Q_{n+1}(x) + R_n \,Q_{n-1}(x) + S_n\, Q_{n-3}(x)~. \label{eq:2mm3tr}
\end{equation}
Notice that for the two-matrix model we have to introduce an additional variable $S_n$. We can only choose a polynomial that has the same parity as $x Q_n(x)$ under $x\leftrightarrow-x$ since we have an even measure. Thus, we exclude $Q_n(x)$ and $Q_{n-2}(x)$. 
Following the same steps leading to $(\ref{eq:hnRn})$ we rewrite $(\ref{eq:2matrixeigenvalueA})$ as
\begin{equation}
\mathcal{T}_N(\alpha,c)=\mathcal{C}_N N! \prod_ {n=0}^{N-1} k_n(\alpha,c) \label{eq:2mminh}~.
\end{equation}
Combining orthogonality and the three-term recurrence relation we obtain, using an integration by parts, three independent equations for the three unknowns $k_n$,  $R_n$ and $S_n$ 
\begin{eqnarray}\label{eq:threeeq}
 \int_{\mathbb{R}^{2}} \dd \mu(a,b)\,Q_{n-1}(a)Q_n(b)  \partial_aW(a,b) &=& 0~, \\ 
 \int_{\mathbb{R}^{2}} \dd \mu(a,b)\, Q_n(a)Q_{n-1}(b) \partial_a W(a,b) &=& \frac{n}{N}\,k_{n-1}(\alpha,c)~,\\ 
 \int_{\mathbb{R}^{2}}\dd\mu(a,b)\,Q_{n-3}(a)Q_n(b) \partial_aW(a,b) &=& 0~.
\end{eqnarray}
Solving the three integrals on the left hand side is straightforward. The only slightly more difficult one is the second integral where we have to apply the three-term recurrence relation multiple times.
It is now convenient to define $z_n(\alpha,c)\equiv 2\alpha k_{n}(\alpha,c)/(ck_{n-1}(\alpha,c))$ for $\alpha\neq 0$ and $z_n(0,c)\equiv 2k_n(0,c)/(c k_{n-1}(0,c))$ for $\alpha =0$. For $\alpha \neq 0$, the above three equations then imply
\begin{align}\label{eq:Scont}
 z_n(\alpha,c)&=4\alpha\, R_n(\alpha,c)\left(1 + 4\alpha\Big(R_{n+1}(\alpha,c) + R_n(\alpha,c) + R_{n-1}(\alpha,c)\Big)\right)^{-1}~,\\ \nonumber
z_n(\alpha,c) &=  \frac{\alpha}{c^2}\,R_n(\alpha,c)\left( 1 + 4\alpha\Big( R_{n+1} (\alpha,c)+ R_n(\alpha,c)+ R_{n-1}(\alpha,c)\Big)\right)~\\  \label{eq:Scont2}
 &+ \frac{4\alpha^2}{c^2}\Big(S_{n+2}(\alpha,c) + S_{n+1}(\alpha,c) + S_n(\alpha,c)\Big)-\frac{\alpha}{c^2}\frac{n}{N}~,
\end{align}
and 
\begin{equation}\label{eq:znSn}
 z_n(\alpha,c) z_{n-1}(\alpha,c) z_{n-2}(\alpha,c) =\frac{4\alpha^2}{c^2} S_n(\alpha,c)~.
\end{equation}
For $\alpha=0$ the three equations become 
\begin{equation}
z_n(0,c)=4R_n(0,c)~,\quad c^2 z_n(0,c)=R_n(0,c)-\frac{n}{N}~,\quad
S_n(0,c) = 0~.
\end{equation}
Upon introducing $\varepsilon=1/N$ and $x=n\varepsilon$, at large $N$ we can express $z_n(\alpha,c)$, $R_n(\alpha,c)$, and $S_n(\alpha,c)$ in terms of continuous functions
\begin{equation}
z_n(\alpha,c) \equiv  z(x,\alpha,c), \quad\quad R_n(\alpha,c) \equiv r(x,\alpha,c), \quad\quad S_{n+1}(\alpha,c)  \equiv s(x,\alpha,c)~. \label{eq:alpha0}
\end{equation}
To leading order in a small $\varepsilon$-expansion the expressions (\ref{eq:Scont}), (\ref{eq:Scont2}), and (\ref{eq:znSn}) are given by
\begin{eqnarray}\label{eq:implf}
z(x,\alpha,c)&=&4\alpha\, r(x,\alpha,c)\left(1+12\alpha\, r(x,\alpha,c)\right)^{-1},\\
z(x,\alpha,c)&=&\frac{\alpha}{c^2}\,r(x,\alpha,c)\big(1+12\alpha \,r(x,\alpha,c)\big)+\frac{12\alpha^2}{c^2}s(x,\alpha,c)-\frac{\alpha x}{c^2}~,\\
z(x,\alpha,c)^3&=&\frac{4\alpha^2}{c^2}\,s(x,\alpha,c)~.
\end{eqnarray} 
Eliminating $r(x,\alpha,c)$ and $s(x,\alpha,c)$ in the above equations we find the following expressions
\begin{equation}
\frac{\omega(z(x,\alpha,c))}{ (1-3z(x,\alpha,c))^2}=0 \quad\quad \text{and} \quad\quad z(x,0,c)= \frac{4x}{(1-4c^2)}~,\label{eq:equationf}
\end{equation}
where we have defined
\begin{equation}\label{eq:quintic}
\omega(z)\equiv -108 c^2 z^5+72 c^2 z^4+24 c^2 z^3 +3(12 \alpha x -8 c^2 )z^2-(24 \alpha x -4 c^2 +1)z+4\alpha x~.
\end{equation}
We have thus boiled down the large $N$ approximation of our problem to solving a quintic polynomial. Generally speaking, quintic polynomials do not admit simple solutions. Nevertheless, for certain special values of $\alpha$ and $c$ the quintic of interest (\ref{eq:quintic}) admits simple solutions. These turn out to  encode a certain critical behaviour in the planar approximation. For instance, when 
\begin{align}\nonumber
\alpha_1 x &= \frac{1}{9}\left(\sqrt{2c}+3c -2c^2\right)~, \quad \alpha_{2}x =-\frac{1}{9}\left(\sqrt{2c}-3 c +2c^2\right)~, \quad \alpha_3 x= -\frac{1}{48}\left(1-\frac{32c^2}{3}\right)~,\\
\alpha_4 x &= \frac{1}{9}\left(\sqrt{-2c}- 3c- 2c^2\right)~, \quad \alpha_5 x = -\frac{1}{9}\left(\sqrt{-2c}+3c +2c^2\right)~.
\end{align}
the discriminant of the quintic (\ref{eq:quintic}) vanishes. Consequently, $\omega(z)$ shares a zero with its derivative $\omega'(z)$. For $\alpha_2 x$  and $\alpha_4 x$ the discriminant of $\omega'(z)$, which now is a polynomial in $c$, vanishes provided $c=\pm 1/8$. A quartic polynomial with a vanishing discriminant contains a multiple root, which for the specific case of $\omega'(z)$ lies at $z=-1/3$. 
  It can be easily checked that (\ref{eq:quintic}) at $c=\pm1/8$ and $\alpha x = -5/288$ can be written as
\begin{align}\label{eq:quinticspecial}
\omega(z)= - \frac{3}{16}\left(z+\frac{1}{3}\right)^3(9z^2-15z+10)~.
\end{align}
To identify the non-analytic structure of interest, it is enough to solve (\ref{eq:equationf}) with $c=\pm 1/8$, in a small $\delta=\left(\alpha x+5/288\right)$ expansion about the $z=-1/3$ solution. To leading order we find 
\begin{equation}\label{eq:ztreelevel}
z(x,\alpha,\pm1/8) = -\frac{1}{3} + \frac{1}{3} \,\sqrt[3]{\frac{5}{6}}\,\sqrt[3]{1+\frac{288}{5}\, \alpha x} + \ldots~.
\end{equation}
The above expansion exhibits the leading non-analyticity we were after. Away from $c=\pm 1/8$ the quintic (\ref{eq:quintic}) has at most a second order zero leading to a different non-analytic structure more akin to that of the single matrix case.
\newline\newline
\textbf{Planar contribution.} As for the study of single matrix integrals, we can apply the orthogonal polynomial technique to evaluate 
$\mathcal{G}_N(\alpha,c)\equiv -\log \mathcal{T}_N(\alpha,c)/\mathcal{T}_N(0,c)$ in a systematic large $N$ expansion. Here $\mathcal{T}_N(\alpha,c)$ is obtained from (\ref{eq:2mminh}) in the same way that (\ref{eq:hnRn}) was obtained for the single matrix integral: 
\begin{align}
\mathcal{T}_{N}(\alpha,c)= \mathcal{C}_NN!\left(\frac{c}{2\alpha}\right)^{N(N-1)/2}k_0^N\,z_1^{N-1}z_{2}^{N-2}\cdots z_{N-1}~.
\end{align}
Performing steps analogous to those leading to (\ref{eq:FEinRn}) and combining the resulting expression for $\mathcal{G}_N(\alpha,c)$ with the Euler-Maclaurin formula $(\ref{eq:eulerMac})$ results in the expression
\begin{equation}\label{eq:F2mm}
\frac{1}{N^2}\,\mathcal{G}_N(\alpha,c)
=-\int_0^1\dd x \,(1-x)\log\frac{z(x,\alpha,c)}{z(x,0,c)}+ \frac{1}{2}\log \alpha +\mathcal{O}\left(N^{-2}\right)~.
\end{equation}
Evaluating $(\ref{eq:F2mm})$ for $c=\pm 1/8$ by using the expansion (\ref{eq:ztreelevel}), we conclude that the non-analytic behaviour is given by 
\begin{equation}\label{2Mcrit}
\lim_{\alpha\to\alpha_c}\partial_\alpha^{(3)}\mathcal{G}^{(0)}_{\mathrm{n.a.}}(\alpha,\pm 1/8)\sim  \left(\alpha-\alpha_c\right)^{-2/3}~,
\end{equation}
where we recall the critical value $\alpha_c=-5/288$. 
\newline\newline
\textbf{Non-planar contributions.} One can continue along these lines and calculate non-planar contributions. As for the single matrix case, one can argue that the non-analytic structure of the non-planar contributions takes the form
\begin{equation}\label{np2m}
\lim_{\alpha\to\alpha_c} \mathcal{G}_{\text{n.a.}}^{(h)} = g_h \left( \alpha-\alpha_c\right)^{7\chi_h/6}~, \quad\quad h \in \mathbb{N}~,
\end{equation}
where the $h=1$ case is understood to be logarithmic. Near criticality, the essential difference \cite{Kazakov:1985ea, Kazakov:1986hu, Staudacher:1989fy} from the single matrix case (\ref{eq:naF}) is that the critical exponent is now $7/6$ instead of $5/4$. 
\newline\newline
\textbf{Double scaling limit.} As for the single matrix mode, one can consider a double scaling limit in which the combination $\kappa^{-1} = N \left(\alpha-\alpha_c \right)^{7/6}$ is kept fixed as $N$ tends to infinity. Though we do not provide the details, it is worth mentioning that the Painlev{\'e} I equation (\ref{diffeq}) uncovered in the double scaling limit of the single matrix model has an avatar for the two-matrix case under consideration. For the curious reader, we state that the analogous equation can be put in the following form \cite{Gross:1989ni, Brezin:1989db, Crnkovic:1989tn}
\begin{equation}\label{painleve2MM}
\delta r(z)^3 -\delta r(z) \delta r''(z) - \frac{1}{2}\delta r'(z)^2 + \frac{2}{27} \,\delta r^{(4)}(z) - z = 0~.
\end{equation}
The assymptotic expansion is given by  \cite{Ginsparg:1991ws} 
\begin{equation}\label{pertsol2MM}
\delta r(z)= z^{1/3}\left(1+\sum_{n=1}^\infty a_k z^{-7n/3}\right)~,
\end{equation}
where the first few coefficients are given by 
\begin{align}\label{eq:expansionnp2mm}
\delta r(z)&= z^{1/3}\Big(1- \frac{1}{18}\,z^{-7/3}- \frac{1925}{26244}\,z^{-14/3}- \frac{509575}{1417176}\, z^{-21/3}- \frac{445712575}{114791256}\, z^{-28/3}+\ldots \Big)~.
\end{align}
As for the single-matrix case, the above expansion admits non-perturbative corrections of the type 
\begin{equation}\label{exponents2MM}
\epsilon(z) = c_1\,z^{-1/4}\,e^{-\frac{18}{7} \, z^{7/6}}+  c_2\, z^{-1/4}\,e^{-\frac{9\sqrt{2}}{7} \, z^{7/6}}~.
\end{equation}
We note that the integration constants can not be fixed by the WKB analysis. The appearance of multiple solutions has been discussed in \cite{Eynard:1992sg}. 

\subsection{A diagrammatic expansion: decorated Riemann surfaces}\label{decorated}

As a final note, we consider how the perturbative expansion in the 't Hooft limit is modified due to the appearance of a second matrix. Taking into account that the quadratic part of the exponent in the integral (\ref{eq:2MI}) contains a mixed term, we have the following set of propagators 
\begin{equation}\label{ABprop}
\langle A_{IK}B_{JL}\rangle=\frac{2c}{N(1-4c^2)}\,\delta_{IL}\delta_{JK}~,\quad \langle A_{IK}A_{JL}\rangle=\langle B_{IK}B_{JL}\rangle= \frac{1}{N(1-4c^2)}\,\delta_{IL}\delta_{JK}~.
\end{equation}
Thus, we have additional structure at each face indicating whether it has edges built from $AA$, $BB$ or $AB$ type propagators 
\begin{figure}[H]
\begin{center}
\begin{tikzpicture}[scale=1]
\draw[->,decorate,decoration={snake,segment length=.4cm}, line width= 0.2mm,magenta] (-3.4,-4) --(-5.2,-4); 
\node[scale= .7] at (-5.55,-3.85) {$A$};
\node[scale= .7] at (-3,-3.88) {$B$~,};
\node[scale= .7] at (-2.25,-3.85) {$A$};
\node[scale= .7] at (-.2,-3.88) {$A$~,};
\node[scale= .7] at (.7,-3.85) {$B$};
\node[scale= .7] at (2.9,-3.85) {$B$~.};
\draw[->,decorate,decoration={snake,segment length=.4cm}, line width= 0.2mm,magenta] (-5.2,-3.7) --(-3.4,-3.7); 
\draw[middlearrow={>},line width=0.15mm] (-.5,-4) --(-2,-4);
\draw[middlearrow={>},line width=0.15mm] (-2,-3.7) --(-.5,-3.7); 
\draw[middlearrow={<},line width=0.15mm] (1,-4) --(2.5,-4); 
\draw[middlearrow={>},line width=0.15mm] (1,-3.7) --(2.5,-3.7); 
\node[scale= .9] at (-4.4,-4.55) {$\sim \frac{2c}{N(1-4c^2)}$};
\node[scale= .9] at (-1.3,-4.55) {$\sim  \frac{1}{N(1-4c^2)}$};
\node[scale= .9] at (1.8,-4.55) {$\sim  \frac{1}{N(1-4c^2)}$};
\end{tikzpicture}
\end{center}
\caption{Various propagators in the two-matrix model.}
\end{figure}

We might envision a continuum limit in which the vertices densely fill the discretised Riemann surface. In such a situation, the additional structure at the vertices may be viewed as an additional field taking two values living on each vertex of the continuous surface. More generally, we could have added an additional coupling weighting the two quartic vertices differently. This would generalise (\ref{quarticW}) to 
\begin{equation}\label{quarticWh}
W(a,b)= \frac{1}{2} (a^2 + b ^2) + \alpha (g \, a^4 + g^{-1} \, b^4) - 2 c \, a \, b~,
\end{equation}
where $g>0$ parametrises the relative weight of the $A$ and $B$ vertices. 
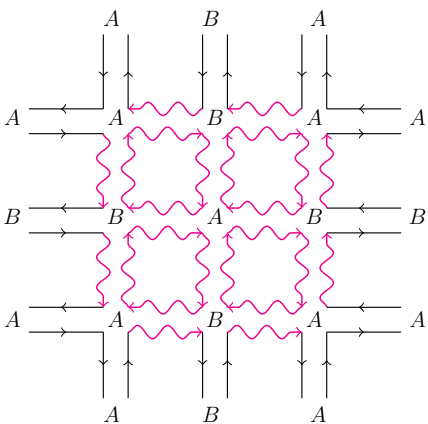
\begin{figure}[H]
\begin{center}
\begin{tikzpicture}[scale=1.1]
\node[scale= .7] at (.4,.15) {$A$};
\node[scale= .7] at (.4,1.4) {$B$};
\node[scale= .7] at (.4,2.6) {$A$};
\node[scale= .7] at (1.65,.15) {$A$};
\node[scale= .7] at (1.65,1.4) {$B$};
\node[scale= .7] at (1.65,2.6) {$A$};
\node[scale= .7] at (2.85,.15) {$B$};
\node[scale= .7] at (2.85,1.4) {$A$};
\node[scale= .7] at (2.85,2.6) {$B$};
\node[scale= .7] at (4.05,.15) {$A$};
\node[scale= .7] at (4.05,1.4) {$B$};
\node[scale= .7] at (4.05,2.6) {$A$};
\node[scale= .7] at (5.3,.15) {$A$};
\node[scale= .7] at (5.3,1.4) {$B$};
\node[scale= .7] at (5.3,2.6) {$A$};
\node[scale= .7] at (1.6,3.8) {$A$};
\node[scale= .7] at (2.8,3.8) {$B$};
\node[scale= .7] at (4.1,3.8) {$A$};
\node[scale= .7] at (1.6,-1) {$A$};
\node[scale= .7] at (2.8,-1) {$B$};
\node[scale= .7] at (4.1,-1) {$A$};
\draw[middlearrow={>},line width=0.15mm] (.6,0) --(1.5,0); 
\draw[middlearrow={<},line width=0.15mm] (.6,.3) --(1.5,.3); 
\draw[->,decorate,decoration={snake,segment length=.4cm}, line width=0.2mm,magenta] (1.5,1.2) --(1.5,.3); 
\draw[->,decorate,decoration={snake,segment length=.4cm}, line width= 0.2mm,magenta] (1.8,.3) --(1.8,1.2); 
\draw[middlearrow={<},line width=0.15mm] (0.6,1.5) --(1.5,1.5); 
\draw[middlearrow={>},line width=0.15mm] (0.6,1.2) --(1.5,1.2); 
\draw[->,decorate,decoration={snake,segment length=.4cm}, line width= 0.2mm,magenta] (1.5,2.4) --(1.5,1.5); 
\draw[->,decorate,decoration={snake,segment length=.4cm}, line width= 0.2mm,magenta] (1.8,1.5) --(1.8,2.4); 
\draw[middlearrow={<},line width=0.15mm] (1.5,2.7) --(1.5,3.6); 
\draw[middlearrow={>},line width=0.15mm] (1.8,2.7) --(1.8,3.6); 
\draw[middlearrow={<},line width=0.15mm] (0.6,2.7) --(1.5,2.7); 
\draw[middlearrow={>},line width=0.15mm] (0.6,2.4) --(1.5,2.4); 
\draw[->,decorate,decoration={snake,segment length=.4cm}, line width= 0.2mm,magenta] (2.7,1.5) --(1.8,1.5); 
\draw[->,decorate,decoration={snake,segment length=.4cm}, line width= 0.2mm,magenta] (1.8,1.2) --(2.7,1.2); 
\draw[->,decorate,decoration={snake,segment length=.4cm}, line width= 0.2mm,magenta] (2.7,2.4) --(2.7,1.5); 
\draw[->,decorate,decoration={snake,segment length=.4cm}, line width=0.2mm,magenta] (3,1.5) --(3,2.4); 
\draw[->,decorate,decoration={snake,segment length=.4cm}, line width= 0.2mm,magenta] ((1.8,2.4) --(2.7,2.4); 
\draw[->,decorate,decoration={snake,segment length=.4cm}, line width= 0.2mm,magenta] (2.7,2.7) --(1.8,2.7); 
\draw[->,decorate,decoration={snake,segment length=.4cm}, line width= 0.2mm,magenta] (1.8,0) --(2.7,0); 
\draw[->,decorate,decoration={snake,segment length=.4cm}, line width= 0.2mm,magenta] (2.7,.3) --(1.8,.3); 
\draw[middlearrow={<},line width=0.15mm] (1.8,0) --(1.8,-.8); 
\draw[middlearrow={>},line width=0.15mm] (1.5,0) --(1.5,-.8); 
\draw[middlearrow={<},line width=0.15mm] (3,0) --(3,-.8); 
\draw[middlearrow={>},line width=0.15mm] (2.7,0) --(2.7,-.8); 
\draw[->,decorate,decoration={snake,segment length=.4cm}, line width= 0.2mm,magenta] (2.7,1.2) --(2.7,.3); 
\draw[->,decorate,decoration={snake,segment length=.4cm}, line width= 0.2mm,magenta] (3,.3) --(3,1.2); 
\draw[->,decorate,decoration={snake,segment length=.4cm}, line width= 0.2mm,magenta] (3,1.2) --(3.9,1.2); 
\draw[->,decorate,decoration={snake,segment length=.4cm}, line width= 0.2mm,magenta] (3.9,1.5) --(3,1.5); 
\draw[->,decorate,decoration={snake,segment length=.4cm}, line width= 0.2mm,magenta] (3.9,.3) --(3,.3); 
\draw[->,decorate,decoration={snake,segment length=.4cm}, line width=0.2mm,magenta] (3,0) --(3.9,0); 
\draw[->,decorate,decoration={snake,segment length=.4cm}, line width= 0.2mm,magenta] (3,2.4) --(3.9,2.4); 
\draw[->,decorate,decoration={snake,segment length=.4cm}, line width= 0.2mm,magenta] (3.9,2.7) --(3,2.7); 
\draw[middlearrow={<},line width=0.15mm] (3.9,2.7) --(3.9,3.6); 
\draw[middlearrow={>},line width=0.15mm] (4.2,2.7) --(4.2,3.6); 
\draw[middlearrow={>},line width=0.15mm] (3,2.7) --(3,3.6); 
\draw[middlearrow={<},line width=0.15mm] (2.7,2.7) --(2.7,3.6); 
\draw[->,decorate,decoration={snake,segment length=.4cm}, line width= 0.2mm,magenta] (3.9,1.2) --(3.9,.3); 
\draw[->,decorate,decoration={snake,segment length=.4cm}, line width=0.2mm,magenta] (4.2,.3) --(4.2,1.2); 
\draw[->,decorate,decoration={snake,segment length=.4cm}, line width= 0.2mm,magenta] (3.9,2.4) --(3.9,1.5); 
\draw[->,decorate,decoration={snake,segment length=.4cm}, line width= 0.2mm,magenta] (4.2,1.5) --(4.2,2.4); 
\draw[middlearrow={>},line width=0.15mm] (3.9,0) --(3.9,-.8); 
\draw[middlearrow={<},line width=0.15mm] (4.2,0) --(4.2,-.8);  
\draw[middlearrow={>},line width=0.15mm] (4.2,0) --(5.1,0); 
\draw[middlearrow={<},line width=0.15mm] (4.2,.3) --(5.1,.3); 
\draw[middlearrow={>},line width=0.15mm] (4.2,1.2) --(5.1,1.2); 
\draw[middlearrow={<},line width=0.15mm] (4.2,1.5) --(5.1,1.5); 
\draw[middlearrow={>},line width=0.15mm] (4.2,2.4) --(5.1,2.4); 
\draw[middlearrow={<},line width=0.15mm] (4.2,2.7) --(5.1,2.7); 
\end{tikzpicture}
\end{center}
\caption{Piece of a planar diagram in the two-matrix model.}
\end{figure}
At this stage, we have no immediate reason to suspect that such a two-dimensional field theory is itself local. Nevertheless, and somewhat remarkably, we will discuss evidence for this in section \ref{LiouvilleSec} where we will also provide an interpretation for the three couplings in (\ref{quarticWh}), namely $\alpha$, $c$, and $g$.
\begin{center} *** \end{center}
Adding more matrices further decorates the Feynman diagrams. One might imagine a limit where we have an infinite chain of matrices such that they effectively carry a continuous label $t$. We will now consider precisely such a situation. 

%

\section{Quantum mechanical matrices$^{\color{magenta} F}$}\label{sec5}

In the previous sections, we considered certain classes of integrals in the limit of a large number of variables. In this section we will consider the quantum mechanical generalisation of those integrals over a single matrix. The original $\mathbb{C}$-number elements of our $N\times N$ Hermitian matrices will be promoted to operators acting on a Hilbert space, and the matrix integral will be naturally replaced by a matrix path integral. Our most important goal in this section will be to characterise the ground state for such systems. 

\subsection{Action and Hamiltonian}

We begin by introducing the following class of classical actions
\begin{equation}\label{action}
S_N[M(t)] = N \, \Tr \int \dd t \left(  \frac{1}{2} \dot{M}(t)^2 - V(M(t)) \right)~.
\end{equation}
Here, $V(M(t))$ is a polynomial potential governing the $N\times N$ Hermitian matrix valued path $M(t)$, and the dot denotes a derivative with respect to time. We can view the above as a class of classical theories comprising $N^2$  degrees of freedom whose dynamical features are governed by the potential $V(M(t))$. Quantum mechanical transition amplitudes are given by the Feynman path integral
\begin{equation} \label{eq:MQMpartf}
\mathcal{A}_N(M_f,M_i)= \int \mathcal{D} M(t) \, e^{\frac{i}{\hbar} S_N[M(t)]}~,
\end{equation}
with boundary conditions $M(t_i)=M_i$ and $M(t_f)=M_f$, and our measure is now given by
\begin{equation}
\mathcal{D}M(t) \equiv \prod_{t\in \mathbb{R}} [\dd M(t) ]~.
\end{equation}
The path integral (\ref{eq:MQMpartf}) serves as the quantum mechanical generalisation of the matrix integrals previously explored. 

As for the ordinary matrix integrals, we can exploit the global $U(N)$ symmetry of $(\ref{action})$ by parametrising $M(t)$ as 
\begin{align} \label{eq:MQMdiagonalize}
M(t)\rightarrow \mathcal{U}(t)\,D_M(t)\,\mathcal{U}(t)^\dagger~, \quad\quad D_M(t)= \mathrm{diag}(\lambda_1(t),\lambda_2(t),...,\lambda_N(t))~.
\end{align}
Here, $\mathcal{U}(t)$ is a time dependent element of $U(N)/U(1)^N$, and the $\lambda_I(t)$ are the real valued time dependent eigenvalues of $M(t)$.
Under $(\ref{eq:MQMdiagonalize})$ the kinetic term in the action $(\ref{action})$ is expressed as
\begin{equation} \label{eq:infMt}
\Tr \,\dot{M}(t)^2
= \Tr\, \dot{D}_M(t)^2 + \Tr [D_M(t), \dot{\mathcal{U}}(t)\mathcal{U}(t)^\dagger]^2~. 
\end{equation}
We note that $\dot{\mathcal{U}}(t)\,\mathcal{U}(t)^\dagger$ is again a Hermitian matrix which can be expressed as 
\begin{align}\label{eq:decomp}
\dot{\mathcal{U}}(t)\,\mathcal{U}(t)^\dagger=  \frac{i}{\sqrt{2}}\sum_{a=1}^{N(N-1)/2} \left(T^{S}_{a} \, \dot{\beta}_{a}(t) + {T}^A_{a} \, \dot{\gamma}_{a}(t) \right)~,
\end{align}
where $T^{S}_{a}$ and ${T}^A_{a}$ are a basis of real symmetric, and pure imaginary antisymmetric $N\times N$ generators of $U(N)/U(1)^N$, with normalisation $\text{Tr} \,T^\epsilon_{a} T^{\epsilon'}_{b} = \delta_{ab}\,\delta^{\epsilon\epsilon'}$ with $\epsilon,\epsilon' \in \{S,A\}$.
 Thus, 
\begin{equation}\label{eq:infMt}
\Tr \,\dot{M}^2
= \Tr\,\dot{D}_M^2 + \frac{1}{2}\sum_{I <J}(\lambda_I-\lambda_J)^2(\dot{\beta}_{IJ}^2+\dot{\gamma}_{IJ}^2)~,
\end{equation}
where for notational convenience we suppressed the explicit time dependence, $\beta_{IJ}$ are the matrix elements of $\sum_a T^{S}_{a} \, \dot{\beta}_{a}$ and analogously $\gamma_{IJ}$ the matrix elements of  $\sum_a T^{A}_{a} \, \dot{\gamma}_{a}$.

Turning to the Hamiltonian formalism, we calculate the canonical momenta for $\lambda$, $\beta_{IJ}$ and $\gamma_{IJ}$ respectively:
 \begin{equation} \label{eq:MQMcanmom}
 \pi_{\small{I}} = N \dot{\lambda}_I, \quad\quad  \pi_{\small{IJ}} = \frac{N}{2} (\lambda_I - \lambda_J)^2\, \dot{\beta}_{IJ}~,  \quad\quad
\tilde{\pi}_{\small{IJ}} =  \frac{N}{2}\,   (\lambda_I - \lambda_J)^2\, \dot{\gamma}_{IJ}~.
 \end{equation}
From these, we can construct the classical Hamiltonian 
 \begin{align} \label{eq:classH}
H = \sum_{I=1}^N\left(\frac{1}{2 N}\,\pi_I^2 + N V(\lambda_I)\right) + \frac{1}{N}\sum_{I<J} \frac{\pi_{IJ}^2 + \tilde{\pi}_{IJ}^2}{(\lambda_I - \lambda_J)^2}~.
 \end{align}
At the classical level, the lowest energy configuration lies at the minimum of $V(\lambda_I)$ with all classical momenta vanishing. As we shall soon see, the quantum mechanical ground state differs considerably from its classical counterpart. 

 \subsection{Quantisation $\&$ free fermions}

It is straightforward to promote $(\ref{eq:classH})$ to a quantum Hamiltonian. Recall that in the eigenvalue basis (\ref{eq:MQMdiagonalize}), the $N^2$ degrees of freedom are living on a curved space. Moreover, this curved space is merely a coordinate transformation of the flat metric on $\mathbb{R}^{N^2}$. Consequently, the quantum kinetic term is nothing other than the Laplace-Beltrami differential operator associated to our curved space.  
The quantum version of the Hamiltonian in $(\ref{eq:classH})$ is consequently given by
 \begin{equation} \label{eq:MQMqH}
\hat{H} = \sum_{I=1}^N \left(  -\frac{1}{2N}\frac{1}{\Delta_N(\lambda)}\frac{\partial^2}{\partial \lambda_I^2}\,\Delta_N(\lambda) + N V(\lambda_I) \right) +  \frac{1}{N}\sum_{I<J} \frac{\hat{\pi}_{IJ}^2 + \hat{\tilde{\pi}}_{IJ}^2}{(\lambda_I - \lambda_J)^2}~,
 \end{equation}
where we have explicitly written the eigenvalue dependence for the Laplacian, we are working in units for which $\hbar=1$ and $\Delta_N(\lambda)$ is the determinant of the Vandermonde matrix introduced in (\ref{eq:vandermondedet}). 
The angular part in $(\ref{eq:MQMqH})$ describes the motion on the compact coset space $U(N)/U(1)^N$.\footnote{It is worth mentioning that for certain values of $\hat{\pi}_{IJ}$ and $\hat{\tilde{\pi}}_{IJ}$ the above Hamiltonian is of the form of a Calogero model  \cite{Kazhdan1978,Polychronakos:2006nz}.}
Due to the compactness of the coset space, the eigenfunctions of $H$ can be annihilated by $\hat{\pi}_{IJ}$ and $\hat{\tilde{\pi}}_{IJ}$ whilst retaining normalisability. Thus, it is natural to study wavefunctions $\Psi(\lambda_I)$ that are functions only of the eigenvalues, but not the coordinates on the compact coset space $U(N)/U(1)^N$. 
Acting on such wavefunctions, the Hamiltonian is reduced to
\begin{equation}
\hat{H}= \sum_{I=1}^N  \left( - \frac{1}{2 N} \frac{1}{\Delta_N(\lambda)}\frac{\partial^2}{\partial \lambda_I^2}\,\Delta_N(\lambda) + N V(\lambda_I) \right)~. \label{eq:matrixH}
  \end{equation}
A subsequent redefinition of the wavefunctions
 \begin{align}\label{reducedPsi}
 \Psi(\lambda_1,\ldots,\lambda_N)= \frac{\psi(\lambda_1,\ldots,\lambda_N)}{\Delta_N(\lambda)}~,
 \end{align}
leads to a Hamiltonian with standard kinetic term. Thus, we have reduced our eigenvalue problem to the following single particle problem
 \begin{align} \label{eq:decH}
 \left(-\frac{1}{2N}\frac{\partial^2}{\partial \lambda^2}+ N V(\lambda)\right)\psi_{\varepsilon_n}(\lambda) = N \varepsilon_n \psi_{\varepsilon_n}(\lambda)~, \quad\quad n = 1,2,\ldots, N~.
 \end{align}
Recall that $\Delta_N(\lambda)$ is antisymmetric under exchange of any two eigenvalues $\lambda_I$. It then follows from (\ref{reducedPsi}) that a general state $\Psi$ invariant under the permutation subgroup $S_N \subset U(N)$ enforces the rescaled wavefunction $\psi$ to be an {\it anti-symmetric} function of the $\lambda_I$. In particular, the problem has been reduced to solving a system of $N$ free particles subject to the Pauli exclusion principle \cite{Brezin:1977sv}. It is already clear at this stage that the quantum mechanical state, particularly at large $N$, differs substantially from its classical counterpart. In other words, quantum effects play a dominant role in the description of the system. 

The ground state of the matrix quantum mechanical systems is described by $N$ free fermions, each governed by the potential $V(\lambda)$.  
The fermions fill up the first $N$ energy levels, all the way up to the Fermi level $\varepsilon_F\equiv \varepsilon_N$. We will be interested in  small energy excitations or ripples above the filled Fermi sea. Rather generally, in the large $N$ limit, the low energy physics near a filled Fermi sea has a linear dispersion relation in two-dimensions \cite{Polchinski:1992ed}.

\subsection{Non-analyticity $\&$ the inverted harmonic oscillator}

We now assume that the potential has a maximum at some value of $\lambda$. Recall that in our study of ordinary matrix integrals, we uncovered non-analytic behaviour for particular choices of the polynomial $V(\lambda)$. For instance, for $V_\alpha(\lambda)=\lambda^2/2+\alpha \lambda^4$ we uncovered non-analytic behaviour for a negative value of $\alpha=\alpha_c<0$. For negative values of $\alpha$, $V_\alpha(\lambda)$ contains a maximum. Thus, in tuning the value of $\alpha$ beyond $\alpha_c$, the confining effect of the quadratic term in the Lagrangian will no longer be strong enough to compete with the repulsion of eigenvalues from both the Vandermonde and  the quartic contribution to $V_\alpha(\lambda)$. Recall further that for the matrix integrals the non-analyticities were important in obtaining a continuum limit with a divergent number of vertices in the discretised Riemann surface.

It is natural, then, to ask what kind of non-analyticities might we expect from the free fermions stemming from quantum mechanical matrices? As a simple test observable, we can consider the Fermi energy for a potential given by 
\begin{equation}
V(\lambda) = \frac{1}{2} \gamma^2 \lambda^2 - \alpha \lambda^4~.
\end{equation}
We have maxima at $\lambda_\pm = \pm \gamma \left(2\sqrt{\alpha}\right)^{-1}$, such that $V(\lambda_\pm) = \gamma^4 (16 \alpha)^{-1}$. Given a fixed number, $N$, of fermions we can consider how the Fermi energy $\varepsilon_F$ behaves as a function of the dimensionless parameter $\tilde{\alpha} \equiv \alpha \, \gamma^{-3}$. As we vary $\tilde{\alpha}$, the Fermi level will begin to approach the maximum of the potential. In the large $N$ limit we can resort to a semi-classical approximation. This approximation can be cast in the form of a Bohr-Sommerfeld condition 
\begin{equation}
\int_{0}^{\lambda_+} \dd\lambda \sqrt{\varepsilon_F-V(\lambda)} = \frac{\pi}{2\sqrt{2}} \left(1 + \frac{1}{2N} \right)~.
\end{equation}
Examining the above integral reveals that upon tuning $\varepsilon_F$ to the maximum value of the potential gives rise to non-analytic behaviour. This stems from the part of the integral near the maxima. The same non-analyticity is captured by the following, simpler, integral which we evaluate in a small $\delta\varepsilon_F$-expansion \cite{Brezin:1989ss, Parisi:1989dka,Gross:1990ay, Ginsparg:1990as}: 
\begin{equation}
 \int_0^{\lambda_+} \dd\lambda \sqrt{\delta\varepsilon_F - V''(\lambda_+)(\lambda-\lambda_+)^2/2} =  -\frac{\delta\varepsilon_F}{4\gamma}\,\log \tilde{\alpha} \frac{\delta\varepsilon_F}{\gamma}+ \,\text{analytic}~.
\end{equation} 
We should view $-\delta\varepsilon_F$ as the difference between the Fermi level and the maximum value of the potential. So long as the Fermi level is close to the top of the potential the detailed features of the potential will not affect the preceding analysis.

For the sake of generality, it is perhaps worth pointing out that tuning the potential to have a non-quadratic maximum would affect the non-analytic behaviour. For instance
\begin{equation}
 \int_0^{\lambda_+} \dd\lambda \sqrt{\delta\varepsilon_F + (\lambda-\lambda_+)^m} = \sqrt{\delta \varepsilon_F } \, \lambda_+ \times {_2F_1}\left(-\frac{1}{2};\frac{1}{m};1+\frac{1}{m};-\frac{ (-\lambda_+)^m}{\delta \varepsilon_F }\right)~,
\end{equation}
from which we extract a non-analytic behaviour of the type $\sim \delta \varepsilon_F^{1/2+1/m}$ for $m\neq2$ \cite{Parisi:1989dka, Ginsparg:1990as}. In what follows we focus on the $m=2$ case.

\subsection{A scattering problem}

A final (albeit somewhat a posteriori) motivation for assuming a local maximum is that we will eventually compare calculations for the matrix quantum mechanical theory to a certain worldsheet string theory. The natural observable from the worldsheet perspective is an $S$-matrix. In the presence of a local maximum, the low energy excitations naturally encode an $S$-matrix. This is most evident, if we consider the physics localised near the maximum, where the model reduces to $N$ fermions in an inverse harmonic well. Let us introduce the following dimensionless quantities
\begin{equation}\label{rescalingQM}
x = 2^{3/4}N^{1/2}\, \left( \lambda-\lambda_+ \right)\,\gamma^{1/2} ~, \quad\quad\quad \nu = \frac{N}{2^{1/2}} \, \left(\frac{\varepsilon_F + \delta \varepsilon -V(\lambda_+)}{\gamma}\right)~,
\end{equation}
and consider excitations with energy near $\varepsilon$ such that at large $N$ we have $\delta \varepsilon/\varepsilon_F \ll 1$. 
The single particle Schr\"odinger equation (\ref{eq:decH}) near the maximum of the potential now becomes
\begin{equation}\label{doublescaling}
\left(- \partial_x^2 - \frac{x^2}{4} \right) \psi_{\nu,p} = \nu \, \psi_{\nu,p}~.
\end{equation}
Note further that the rescaling (\ref{rescalingQM}) involves $N$ in such a way that the range of $x$ and $\nu$ becomes effectively infinite as $N$ tends to infinity. The parity index in $\psi_{\nu,p}$, $p\in\{\pm\}$, reflects the behaviour of the wavefunction under the discrete symmetry $x\to-x$, namely $\psi_{\nu,p}(-x) =  p \, \psi_{\nu,p}(x)$.  

When considering the scattering problem, we take the Fermi level $\nu_F$ of our new Schr{\"o}dinger problem (\ref{doublescaling}) to be negative, such that we can send in an incoming wave from negative $x$, let it bounce against the wall, and compute the reflected wavefunction. The Schr\"odinger equation (\ref{doublescaling}) admits solutions whose explicit form is given by the parabolic cylinder functions $D_a(z)$. Explicitly, 
\begin{equation}\label{solnsWE}
\Psi_{\nu,\pm}(x) = e^{-i\pi(1/2-i \nu)/4}  \left( D_{ i (\nu + i/2)}\left(e^{\frac{3\pi i}{4}} x\right)\pm D_{ i (\nu + i/2)}\left(e^{ -\frac{\pi i}{4}} x\right) \right)~.
\end{equation}
It will also prove convenient to express the above wavefunctions in a basis of incoming/outgoing waves near large and negative values of $x$. These can be expressed as follows
\begin{equation}
\psi^{(out)}_{\nu}(x) =\frac{1}{2}\left( \Psi_{\nu,+}(x) + \Psi_{\nu,-}(x)\right)~, \quad\quad \psi^{(in)}_{\nu}(x) = \left(\psi^{(out)}_{\nu}(x) \right)^*~.
\end{equation}
The outgoing wavefunctions admit an asymptotic expansion of the following form
\begin{equation}\label{asy}
\lim_{x\to-\infty} \psi^{(out)}_{\nu}(x) =   e^{i x^2/4} (-x)^{-1/2+i\nu} + \ldots~.
\end{equation}
Notice that in addition to the energy, $\nu$ parameterises the wavelength of the spatial oscillations. It is easiest to see this by considering a coordinate transformation to the spatial coordinate $x = -e^{-u}$.
The Wronskian for the above wavefunctions is given by
\begin{equation}\label{wronskian}
W(\nu) = \psi^{(out)}_{\nu}(x) \partial_x \psi^{(out)}_{\nu}(-x) - \psi^{(out)}_{\nu}(-x) \partial_x \psi^{(out)}_{\nu}(x) =  i\frac{\sqrt{2\pi}}{\Gamma\left(1/2-i\nu \right)} \, e^{-\pi\nu/2}~.
\end{equation}
As expected, $W(\nu)$ is independent of $x$. We can also express the parity eigenstates in terms of $\psi^{(out)}(x)$ as follows
\begin{equation}\label{eq;paritystates}
\psi_{\nu,+}(x) \equiv \frac{1}{\sqrt{2}}\,\frac{\psi^{(out)}_{\nu}(x) + \psi^{(out)}_{\nu}(-x)}{W(\nu)}~, \quad\quad \psi_{\nu,-}(x) \equiv \frac{1}{\sqrt{2}}\,\frac{\psi^{(out)}_{\nu}(x) - \psi^{(out)}_{\nu}(-x)}{W(\nu)^*}~,
\end{equation}
where we have rescaled (\ref{solnsWE}) with respect to the Wronskian. 

By virtue of the discrete symmetry $x\leftrightarrow-x$, $\psi_\nu^{(out)}(-x)$ and its complex conjugate also describe solutions to (\ref{doublescaling}). These are waves that are purely incoming or outgoing on the other side of the potential barrier.

\subsection{Releasing the particle number}\label{secondQ}

So far we kept the particle number $N$ fixed. It will be convenient to also allow the particle number to vary, thus constructing a second quantised picture of the quantum mechanical theory. To do so, we  introduce lowering and raising operators, $\hat{a}_{\nu,p}$ and $\hat{a}^\dagger_{\nu,p}$ 
satisfying the standard anti-commutation relations
\begin{equation}\label{AC}
\{\hat{a}_{\nu,p}, \hat{a}_{\nu',p'} \} =0~, \quad\quad \{\hat{a}_{\nu,p}, \hat{a}^\dagger_{\nu',p'} \} =2\pi\,\delta_{\nu,\nu'} \delta_{p,p'}~.
\end{equation}
The ground state is filled until the Fermi level and we introduce the state $|\text{Fermi}\rangle$ corresponding to a filled Fermi sea by
\begin{eqnarray}\nonumber
\hat{a}_{\nu,p} \, |\text{Fermi}\rangle &=& 0~, \quad\quad \nu > \nu_F~\quad\quad \& \quad\quad p \in \{ \pm \}~, \\ \label{eq:raisingloweringop}
\hat{a}^\dag_{\nu,p} \, |\text{Fermi}\rangle &=& 0~, \quad\quad \nu  < \nu_F~\quad\quad \& \quad\quad p \in \{ \pm \}~,
\end{eqnarray}
where we recall that $\nu_F<0$. The above Fermi sea is filled on both sides of the potential.\footnote{One can also consider filling one of the two sides, or perhaps filling the two sides with Fermi seas with different levels. The case where both sides are filled equally was considered in \cite{Takayanagi:2003sm,Douglas:2003up}.} 
We can create multi-particle energy eigenstates on top of $|\text{Fermi}\rangle$ by acting with $\hat{a}^\dag_{\nu,p}$ with $\nu>\nu_F$, and hole type excitations by acting with $\hat{a}_{\nu,p}$ with $\nu < \nu_F$.

We can thus introduce field operators for the creation and annihilation of wavefunctions at $x$:
\begin{equation}\label{eq:PsiPsidFS}
\hat{\Psi}_p(t,x)= \int_{\mathbb{R}} \frac{\dd \nu}{2\pi} \,  e^{-i\nu t} \, \hat{a}_{\nu,p} \, \psi^*_{\nu,p}(x)~, \quad\quad \hat{\Psi}^\dagger_p(t,x) = \int_{\mathbb{R}} \frac{\dd \nu}{2\pi} \,e^{i\nu t}\, \hat{a}^\dag_{\nu,p}  \, \psi_{\nu,p}(x)~,
\end{equation}
where the integration is over the continuous energy levels $\nu$ and $\psi_{\nu,p}^*(x)$ and $\psi_{\nu,p}(x)$ are the parity eigenstates introduced in (\ref{eq;paritystates}). In the expression above $*$ denotes the complex conjugation of $\mathbb{C}$-numbers, while $\dagger$ denotes the complex conjugation of quantum operators. Given the above fermionic operators, we can define a number density operator
\begin{equation}\label{density}
\hat{n}(t,x) = \hat{\Psi}^\dag(t,x)\hat{\Psi}(t,x)~.
\end{equation}
In the above, we have defined $\hat{\Psi}(t,x) \equiv \hat{\Psi}_+(t,x) + \hat{\Psi}_-(t,x)$ and its conjugate, satisfying 
\begin{equation}
\{ \hat{\Psi}(t,x), \hat{\Psi}^\dag(t,x')  \} = 2\,\delta(x-x')~.
\end{equation}
The operator $\hat{\Psi}^\dag(t,x)$ creates states that can be perceived as single particle states on top of the filled Fermi sea when viewed near the left boundary at large negative values of $x$. 

\subsubsection{Feynman propagator}

Given a second quantised theory we can construct various propagators for the fermions. For instance, the Feynman propagator is given by 
\begin{equation}\label{Fpropagator}
S_F(t,x;t',x') =   - i \,\langle \text{Fermi}| \, T _F \, \hat{\Psi}(t,x) \hat{\Psi}^\dagger(t',x')  |\text{Fermi}\rangle~,
\end{equation}
where $T_F$ denotes Fermionic time-ordering. It satisfies 
\begin{equation}\label{Seqn}
\left(-\partial_{x}^2 - \frac{x^2}{4} - i \partial_{t} \right)  S_F(t,x;t',x') = \delta(t-t')\delta(x-x')~.
\end{equation}
More explicitly, we can express $S_F(t,x;t',x')$ as
\begin{align}\label{SFexplicit}
i\,S_F(t,x;t',x') =  \Theta(s)  \int_{\nu_F}^{\infty} \frac{\dd\nu}{2\pi}\,e^{-i\nu s}\, \psi^*_{\nu,p}(x) \psi_{\nu,p}(x')-  \Theta(-s)   \int_{-\infty}^{\nu_F}  \frac{\dd\nu}{2\pi}\,e^{-i\nu s}\, \psi^*_{\nu,p}(x) \psi_{\nu,p}(x')~,
\end{align}
where we sum over the parity index, and for notational simplicity we have defined $s\equiv t-t'$. 
We can also write down the Fourier transform of the Fermion propagator with respect to $s$. For this we use the contour integral representation of the Heaviside function 
\begin{align}
\Theta(s)=\lim_{\varepsilon\rightarrow 0^+}\,\int_{\mathbb{R}} \frac{\dd \Omega}{2\pi i} \frac{e^{i\Omega s}}{\Omega - i\varepsilon}~.
\end{align}
The Fourier transform then reads
\begin{align}\label{eq:Lehmann}
\tilde{S}_F(x,x';\omega)
&= -\lim_{\varepsilon\rightarrow 0^+}\,\int_{\mathbb{R}} \frac{\dd\nu}{2\pi}\, \psi^*_{\nu,p}(x) \psi_{\nu,p}(x')\left(~\frac{\Theta(\nu-\nu_F)}{\nu-\omega-i \varepsilon} + \frac{\Theta(\nu_F-\nu)}{\nu-\omega+i \varepsilon}~\right)~,
\end{align}
and we use the following conventions for the Fourier transform:
\begin{align}
f(t)= \int_{\mathbb{R}}\frac{\dd \omega}{2\pi}\,e^{-i\omega t}\tilde{f}(\omega)\quad\quad \& \quad \quad \tilde{f}(\omega)= \int_{\mathbb{R}}\dd t\, e^{+i\omega t}\, f(t)~.
\end{align}
Complex conjugation at the level of the Fourier transform mirrors itself in $x\leftrightarrow x'$ and a particle hole exchange. 
Finally, we note that for large negative values of $x$ and $x'$ the parity states behave as
\begin{multline}\label{eq:FeynmanFS}
\lim_{x,x'\rightarrow -\infty}\left(\psi_{+,\nu}^*(x)\psi_{+,\nu}(x')+\psi_{-,\nu}^*(x)\psi_{-,\nu}(x') \right) = \\ \frac{1}{\sqrt{xx'}}\left(e^{-i(x^2-x'^2)/4-i\nu \log x/x'}+R_\nu^*\,e^{-i(x^2+x'^2)/4-i\nu \log xx'} \right)+ \text{h.c.}~,
\end{multline}
where we defined 
\begin{equation}\label{eq:Rpurephase}
R_\nu \equiv- i\,\frac{\Gamma(1/2-i\nu)}{\sqrt{2\pi}} \,e^{-\pi\nu/2}~.
\end{equation}
We will provide a physical interpretation of the  definition of $R_\nu$ in the next section. 
For large negative values of $x$ and $x'$, with $x>x'$, we then find the asymptotic form for (\ref{eq:FeynmanFS}):
\begin{align}
\tilde{S}_F(x,x';\omega)&\approx
-\frac{i}{\sqrt{xx'}}\left(e^{i(x^2-x'^2)/4 + i\omega\log x/x'}+R_{\omega}\,e^{i(x^2+x'^2)/4+ i\omega\log xx'}\right)\Theta(\omega-\nu_F)\cr
&+\frac{i}{\sqrt{xx'}}\left(e^{-i(x^2-x'^2)/4- i\omega\log x/x' }+R^*_{\omega}\,e^{-i(x^2+x'^2)/4- i\omega\log xx'}\right)\Theta(\nu_F-\omega)\,.\label{eq:SFFourier}
\end{align}
For $x'>x$, both large and negative we find
\begin{align}
\tilde{S}_F(x,x';\omega)&\approx
-\frac{i}{\sqrt{xx'}}\left(e^{-i(x^2-x'^2)/4- i\omega\log x/x' }+R_{\omega}\,e^{i(x^2+x'^2)/4+ i\omega\log xx'}\right)\Theta(\omega-\nu_F)\cr
&+\frac{i}{\sqrt{xx'}}\left(e^{i(x^2-x'^2)/4+ i\omega\log x/x'}+R^*_{\omega}\,e^{-i(x^2+x'^2)/4- i\omega\log xx'}\right)\Theta(\nu_F-\omega)~.\label{eq:SFFourier2}
\end{align}
We will now put all these expressions to good use, as they have provided us the basic building blocks to construct an $S$-matrix for our quantum mechanical matrices. 


\section{Scattering from quantum mechanical matrices$^{\color{magenta}F}$}\label{sec6}

As briefly mentioned in the previous section, one can setup an $S$-matrix problem for the fermions propagating near the filled Fermi sea. These excitations reflect off the region near the maximum of the potential. In this section, we provide some details for the scattering amplitude introduced in \cite{Moore:1991sf, Moore:1991zv}
of the number density operator $\hat{n}(t,x)$ defined in (\ref{density}). These are simple observables in our theory. 

\subsection{Reflection coefficient \& Green's function}\label{secRcGf}
\begin{figure}[H]\label{reflection}
\centering
\begin{tikzpicture}[scale=0.8]
\begin{axis}[
    axis lines = none,
    xlabel = $x$,
    ylabel = {$f(x)$},
]
\addplot [name path=A,
    domain=-8:8, 
    samples=100, 
    color=black,line width=.3mm
]
{-x^2 };
\draw[->,decorate,decoration={snake,segment length=.5cm},magenta, line width= 0.3mm] (3.5,-8) --(7.8,-8); 
\draw[->,decorate,decoration={snake,segment length=.5cm},orange, line width= 0.3mm] (-8.6,-11) --(-3.5,-11); 
\draw[->,decorate,decoration={snake,segment length=.5cm},teal, line width= 0.3mm] (-3.5,-6) --(-8.6,-6); 
\addplot [name path=L,NavyBlue, domain=-8:-3.95,line width=.3mm] {-x^2};
\addplot[name path=E, NavyBlue,line width=.2mm] coordinates {(-4,-16.2) (-10,-16.2)};
\addplot[opacity=0, line width=.2mm] coordinates {(-4,-16.2) (-10,-16)};
\addplot[name path=E, NavyBlue,line width=.2mm] coordinates {(-10,-16.2) (-10,-64)};
\node[orange] at (-10, -11.5)   {$I_\nu$ };
\node[teal] at (-10, -3.5)   {$R_\nu$ };
\node[magenta] at (8.9, -8)   {$T_\nu$ };
\node[NavyBlue] at (-2.4, -18)   {${\small{\nu_F}}$ };
\addplot[RoyalBlue!50] fill between[of=L and E];
\end{axis}
\end{tikzpicture}
\caption{We consider an incoming wave from negative infinity (orange), part of which is reflected (teal), and part of which is transmitted (magenta) from the inverted harmonic oscillator potential. The ground state configuration fills the Fermi sea up to $\nu_F$.}
\label{inreftrans}
\end{figure}
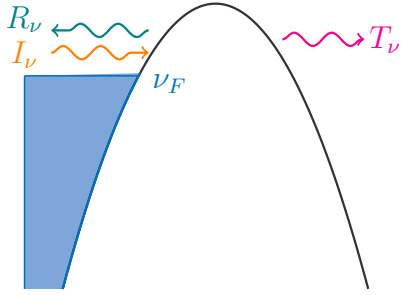

Recall the Hamiltonian (\ref{doublescaling}) we obtained at the end of the previous section
\begin{equation}\label{Hamiltonian}
\hat{H} = -\partial_x^2 - \frac{x^2}{4}~.
\end{equation}
A natural object to consider is given by the reflection coefficient, $R_\nu$, of a wave of frequency $\nu$ coming in from the asymptotic region at large and negative values of $x$ and reflecting back, as depicted in figure \ref{inreftrans}. 
\newline\newline

As our boundary condition we impose that the incoming flux from large positive values of $x$ vanishes. In this regard, it is convenient to note that $\psi_\nu^{(out)}(-x)$
is a wave that is purely outgoing for large positive values of $x$. Expanding for large negative values of $x$ we find a superposition of an incoming and reflected wave 
\begin{equation}\label{eq:inref}
\frac{\psi^{(out)}_\nu(-x)}{W(\nu)} \approx  \frac{e^{-ix^2/4-i\nu\log(-x)}}{\sqrt{-x}}-i\, \frac{\Gamma(1/2-i\nu)}{\sqrt{-2\pi x}} e^{-\pi\nu/2} e^{ix^2/4+i\nu\log(-x)}  + \ldots~,
\end{equation}
where we have normalised the wavefunction such that the incoming (first) part has unit coefficient.
To obtain the reflection and transmission coefficient we calculate the probability current
\begin{align}\label{eq:propcurrent}
\mathcal{J}(t,x) =-i\left( \psi^* \partial_x\psi-\psi \partial_x\psi^*\right)~, \quad\quad \partial_t \rho + \partial_x \mathcal{J} = 0~.
\end{align}
From the probability current one can readily obtain the reflection and transmission coefficients
\begin{align}\label{eq:RT}
|R_\nu|^2=\frac{|\mathcal{J}_{ref|}}{|\mathcal{J}_{in}|}= \frac{1}{1+e^{2\pi\nu}}~,\quad |T_\nu|^2=\frac{|\mathcal{J}_{{trans}}|}{|\mathcal{J}_{{in}}|}= \frac{1}{1+ e^{-2\pi\nu}} = 1-|R_\nu|^2~,
\end{align}
where we obtain $\mathcal{J}_{{in}}$ from $(\ref{eq:propcurrent})$ by using the purely incoming part of $(\ref{eq:inref})$ and similarly the purely reflected and transmitted wave for $\mathcal{J}_{{ref}}$ and $\mathcal{J}_{{trans}}$ respectively. For negative values of $\nu$, which is the situation we are most interested in, the transmission coefficient is exponentially suppressed. In fact, the expression $(\ref{eq:inref})$ encodes more information. For instance, it encodes the reflection coefficient
\begin{equation}\label{eq:Rpurephase}
R_\nu =- i\,\frac{\Gamma(1/2-i\nu)}{\sqrt{2\pi}} \,e^{-\pi\nu/2}~,
\end{equation}
which can be viewed as an $S$-matrix element for the scattering of a single wave against the potential barrier. Perturbative unitarity of the $S$-matrix is given by the fact that $R_\nu$ is a pure phase up to $e^{-\pi\nu/2}$ corrections. 
Using the Gamma function identity 
\begin{align}
|\Gamma(1/2-i\nu)|^2=\frac{\pi}{\cosh(\pi\nu)}~,
\end{align}
the absolute value of $(\ref{eq:Rpurephase})$ reduces to $(\ref{eq:RT})$.

The Green's function is given by evaluating the matrix elements of the resolvent
\begin{equation}
G(x,x';z) = \langle x'| \left({\hat{H}-z}\right)^{-1}|x\rangle~, \quad\quad \left( G(x,x';z) \right)^* = G(x',x;z^*)~,
\end{equation}
where $z$ takes values across the whole complex plane, modulo the spectrum of $\hat{H}$. The Green's function satisfies the equation
\begin{equation}
\left(-\partial_x^2 - \frac{x^2}{4} -z\right) G(x,x';z) = \delta(x-x')~.
\end{equation}
It is often convenient to express $G(x,x';z)$ in terms of eigenfunctions of $\hat{H}$ with complex eigenvalue $z$. Concretely, 
\begin{align}\label{propagator}
G(x,x';z)= \frac{1}{W(z)}\left(\psi^{R}_z(x)\psi^{L}_z(x')\Theta(x'-x)+\psi^{R}_z(x')\psi^{L}_z(x)\Theta(x-x')\right)~,
\end{align}
where $\psi^{L}_z(x)$ decays at $x=-\infty$, and $\psi^{R}_z(x)$ decays at $x=\infty$, and $W(z)$ is the Wronskian (\ref{wronskian}). As an example, when $z$ takes values in the upper half of the complex plane we can express $\psi^{L,R}_z(x)$ as 
\begin{equation}\label{eq:wavefunctionsuph}
\psi^L_z(x) = \psi_z^{(out)}(x)~, \quad\quad \psi^R_z(x) =  \psi_z^{(out)}(-x)~.
\end{equation} 
For large negative and large positive values of $x$ these wave functions scale as
\begin{align}
\lim_{x\rightarrow -\infty}\psi^L_z(x) =  \frac{1}{\sqrt{-x}}\,e^{ix^2/4+iz\log(-x)}\qquad \mathrm{and}\qquad \lim_{x\rightarrow \infty}\psi^R_z(x) = \frac{1}{\sqrt{x}}\,e^{ix^2/4+iz\log x}~.
\end{align}
Using the expressions for the wavefunctions in $(\ref{eq:inref})$, we can obtain an asymptotic expansion of $G(x,x';z)$ for large and negative values of $x$ and $x'$. For $z$ in the upper half plane we obtain
\begin{align}\label{eq:GreensMQM}
G(x,x';z) \approx &-\frac{i}{\sqrt{xx'}}\left(e^{-i(x^2-x'^2)/4- iz\log x/x' }+R_{z}\,e^{i(x^2+x'^2)/4+ i z\log xx'}\right)\Theta(x'-x)\cr
&-\frac{i}{\sqrt{xx'}}\left(e^{i(x^2-x'^2)/4+ iz\log x/x'}+R_{z}\,e^{i(x^2+x'^2)/4+ i z\log xx'}\right)\Theta(x-x')~.
\end{align}
A similar expression can be derived for $z$ in the lower half plane. 
The reflection coefficient $R_\nu$ given in ($\ref{eq:Rpurephase}$) and the above Green's function are the basic building blocks for the scattering problem of interest -- namely, the scattering of the probability density $\rho(t,x) = \psi^*(t,x)\psi(t,x)$ about the filled Fermi sea. We note that $\rho(t,x)$ is a quantum mechanical analogue of the eigenvalue density $\rho(\lambda)$ introduced in (\ref{eq:CH}). 


As a final remark, before embarking on the calculation of various scattering processes we would like to relate the Green's function introduced in (\ref{propagator}) to the Feynman propagator (\ref{SFexplicit}) discussed in the previous section. These obey the same equation. Comparing the first line in (\ref{eq:SFFourier}) with the lower line of the Green's function in (\ref{eq:GreensMQM}), we note that the two expressions take the same form. The difference lies in the Heaviside function appearing in $\tilde{S}_F(x,x';\omega)$. This is a simple manifestation of the fact that $G(x,x';z)$ stems from a first quantised setup, whereas $\tilde{S}_F(x,x';\omega)$ stems from a `second quantised' picture allowing for particle creation and annihilation.

\subsection{Multi-particle scattering}\label{multiparticlesection}

To ensure a large range of admissible scattering frequencies we take $\nu_F^2 \gg1$. This condition will also ensure that the transmitted wave is exponentially suppressed, such that we can aptly refer to scattering processes leaking across the barrier as non-perturbative phenomena. The physical problem we are after consists of sending in an incoming density wave $\rho_\omega(x)$ at some incoming frequency $\omega$ near the spatial boundary at large and negative values of $x$, and measuring the scattered wave at some outgoing frequency $\omega'$, again at large negative values of $x$. Since we are scattering a composite object, the elements of our perturbative $S$-matrix will be given by certain convolutions of the reflection coefficients $R_{\nu}$. In what follows, we will often express results in terms of the spatial coordinate 
\begin{equation}
u = -\log(-x)~, \quad\quad x<0~.
\end{equation}
Notice that for large negative values of $x$, $u$ is also large and negative and $u$ increases monotonically with $x$. 

To compute scattering processes for multiple incoming and outgoing waves\footnote{The $S$-matrix in a two-dimensional world is a rather subtle object due to the absence of an infinitely large celestial sphere at null infinity. This implies we cannot have parametrically separated asymptotic states. Instead, in two-dimensions the celestial sphere collapses to a celestial point. 
} it is convenient to resort to the multi-particle picture described in section \ref{secondQ}.
We begin by considering the $n$-point function 
\begin{equation}\label{eq:qcf}
\mathcal{Q}\left( \{ t_k, x_k \} \right) =  \langle \text{Fermi} |T_B    \prod_{k=1}^{n} \hat{n}(t_k, x_k) | \text{Fermi} \rangle~,
\end{equation}
where now $T_B$ indicates bosonic time ordering and we defined $\hat{n}(t_k,x_k)$ in (\ref{density}). Given that our theory is non-interacting, we can calculate $\mathcal{Q}$ via multiple Wick contractions using (\ref{SFexplicit}). We note that the above correlation function is invariant under $(t_i,x_i)\leftrightarrow (t_j,x_j)$.
\newline\newline
\textbf{General approach.} Before delving into concrete examples, it is worth explaining the general strategy. Starting with the quantum correlation function (\ref{eq:qcf}) we perform all possible Wick contractions. Keeping the fully connected part  we obtain 
\begin{equation}\label{eq:qcf1tn}
\mathcal{Q}_c\left(\{t_k,x_k\}\right)=-\frac{i^n}{n}\sum_{\sigma\in \mathrm{Perm}(n)}\prod_{k=1}^n S_F(t_{\sigma_k},x_{\sigma_k};t_{\sigma_{k+1}},x_{\sigma_{k+1}})~,
\end{equation}
where  $\mathrm{Perm}(n)$ denotes the set of order $n!$ permuting the elements in $\{1,2,\ldots ,n\}$. We also compute indices modulo $n$ (e.g. $\sigma_{n+1}\equiv \sigma_1$ and so on). The subscript $c$ is a reminder that we are considering the fully connected part. Notice that (\ref{eq:qcf1tn}) is invariant under $(t_i,x_i)\leftrightarrow (t_j,x_j)$. Going to Fourier space the above leads to an expression manifestly invariant under $(\omega_i,x_i)\leftrightarrow (\omega_j,x_j)$. From this expression we extract the fully connected part $\tilde{\mathcal{Q}}_c(\{\omega_i,x_i\})$ and obtain the one-to-$n$ scattering amplitude $S(\omega_1,\dots,\omega_{n-1}|\omega_n)$ defined as the part of the correlation function taking the form
\begin{align}
\delta\left(\omega_1+\dots+\omega_n\right)\, \exp \left( {i\omega_n u_n-i \sum\limits_{k=1}^{n-1}\omega_ku_k} \right) \, S(\omega_1,\dots,\omega_{n-1}|\omega_n)~.\label{eq:amplitudegeneral}
\end{align}
Outgoing frequencies $\omega_k$ are positive for $k=1,\ldots,n-1$ and as a result, the frequency of the incoming wave $\omega_n$ must be negative.
Since the original correlation function (\ref{eq:qcf}) is invariant under $(t_i,x_i)\leftrightarrow (t_j,x_j)$ we impose a specific ordering of the spatial coordinates, namely $x_1>x_2 >\ldots >x_n$, to avoid overcounting.  For $m>$1 we can similarly define the $m$-to-$n$ amplitude. We now proceed to study some concrete examples. 
\newline\newline
\textbf{One-to-one scattering.} As a warm up exercise, we compute the one-to-one scattering problem in the second quantised picture. We wish to calculate the Fourier transform of the connected part of
\begin{equation}\label{eq:qcf1t1}
\mathcal{Q}\left( \{ t_i, x_i \}\right) = \langle \text{Fermi} |\,T_B\, \hat{n}(t_1, x_1) \, \hat{n}(t_2, x_2)  | \text{Fermi} \rangle~,
\end{equation}
where now $T_B$ corresponds to bosonic time ordering. 
Keeping track of the various Wick contractions 
and using the Fermion propagator (\ref{SFexplicit}), we find 
\begin{equation}\label{eq:con1to1}
\mathcal{Q}_c( t_1, x_1;t_2,x_2) 
=\frac{1}{2} \sum_{\sigma\in \mathrm{Perm}(2)}\prod_{k=1}^2S_F(t_{\sigma_k},x_{\sigma_k};t_{\sigma_{k+1}},x_{\sigma_{k+1}})~.
\end{equation}
In appendix \ref{Wickcont} we provide some steps leading to (\ref{eq:con1to1}). 
Going to Fourier space we obtain 
\begin{equation}
\tilde{\mathcal{Q}}_c(\omega_1,x_1;\omega_2,x_2) =\int_{\mathbb{R}^2} \dd t_1 \dd t_2 \, e^{+i\omega_1 t_1+i\omega_2 t_2} \mathcal{Q}_c( t_1, x_1;t_2,x_2)~.
\end{equation}
Performing the integrals over $t_1$ and $t_2$, we are led to the following expression
\begin{multline}\label{eq:1to2qcf}
\tilde{\mathcal{Q}}_c(\omega_1,x_1;\omega_2,x_2)= 
\frac{2\pi}{2}\delta(\omega_1+\omega_2) \times
\\ \int_\mathbb{R} \frac{\dd \upsilon}{2\pi} \left(\tilde{S}_F(x_1,x_2;\upsilon +\omega_1)\tilde{S}_F(x_2,x_1;\upsilon)+ \tilde{S}_F(x_2,x_1;\upsilon+\omega_2)\tilde{S}_F(x_1,x_2;\upsilon)\right)\, .
\end{multline}
We can expand the above expression at large and negative values of $x_1$ and $x_2$. There are various pieces in this regime. Of these, we would like to keep those terms corresponding to particles that have been scattered by the barrier, and moreover do not exhibit large oscillations of the type $\sim e^{\pm i x^2/4}$. Expressing the result in terms of the spatial coordinate $u = -\log(-x)$ we find
\begin{align}
\tilde{\mathcal{Q}}_c(\omega_1,u_1;\omega_2,u_2)= 2\pi {\delta(\omega_1+\omega_2)}\left(\omega_2\,\frac{e^{i\omega_1u_1 -i\omega_2 u_2}}{2\pi}+ \frac{e^{-i\omega_1u_1 +i\omega_2 u_2}}{2\pi}\,S(\omega_1 |\omega_2)\right)~.
\end{align}
and $S(\omega_1|\omega_2)$ is defined as
\begin{equation}\label{eq:Sm1to1}
S(\omega_1 |\omega_2)\equiv e^{-i\omega_2\log(-\nu_F)}\,\int_0^{\omega_1} \dd \upsilon \, R^*_{\nu_F-\upsilon}R_{\nu_F-\upsilon-\omega_2}~.
\end{equation}
The phase $e^{-i\omega_2\log(-\nu_F)}$ subtracts the phase appearing in the perturbative expansion of the reflection coefficient (\ref{eq:pertR}). (One can also remove this phase by a redefinition of the $u$-coordinate.) Notice that the frequency of $R_\nu$ is always above the Fermi level since it corresponds to particle scattering, whereas the frequency of $R^*_\nu$ is always below the Fermi level since it corresponds to the scattering of a hole. 
Performing a large $\nu_F$ expansion, we obtain 
\begin{equation}\label{onetoone}
S(\omega_1 |\omega_2)
=\omega_1+\frac{1}{24\nu_F^2}\left(i\omega_1^2 -\omega_1^4(\omega_1-2i) \right)+ \mathcal{O}\left(\nu_F^{-4}\right)~.
\end{equation}
It is worth noting that the leading piece of $S(\omega_1|\omega_2)$, including the energy conserving delta-function, is the one-to-one scattering amplitude we would have obtained for a theory of massless fields in two-spacetime dimensions in a theory subject to Poincar{\'e} invariance. The terms subleading in $1/\nu_F$ are corrections which indicate any such Poincar{\'e} invariance is ultimately broken. We will comment on this from the perspective of a continuum description in the next section.
Finally dimensional analysis may raise concerns about the above expression. We must remember, however, that in going to the Hamiltonian (\ref{Hamiltonian}) we have set several dimensionful parameters to unity. Upon restoring them, we can also restore our faith in dimensional analysis.
\newline\newline
\textbf{One-to-two scattering.} We now move on to the case of one-to-two scattering. This scattering amplitude encodes the type of interactions present in the theory.  
The correlation function of interest is
\begin{equation}
\mathcal{Q}\left( \{ t_i, x_i \}\right) = \langle \text{Fermi} | \,T_B\,\hat{n}(t_1,x_1) \hat{n}(t_2,x_2) \hat{n}(t_3,x_3)  | \text{Fermi} \rangle~,
\end{equation}
The resulting expression for the connected part reads
\begin{equation}
\mathcal{Q}_c\left( \{ t_i, x_i \}\right)
=\frac{i}{3}\sum_{\sigma\in \mathrm{Perm}(3)}\prod_{k=1}^3 S_F(t_{\sigma_k},x_{\sigma_k};t_{\sigma_{k+1}},x_{\sigma_{k+1}})~.
\end{equation}
Fourier transforming with respect to the time coordinates, we obtain 
\begin{equation}\label{eq:qcf1t3}
\tilde{\mathcal{Q}}_c(\omega_1,x_1;\omega_2,x_2;\omega_3,x_3) =\int_{\mathbb{R}^3} \dd t_1 \dd t_2 \dd t_3\, e^{i\omega_1 t_1+i\omega_2 t_2+i\omega_3 t_3} \mathcal{Q}_c( t_1, x_1;t_2,x_2; t_3,x_3)~.
\end{equation}
Due to time-translation invariance, we obtain an overall delta function imposing the conservation of energy. Combining (\ref{eq:amplitudegeneral}) with (\ref{eq:qcf1t3})
we obtain
\begin{align}\label{eq:1to2final}
S(\omega_1,\omega_2 |\,\omega_3)
&\equiv e^{-i\omega_3\log(-\nu_F)}\,\left[ \int_{0}^{\omega_1}\dd \upsilon \, R^*_{\nu_F-\upsilon}R_{\nu_F-\upsilon-\omega_3}-\int_{\omega_2}^{\omega_1+\omega_2} \dd\upsilon \, R^*_{\nu_F-\upsilon}R_{\nu_F-\upsilon-\omega_3}~\right]~.
\end{align}
We can express the above expression in a large $\nu_F$ expansion using (\ref{eq:pertR})
\begin{multline}\label{eq:1to2final}
S(\omega_1,\omega_2 |\,\omega_3)=\frac{i}{\nu_F} \, \omega_1 \omega_2\omega_3  -\frac{i}{24\nu_F^3}\omega_1\omega _2 \omega _3 \left(\omega_3+i\right) \left(\omega_3+2 i\right)  \\ 
\times \left(\omega _1 \left(\omega _1-i\right)+\omega _2 \left(\omega _2-i\right)+1\right)+\mathcal{O}\left(\nu_F^{-5}\right)\,.
\end{multline}
Note that the alternating signs arise from the alternating signs in (\ref{eq:SFFourier}) and we take the frequencies of the scattered waves $\omega_1, \omega_2 >0$. 
 \newline\newline
\textbf{One-to-three scattering.} In direct analogy to the one-to-one and one-to-two case, we can work out the one-to-three scattering amplitude. After obtaining the possible Wick contractions of 
\begin{equation}
\mathcal{Q}\left( \{ t_k, x_k \}\right)  = \langle \text{Fermi} | \,T_B \, \hat{n}(t_1,x_1) \hat{n}(t_2,x_2)\hat{n}(t_3,x_3) \hat{n}(t_4,x_4)  | \text{Fermi} \rangle~,
\end{equation}
we can express the connected piece as
\begin{align}\label{eq:qcf1to3}
\mathcal{Q}_c\left( \{ t_k, x_k \}\right)=-\frac{1}{4}\sum_{\sigma\in \mathrm{Perm}(4)}\prod_{k=1}^4 S_F(t_{\sigma_k},x_{\sigma_k};t_{\sigma_{k+1}},x_{\sigma_{k+1}})~.
\end{align}
Fourier transforming, extracting the term without the $e^{\pm i x^2/4}$ oscillations,
and using (\ref{eq:amplitudegeneral}) we end up with
\begin{multline}
S(\omega_1,\omega_2,\omega_3\,|\,\omega_4)\, \equiv e^{-i\omega_4\log(-\nu_F)}\,\Bigg[\int_{\omega_2+\omega_3}^{\omega_1+\omega_2+\omega_3}\dd \upsilon\, R_{\nu_F-\upsilon}^*R_{\nu_F-\upsilon-\omega_4}-\int_{\omega_2}^{\omega_1+\omega_2}\dd \upsilon\, R_{\nu_F-\upsilon}^*R_{\nu_F-\upsilon-\omega_4}\cr
-\int_{\omega_3}^{\omega_1+\omega_3}\dd \upsilon\,  R_{\nu_F-\upsilon}^*R_{\nu_F-\upsilon-\omega_4}+\int_{0}^{\omega_1}\dd \upsilon\, R^*_{\nu_F-\upsilon}R_{\nu_F-\upsilon-\omega_4}\Bigg]~.
\end{multline}
Note that there is an overall minus sign arising from our original definition of the scattering amplitude (\ref{eq:amplitudegeneral}) and (\ref{eq:qcf1to3}).
Expanding the above expression at large $\nu_F$ by making use of (\ref{eq:pertR}) we find
\begin{multline}
S(\omega_1,\omega_2,\omega_3\,|\,\omega_4)=-\frac{1}{\nu_F^2}\,\omega_1\omega_2\omega_3\omega_4 (\omega_4+i)  +\frac{1}{24\nu_F^4}\,\omega_1\omega _2 \omega _3 \omega _4\left(\omega_4 +i\right) \left(\omega_4+2i\right) \left(\omega_4+3 i\right) \\ \times \big(\omega _1 \left(\omega _1-i\right)
+\omega _2
   \left(\omega _2-i\right)+\omega _3 \left(\omega _3-i\right)+1\big)+ \mathcal{O}\left(\nu_F^{-6}\right)~.
\end{multline}
{\textbf{One-to-$(n-1)$ scattering.}} From the above scattering amplitudes we observe 
\begin{equation}
S(\omega_1,\ldots, \omega_{n-1}|\,\omega_n)= \sum_{L=0}^{\infty}\nu_F^{n-1+2L}S^{(L)}(\omega_1,\ldots,\omega_{n-1}|\,\omega_n)~, 
\end{equation}
where for the case of one-to-(n-1) scattering the left hand side is given by \cite{Moore:1991zv}
\begin{equation}
S(\omega_1,\ldots, \omega_{n-1}|\,\omega_n)\equiv e^{-i\omega_n\log(-\nu_F)}\,\sum_{\Omega\, \subseteq \{\omega_1,..,\,\omega_{n-1}\}}\,(-1)^{|\Omega|+1}\,\int_0^{\omega(\Omega)}\dd \upsilon\, R^*_{\nu_F-\upsilon}R_{\nu_F-\upsilon-\omega_n}~.
\end{equation}
where $\omega(\Omega)$ is the sum of all elements in $\Omega$.
Furthermore, for the leading term in a large $\nu_F$ expansion of the one-to-$(n-1)$ scattering amplitude, we observe
\begin{align}
S^{(0)}(\omega_1,\ldots,\omega_{n-1}|\,\omega_n)= -i^n\,\omega_1\cdots \omega_{n-1}\prod_{k=0}^{n-3}\, (\omega_n+  i k)~, \quad\quad n\geq 3~.
\end{align}
For the the first sub-leading term and $n\geq 3$ we find 
\begin{equation}
S^{(1)}(\omega_1,\ldots,\omega_{n-1}|\,\omega_n)=(i)^n\,\frac{1}{24}\,\omega_1\cdots\omega_{n-1}\prod_{k=0}^{n-1}\,(\omega_n+ k i)\left(\sum_{\ell=1}^{n-1}\omega_\ell(\omega_\ell-i)+1\right)~, 
\end{equation}
where $\omega_n=-\sum_{k=1}^{n-1}\omega_k$. For both expressions we made use of the large $\nu_F$ expansion (\ref{eq:pertR}). In appendix \ref{higherscat} we obtain some higher-point scattering amplitudes that confirm the above expressions, at least for low enough particle number. 
\newline\newline
{\textbf{Two-to-two scattering.}} 
To obtain the two-to-two amplitude, instead of looking for terms of the form (\ref{eq:amplitudegeneral}) we must extract the term proportional to
\begin{equation}\label{eq:2to2}
e^{-i\omega_1u_1-i\omega_2u_2+i\omega_3 u_3+i\omega_4 u_4}\,\delta(\omega_1+\omega_2+\omega_3+\omega_4)\, S(\omega_1,\omega_2\,|\,\omega_3,\omega_4)~
\end{equation}
from the connected part of the four point correlation function (\ref{eq:qcf1to3}). This defines for us the two-to-two amplitude $S(\omega_1,\omega_2\,|\,\omega_3,\omega_4)$.
In the above expression we assume that  $\omega_1$ and $\omega_2$ are the positive frequencies of the outgoing waves, while $\omega_3$ and $\omega_4$ are the negative frequencies of the two incoming waves.  Additionally we have to order the frequencies. We choose
\begin{equation}\label{eq:assump}
\omega_2>\omega_1~, \quad \omega_4>\omega_3~, \quad \omega_2>-\omega_3~, \quad \omega_2>-\omega_4~,
\end{equation}
such that $\omega_2= \mathrm{max}\{\omega_1,\omega_2, |\omega_3|,|\omega_4|\}$. Combining with (\ref{eq:2to2}) and (\ref{eq:qcf1to3}) we obtain \cite{Moore:1991zv, Mandal:1991ua, Polchinski:1991uq}
\allowdisplaybreaks
\begin{align}
&S(\omega_1,\omega_2 |\omega_3,\omega_4)\equiv e^{-i(\omega_3+\omega_4)\log(-\nu_F)}\Bigg[\int_{\omega_3}^{\omega_1+\omega_3}\dd \upsilon R_{\nu_F-\upsilon+\omega_3}^*R_{\nu_F-\upsilon-\omega_4}\cr
&+\int_{\omega_2}^{\omega_1+\omega_2}\dd \upsilon R_{\nu_F-\upsilon}^*R_{\nu_F-\upsilon+\omega_1+\omega_2}
-\int_{\omega_3}^{\omega_1+\omega_3}\dd \upsilon R_{\nu_F-\upsilon}R_{\nu_F-\upsilon+\omega_3}^*R_{\nu_F-\upsilon+\omega_1+\omega_3}R_{\nu_F-\upsilon-\omega_2}^*\cr
&-\int_{\omega_4}^{\omega_1+\omega_4}\dd \upsilon R_{\nu_F-\upsilon}R_{\nu_F-\upsilon+\omega_4}^*R_{\nu_F-\upsilon+\omega_1+\omega_4}R_{\nu_F-\upsilon-\omega_2}^*\Bigg]\cr
&=-\frac{1}{\nu_F^2}\omega_1\omega_2\omega_3\omega_4(\omega_2-i)+ \mathcal{O}(\nu_F^{-4})~,
\end{align}
as the two-to-two scattering amplitude.
\newline\newline
{\textbf{Two-to-three scattering.}}
To obtain the two-to-three amplitude we consider the connected correlation function (\ref{eq:connected5point}).  We are then looking for terms in proportional to
\begin{align}\label{eq:2to3}
 e^{-i\omega_1u_1-i\omega_2u_2-i\omega_3u_3+i\omega_4u_4+i\omega_5 u_5}\delta(\omega_1+\omega_2+\omega_3+\omega_4+\omega_5)\,S(\omega_1,\omega_2,\omega_3\,|\, \omega_4,\omega_5)
\end{align}
with $\omega_1, \omega_2, \omega_3$ outgoing and positive and $\omega_4, \omega_5$ incoming negative. The above expression defines for us the scattering amplitude for two-to-three scattering. Additionally we assume 
\begin{equation}\label{eq:assump2to3}
\omega_3>\omega_2>\omega_1~,\quad  \omega_5>\omega_4~, \quad  \omega_3>-\omega_4~, \quad \omega_2<-\omega_5~, \quad \omega_1<-\omega_5~,
\end{equation}
and $x_1>x_2>x_3>x_4>x_5$.
In particular (\ref{eq:assump2to3}) implies that $\omega_3= \mathrm{max}\{\omega_1,\omega_2,\omega_3,|\omega_4|,|\omega_5|\}$. 
We then find : 
\allowdisplaybreaks
\begin{align}
&S(\omega_1,\omega_2,\omega_3\,|\, \omega_4,\omega_5)\equiv e^{-i(\omega_4+\omega_5)\log(-\nu_F)}\Bigg[\int_{0}^{\omega_1}\dd \upsilon \,R_{\nu_F-\upsilon-\omega_2-\omega_4} R_{\nu_F-\upsilon+\omega_1+\omega_5}^*R_{\nu_F-\upsilon+\omega_1}R_{\nu_F-\upsilon-\omega_2}^*\cr
-&\int_{\omega_4}^{\omega_1+\omega_4}\dd \upsilon \, R_{\nu_F-\upsilon-\omega_3-\omega_5}R_{\nu_F-\upsilon+\omega_4}^*R_{\nu_F-\upsilon}R_{\nu_F-\upsilon-\omega_3}^*-\int_{\omega_2+\omega_3}^{\omega_1+\omega_2+\omega_3}\dd \upsilon \, R_{\nu_F-\upsilon-\omega_4-\omega_5}R_{\nu_F-\upsilon}^*\cr
-&\int_{\omega_5}^{\omega_1+\omega_5}\dd \upsilon \, R_{\nu_F-\upsilon-\omega_3-\omega_4}R_{\nu_F-\upsilon+\omega_5}^*R_{\nu_F-\upsilon}R_{\nu_F-\upsilon-\omega_3}^*+\int_{\omega_3}^{\omega_1+\omega_3}\dd \upsilon\, R_{\nu_F-\upsilon-\omega_4-\omega_5}R_{\nu_F-\upsilon}^*\cr
+&\int_{0}^{\omega_1}\dd \upsilon \,R_{\nu_F-\upsilon-\omega_2-\omega_5} R_{\nu_F-\upsilon+\omega_1+\omega_4}^*R_{\nu_F-\upsilon+\omega_1}R_{\nu_F-\upsilon-\omega_2}^*+\int_{0}^{\omega_1}\dd \upsilon\, R_{\nu_F-\upsilon-\omega_4-\omega_5}R_{\nu_F-\upsilon}^*\cr
-&\int_{\omega_2}^{\omega_1+\omega_2}\dd \upsilon \,R_{\nu_F-\upsilon-\omega_4-\omega_5}R_{\nu_F-\upsilon}^*\Bigg]= \frac{i}{\nu_F^3}\omega_1\omega_2\omega_3\omega_4\omega_5(\omega_3-i)(\omega_1+\omega_2+\omega_3-2i)+ \mathcal{O}\left(\nu_F^{-5}\right)~.
\end{align}
We note that any two-to-$n$ amplitude contains factors of either two or four reflection coefficients. 

In such a way we can continue to compute higher amplitudes from our basic building blocks. We thus observe the emergence of a perturbative $S$-matrix admitting a small $1/\nu_F$ expansion. 
So far, we obtained the $S$-matrix directly from the large $N$ quantum mechanical matrix model. It is natural to ask whether the same $S$-matrix can be obtained directly from some weakly coupled system. Before turning to this we summarise the results of the $(n-m)$-to-$m$ scattering amplitudes $S(\omega_1,\dots,\omega_m\,|\,\omega_{m+1},\dots,\omega_n)$ calculated in this section and appendix \ref{higherscat}. 
\begingroup
\renewcommand{\arraystretch}{2}
\begin{table}[H]
\begin{center}
\begin{tabular}{ cV{1.8}c | c| c|} 
$*$ & $S(\omega_1\,|\,\omega_2)$ & $S(\omega_1,\omega_2\,|\,\omega_3)$ & $S(\omega_1,\omega_2,\omega_3\,|\,\omega_4)$ \\ \Xhline{3\arrayrulewidth}
$\nu_F^{0}$ &$\omega_1$ & 0 & 0  \\ \hline
 $\nu_F^{-1}$ &0 & $i\omega_1\omega_2\omega_3$ & 0  \\ \hline
  $\nu_F^{-2}$ & $\frac{1}{24}\left(i\omega_1^2 -\omega_1^4(\omega_1-2i) \right)$ &0  & $-\omega_1\omega_2\omega_3\omega_4(\omega_4+i)$ \\ \hline
  $\nu_F^{-3}$ & 0  &  \makecell{$-\frac{i}{24}\,\omega_1\omega_{2}\prod_{k=0}^{2}\,(\omega_3+ k i)$\\ $\times \left(\sum_{\ell=1}^{2}\omega_\ell(\omega_\ell-i)+1\right)$} & 0 \\ \hline
  $\nu_F^{-4}$ & x & 0 &  \makecell{$\frac{1}{24}\,\omega_1\omega_2\omega_{3}\prod_{k=0}^{3}\,(\omega_4+ k i)$\\ $\times \left(\sum_{\ell=1}^{3}\omega_\ell(\omega_\ell-i)+1\right)$} \\ \hline
  \multicolumn{1}{c:}{} & \multicolumn{1}{c:}{0} & \multicolumn{1}{c:}{x} & \multicolumn{1}{c:}{0}  \\ 
            \multicolumn{1}{c:}{} & \multicolumn{1}{c:}{x} & \multicolumn{1}{c:}{0} & \multicolumn{1}{c:}{x}  \\ 
\end{tabular}
\end{center}
\caption{Summary of the scattering amplitudes calculated in this section and appendix \ref{higherscat}. Each box contains the contribution to the scattering amplitude at the inverse power of $\nu_F$ in the leftmost box. An x indicates a nonzero result easily obtained by expanding the reflection coefficients to higher order in $1/\nu_F$. Within $S(\omega_1,\dots,\omega_{n-1}\,|\,\omega_n)$,  $\omega_n$ denotes the incoming (negative) frequency, whereas $\{\omega_1,\dots,\omega_{n-1}\}$ denote the $n-1$ outgoing (positive) frequencies.}
\end{table}
\endgroup

\begingroup
\renewcommand{\arraystretch}{2}
\begin{table}[H]
\begin{center}
\begin{tabular}{ cV{1.8}c | c | c| c|} 
 $*$& $S(\omega_1,\dots,\omega_4\,|\,\omega_5)$  &$S(\omega_1,\dots,\omega_5\,|\,\omega_6)$ & $S(\omega_1,\dots,\omega_6\,|\,\omega_7)$  \\ \Xhline{3\arrayrulewidth}
   \multicolumn{1}{c:}{} & \multicolumn{1}{c:}{0} & \multicolumn{1}{c:}{0} & \multicolumn{1}{c:}{0}  \\ \hline
$\nu_F^{-3}$ & \makecell{$-i\omega_1\omega_2\omega_3\omega_4\omega_5$\\ $\times (\omega_5+i)(\omega_5+2i)$} & 0 & 0  \\ \hline
 $\nu_F^{-4}$ &0 & \makecell{$\omega_1\cdots\omega_6$\\ $\times (\omega_6+i)(\omega_6+2i)(\omega_6+3i)$}  & 0  \\ \hline
  $\nu_F^{-5}$ & \makecell{$\frac{i}{24}\,\omega_1\cdots\omega_{4}\prod_{k=0}^{4}\,(\omega_5+ k i)$\\ $\times\left(\sum_{\ell=1}^{4}\omega_\ell(\omega_\ell-i)+1\right)$} & 0 &  \makecell{$i\omega_1\cdots\omega_7$\\ $\times \prod_{k=0}^4(\omega_7+ki)$}    \\ \hline
  $\nu_F^{-6}$ & 0  &  \makecell{$-\frac{1}{24}\,\omega_1\cdots\omega_{5}\prod_{k=0}^{5}\,(\omega_6+ k i)$\\ $\times\left(\sum_{\ell=1}^{5}\omega_\ell(\omega_\ell-i)+1\right)$} & 0 \\ \hline
  $\nu_F^{-7}$ & x & 0 &   \makecell{$-\frac{i}{24}\,\omega_1\cdots\omega_{6}\prod_{k=0}^{6}\,(\omega_7+ k i)$\\ $\times\left(\sum_{\ell=1}^{6}\omega_\ell(\omega_\ell-i)+1\right)$} \\ \hline
     \multicolumn{1}{c:}{} & \multicolumn{1}{c:}{0} & \multicolumn{1}{c:}{x} & \multicolumn{1}{c:}{0}  \\ 
          \multicolumn{1}{c:}{} & \multicolumn{1}{c:}{x} & \multicolumn{1}{c:}{0} & \multicolumn{1}{c:}{x}  \\ 
\end{tabular}
\end{center}
\caption{Summary of the scattering amplitudes calculated in this section and appendix \ref{higherscat}.  Each box contains the contribution to the scattering amplitude at the inverse power of $\nu_F$ in the leftmost box. An x indicates a nonzero result easily obtained by expanding the reflection coefficients to higher order in $1/\nu_F$.  Within $S(\omega_1,\dots,\omega_{n-1}\,|\,\omega_n)$,  $\omega_n$ denotes the incoming (negative) frequency, whereas $\{\omega_1,\dots,\omega_{n-1}\}$ denote the $n-1$ outgoing (positive) frequencies.}
\end{table}
\endgroup

\begingroup
\renewcommand{\arraystretch}{2}
\begin{table}[H]
\begin{center}
\begin{tabular}{ cV{1.8}c | c | c|} 
$*$ & $S(\omega_1,\omega_2\,|\,\omega_3,\omega_4)$    & $S(\omega_1,\omega_2\,|\,\omega_3,\omega_4,\omega_5)$ & $S(\omega_1,\omega_2,\omega_3\,|\,\omega_4,\omega_5)$  \\ \Xhline{3\arrayrulewidth}
       \multicolumn{1}{c:}{} & \multicolumn{1}{c:}{0} & \multicolumn{1}{c:}{0} & \multicolumn{1}{c:}{0}  \\ \hline
$\nu_F^{-2}$ & $-\omega_1\omega_2\omega_3\omega_4(\omega_2-i)$ & 0 & 0  \\ \hline
 $\nu_F^{-3}$ &0 & \makecell{$-i\omega_1\omega_2\omega_3\omega_4\omega_5$\\ $\times (\omega_2-i)(\omega_2-2i)$} &\makecell{$ i\omega_1\omega_2\omega_3\omega_4\omega_5$ \\$\times (\omega_3-i)(\omega_1+\omega_2+\omega_3-2i)$}  \\ \hline
  $\nu_F^{-4}$ &x  &0  & 0 \\ \hline
  $\nu_F^{-5}$ & 0  & x& x \\ \hline
 \multicolumn{1}{c:}{} & \multicolumn{1}{c:}{x} & \multicolumn{1}{c:}{0} & \multicolumn{1}{c:}{0}  \\ 
 \multicolumn{1}{c:}{} & \multicolumn{1}{c:}{0} & \multicolumn{1}{c:}{x} & \multicolumn{1}{c:}{x}  \\ 
\end{tabular}
\end{center}
\caption{Summary of the scattering amplitudes calculated in this section and appendix \ref{higherscat}. Each box contains the contribution to the scattering amplitude at the inverse power of $\nu_F$ in the leftmost box. An x indicates a nonzero result easily obtained by expanding the reflection coefficients to higher order in $1/\nu_F$. Within $S(\omega_1,\dots,\omega_m\,|\,\omega_{m+1},\dots,\omega_n)$, $\{\omega_{m+1},\ldots,\omega_n\}$ denote the $n-m$ incoming (negative) frequencies and $\{\omega_1,\ldots, \omega_m\}$ the $m$ outgoing (positive) frequencies. Note in particular that we choose an ordering of the frequencies which implies that $\omega_2$ is the maximum for the two-to-two and three-to-two amplitude, whereas $\omega_3$ is the maximum for the two-to-three amplitude. 
}
\end{table}
\endgroup


\section{A glimpse into the continuum$^{\color{magenta}{G}}$}\label{LiouvilleSec}

Up until this point, we have focused on the large $N$ limit of a variety of systems. We have mentioned on several instances that upon taking the large $N$ limit, and further tuning certain coefficients, the systems often exhibit certain critical behaviour. In the case of large $N$ matrices, one might then imagine the emergence of a continuous theory residing on a genus $h$ Riemann surface $\Sigma_h$. The goal of this section is to briefly elaborate on these two-dimensional continuum theories. 

\subsection{Random pure geometry in two-dimensions}


We begin our discussion by considering a theory of pure geometry in two-dimensions. The question of interest is as follows. Let us endow a compact genus $h$ Riemann surface, $\Sigma_h$, with a metric $g_{ij}$. The most general local, diffeomorphism invariant, two-derivative action is given by
\begin{equation}
S_{\mathrm{grav}}[g_{ij};\Sigma_h] = - \frac{\vartheta}{4\pi}  \int_{\Sigma_h} \dd^2x \sqrt{g} \, R + \Lambda \int_{\Sigma_h} \dd^2x \sqrt{g}~,
\end{equation}
where $\vartheta$ and $\Lambda$ are real parameters and $R$ is the Ricci scalar.
By the Gauss-Bonnet theorem, the first term is a topological invariant proportional to the Euler characteristic $\chi_h$. The second term, which is the cosmological constant term, computes the area of the surface. Motivated by the continuum limit of the discrete picture discussed in section \ref{singlematrix} and the resulting expression (\ref{critgen}), we are prompted to study the path integral
\begin{align}\label{eq:liouvillestart}
\mathcal{Z}_{\Sigma_h} [\vartheta;\Lambda]=  e^{\vartheta \, \chi_h} \, \int\frac{ \mathcal{D} {g}_{ij}}{\mathrm{vol \, diff}} \, \exp \left( {- \Lambda \int_{\Sigma_h} \dd^2x \sqrt{g}} \right)~.
\end{align}
At this stage, it is useful to recall that a two-dimensional metric is diffeomorphic, at least locally, to the following
\begin{equation}\label{Weylfix}
g_{ij}(x) = e^{\varphi(x)} \tilde{g}_{ij}(x)~,
\end{equation}
where $\tilde{g}_{ij}$ is a fixed reference metric, often referred to as the fiducial metric, and $\varphi(x)$ is a local Weyl factor that remains unfixed. The parameterization (\ref{Weylfix}) leaves a particular subgroup of diffeomorphisms unfixed. Indeed, if we consider the set of coordinate transformations that map $\tilde{g}_{ij}$ to itself times a local Weyl factor $e^{\sigma(x)}$, we can return to the original expression by shifting $ \varphi(x) \to \varphi(x) - \sigma(x)$.
That the group of residual diffeomorphisms in the gauge (\ref{Weylfix}) is the two-dimensional conformal group will play an important role in what comes. In fact, the transformation
\begin{equation}\label{redund}
\tilde{g}_{ij}(x) \to e^{\tilde{\sigma}(x)} \tilde{g}_{ij}(x)~, \quad\quad \varphi(x) \to \varphi(x) - \tilde{\sigma}(x)~
\end{equation}
for general $\tilde{\sigma}(x)$ is a redundancy of the parameterization (\ref{Weylfix}). As such, the resulting theory governing $\varphi(x)$ and the accompanying Fadeev-Popov ghost fields should be invariant under the more general Weyl transformations (\ref{redund}).
In the gauge (\ref{Weylfix}), the ghost theory is described by a conformal field theory of $b c$-ghosts in a fixed background $\tilde{g}_{ij}$ whose central charge is $c_{g}=-26$. The resulting theory consisting of the $b c$-ghosts and the $\varphi$-field should be invariant under the redundancy (\ref{redund}). Thus, the theory governing the Weyl mode $\varphi(x)$ must be a two-dimensional conformal field theory with central charge $c_L = 26$ such that the net conformal anomaly vanishes.
If we further assume the $\varphi$-sector admits a local description, one can postulate an action of the following type \cite{David:1988hj,Distler:1988jt}
\begin{equation}\label{SL}
S_L[\varphi,\tilde{g}_{ij}] =  \frac{1}{4\pi} \int_{\Sigma_h} \dd^2 x \sqrt{\tilde{g}} \left(\tilde{g}^{ij} \partial_i \varphi \partial_j \varphi + Q \tilde{R} \varphi + 4\pi \Lambda \, e^{2 b \varphi}  \right)~.
\end{equation}
The above action is known as the Liouville action and has been studied extensively \cite{Seiberg:1990eb, Teschner:2001rv,Nakayama:2004vk}.
In order for the kinetic term to have a standard normalisation, we have allowed for a constant rescaling of $\varphi$ by $2b$, such that the resulting theory should be invariant under the following redundancy
\begin{equation}\label{redundii}
\tilde{g}_{ij}(x) \to e^{\tilde{\sigma}(x)} \tilde{g}_{ij}(x)~, \quad\quad \varphi(x) \to \varphi(x) - {\tilde{\sigma}(x)}/{2b}~.
\end{equation}
When $Q = b+1/b$, the Liouville action describes a two-dimensional conformal field theory. It has some unusual features such as a continuous spectrum and the absence of a normalisable vacuum state. Nevertheless, it admits a consistent quantisation and many of its properties are known explicitly. For instance, its central charge is given by $c_L = 1+6 Q^2$, and the theory contains spinless primary operators $\mathcal{O}^{(\alpha)} = e^{2\alpha \varphi}$ whose conformal dimensions are
\begin{equation}
\Delta_\alpha  = \alpha\left(Q-{\alpha}\right)~,\quad\quad \bar{\Delta}_\alpha =\alpha\left(Q-{\alpha}\right)~.
\end{equation}
These conformal dimensions can be used to fix $b$ in (\ref{SL}), since conformal invariance requires $\mathcal{O}^{(b)}= e^{2b \varphi}$ to have conformal dimensions $(\Delta_b,\bar{\Delta}_b) = (1,1)$. One thus finds
\begin{equation}\label{gamma}
b =\frac{Q}{2}\pm \sqrt{\frac{Q^2}{4}-1}~.
\end{equation}
Requiring $c_L + c_g = 0$ fixes $Q= \pm 5/\sqrt{6}$. Given the discrete symmetry $Q \to -Q$, $b \to -b$, and $\varphi \to -\varphi$, we can pick $Q>0$ without loss of generality. One then finds the two solutions $b = b_\pm$ with $b_+ = \sqrt{3/2}$ and $b_- = \sqrt{2/3}$. 
If one further requires the existence of a semi-classical limit, i.e. a limit where $Q \to \infty$ gives rise to a sensible saddle-point approximation, one is forced to pick the negative root in (\ref{gamma}). We will consider the negative root in what follows even in the absence of a semiclassical limit.

In this way we obtain a description of the path integral for a theory of pure two-dimensional geometry in terms of Liouville theory.
A natural collection of observables is given by the set of diffeomorphism invariant functionals of $g_{ij}$. In the Weyl gauge, this is given by the set of conformal primaries built from $\varphi$ with conformal dimensions $\left(\Delta,\bar{\Delta} \right) = (1,1)$. A particularly important one is the area operator
\begin{equation}\label{Aop}
\mathcal{A}_h = \int_{\Sigma_h} \dd^2 x \sqrt{g} = \int_{\Sigma_h} \dd^2 x \sqrt{\tilde{g}} \, e^{2b \varphi}~,
\end{equation}
where $b$ is given by (\ref{gamma}). Expectation values of $\mathcal{A}_h$ are calculated by taking derivatives of $-\log \mathcal{Z}_{\Sigma_h} [\vartheta;\Lambda]$ with respect of $\Lambda$.

\subsection{Sprinkling matter}\label{sprinkling}

Generally speaking a local quantum field theory can be placed on an arbitrary curved background. If the theory has no diffeomorphism anomaly, we can further gauge the diffeomorphism group. Thus, it would seem we can couple arbitrary matter content to two-dimensional gravity. Following this reasoning, and going to the Weyl gauge (\ref{Weylfix}), a consistent theory of matter coupled to two-dimensional gravity is described by the following action
\begin{equation}\label{2dtheory}
S_{\mathrm{eff}} = S_L[\varphi,\tilde{g}_{ij}] + S_{\mathrm{ghost}}[b,c,\tilde{g}_{ij}] + S_{\mathrm{matter}}[X,\varphi,\tilde{g}_{ij}]~.
\end{equation}
Invariance under the redundancy (\ref{redund}) and the residual diffeomorphisms implies that the above theory must be a two-dimensional conformal field theory with vanishing central charge. This statement must be true, curious as it may sound, even for a massive matter theory. Somehow, the Liouville mode must `dress' the matter theory in such a way as to make the resulting system conformal. Of course, given that the original matter theory is not conformal, the resulting conformal theory obtained upon coupling to the Liouville mode must be strongly interacting even if the original matter theory is not \cite{Zamolodchikov:2005jb}. 

In what follows we will consider the simpler situation, namely, a matter theory which is itself a two-dimensional conformal field theory with central charge $c_m$. In this case, the matter action $S_{\mathrm{matter}}$ becomes solely a function of the matter fields and $\tilde{g}_{ij}$. However, the path integration measures for the matter and ghost fields produce the Liouville action in the Weyl gauge \cite{Polyakov:1981rd}.\footnote{More generally, there is no reason to insist on the existence of a matter action. The relevant point is that the conformal anomaly of a two-dimensional conformal field theory on the background (\ref{Weylfix}) is governed by the Liouville theory.} Thus, one can postulate that (\ref{SL}) continues to govern the $\varphi$ sector, except now we must fix the parameters $Q$ and $b$ such that $c_L + c_m -26 =0$. Proceeding along similar lines to the previous sub-section, we end up with the expressions
\begin{equation}\label{Qgamma}
Q = \sqrt{\frac{25-c_m}{6}}~, \quad\quad b = \frac{\sqrt{25-{c_m}} \pm \sqrt{1-{c_m}}}{2 \sqrt{6}}~.
\end{equation}
For $c_m=0$, we recover (\ref{gamma}). 

The expressions (\ref{Qgamma}) demonstrate an important point. The resulting theory is highly sensitive to $c_m$. For $c_m \le 1$, we have that $b$ is real and positive. For $c_m \in (1,25)$ the parameter $b$ becomes complex with non-vanishing real and imaginary parts, while $Q$ remains real. Finally, for $c_m \ge 25$, we have that $b$ and $Q$ become pure imaginary. If we take $c_m = 25$ and tune $\Lambda = 0$, we observe that (\ref{SL}) reduces to the action of a free scalar, such that our resulting theory is nothing more than the critical bosonic string. All other cases describe non-critical string theories. 

In the presence of matter, we can extend the space of admissible observables. For instance, a spinless weight $(\Delta,\Delta)$ conformal primary $\mathcal{O}_{\Delta}$ of the matter theory can be combined with the Liouville exponential operator to produce a `dressed' operator $\mathcal{\tilde{O}}_\Delta = e^{2\sigma \varphi} \mathcal{O}_\Delta$ with
\begin{equation}\label{sigma}
\sigma =  \frac{\sqrt{25-c_m} \pm \sqrt{24 \Delta +1-c_m}}{2 \sqrt{6}}~.
\end{equation}
In this way, we can obtain a significant collection of observables. It is worth noticing that for certain values of $c_m$ and $\Delta$, $\sigma$ may become complex. For such situations, $\mathcal{O}^{(\sigma)} = e^{2\sigma\varphi}$ becomes a complex (rather than Hermitian) operator. To render it Hermitian we can consider linear combinations of $\mathcal{O}^{(\sigma)}$ with its complex conjugate. 
\newline\newline
{\textbf{Small and/or negative central charge.}} The most understood situation occurs when $c_m < 1$. For example, we could take the matter conformal field theory to be one of the minimal models  \cite{Friedan:1986qt, Polyakov:1984yq} with central charge
\begin{equation}\label{cminimalmodel}
c^{(p,q)}_m = 1 - \frac{6(p-q)^2}{p q}~, \quad\quad (p,q)~\, \text{coprime}~,
\end{equation}
and operators of conformal dimension 
\begin{equation}
\Delta_{r,s}= \frac{(rq-sp)^2- (p-q)^2}{4pq}~, \quad r=1,\ldots, p-1~, \quad s=1,\ldots , q-1~.
\end{equation}
The limit $c_m \to -\infty$ leads to a parameterically large $Q$ and, for the negative branch, a parametrically small $b$. In this limit, the system is driven to a semiclassical regime where saddle point techniques may be employed. Though there are several known minimal models with large and negative $c_m$, they are all non-unitary.\footnote{Nevertheless, upon coupling to gravity many of the states in the original matter theory are removed from the resulting Hilbert space possibly rendering the non-unitarity less severe.}
\newline\newline
{\textbf{Large central charge.}} For $c_m > 25$, we could imagine a contour rotation $\varphi \to i \varphi$ that yields the action (\ref{SL}) real. This comes at the price (or perhaps reward) of changing the sign of the kinetic term of $\varphi$. In critical string theory, we would interpret a wrong sign scalar as a time-like direction in the target space. Perhaps the same is true for $c_m>25$, in which case non-critical string theory might provide an interesting window into time-dependent backgrounds in string theory. In such circumstances caution must be exercised, as many of the techniques we understand require rotating time-like target space coordinates back to their spacelike values, which for $c_m>25$ would take us back to a complex action. 
\newline\newline
{\textbf{Intermediate central charge.}} For $c_m \in (1,25)$ the situation is far less understood. In this case, the action is intrinsically complex, and new methods are required to deal with it. Curiously, and perhaps interestingly, this case includes a matter conformal field theory with $c_m=3$ and $c_m=4$.

\begin{center}\it{Matter matters} \end{center}

The regime $c_m \in (1,25)$ teaches us an important lesson. The ability to couple a particular matter theory to gravity at the classical level does not guarantee that the combined system makes sense at the quantum level. Even if sense can be made of the path integral for a matter theory with $c_m \in (1,25)$, it seems the resulting theory will not resemble an ordinary theory of two-dimensional geometry. The difficulties surrounding the $c_m>1$ regime have been dubbed the $c_m>1$ {\it barrier} in the literature. 

In the next section we will discuss several contexts in which this barrier has been surpassed. As a general rule, addressing which matter theories can indeed be consistently coupled to gravity at the quantum level, particularly in higher-dimensions, seems to require a more complete understanding of the theory in the deep ultraviolet.  
%

\subsection{Critical exponents $\&$ the area operator}\label{Acrit}

At this stage we have equipped ourselves with the necessary tools to compute the following quantity 
\begin{equation}\label{eq:partfuncA}
\tilde{\mathcal{Z}}_{\Sigma_h}[\upsilon;\Lambda]= \int  \frac{\mathcal{D}g_{ij}}{\text{vol} \,  \text{diff}}  \, \mathcal{D}_{g_{ij}}X  \, e^{-S_{\mathrm{grav}}[g_{ij}] - S_{\mathrm{matter}}[X,g_{ij}]} \, \delta \, \left(\int_{\Sigma_h} \dd^2 x\sqrt{{g}}-\upsilon\right)~. 
\end{equation}
The $\delta$-function in the above expression fixes the area to the particular value $\upsilon$. By fixing the area operator (\ref{Aop}) to take a fixed value, the fixed area partition function (\ref{eq:partfuncA}) tames any potential divergences in the original path integral which can subsequently be analysed in a clearer fashion.

Once again, resorting to the Weyl gauge (\ref{Weylfix}) and keeping in mind the rescaling of $\varphi$, we can make progress in evaluating (\ref{eq:partfuncA}). We obtain the path integral expression
\begin{equation}\label{Zmatter}
\tilde{\mathcal{Z}}_{\Sigma_h}[\upsilon;\Lambda]=   e^{\vartheta \, \chi_h} \, \int  \frac{\mathcal{D}\varphi}{\text{vol} \, \mathcal{G}} \, e^{-S_L[\varphi,\tilde{g}_{ij}]}  \, \delta \, \left(\int_{\Sigma_h} \dd^2 x\sqrt{{\tilde{g}}}\, e^{2b\varphi}-\upsilon\right)~. 
\end{equation}
Any local ultraviolet divergences arising upon integrating out the matter and ghost fields are absorbed in the bare couplings of the gravitational theory. 
The group $\mathcal{G}$ is the residual gauge group upon fixing the Weyl gauge. For instance, when $h=0$ we could fix $\tilde{g}_{ij}$ to be the round metric on the two-sphere in which case $\mathcal{G} = PSL(2,\mathbb{C})$. In contrast to the critical string, dividing by the potentially divergent volume of $\mathcal{G}$ does not mean that the partition function vanishes. This is due to the fact that $\varphi$ transforms non-trivially under the residual gauge transformations.

Given (\ref{Zmatter}), the approach is to consider a constant shift $\varphi \to \varphi + \log \upsilon^{1/2b}$ that allows us to remove $\upsilon$ from the $\delta$-function while only affecting the remaining Liouville theory in a simple way \cite{David:1988hj,Distler:1988jt}. Using $\delta(\zeta x)= \delta(x)/|\zeta|$ and (\ref{SL}) it is readily found that 
\begin{equation}\label{AreaZ}
{\tilde{\mathcal{Z}}_{\Sigma_h}[\upsilon;\Lambda]} =  \mathcal{N}_h \, e^{\vartheta \chi_h} \, e^{-\Lambda \upsilon} \, {\upsilon}^{-{Q} \chi_h/{2b} -1}~,
\end{equation}
where $\mathcal{N}_h$ is a normalisation constant that may depend on the genus $h$ but is independent of $\upsilon$, $\vartheta$, and $\Lambda$. We recall that $b$ is given by (\ref{Qgamma}). The continuum partition function can be recovered by integrating over $\upsilon$ as follows
\begin{equation}\label{Flambda}
\mathcal{Z}_{\Sigma_h} [\vartheta;\Lambda] = e^{\vartheta \chi_h} \int^\infty_0 \dd\upsilon \, \tilde{\mathcal{Z}}_{\Sigma_h}[\upsilon;\Lambda] = \mathcal{N}_h\,e^{\vartheta \chi_h}  \, \Lambda^{\frac{Q \chi_h}{2b}} \, \Gamma \left(-Q \chi_h/2b \right)~, \quad\quad h \neq 1~, 
\end{equation}
where we have regularised the integral near $\upsilon = 0$ by analytic continuation. 
When $h=1$ the resulting expression is logarithmic in $\Lambda$, regardless of the value of $Q/2b$.

It is tempting to view (\ref{Flambda}) from the perspective of critical systems near a phase transition, with $\Lambda$ being a tuneable parameter driving the system towards criticality. To this end, it is customary in the literature to define a critical exponent $\Gamma_{\text{str}} \equiv 2-Q/b$, often referred to as the string susceptibility. Combining $Q$ and $b$ in (\ref{Qgamma}) and choosing the negative branch for $b$, we obtain the KPZ relation \cite{Knizhnik:1988ak}
\begin{equation}
\Gamma_{\mathrm{str}}= \frac{1}{12}\left(c_m-1-\sqrt{(c_m-1)(c_m-25)}\right)~.
\end{equation}
Some examples are given by  \cite{Ambjorn:1985az, David:1985nj,Kazakov:1985ea, Kazakov:1986hu, Boulatov:1986sb, Kazakov:1988ch}
\begin{align}\label{tablecrit}
\big(\Gamma_{\mathrm{str}} , {Q}/{2b} \big) = 
\begin{cases}
(-1/2, 5/4 ) \quad\quad &\mathrm{for} \quad\quad c_m= 0~,\\
(-1/3, 7/6) \quad\quad &\mathrm{for} \quad\quad c_m= 1/2~,\\
(0, 1) \quad\quad &\mathrm{for} \quad\quad  c_m=1~.
\end{cases}
\end{align}
At large negative $c_m$ we have
\begin{equation}\label{curious}
\mathcal{Z}_{\Sigma_h} [\vartheta;\Lambda] = \tilde{\mathcal{N}}_h \, e^{\vartheta\chi_h}  \left(\frac{1}{\sqrt{\Lambda}}\right)^{\frac{c_m \chi_h}{6}}~,
\end{equation}
where $\tilde{\mathcal{N}}_h$ is again independent of $\Lambda$. For genus zero (\ref{curious}) is reminiscent of the partition function of a CFT of central charge $c_m$ on a round two-sphere with Ricci scalar $\tilde{R} = 2\Lambda$ \cite{Zamolodchikov:2001dz}, and consequently an entanglement entropy \cite{Casini:2011kv,Holzhey:1994we,Calabrese:2004eu}. For a genus $h$ Riemann surface (\ref{curious}) is reminiscent of the partition function obtained from Euclidean AdS$_3$/CFT$_2$ considerations at large positive central charge (see for instance \cite{Krasnov:2000zq}).

\begin{center}{\it A gravitational conformal weight}\end{center}

When coupling a conformal field theory to gravity, one is naturally led to the question of what replaces the conformal weights of a spinless conformal primary $\mathcal{O}_\Delta$. One way to quantify this is by computing the following quantity
\begin{multline}\label{areaoperator}
\langle \mathcal{O}_\Delta \rangle_{\upsilon} = \frac{1}{\tilde{\mathcal{Z}}_{\Sigma_h}[\upsilon;\Lambda]}  \\ \times {e^{\vartheta \, \chi_h}} \, \int  \frac{\mathcal{D}\varphi}{\text{vol} \, \mathcal{G}} \mathcal{D} X \, e^{-S_L[\varphi,\tilde{g}_{ij}] - S_{\mathrm{matter}}[X,\tilde{g}_{ij}]}  \, \delta \, \left(\int_{\Sigma_h} \dd^2 x\sqrt{{\tilde{g}}} \, e^{2b\varphi}-\upsilon\right) \int_{\Sigma_h} \dd^2 x \sqrt{\tilde{g}} \, e^{2\sigma \varphi} \mathcal{O}_\Delta~,
\end{multline}
where $\sigma$ was given in (\ref{sigma}) and $b$ in (\ref{Qgamma}). Once again, using the technique of \cite{David:1988hj,Distler:1988jt}, we can shift $\varphi \to \varphi + \log \upsilon^{1/(2b)}$ to conclude
\begin{equation}
\langle \mathcal{O}_\Delta \rangle_{\upsilon} = \mathcal{N}_h \, \upsilon^{\sigma/b}~,
\end{equation}
with a $\upsilon$-independent normalisation constant $\mathcal{N}_h$. Rescaling the area parameter $\upsilon \to k^2 \upsilon$ we find $\langle \mathcal{O}_\Delta \rangle_{\upsilon} \to k^{2\sigma/b} \langle \mathcal{O}_\Delta \rangle_{\upsilon}$. In an ordinary two-dimensional conformal field theory, an integrated spinless conformal operator with weights $(\tilde{\Delta},\tilde{\Delta})$ would scale as $k^{2(1-\tilde{\Delta})}$. We are thus prompted to define the quantity $\Delta_{\mathrm{grav}} \equiv 1-\sigma/b$ as the gravitational analogue of the conformal weight. Explicitly
\begin{equation}\label{eq:gravConfweight}
\Delta_{\mathrm{grav}} = \frac{\sqrt{24 \Delta +1-c_m}-\sqrt{1-c_m}}{\sqrt{25-c_m}-\sqrt{1-c_m}}~,
\end{equation}
 which is our second KPZ relation. For the identity operator both $\Delta=0$ and $\Delta_{\mathrm{grav}}=0$. We also have that $\Delta_{\mathrm{grav}} = 1$ when $\Delta=1$. For large and negative $c_m$ we have $\Delta_{\mathrm{grav}} = \Delta + \mathcal{O}(1/c_m)$. More generally, the two differ. 

\subsection{Large $N$ matrices $\&$ the continuum}

We are now in a position to connect the continuum description developed in this section to the matrix models we examined throughout the previous ones. 
\newline\newline
{\textbf{Single matrix integrals.} Let us consider first the integrals over single matrices considered in section \ref{singlematrix}. There, we argued for the existence of a continuum limit in which the expectation number $\langle n_h \rangle$ defined in (\ref{eq:continuumlimit}) diverges as we approach a critical coupling. We presented evidence that near criticality and at large $N$
\begin{equation}
\mathcal{F}_N( \alpha ) \approx \sum_{h=0}^\infty f_h \, e^{\chi_h \log N} \left( \alpha-\alpha_c \right)^{5 \chi_h /4}~.
\end{equation}
How should we compare the above to our continuum expressions? We might expect that the sum (\ref{eq:connecteddiagrams}) should be viewed as the discrete version of the pure geometry path integral (\ref{eq:liouvillestart}), with $\vartheta = \log N$. In such a case, we should compare the critical limit (\ref{eq:naF}) to (\ref{Flambda}). Using the $c_m =0$ result for $Q/2b$, and recalling (\ref{eq:naF}), we conclude that the cosmological constant of the continuum theory $\Lambda$ is propotional to the deviation from criticality, i.e. $\Lambda \propto (\alpha-\alpha_c)$. Notice that we recover the logarithmic behaviour in (\ref{eq:sumF1}) for $h=1$. 

That the continuum cosmological constant is defined as a deviation from the critical coupling $\alpha_c$ is part of a recurring theme. It is the same theme that led us to consider physics just above the filled Fermi sea. 
\newline\newline
{\textbf{Multicritical matrix model.}} 
It has been argued \cite{Staudacher:1989fy,Brezin:1989db, Gross:1989vs} that multicritical matrix models correspond to two-dimensional gravity coupled to a non-unitary minimal CFT (\ref{cminimalmodel}) with ($p,q$)= $(2,2m-1)$ and string susceptibility $\Gamma_{\mathrm{str}}= -{1}/{m}$~. For instance, the multicritical model discussed in (\ref{eq:multcritmodel}) corresponds to $m=3$: i.e. two-dimensional gravity coupled to the Lee-Yang minimal model with $c_m =-22/5$. For these models, one has to use a slightly modified version \cite{Brezin:1989db} of the KPZ relation due to the presence of negative weight operators in the spectrum of the  matter CFT. It is remarkable that tuning the parameters of $V(M)$ for a {\textit{single}} matrix leads to a continuum description of gravity coupled to conformal field theories with varying central charges. 
\newline\newline
{\textbf{Double matrix integrals.} Given (\ref{np2m}) and using $\Lambda \propto (\alpha-\alpha_c)$ one is led to a theory with $Q/(2b) = 7/6$. Looking at (\ref{tablecrit}) we see that this occurs for $c_m=1/2$. It is natural to propose \cite{Kazakov:1986hu} that the two matrix model corresponds to $c_m = 1/2$ matter, and in particular two-dimensional gravity coupled to a free massless fermion. 
The free massless fermion in two dimensions is a minimal model which has three spinless conformal primaries, namely the identify operator with $\Delta_0=0$, the $\sigma$-operator with $\Delta_\sigma =1/16$, and the energy operator with $\Delta_\epsilon = 1/2$. These operators are sourced by the cosmological constant $\Lambda$, the magnetic field $H$ and the temperature $\beta$. These sources make their appearance in the matrix model as simple functions of the couplings $\alpha$, $g$, and $c$ of the potential (\ref{quarticWh}).
Using (\ref{eq:gravConfweight}) we infer the gravitational conformal weight of these primaries to be
$\Delta_{\mathrm{grav},0}=0$, $\Delta_{\mathrm{grav},\sigma}={1}/{6}$, and  $\Delta_{\mathrm{grav},\epsilon}={2}/{3}$. These were obtained from considerations of the generalised two-matrix model (\ref{quarticWh}) in \cite{Brezin:1989db} where it was found that there are certain non-analyticities of the large $N$ two-matrix integral with potential (\ref{quarticWh}) upon tuning $\alpha$, $c$, and $g$. The critical coefficients stemming from these non-analyticities are expressed in terms of the gravitational conformal weights $\Delta_{\mathrm{grav},0}$, $\Delta_{\mathrm{grav},\epsilon}$, and $\Delta_{\mathrm{grav},\sigma}$ via the scaling relations of the two-dimensional Ising model. Concretely, the critical exponents associated to $c$ and $g$, which we denote by $\boldsymbol{\alpha}$ and $\boldsymbol{\beta}$ respectively, are related to the gravitational conformal weights as
\begin{equation}
\boldsymbol{\alpha} = \frac{1-2\Delta_{\mathrm{grav},\epsilon}}{1-\Delta_{\mathrm{grav},\epsilon}} = -1~, \quad\quad \boldsymbol{\beta} = \frac{\Delta_{\mathrm{grav},\sigma}}{1-\Delta_{\mathrm{grav},\epsilon}} = \frac{1}{2}~.
\end{equation}

It might be worth recalling that a free massless fermion describes the critical behaviour of the two-dimensional Ising model. The Ising model has two states at every point on the lattice. This may remind us, somewhat, of the decorated Feynman diagrams discussed in section \ref{decorated}. This bears some truth, but caution should be exercised before decorating the Feynman diagrams too elaborately. For instance, the continuum limit of mulitcritical single matrix models corresponds to non-unitary minimal models with $c_m<0$ coupled to gravity. It has been further argued \cite{tada,Douglas:1990pt} that the whole remaining family of minimal models can be obtained from particularly selected two-matrix models. 

Upon coupling a minimal model to two-dimensional quantum gravity one obtains a new set of dressed and integrated operators generalising to the area operator in (\ref{areaoperator}). In addition to their gravitional conformal weights (\ref{eq:gravConfweight}), one can also study the correlation functions of these operators and compare them to quantities obtained from matrix integral calculations.
Work in this direction includes \cite{Belavin:2005jy,Belavin:2006ex,Belavin:2008kv,Belavin:2010ba,Belavin:2010sr}. 
\newline\newline
{\textbf{Non-perturbative features from the continuum.}} In our discussion of large $N$ matrices we touched upon the possibility of non-perturbative corrections to the planar expansion. In the double scaling limit, these manifest themselves in terms of small WKB type corrections of the series expansion solutions to the various non-linear differential equations, such as the Painlev{\'e} I equation. We are naturally led to the question of whether such non-perturbative terms can be recovered in the continuum picture. 

In \cite{Zamolodchikov:2001ah} it was shown that Liouville theory can be studied on the disk topology, which is a manifold with a boundary.  In this case, the Liouville field diverges near the boundary of the disk. Taking the metric $\tilde{g}_{ij}$ on the disk to be $\dd s^2 = \dd\rho^2 + \rho^2 \dd\theta^2$ with $\rho \in (0,1)$, in the semiclassical $b\to0$ limit the solution of $\varphi$ is given by
\begin{equation}
e^{2b\varphi_{cl}} = \frac{1}{\pi b^2 \Lambda} \frac{1}{(1-\rho^2)^2}~,
\end{equation}
such that the physical metric $g_{ij} = e^{2b\varphi}\tilde{g}_{ij}$ is the Poincar{\'e} disk. The authors of \cite{Zamolodchikov:2001ah} showed the existence of a family of boundary states labelled by two integers $(m,n)$. These configurations are known as ZZ-branes or instantons. Processes involving Liouville theory on the disk have been argued \cite{Klebanov:2003km,Martinec:2003ka,McGreevy:2003kb} to be related to non-perturbative features in the matrix models. For instance, \cite{Alexandrov:2003nn} recovered the exponent (\ref{epsilonPainleve}) and (\ref{exponents2MM}) of the non-perturbative corrections in the double scaling limit by calculating the Liouville disk partition function. In \cite{Seiberg:2004at}, the ZZ-instantons were interpreted as single eigenvalues sitting on critical points of the potential away from the dense eigenvalue distribution. 

There is another type of boundary condition that arises in the study of Liouville theory on a disk topology. Here, one adds a boundary interaction \cite{Fateev:2000ik, Teschner:2000md}
\begin{equation}
S_{\mathrm{bdy}} =   \int_{S^1} \dd u \sqrt{h} \left( \frac{Q}{2\pi}  K \varphi + \Lambda_B e^{b\varphi} \right)
\end{equation}
to the Liouville action (\ref{SL}). The boundary cosmological constant $\Lambda_B$ is a new continuous parameter labelling the states in this theory. The induced metric at the $S^1$ boundary of the disk is $h$ and $K$ is the extrinsic curvature at the boundary. $\Lambda_B$ can be viewed as a chemical potential for the size of the boundary of the physical metric on a disk topology. For $\Lambda_B>0$, the boundary action $S_{\mathrm{bdy}}$ suppresses configurations at large values of $\varphi$. These configurations are known as FZZT branes  and can be viewed as extended across the weakly coupled region $\varphi \lesssim -(\log \Lambda_B)/b$. From the matrix model perspective, an interpretation of the FZZT brane should be an object that is parameterised by a continuous variable. Once again, one may ask whether FZZT branes are related to non-perturbative features of the matrix integral.  In \cite{Seiberg:2004at}, the FZZT branes were argued to be described by the single trace $\Tr \log \left(\Lambda_B-M\right)$. Indeed, the relation of $\Tr \log \left(\Lambda_B-M\right)$ to triangulated Riemann surfaces with a boundary was explored in section \ref{Riemannsurfaces}.
\newline\newline
{\textbf{Quantum mechanical matrices.} In the case of quantum mechanical matrices, we uncovered a rich structure in the form of a two-dimensional $S$-matrix. We now turn to its realisation in the continuum picture. 

\subsection{Scattering from the continuum}


We now discuss how to recover the scattering discussed in the previous section from the continuum theory, in particular the one-to-one and one-to-two amplitudes. In order to do so, we must first understand what is being scattered in the continuum picture. For instance the two-dimensional theory (\ref{2dtheory}) lives on a Euclidean compact surface $\Sigma_h$ which has no asymptotic regions to scatter to and from. If, on the other hand, we view the two-dimensional theory as the worldsheet theory of a string propagating some target space, there may be room for a scattering amplitude in the assymptotia of the target space. To this end we take the $c_m=1$ theory to be a free scalar field denoted by $X^0$. In particular for $c_m=1$ we conclude using (\ref{Qgamma}) that $b=1$ and $Q=2$. Our worldsheet theory is then given by
\begin{equation}\label{NCworldsheet}
S_{\mathrm{w.s.}} =  -\frac{1}{{4}\pi} \int_{\Sigma_h} \dd^2 x \sqrt{\tilde{g}} \, \tilde{g}^{ij} \, \partial_i X^0 \partial_j X^0 + \frac{1}{4\pi} \int_{\Sigma_h} \dd^2 x \sqrt{\tilde{g}} \left(\tilde{g}^{ij} \partial_i \varphi \partial_j \varphi + 2 \tilde{R}\, \varphi + 4\pi \Lambda\, e^{2 \varphi}  \right)~. 
\end{equation}
The free scalar $X^0$ encodes the time coordinate of the target space, and as such comes with a kinetic term with the wrong sign. The Liouville mode $\varphi$ is then associated to the spatial coordinate of the target space, such that the target space is two-dimensional. Taking the contribution of the ghost conformal field theory into account, the net theory has vanishing central charge.\footnote{Said otherwise, given that the central charge of a worldsheet matter theory consisting of a free scalar and the Liouville conformal field theory can be tuned to equal $26$, we can regard the system as a bosonic critical string.} A string propagating in two-dimensions has no transverse excitations. Consequently, its spectrum is given by a single scalar field encoding the target space centre of mass position of the string. The natural asymptotic region in the target space is given by the region $\varphi \to -\infty$, where the Liouville interaction is switched off. 

We can now compute the $S$-matrix elements as is done for more familiar worldsheet theories. We calculate the expectation value of several vertex operator insertions. For a free scalar we can construct conformal operators by exponentiating $X^0$. It follows from our discussion in section \ref{sprinkling}, that we can construct the dressed integrated vertex operator
\begin{equation}\label{tachyon}
\mathcal{V}^\pm_\omega = g_s \int_{\Sigma_h} \dd^2 x \sqrt{\tilde{g}} \, : e^{\pm i \omega X^0} :  \mathcal{O}({\omega})~, \quad\quad \omega \ge 0~.
\end{equation}
In the above, $\mathcal{O}({\omega})$ is taken to be a Hermitian Liouville operator of the form 
\begin{equation}
\mathcal{O}({\omega}) \equiv S(\omega) e^{(2+i\omega) \varphi} + S(\omega)^{*} e^{(2-i\omega) \varphi}~, \quad\quad\quad  \omega \ge 0~.
\end{equation}
The phase factors have been chosen such that the two-point function of the Liouville operator $\mathcal{O}({\omega};z,\bar{z})$ has standard $\delta$-function normalisation
\begin{equation}\label{2ptfunction}
\langle \mathcal{O}({\omega}_1;z_2,\bar{z}_2) \mathcal{O}({\omega}_2;z_2,\bar{z}_2) \rangle =2\pi  \delta(\omega_1-\omega_2) \times \frac{1}{|z_1-z_2|^{4+\omega_1^2}}~,
\end{equation}
where $z$ and $\bar{z}$ label points on $\mathbb{C}$, and we adhere to the conventions of \cite{PolchinskiBook} with $\alpha'=1$. The phase factor is given by \cite{Seiberg:1990eb} 
\begin{equation}\label{eq:phasefactor}
S(\omega)= \sqrt{\frac{\Gamma(i\omega)}{\Gamma(1-i\omega)}\frac{\Gamma(1+i\omega)}{\Gamma(-i\omega)}}~.
\end{equation}
To compute target space $S$-matrix elements we must calculate expectation values for the above vertex operators. The parameter $\omega$ corresponds to the energy of the asymptotic state, and the sign determines whether it is incoming (negative) or outgoing (positive).\footnote{This is a different convention from our previous section where $\omega$ could take either sign. We choose it to comply with recent literature on Liouville theory.}\newline\newline
{\textbf{One-to-one from the continuum.} We begin by considering the one-to-one amplitude in the continuum picture. For this, we require the expectation value for two vertex operator insertions on the two-sphere. Given the two-point function (\ref{2ptfunction}), and following the discussion in \cite{Erbin:2019uiz}, one finds
\begin{equation}
A_{S^2}(\omega_1|\omega_2) = 2 \pi  \omega_1 \delta(\omega_1-\omega_2) \times \left( \frac{g_s^2 c_{S^2}}{4\pi} \right)~,
\end{equation}
where $c_{S^2}$ is a coefficient which we will soon relate to the one-to-two amplitude. Comparing to (\ref{onetoone}) we fix $c_{S^2} = 4\pi/g_s^2$. 
\newline\newline
{\textbf{One-to-two from the continuum.} Next we consider three insertions on a genus zero surface
\begin{equation}\label{onetwoL}
A_{S^2}(\omega_1,\omega_2\,|\,\omega_3) \equiv \langle c\tilde{c} \mathcal{V}^+_{\omega_1}\, c\tilde{c} \mathcal{V}^+_{\omega_2}\, c\tilde{c} \mathcal{V}^-_{\omega_3} \rangle_{S^2}~.
\end{equation}
We follow the presentation in \cite{Balthazar:2017mxh}.
To evaluate the above expression requires knowledge of the three-point function for the Liouville operator $\mathcal{O}(\omega;z,\bar{z})$
\begin{equation}\label{eq:threepoint}
\langle \mathcal{O}(\omega_1; z_1,\bar{z}_1)\, \mathcal{O}(\omega_2; z_2,\bar{z}_2)\, \mathcal{O}(\omega_3; z_3,\bar{z}_3)  \rangle = \frac{\mathcal{C}(\omega_1,\omega_2,\omega_3)}{|z_{12}|^{2+(\omega_1^2+\omega_2^2-\omega_3^2)/2}|z_{23}|^{2+(\omega_2^2+\omega_3^2-\omega_1^2)/2}|z_{31}|^{2+(\omega_3^2+\omega_1^2-\omega_2^2)/2}}~.
\end{equation}
This quantity was studied  by Dorn-Otto \cite{Dorn:1994xn} and Zamolodchikov-Zamolodchikov \cite{Zamolodchikov:1995aa} which gives it the name DOZZ formula. The structure constant $\mathcal{C}(\omega_1,\omega_2,\omega_3)$ is given by
\begin{align}\label{eq:generalDOZZ}
\mathcal{C}(\omega_1,\omega_2,\omega_3)&\equiv \left(S(\omega_1)S(\omega_2)S(\omega_3)\right)^{-1}C(\omega_1,\omega_2,\omega_3)~,
\end{align}
with the DOZZ coefficient
\begin{multline}\label{eq:DOZZunnormalized}
C(\omega_1,\omega_2,\omega_3)= \,\frac{\Upsilon'_1 (0)}{\Upsilon_1(1+i(\omega_3+\omega_2+\omega_1)/2)}\frac{\Upsilon_1(2+i\omega_1)}{\Upsilon_1(1+i(\omega_1+\omega_2-\omega_3)/2)}\cr
\times\,\frac{\Upsilon_1(2+i\omega_2)}{\Upsilon_1(1+i(\omega_2+\omega_3-\omega_1)/2)}\frac{\Upsilon_1(2+i\omega_3)}{\Upsilon_1(1+i(\omega_3+\omega_1-\omega_2)/2)}~.
\end{multline}
In the literature the three-point function contains an additional phase has been absorbed in a shift of the Liouville field. The function $\Upsilon_b(z)$ is a holomorphic function that admits an integral expression as
\begin{equation}
 \log\Upsilon_b (z)= \int_0^\infty\frac{\dd t}{t}\left[\left(\frac{Q}{2}-z\right)^2e^{-t}- \frac{\sinh^2\left[\left(\frac{Q}{2}-z\right)\frac{t}{2}\right]}{\sinh \frac{tb}{2}\sinh\frac{t}{2b}}\right]~,
\end{equation}
where the real part of $z$ is restricted to the interval $z \in (0, Q)$ and $b+b^{-1}=Q$. To evaluate $\Upsilon_b (z)$ at complex values of $z$ we must analytically  continue the above expression. It is useful to note that we can express $\Upsilon_b(z)$  in terms of the Barnes double Gamma-function $\Gamma_2(z\,|\, b_1,b_2)$:
\begin{align}\label{eq:UpsilondoubleGamma}
\Upsilon_b(z)= \frac{1}{\Gamma_b(z)\Gamma_{b}(Q-z)}~,\quad\quad \Gamma_b(z)\equiv \frac{\Gamma_2(z\,|\, b, b^{-1})}{\Gamma_2(Q/2\,|\, b,b^{-1})}~.
\end{align}
In particular we have $\Gamma_2(z\,|\, 1,1)= (2\pi)^{z/2}/G(z)$ with $G(z)$ being the Barnes $G$-function. $\Upsilon_b(z)$ also satisfies
\begin{equation}\label{eq:propertiesUpsilon}
\Upsilon_b(z+b)= \gamma(bz) \, b^{1-2b z}\, \Upsilon_b(z)~,\quad \Upsilon_b(z)= \Upsilon_{b^{-1}}(z)~,\quad \Upsilon_b(Q-z)= \Upsilon_b(z)~,
\end{equation}
with $\gamma(z)\equiv \Gamma(z)/\Gamma(1-z)$. The phase $S(\omega)$ in (\ref{eq:phasefactor}) can be expressed as $S(\omega)= \gamma(i\omega)^{1/2}\gamma(1+i\omega)^{1/2}$. 

Additionally (\ref{tachyon}) contains a piece stemming from the operator $e^{\pm i \omega X^0}$. The structure constants for this piece (as may be familiar from the critical string) will not produce non-trivial functions of the $\omega_i$. Nevertheless, due to the symmetry $X^0 \to X^0 + a$, it yields a delta-function imposing the conservation of target space energy. 

At this point we have asembled the tools required to evaluate  (\ref{onetwoL}). Using the relations (\ref{eq:propertiesUpsilon}) for $(Q,b)=(2,1)$ we can rewrite $\Upsilon_1(2+i\omega_k)$ as 
\allowdisplaybreaks
\begin{equation}
\Upsilon_1(2+i\omega_k)= \Upsilon_1(2+i\omega_k)^{1/2}\Upsilon_1(2+i\omega_k)^{1/2}= \Upsilon_1(-i\omega_k)^{1/2}\Upsilon_1(i\omega_k)^{1/2}\gamma(i\omega_k)^{1/2}\gamma(-i\omega_k)^{-1/2}~,
\end{equation}
with $\gamma(x)$ defined below (\ref{eq:propertiesUpsilon}) satisfying in particular $\gamma(1+i\omega_k)= \gamma(-i\omega_k)^{-1}$. 
This way we obtain for (\ref{eq:generalDOZZ})
\begin{align}
\mathcal{C}(\omega_1,\omega_2,\omega_3)&=\frac{1}{\Upsilon_1(1+i(\omega_1+\omega_2+\omega_3)/2)} \left(\frac{ \Upsilon_1(i\omega_1)^{1/2}\Upsilon_1(-i\omega_1)^{1/2}}{\Upsilon_1(1+i(\omega_1+\omega_2-\omega_3)/2)}\times {\mathrm{2~perm.}}\right) \cr
&=\frac{1}{\Upsilon_1(1+i(\omega_1+\omega_2+\omega_3)/2)} \left(\frac{ \omega_1 \Upsilon_1(1+i\omega_1)}{\Upsilon_1(1+i(\omega_1+\omega_2-\omega_3)/2)}\times {\mathrm{2~perm.}}\right)~,
\end{align}
where we made use of (\ref{eq:UpsilondoubleGamma}) to evaluate $\Upsilon_1'(0)=1$ and in going to the second line we used
\begin{align}
\Upsilon_1(-i\omega_k)= \Upsilon_1(2+i\omega_k)= \gamma(1+i\omega_k)\Upsilon_{1}(1+i\omega_k)~.
\end{align}
Putting everything together, and in particular applying the energy conservation obtained from the delta function as well as $\Upsilon_1(1)=1$, we have that   
\begin{equation}
A_{S^2}(\omega_1, \omega_2\,|\,\omega_3) = i\,c_{S^2} \, g_s^3 \, \delta(\omega_1 + \omega_2 - \omega_3) \mathcal{C}(\omega_1,\omega_2,\omega_3)=  i   \, c_{S^2}  \, g_s^3 \, \delta(\omega_1 + \omega_2 - \omega_3)\,\omega_1\omega_2\omega_3~,
\end{equation}
where $c_{S^2}$ is a non-vanishing constant. Comparing to (\ref{eq:1to2final}), we see that the amplitude matches the leading term in the large $\nu_F$ expansion \cite{Klebanov:1991qa, Ginsparg:1993is, Polchinski:1994mb}, so long as we identify $1/|\nu_F| = c_{S^2} g_s^3 = 4\pi g_s$. 

The terms (\ref{eq:1to2final}) which are subleading in $1/\nu_F$ are predictions about higher genus contributions to the one-to-two scattering process in the continuum picture. We note that this identification is consistent with the powers of $\nu_F$ appearing in the various scattering amplitudes computed in section \ref{multiparticlesection}. A scattering amplitude on a Riemann surface of genus $h$ and $n$ vertex operator insertions scales as $g_s^{2h-2+n}$. 
A similar picture holds for other perturbative scattering amplitudes. 

Further remarkable comparisons between the quantum mechanical matrix model $S$-matrix and Liouville theory have been obtained more recently in \cite{Balthazar:2017mxh}. It is interesting to note that the whole structure of the perturbative string amplitudes is encoded in the reflection coefficient (\ref{eq:Rpurephase}). Though manifest from the quantum mechanical matrix model, the explicit form of $R_\nu$ is far from obvious from the Liouville perspective.
\begin{center}\it{Two-dimensional target space picture} \end{center}

To end the section, we would like to make some remarks about the target space picture of the non-critical string theory (\ref{NCworldsheet}). For $\varphi\to-\infty$ the exponential interaction of the Liouville mode is switched off, and the theory is approximated by the combination of a free scalar and a linear dilaton CFT. From this perspective, at least in the limit of large and negative $\varphi$, we have a two-dimensional target space with a running dilaton $\Phi$. The free scalar encodes the time direction of the target space, whereas $\varphi$ encodes the spatial direction. The target space metric and dilaton field are given by
\begin{equation}\label{target}
\frac{\dd s^2}{\alpha'} = -(\dd X^0)^2 + \dd\varphi^2~, \quad\quad\quad \Phi(X^0,\varphi) = 2\varphi~,
\end{equation}
such that the space-dependent string coupling $g_s = e^{\Phi}$ becomes parameterically weak as $\varphi\to-\infty$ and increasingly strong in the positive $\varphi$ region. We also have reinstated the string tension $\alpha'$. 

Due to the space-dependence, the vacuum is not Poincar{\'e} invariant. As $\varphi$ increases, so does the string coupling, and we moreover enter a regime where the Liouville interaction becomes significant. This regime is often referred to as the {\it Liouville wall} in the literature. As was mentioned, the spectrum of a string embedded in a two-dimensional target space consists of a single propagating degree of freedom, which is the `tachyon field' $\tau$. It is the target space field corresponding to the vertex operator (\ref{tachyon}). In two-dimensions, $\tau$ is in fact a massless field, so tachyon is somewhat of a misnomer. From the target space perspective, the $S$-matrix is given by sending an incoming tachyon wave from $\varphi \to -\infty$, which reflects off the Liouville wall, finally heading back to $\varphi \to -\infty$. Finally, we note that the background (\ref{target}) is intrinsically stringy since the slope of $\Phi$ is the string length.

\section{Further Developments}

The goal of this section is to mention some further examples in the multiverse of large $N$ theories. The section will be brief and the selection of topics is in no sense comprehensive. 

\subsection{Supersymmetric quantum mechanical matrices$^{\color{magenta}{I}}$}

One of the main tools that has allowed for tremendous progress in many corners of modern theoretical physics, ranging from quantum field theory and mathematics all the way to string theory and black hole physics, is supersymmetry. Here we mention the supersymmetric generalisation of some quantum mechanical matrix models. A supersymmetric quantum mechanical theory has a Hamiltonian of the following form
\begin{equation}
\{  \hat{\mathcal{Q}}^A,  \hat{\mathcal{Q}}^\dag _B \} = 2 {\delta^A}_B \hat{H}_{\text{SUSY}}~, \quad\quad \{  \hat{\mathcal{Q}}^A,  \hat{\mathcal{Q}} _B \}  = 0~,
\end{equation}
where the $\hat{\mathcal{Q}}^{A}$ are supercharges, with $A$ and $B$ being indices denoting additional symmetries or structures. In line with our previous considerations of matrix models, we further assume that the models have an $SU(N)$ symmetry, which we may or may not gauge. We take the bosonic elements of the theory to be as in our previously studied example, namely a collection of Hermitian $N\times N$ matrices $\hat{M}^a_{IJ}$, with $a = 1,2,\ldots,n_B$. In a supersymmetric theory, these are accompanied by adjoint valued fermionic operators $\hat{\Theta}^\alpha_{IJ}$, with $\alpha = 1,2,\ldots,n_F$, obeying the operator algebra
\begin{equation}
\{ \hat{\Theta}^\alpha_{IJ} , \hat{\Theta}^\beta_{KL} \} = \delta_{IK} \delta_{JL} \delta^{\alpha\beta}~.
\end{equation}
As a reality condition we consider $(\hat{\Theta}_{IJ}^\alpha)^\dag = \hat{\Theta}_{JI}^\alpha$. From the perspective of an action principle, the fermionic operators translate to Grassmann valued functions of time $\Theta_{IJ}(t)$ satisfying $(\Theta_{IJ}(t))^* = \Theta_{JI}(t)$. 
\newline\newline
{\textbf{Marinari-Parisi models.} The simplest class of models for a supersymmetric theory of quantum mechanical matrices was introduced by Marinari and Parisi \cite{Marinari:1990jc}. They are built from the off-shell field content $\mathcal{M} = \{{M}_{IJ},{\Psi}_{IJ},\bar{\Psi}_{IJ},F_{IJ}\}$, where $M_{IJ}$ and $F_{IJ}$ are Hermitian matrices, while $\Psi_{IJ} = \Theta_{IJ}^1 + i \Theta^2_{IJ}$ and $\bar{\Psi}_{IJ} = \Theta_{IJ}^1 - i \Theta^2_{IJ}$ are Grassmann valued complex matrices. The action reads 
\begin{equation}\label{SUSYaction}
S_{\text{MP}}[M(t)] = N \, \Tr \int_{\mathbb{R}} \dd t \left(  \frac{1}{2} \dot{M}^2 + i \bar{\Psi} \dot \Psi - |W(M)|^2 -  \bar{\Psi} \,\partial_M W(M) \, \Psi \right)~,
\end{equation}
where we have integrated out the auxiliary field $F_{IJ}$, and $W(M)$ is the superpotential, which is itself a Hermitian matrix. 

The models (\ref{SUSYaction}) are distinguished from those built from a single matrix $M_{IJ}$ in an important way. Recall that the single matrix models admitted a description for the ground state sector that could be reduced to an $N$ variable (rather than $N^2$) problem.  Moreover, had we considered gauging the $SU(N)$ symmetry of the single matrix models we would end up again with $N$ degrees of freedom. In the case at hand, $\Psi_{IJ}$ adds additional degrees of freedom, and theories of the type (\ref{SUSYaction}) can no longer be mapped to a simple eigenvalue problem even upon gauging the $SU(N)$. There are not enough symmetries. This conclusion remains true for essentially any model comprised of two or more interacting matrices and a single $SU(N)$ symmetry, regardless of whether they are fermionic or bosonic. Thus, new techniques are required to understand such a situation. 

To our knowledge, a complete description for the worldsheet theory describing the continuum limit of Marinari-Parisi models remains unknown. We now discuss a supersymmetric quantum mechanical matrix model for which a clearer picture of the emergent worldsheet is available. 
\newline\newline
{\textbf{D0-brane quantum mechanical matrices.} An important supersymmetric quantum mechanical matrix model is the D0-brane quantum mechanical model. The model enjoys an $SO(9)$ global symmetry and its field content is given by an $SU(N)$ adjoint valued vector multiplet $\mathcal{V} = \{ M^a_{IJ}, \Theta^\alpha_{IJ}, A_{IJ} \}$. The index $a=1,2,\ldots,9$ is an $SO(9)$ vector index, while $\alpha=1,\ldots,16$ is a spinor index of the 16-dimensional representation of $SO(9)$. The field $A_{IJ}$ is a gauge field for the $SU(N)$ symmetry. The role of $A_{IJ}$ is solely to impose the gauge constraint, and carries no dynamics of its own. The system of interest has an interpretation stemming from considerations of type IIA superstring theory. It describes the low energy theory governing the open string sector of a collection of $N$ D0-branes \cite{Banks:1996vh}. The theory is governed by the action 
\begin{equation}\label{BFSS}
S_{\text{D0}} =  \frac{N}{\lambda^2} \, \text{Tr} \int_{\mathbb{R}} \dd t \left( \frac{1}{2} \dot{M}^a \dot{M}^a + \frac{i}{2} \Theta^T\dot{\Theta} + \frac{1}{4} [M^a, M^b]^2 -  \frac{1}{2} \Theta^T \Gamma^a [\Theta, M^a]\right)~,
\end{equation}
where we have set $A_{IJ}=0$, and must recall the gauge singlet constraint. The $\Gamma^a$ are 16$\times$16 dimensional matrices satisfying the Clifford algebra 
\begin{equation}
\{\Gamma^a, \Gamma^b\} = 2 \, \delta^{ab}~. 
\end{equation}
In addition to the global $SO(9)$ symmetry, the above model has sixteen supercharges. The dimensionful parameter $\lambda$ is the 't Hooft coupling of the theory. The theory is effectively weakly coupled at high energies, and strongly coupled at low energies. 

There is a remarkable proposal originally developed by Banks, Fischler, Shenker, and Susskind (BFSS) relating the above theory in the large $N$ limit to quantum gravity in the ten-dimensional geometry residing in the vicinity of a stack of $N$ D0-branes \cite{Banks:1996vh}. The string frame metric and dilaton of this geometry are described by the $SO(9)$ invariant solution
\begin{equation}\label{D0metric}
\frac{\dd s^2}{\alpha'} = - \frac{u^{7/2} \dd t^2 }{\sqrt{a\lambda}} + \sqrt{a\lambda} \left( \frac{\dd u^2}{u^{7/2}} + \frac{\dd\Omega_8^2}{u^{3/2}} \right) ~, \quad\quad e^\Phi = \frac{(2\pi)^2 a^{3/4}}{N} \left( {u}\, {\lambda^{-1/3}}\right)^{-21/4}~,
\end{equation}
with $a = 240 \pi^5$. In addition to the metric and dilaton, the solution contains $N$ units Ramond-Ramond two-form flux sourced by the D0-branes. It is also worth noting there have been several variations of the D0-brane model, such as the BMN model \cite{Berenstein:2002jq} which describes a massive deformation of (\ref{BFSS}), whereby the flat direction of mutually commuting $M^a$'s is lifted.  

A particular incarnation of the BFSS proposal \cite{Itzhaki:1998dd}, which is in the same spirit as AdS/CFT, claims that quantum gravity in the  background (\ref{D0metric}) is dual to the D0-brane quantum mechanics theory. If true, the proposal surpasses the $c>1$ barrier in a remarkable way. Unravelling the details of the BFSS proposal, by whatever means need be \cite{Anagnostopoulos:2007fw,Berenstein:2004kk,Wiseman:2013cda,Lin:2014wka,Anous:2017mwr,Filev:2015hia}, remains a fascinating challenge in string theory. 
\newline\newline
{\textbf{$d$-dimensional models?} 
One might ask about simple supersymmetric generalisations of (\ref{BFSS}) which have an $SO(d)$ symmetry. Let $\Gamma^a$ be the $s_d\times s_d$ dimensional generators of the $s_d$ dimensional, real representation of the Clifford algebra satisfying the Clifford relation 
\begin{align}\label{eq:clifford}
\{\Gamma^a,\Gamma^b\} = 2\, \delta^{ab}~, \quad\quad\quad a,b =  1,..,d~. 
\end{align} 
The $d$-dimensional Clifford algebra contains irreducible representations with dimension 
\begin{align}
s_d = \begin{cases} 2^{\left\lfloor \frac{d}{2}\right\rfloor}~, \quad\quad d= 0,1,2 \; \mathrm{mod}\; 8 \\
2^{\left\lfloor \frac{d}{2}\right\rfloor+1}~, \;\;\;\; \mathrm{otherwise} \label{eq:dimirrepClif}
\end{cases}
\end{align}
A simple degree of freedom counting informs us that any supersymmetric theory built from a vector multiplet $\mathcal{V} = \{ M^a_{IJ}, \Theta^\alpha_{IJ}, A_{IJ} \}$ requires $s_d = 2(d-1)$. When combined with (\ref{eq:dimirrepClif}), this implies that only $d=2,3,5,9$, and correspondingly $s_d=2,4,8,16$ are valid dimensions \cite{Baake:1984ie, Frohlich:1999zf}. Perhaps the difficulty in constructing models with $d>9$ is related to the absence of perturbative superstring theories in larger than ten spacetime dimensions.

\subsection{Two-dimensional Yang-Mills theory$^{\color{magenta}{J}}$}

Another natural class of matrix models consists of matrices living on a spacetime manifold $\mathcal{M}$, as opposed to being merely functions of time. An example of such a theory is Yang-Mills theory with $SU(N)$ gauge group. 

Perhaps the simplest example of such a theory is Yang-Mills theory in two spacetime dimensions, described by the Euclidean action
\begin{equation}
S_{\text{YM}} = \frac{1}{4g^2} \int_{\mathcal{M}} \dd^2x \sqrt{g} \,F_{ij} F^{ij}~,
\end{equation}
where $F_{ij}$ is the field strength for the gauge field $A_i$ and $\mathcal{M}$ is a two-dimensional manifold endowed with metric $g_{ij}$. The theory carries no local degrees of freedom, but it has a non-vanishing Hilbert space. The large $N$ loop equations for two-dimensional Yang-Mills theory have been studied in \cite{Kazakov:1983fn}.

If one quantises the theory on a spatial $S^1$, it can be shown that the theory is mapped to a quantum mechanical matrix model, but in this case the matrices are {\it unitary} rather than Hermitian \cite{Makeenko:1979pb, Migdal:1975zg, Rusakov:1990rs,Minahan:1993np}. Such unitary matrices admit a treatment similar to that of Hermitian matrices, except the resulting fermion theory now lives on the circle rather than the half-line. 

It has been proposed by Gross and Taylor \cite{Gross:1992tu} that two-dimensional Yang-Mills theory is described by a worldsheet theory in the large $N$ limit. Although much remains to be understood about this worldsheet theory, it seems clear that it is not of the Liouville type. 

\subsection{Chern-Simons theory $\&$ topological strings$^{\color{magenta}{J}}$}
Perhaps the second simplest quantum field theory built from adjoint matrices is the Chern-Simons theory in three-dimensions. The action of this theory is given by
\begin{equation}
S_{\text{CS}} = \frac{k}{4\pi} \text{Tr} \int_{\mathcal{M}} \left( A \wedge \dd A + \frac{2}{3} A\wedge A \wedge A \right)~.
\end{equation}
The above theory has the remarkable property of being a topological theory, insensitive to the geometric features of $\mathcal{M}$. When quantised on a compact spatial two-manifold the above theory has a finite dimensional Hilbert space. For concreteness we consider an $U(N)$ gauge group. 

The standard perturbative expansion of Chern-Simons theory is given by taking the level $k \in \mathbb{Z}$ to be large while keeping $N$ fixed. On the other hand, it is natural to explore the behaviour of Chern-Simons theory in the 't Hooft limit for which we take $N$ large while keeping $\lambda = N/(N+k)$ fixed. Indeed, as argued by Gopakumar and Vafa \      \cite{Periwal:1993yu,Gopakumar:1998ki}, the 't Hooft limit of Chern-Simons theory on a three-sphere with $U(N)$ gauge group is captured by the topological closed string on a resolved conifold background. The target space of the worldsheet theory has a Ramond-Ramond two-form $B$, a K{\"a}hler two-form $J$, and string coupling $g_s$.  These are related to the Chern-Simons parameters by 
\begin{equation}
\lambda = -\frac{1}{2\pi} \int_{S^2} \left( B + i J \right)~, \quad\quad g_s = \frac{2\pi \lambda}{N}~. 
\end{equation}
Notice that although $\lambda$ was originally a real number, the above identifications require extending its domain to parts of the complex plane. It is also worth noting that the partition function of Chern-Simons theory can be mapped to an ordinary matrix intergral \cite{Marino:2002fk}, rather than a quantum mechanical matrix model.

It is worth noting that the topological string is substantially different than the Liouville string discussed in the previous section. Moreover, the duality between Chern-Simons and the topological string does not require tuning the 't Hooft coupling $\lambda$ to some critical value. 

\subsection{Fermionic quantum mechanical matrices$^{\color{magenta}{K}}$}

When discussing supersymmetric models we introduced a new set of fermionic operators $\hat{\Psi}^\alpha_{IJ} = \hat{\Theta}^{\alpha}_{IJ}+i \, \hat{\Upsilon}^{\alpha}_{IJ}$. This leads to yet another class of quantum mechanical matrix models built entirely from the fermions. As for Chern-Simons theory quantised on a compact Riemann surface, the Hilbert space of these modes is finite dimensional for any finite value of $N$ -- namely $\dim \mathcal{H}_N = 2^{N^2}$. The finiteness of the Hilbert space distinguishes these models from their bosonic and supersymmetric cousins. A simple class of models \cite{Anninos:2015eji,Anninos:2016klf,Tierz:2017nvl,Azeyanagi:2017drg,Gaitan:2020zbm} are described by a quartic Hamiltonian of the general type
\begin{equation}\label{FM}
\hat{H}_{\text{F}} = \frac{N}{\lambda} \, \text{Tr} \, \hat{\Psi}^\dag \Gamma^a \hat{\Psi} \, \hat{\Psi}^\dag \Gamma^a \hat{\Psi}~,
\end{equation}
where as before the $\Gamma^a$ are the $s_d \times s_d$ dimensional $\Gamma^a$ matrices. 

Although the wordsheet description capturing the 't Hooft limit of the above models remains unknown, simple versions of the above model exhibit features that bear some similarity to those encountered in Chern-Simons theory, as well as two-dimensional Yang-Mills theory. The models (\ref{FM}) have been considered both for a gauged as well as ungauged  $SU(N)$ symmetry.


\subsection{Melonic expansions: SYK $\&$ tensor models$^{\color{magenta}{M}}$}

Recently, a novel type of large $N$ limit has been exploited to analyse systems which have several features of interest  from the perspective of the physics of black holes, as well as condensed matter systems. One unusual ingredient in these models, known as quenched disorder, is most often used in the context of glassy physics. Quenched disorder treats the couplings of the theory as random variables drawn from an ensemble. The simplest class of Hamiltonian's considered \cite{Kitaev:2017awl,Polchinski:2016xgd,Sachdev:2015efa,Maldacena:2016hyu} are again built purely from real quantum mechanical fermions $\hat{\Theta}_I$ satisfying the algebra $\{\hat{\Theta}_I,\hat{\Theta}_J \} = \delta_{IJ}$, with $I=1,\ldots,N$. They are
\begin{equation}\label{SYKH}
\hat{H}_{\text{SYK}} = \frac{i^{q/2}}{q!} \mathcal{J}_{I_1\ldots I_q} \hat{\Theta}_{I_1} \hat{\Theta}_{I_2} \ldots \hat{\Theta}_{I_q}~,
\end{equation}
with totally anti-symmetric real couplings $\mathcal{J}_{I_1\ldots I_q}$ drawn from a random Gaussian ensemble with variance 
\begin{equation}\label{probJ}
\langle \mathcal{J}_{I_1\ldots I_q} \mathcal{J}_{I'_1\ldots I'_q} \rangle = \frac{\mathcal{J}^2 (q-1)!}{N^{q-1}} \times \delta_{I_1\ldots I_q,I'_1\ldots I'_q}\,~.
\end{equation}
Notice that the theory is built from vector-like constituents, and the Hilbert space is $2^N$ dimensional. As such, the models do not exhibit a `t Hooft expansion, at least in the standard sense. Nevertheless, the perturbative structure exhibits an interesting structure at large $N$, often referred to as a {\it melonic} expansion. These theories have no continuous global symmetries, and as such bear some relation to matrix models with the $SU(N)$ ungauged. 

We can analyse the diagrammatic expansion most easily by expressing the theory in terms of an action. For any given realisation of the couplings, the action is given by
\begin{equation}
S_{\text{SYK}}[\Theta_I;\mathcal{J}_{I_1\ldots I_q}] = \int_{\mathbb{R}} \dd t \left(  \frac{i}{2} \Theta_I \dot{\Theta}_I  - \frac{i^{q/2}}{q!}  \mathcal{J}_{I_1\ldots I_q} {\Theta}_{I_1} {\Theta}_{I_2} \ldots {\Theta}_{I_q} \right)~.
\end{equation}
To build the diagrams for the theory with randomised couplings, one resorts to the replica trick whereby the theory is replicated to $n \in \mathbb{Z}$ copies and $n$ is then allowed to take real values. This adds an additional `replica' index to the degrees of freedom. Assuming that the symmetry permuting the replicas is unbroken at large $N$, one can simply treat the $\mathcal{J}_{I_1\ldots I_q}$ as additional non-dynamical fields whose propagator is dictated by the Gaussian probability distribution (\ref{probJ}). At large $N$, a particular class of diagrams dominate. This class is displayed in the figure below. 
\newline
\begin{figure}[H]
\begin{center}
{
\begin{tikzpicture}[scale=1.2, rotate=90]
\draw (0,0) circle [radius=0.5cm];
\draw (0,-0.5) to[out=150,in=210] (0,0.5);
\draw [fill] (0,.5) circle [radius=0.03];
\draw [fill] (0,-.5) circle [radius=0.03];
\draw (0,-0.5) to[out=35,in=-35] (0,0.5);
\end{tikzpicture}
}
{}
{
\begin{tikzpicture}[scale=1.2, rotate=90]
\draw (0,0) circle [radius=0.5cm];
\draw (0.5,0) circle [radius=0.18cm];
\draw (-0.5,0) circle [radius=0.18cm];
\draw [fill] (0.46,.18) circle [radius=0.03];
\draw [fill] (0.46,-.18) circle [radius=0.03];
\draw [fill] (-0.46,.18) circle [radius=0.03];
\draw [fill] (-0.46,-.18) circle [radius=0.03];
\end{tikzpicture}
}
{}
{
\begin{tikzpicture}[scale=1.2, rotate=90]
\draw (0,0) circle [radius=0.5cm];
\draw (0.5,0) circle [radius=0.18cm];
\draw (-0.5,0) circle [radius=0.18cm];
\draw (0.7,0) circle [radius=0.12cm];
\draw [fill] (0.65,.10) circle [radius=0.03];
\draw [fill] (0.65,-.10) circle [radius=0.03];
\draw [fill] (0.46,.18) circle [radius=0.03];
\draw [fill] (0.46,-.18) circle [radius=0.03];
\draw [fill] (-0.46,.18) circle [radius=0.03];
\draw [fill] (-0.46,-.18) circle [radius=0.03];
\end{tikzpicture}
}
{}
{
\begin{tikzpicture}[scale=1.2, rotate=90]
\draw (0,0) circle [radius=0.5cm];
\draw (0,-0.5) to[out=150,in=210] (0,0.5);
\draw [fill] (0,.5) circle [radius=0.03];
\draw [fill] (0,-.5) circle [radius=0.03];
\draw (0,-0.5) to[out=35,in=-35] (0,0.5);
\draw (0.5,0) circle [radius=0.18cm];
\draw (-0.5,0) circle [radius=0.18cm];
\draw [fill] (0.46,.18) circle [radius=0.03];
\draw [fill] (0.46,-.18) circle [radius=0.03];
\draw [fill] (-0.46,.18) circle [radius=0.03];
\draw [fill] (-0.46,-.18) circle [radius=0.03];
\end{tikzpicture}
}
{}
{
\begin{tikzpicture}[scale=1.2]
\draw (0,0) circle [radius=0.5cm];
\draw (0.45,0.2) circle [radius=0.18cm];
\draw (-0.45,0.2) circle [radius=0.18cm];
\draw (0,-0.5) circle [radius=0.18cm];
\draw [fill] (0.36,.36) circle [radius=0.03];
\draw [fill] (-0.36,.36) circle [radius=0.03];
\draw [fill] (-0.5,.03) circle [radius=0.03];
\draw [fill] (0.5,.03) circle [radius=0.03];
\draw [fill] (0.18,-.46) circle [radius=0.03];
\draw [fill] (-0.18,-.46) circle [radius=0.03];
\end{tikzpicture}
}
{}
{
\begin{tikzpicture}[scale=1.2, rotate=90]
\draw (0,0) circle [radius=0.5cm];
\draw (0,-0.5) to[out=150,in=210] (0,0.5);
\draw [fill] (0,.5) circle [radius=0.03];
\draw [fill] (0,-.5) circle [radius=0.03];
\draw (0,-0.5) to[out=35,in=-35] (0,0.5);
\draw (0.45,0.2) circle [radius=0.18cm];
\draw (-0.45,0.2) circle [radius=0.18cm];
\draw (0.17,-0.25) circle [radius=0.15cm];
\draw [fill] (0.22,-.12) circle [radius=0.03];
\draw [fill] (0.12,-.38) circle [radius=0.03];
\draw [fill] (0.36,.36) circle [radius=0.03];
\draw [fill] (-0.36,.36) circle [radius=0.03];
\draw [fill] (-0.5,.03) circle [radius=0.03];
\draw [fill] (0.5,.03) circle [radius=0.03];
\end{tikzpicture}
}
{}
{
\begin{tikzpicture}[scale=1.2]
\draw (0,0) circle [radius=0.5cm];
\draw (0.4,0.3) circle [radius=0.18cm];
\draw (-0.4,0.3) circle [radius=0.18cm];
\draw (0.4,-0.3) circle [radius=0.18cm];
\draw (-0.4,-0.3) circle [radius=0.18cm];
\draw [fill] (0.27,.43) circle [radius=0.03];
\draw [fill] (-0.27,.43) circle [radius=0.03];
\draw [fill] (-0.47,.13) circle [radius=0.03];
\draw [fill] (0.47,.13) circle [radius=0.03];
\draw [fill] (0.27,-.43) circle [radius=0.03];
\draw [fill] (-0.27,-.43) circle [radius=0.03];
\draw [fill] (-0.47,-.13) circle [radius=0.03];
\draw [fill] (0.47,-.13) circle [radius=0.03];
\end{tikzpicture}
}
\end{center}
\caption{Melon diagrams for $q=4$.}
\end{figure}
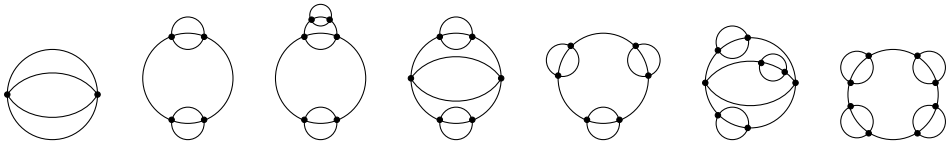
The leading $N$ contribution to the correlation functions can be recast in the form of a Schwinger-Dyson equation which, in turn, exhibits a one-dimensional conformal symmetry at sufficiently large time separations. The conformal symmetry is the space of maps from $\mathbb{R}$ (or $S^1$ when the theory is placed at finite temperature) to itself. It is suggestive of a two-dimensional anti-de Sitter spacetime black hole holographic dual. This interpretation is further supported by considerations of the symmetry breaking low energy sector which is governed by a Schwarzian mode \cite{Maldacena:2016upp}.
It turns out that precisely the same diagrams are those that organise a class of models known as tensor models, which are built from objects with multiple indices \cite{Witten:2016iux, Klebanov:2016xxf}. Tensor models have been previously introduced \cite{Gurau:2010ba,Bonzom:2011zz} as an interesting generalisation to matrix models. The quantum mechanical tensor models do not have random coupling constants. 

The SYK and tensor models are neither matrix models nor vector models. Rather, they are an entirely different class of large $N$ theories. Unlike most quantum mechanical matrix models, they can be dealt with without imposing any gauge singlet constraint. This is perhaps related to their connection to black hole physics. Though not matrix models, it is interesting to note that both matrix integrals as well as two-dimensional gravity seem to stem from detailed considerations of these non-matrix models \cite{Cotler:2016fpe}.

\section{Outlook and speculative remarks}

We would like to end our discussion with a brief outlook, some questions, and speculative remarks. 
\newline\newline
\textbf{AdS/CFT $\&$ criticality.} Over the last two-decades there has been tremendous progress in the development of the AdS/CFT correspondence. This has led to wonderful insights into the nature of quantum gravity and black holes. At the same time, it is important to keep in mind how the various large $N$ models leading to gravity and string theories differ in their detailed form. For instance, many of the large $N$ models considered throughout this work lead to a continuum picture only upon tuning the parameters to certain critical values. At the critical point the Feynman diagrams in the `t Hooft expansion become densely filled Riemann surfaces. This type of criticality seems to be of a different nature than the underlying mechanism in the known examples of AdS/CFT. For instance in $\mathcal{N}=4$ SYM theory at large $N$ we have a continuum worldsheet picture for all values of the `t Hooft coupling $\lambda$. In this picture, different values of $\lambda$ correspond to different values of the AdS length in string units. It remains possible that upon extending $\lambda$ to the complex plane, one may encounter critical behaviour akin to that exhibited by the matrix models we have studied. Interestingly, complex values of $\lambda$ would correspond to complex values of the IIB string coupling. 

In the D0-brane picture the string coupling is space dependent and in some region becomes strongly coupled. This is reminiscent of the running dilaton in the two-dimensional target space picture of the non-critical string. From the D0-brane quantum mechanics perspective, this is understood from the fact that the `t Hooft coupling is dimensionful and thus flows. As such it cannot be tuned to some critical value. It seems less clear for the case of the D0-brane quantum mechanics theory that there is an analogue for tuning the level of the Fermi sea we saw for the quantum mechanical matrix model. 
\newline\newline
$\mathcal{Q}$: Does such critical behaviour of the microscopic model at large $N$ and the emergence of a continuum picture play a role for more general considerations of the emergence of spacetime or is it an artefact of the lower dimensional models?
\newline\newline
\textbf{Wigner's black holes.} In the introduction we mentioned Wigner's hypothesis that complex systems can be treated in statistical fashion. Though heavy nuclei were Wigner's main concern, from the perspective of gravity a natural candidate seems to be a black hole. Indeed, recent ideas motivated by the chaotic nature of the SYK system and its randomised couplings, as well as other considerations \cite{Saad:2019lba,Anninos:2013nra} might provide novel avenues to explore the idea of a Wignerian black hole. 

Another place where matrix integrals have recently played an important role in the context of black hole physics comes from applying ideas of supersymmetric localisation towards calculations of supersymmetric black hole partition functions \cite{Benini:2015eyy,Cabo-Bizet:2019eaf}. This context (at least naively \cite{Anninos:2016szt}) is conceptually different from the Wignerian case described above. Nevertheless, it is interesting to note yet again the appearance of random matrices.
\newline\newline
$\mathcal{Q}$: Is there a role for Wigner's hypothesis, namely that complex systems are well approximated by averaging ensembles of theories, in the theory of black holes?
\newline\newline
\textbf{Emergence of target space equations.} Throughout our discussion we observed how the large $N$ limit of certain matrix models admits a continuum description in terms of a non-critical worldsheet theory. When $c_m=1$, we noted that there is also a two-dimensional target space description. Das and Jevicki \cite{Das:1990kaa} as well as other authors \cite{Sengupta:1990bt, Gross:1990st, Polchinski:1990mf} have considered a construction mapping the degrees of freedom of the quantum mechanical matrix model directly to the target space picture. In such a picture, the target space equations, which include a two-dimensional version of the equations of general relativity coupled to matter, emerge directly from the matrix theory. From a different perspective, one can obtain the target space equations of motion by turning on background fields on the worldsheet and requiring conformality of the deformed theory \cite{Callan:1985ia}. 

There have been various attempts \cite{Lashkari:2013koa} to derive the emergent target space equations of asymptotically AdS spacetimes from a considerations of entanglement entropy, or more general thermodynamic ideas such as Jacobson's picture of the Einstein's equations as a local form of the first law of horizon thermodynamics \cite{Jacobson:1995ab}. 
\newline\newline
$\mathcal{Q}$: Is there a relation among the different approaches for obtaining the target space equations?
\newline\newline
\textbf{New probes for old matrix models.$^{\color{magenta}{L}}$} Recently, there have been several new probes used to assess properties of black holes and spacetime. Several of these has focused on the amount of entanglement between two regions of spacetime  \cite{Ryu:2006bv}. At least in asymptotically anti-de Sitter spacetimes, there are clear indications that even for a piece of spacetime in the vacuum, the amount of entanglement between two regions scales in a similar way to the Bekenstein-Hawking entropy. This is a vast amount of entanglement which cannot be captured by the semi-classical Hilbert space of gravity. 
\newline\newline
$\mathcal{Q}$: Are these probes pertinent to the two-dimensional non-critical string and black hole \cite{Das:1995vj, Hartnoll:2015fca, Donnelly:2019zde}? Is there a generalisation/replacement of the Ryu-Takayanagi entropy for the holographic dual of the D0-brane quantum mechanics \cite{Anous:2019rqb,1789261} or more general non-AdS spacetimes?
\newline\newline
\textbf{Non-perturbative phenomena: Matrix integrals.$^{\color{magenta}{C, H}}$} The matrix models discussed throughout exhibit interesting non-perturbative features. For instance, the perturbative expansion of solutions to the non-linear differential equations discussed in section \ref{singlematrixii} and \ref{section2MM} may contain exponentially suppressed terms capturing non-perturbative behaviour. In \cite{Eynard:1992sg} an equation determining the coefficients in the non-perturbative exponent for general multicritical matrix models has been presented. This equation implies, in particular, Borel summablility for some of the perturbative expansions. 
\newline\newline
$\mathcal{Q}$: The relation between non-perturbative effects of matrix models and the corresponding Liouville theory coupled to unitary minimal models has been explored in \cite{Alexandrov:2003nn}. Can the non-perturbative effects observed for multicritical matrix models be related to the corresponding  Liouville theory coupled to a non-unitary minimal model, particularly in the limit of large and negative central charge? 
\newline\newline
\textbf{Non-perturbative phenomena \& the non-singlet sector: MQM$.^{\color{magenta}{H, L}}$} 
The quantum mechanical matrix model discussed in sections \ref{sec5} and \ref{sec6} also exhibits interesting non-perturbative features. For instance, depending on how we fill the Fermi sea on the other side of the barrier, the perturbative $S$-matrix may admit different non-perturbative completions. At a more practical level, it is interesting to explore non-perturbative corrections to the perturbative $S$-matrix from the continuum string perspective. Although these corrections may disrupt the unitarity of the perturbative $S$-matrix, they may do so in a calculable form. For instance, as recently explored in \cite{Balthazar:2019rnh,Balthazar:2019ypi,Sen:2020oqr}, ZZ-instantons may capture certain non-perturbative contributions. Moreover, when the eigenvalue potential is not infinitely deep, the half-filled Fermi sea will be metastable. Any precise microscopic description of a `metastable string vacuum' might be viewed as encouraging given that many vacua of interest in more realistic string theories may well be metastable \cite{Silverstein:2001xn,Kachru:2003aw} (see also \cite{Zamolodchikov:2006xs}). 

Finally, in \ref{sec5} and \ref{sec6} we have only focused on the singlet sector of the quantum mechanical matrix model. 
Much less is known about the non-singlet sector,  although it has been argued \cite{Maldacena:2005hi} and recently re-examined \cite{Balthazar:2018qdv} that the adjoint sector in the matrix quantum mechanics is dual to a long string with specific boundary conditions corresponding to FZZT branes. More generally, one could study the implications of releasing other non-singlet sectors, particularly with regard to the problem of black holes \cite{Witten:1991yr,Mandal:1991tz,Kazakov:2000pm}.
\newline\newline
$\mathcal{Q}$: Can the metastability of the quantum mechanical matrix model and the instantons mediating it sharpen our holographic understanding of the decay of D-branes \cite{Sen:2002nu} and more realistic metastable string vacua? Is there a target picture of the general non-singlet sector?
\newline\newline
\textbf{Quantum mechanics $\&$ worldline holography.} Throughout our discussion we focused on large $N$ matrix integrals and quantum mechanical matrix models rather than large $N$ quantum field theories. There are several differences between quantum mechanical and quantum field theoretic models. For instance, a quantum mechanical model has a finite number of (non-locally) interacting degrees of freedom, and potentially even a finite dimensional Hilbert space. Aside from topological field theories, continuum quantum field theory has an infinite dimensional Hilbert space and number of degrees of freedom. In quantum mechanics, gauging a symmetry is nothing more than imposing a constraint on the space of states, whereas in quantum field theory one generally introduces additional degrees of freedom. As has been recently emphasised \cite{Anninos:2017hhn,Anninos:2018svg,Gross:2019ach,Gross:2019uxi}, quantum mechanical models can be coupled to a worldine gravity and the quantisation of the resulting `gravitational' theory remains straightforward. This might still be so in two-dimensional quantum field theory, but becomes increasingly complicated in higher dimensions. Finally, quantum mechanical theories are distinguished from quantum field theories in the sense that the degrees of freedom are not required to interact in a local/nearest neighbor type way. From a Wilsonian perspective, in the absence of locality we are confronted with the challenge of which degrees of freedom to integrate out in large $N$ quantum mechanical models to capture the low energy effective theory. 
\newline\newline
$\mathcal{Q}$: Are there important differences from the perspective of the emergent worldsheet theory/ holographic dual when the microscopic theory is quantum mechanical rather than quantum field theoretic?
\newline\newline
\textbf{Finiteness $\&$ cosmology?$^{\color{magenta}{N}}$} One of our motivations for considering these topics is related to cosmological spacetimes, and in particular ones with positive cosmological constant \cite{Anninos:2012qw}. In cosmology one often considers spacetimes with compact Cauchy surfaces, and consequently no asymptotic spatial boundary. 
From this perspective, the two-dimensional models of gravity coupled to minimal models seem to have the desired property of being more `finite' examples of quantum gravity, albeit in a two-dimensional world. They can be quantised on a compact Cauchy surface and the constraints of gravity impose severe restrictions on the allowed space of states, rendering it essentially finite.

Moreover, the two-dimensional quantum gravity path integrals discussed in \ref{LiouvilleSec}  make sense for positive (rather than negative) cosmological constant. From the large $N$ matrix model perspective, this manifests itself in the appearance of branch cut ambiguities as we cross from $\alpha>\alpha_c$ to $\alpha < \alpha_c$. It is interesting to note that from the perspective of the matrix model at fixed $N$, tuning $\Lambda\propto (\alpha-\alpha_c)$ corresponds to tuning the number of vertices of the triangulated Riemann surface. As $\Lambda$ increases, the number goes down. This bears some resemblance to the de Sitter horizon entropy \cite{Gibbons:1977mu} being inversely related to the cosmological constant. Interestingly, the Euclidean continuation of a two-dimensional de Sitter universe continues to the Euclidean two-sphere.  (It has been suggested that any microscopic description of de Sitter will involve a finite number of degrees of freedom or, more extremely, even a finite dimensional Hilbert space \cite{Goheer:2002vf}.)
Perhaps, then, the quantum gravity path integrals studied in section \ref{LiouvilleSec} on an $S^2$ topology are of some relevance  \cite{Polyakov:2007mm,Martinec:2003ka,Bautista:2015wqy, Betzios:2016lne,1789241}. We saw that the $S^2$ path integral of gravity coupled to conformal matter with $c_m < 1$ is dual to a large $N$ matrix integral.\footnote{Matrix integrals have also appeared in other considerations of quantum de Sitter. For instance, it was shown in \cite{Anninos:2017eib} that a matrix integral captures gauge invariant correlation functions in a four-dimensional de Sitter theory with an infinite tower of interacting massless higher spin fields. In \cite{Cotler:2019nbi,Cotler:2019dcj,Maldacena:2019cbz} matrix integrals were discussed in relation to a two-dimensional de Sitter solution of Jackiw-Teitelboim gravity.} It would be interesting to understand the Lorentzian interpretation (if any) of the matrix integral in terms of a Hilbert space and a collection of operators  directly from the matrix picture. One possible avenue may be to consider inserting a macroscopic loop operator $W_\ell$ introduced in section \ref{loopsec}. In the continuum limit, this creates a hole in the Riemann surface leading to a picture similar to that of Hartle and Hawking's wavefunction \cite{Hartle:1983ai}.\footnote{One might even imagine inserting multiple $W_\ell$'s creating multiple boundaries whose interpretation  \cite{Coleman:1988tj,Giddings:1988wv} seems less clear.} A cosmological time emerging from an underlying statistical/Euclidean model remains an elusive but fascinating idea \cite{Ishibashi:1996xs,Strominger:2001gp}.

As a final remark, connecting to the beginning of our discussion, we return to Wigner. There seems to be a certain degree of complexity in constructing de Sitter vacua in string theory. Perhaps this should be taken as a starting point, in the spirit of Wigner, and we could attempt to build a framework for de Sitter based on an ensemble of theories \cite{Denef:2011ee,Anninos:2011kh}.

\section*{Acknowledgements}

It is a great pleasure to acknowledge Jan Ambj{\o}rn, Tarek Anous, Panos Betzios, Joren Brunekreef, Frederik Denef, Dami\'{a}n Galante, Umut G{\"u}rsoy, Nick Halmagyi, Masanori Hanada, Sean Hartnoll, Diego Hofman, Zohar Komargodski, Alex Maloney, Olga Papadoulaki, Antonio Rotundo, Edgar Shaghoulian, Guillermo Silva, Eva Silverstein, Jeremy van der Heijden, Erik Verlinde, G{\'e}rard Watts, Peter West, and Xi Yin for many illuminating discussions. We are also very grateful to Vladimir Kazakov and Spenta Wadia for helpful comments. B.M. would like to thank the theoretical physics group at King's College London for their kind hospitality during the completion of this work. The work of B.M. is part of the research programme of the Foundation for Fundamental Research on Matter (FOM), which is financially supported by the Netherlands Organisation for Science Research (NWO). D.A. is funded by the Royal Society under the grant The Atoms of a deSitter Universe.

{\appendix



\section{$R_n$ staircase}\label{Rnstaircase}

In this appendix we discuss a method to relate the $R_n$ coefficients for the orthogonal polynomial problem to the $h_n$ coefficients. Explicitly we demonstrate how to obtain the integral on the right hand side of (\ref{eq:recursionRn}) in a graphical way. 

Obtaining all possible $R_n$ leading to an expression proportional to $h_n$ as outlined for the quartic potential in $\ref{nonplanar}$ can be tedious. 
For the case of an even degree polynomial there exists a graphical way of computing the $R_n$. Starting with a potential of degree $2p+2$ we have to combine powers of $\lambda$ of degree less or equal to $2p+1$ with polynomials $P_n$ and $P_{n-1}$ in all possible ways to get two degree $n$ polynomials. As a first step one identifies a certain level as the level $n-1$ (corresponding to the degree of $P_{n-1}$). Now we can perform exactly $p+1$ steps in an appropriate unit upwards -- conveniently chosen along 45 degrees -- followed by $p$ steps downwards in a right angle with respect to the highest point. The final level corresponds to level $n$.  
Going downwards mimics the three term recurrence relation and for each step downwards from a level $n$ to a level $n-1$ we get a factor $R_n$.
In total there are $\binom{2p +1}{p}$ different possibilities to start at level $n-1$ and end up at level $n$. Each combination is a product of $p$ $R_n$. 
\newline\newline
{\textbf{Example.}} We illustrate the procedure along the example $V(\lambda)\sim \lambda^6$. Here $p$=2 and we get $\binom{5}{2}=10$ combinations, each a product of two $R_n$.

\begin{figure}[H]
\subfloat[$R_{n+1}R_{n}$]
{
\begin{tikzpicture}[scale=0.7]
  \filldraw (-1,-1) circle[radius=1.5pt];
\filldraw (-0.5,-0.5) circle[radius=1.5pt];
\filldraw (0,0) circle[radius=1.5pt];
\filldraw (.5,-.5) circle[radius=1.5pt];
\filldraw (1,-1) circle[radius=1.5pt];
\filldraw (1.5,-0.5) circle[radius=1.5pt];
\node[left, outer sep=2pt, fill=white,scale=.8] at (-1,-1) {$n-1$};
\node[left, outer sep=2pt, fill=white,scale=.8] at (-0.5,-0.5) {$n$};
\node[left, outer sep=2pt, fill=white,scale=.8] at (0,0) {$n+1$};
\draw[black] (-1,-1) -- (-0.5,-0.5);
\draw[black] (-0.5,-0.5) -- (0,0);
\draw[black] (0,0) -- (.5,-.5);
\draw[black] (.5,-.5) -- (1,-1);
\draw[black] (1,-1) -- (1.5,-0.5); 
\end{tikzpicture}
}
\subfloat[$R_{n}R_{n-1}$]
{
\begin{tikzpicture}[scale=0.7]
\filldraw (-1,-1) circle[radius=1.5pt];
\filldraw (-0.5,-0.5) circle[radius=1.5pt];
\filldraw (0,-1) circle[radius=1.5pt];
\filldraw (.5,-1.5) circle[radius=1.5pt];
\filldraw (1,-1) circle[radius=1.5pt];
\filldraw (1.5,-0.5) circle[radius=1.5pt];
\node[left, outer sep=2pt, fill=white,scale=.8] at (-1,-1) {$n-1$};
\node[left, outer sep=2pt, fill=white,scale=.8] at (-0.5,-0.5) {$n$};
\node[left, outer sep=2pt, fill=white,scale=.8] at (.5,-1.5) {$n-2$};
\draw[black] (-1,-1) -- (-0.5,-0.5);
\draw[black] (-0.5,-0.5) -- (0,-1);
\draw[black] (0,-1) -- (.5,-1.5);
\draw[black] (.5,-1.5) -- (1,-1);
\draw[black] (1,-1) -- (1.5,-0.5);
\end{tikzpicture}
}
\subfloat[$R_{n}R_{n-1}$]
{
\begin{tikzpicture}[scale=0.7]
\filldraw (-1,-1) circle[radius=1.5pt];
\filldraw (-0.5,-1.5) circle[radius=1.5pt];
\filldraw (0,-1) circle[radius=1.5pt];
\filldraw (.5,-.5) circle[radius=1.5pt];
\filldraw (1,-1) circle[radius=1.5pt];
\filldraw (1.5,-0.5) circle[radius=1.5pt];
\node[left, outer sep=2pt, fill=white,scale=.8] at (-1,-1) {$n-1$};
\node[left, outer sep=2pt, fill=white,scale=.8] at (-0.5,-1.5) {$n-2$};
\node[left, outer sep=2pt, fill=white,scale=.8] at (.5,-.5) {$n$};
\draw[black] (-1,-1) -- (-0.5,-1.5);
\draw[black] (-0.5,-1.5) -- (0,-1);
\draw[black] (0,-1) -- (.5,-.5);
\draw[black] (.5,-.5) -- (1,-1);
\draw[black] (1,-1) -- (1.5,-0.5);
\end{tikzpicture}
}
\subfloat[$R_{n-1}R_{n+1}$]
{
\begin{tikzpicture}[scale=0.7]
\filldraw (-1,-1) circle[radius=1.5pt];
\filldraw (-0.5,-1.5) circle[radius=1.5pt];
\filldraw (0,-1) circle[radius=1.5pt];
\filldraw (.5,-.5) circle[radius=1.5pt];
\filldraw (1,0) circle[radius=1.5pt];
\filldraw (1.5,-0.5) circle[radius=1.5pt];
\node[left, outer sep=2pt, fill=white,scale=.8] at (-1,-1) {$n-1$};
\node[left, outer sep=2pt, fill=white,scale=.8] at (-0.5,-1.5) {$n-2$};
\node[left, outer sep=2pt, fill=white,scale=.8] at (.5,-.5) {$n$};
\node[left, outer sep=2pt, fill=white,scale=.8] at (1,0) {$n+1$};
\draw[black] (-1,-1) -- (-0.5,-1.5);
\draw[black] (-0.5,-1.5) -- (0,-1);
\draw[black] (0,-1) -- (.5,-.5);
\draw[black] (.5,-.5) -- (1,0);
\draw[black] (1,0) -- (1.5,-0.5);
\end{tikzpicture}
}
\subfloat[$R_{n-1}R_{n-2}$]
{
\begin{tikzpicture}[scale=0.7]
\filldraw (-1,-1) circle[radius=1.5pt];
\filldraw (-0.5,-1.5) circle[radius=1.5pt];
\filldraw (0,-2) circle[radius=1.5pt];
\filldraw (.5,-1.5) circle[radius=1.5pt];
\filldraw (1,-1) circle[radius=1.5pt];
\filldraw (1.5,-0.5) circle[radius=1.5pt];
\node[left, outer sep=2pt, fill=white,scale=.8] at (-1,-1) {$n-1$};
\node[left, outer sep=2pt, fill=white,scale=.8] at (-0.5,-1.5) {$n-2$};
\node[left, outer sep=2pt, fill=white,scale=.8] at (0,-2) {$n-3$};
\node[left, outer sep=2pt, fill=white,scale=.8] at (1.5,-.5) {$n$};
\draw[black] (-1,-1) -- (-0.5,-1.5);
\draw[black] (-0.5,-1.5) -- (0,-2);
\draw[black] (0,-2) -- (.5,-1.5);
\draw[black] (.5,-1.5) -- (1,-1);
\draw[black] (1,-1) -- (1.5,-0.5);
\end{tikzpicture}
}
\end{figure}

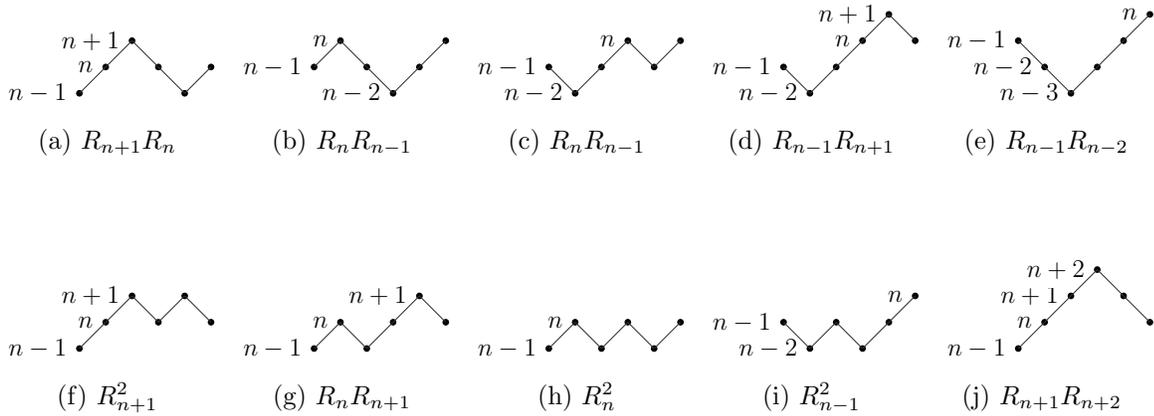
\begin{figure}[H]
\subfloat[$R_{n+1}^2$]
{
\begin{tikzpicture}[scale=0.7]
  \filldraw (-1,-1) circle[radius=1.5pt];
\filldraw (-0.5,-0.5) circle[radius=1.5pt];
\filldraw (0,0) circle[radius=1.5pt];
\filldraw (.5,-.5) circle[radius=1.5pt];
\filldraw (1,0) circle[radius=1.5pt];
\filldraw (1.5,-0.5) circle[radius=1.5pt];
\node[left, outer sep=2pt, fill=white,scale=.8] at (-1,-1) {$n-1$};
\node[left, outer sep=2pt, fill=white,scale=.8] at (-0.5,-0.5) {$n$};
\node[left, outer sep=2pt, fill=white,scale=.8] at (0,0) {$n+1$};
\draw[black] (-1,-1) -- (-0.5,-0.5);
\draw[black] (-0.5,-0.5) -- (0,0);
\draw[black] (0,0) -- (.5,-.5);
\draw[black] (.5,-.5) -- (1,0);
\draw[black] (1,0) -- (1.5,-0.5); 
\end{tikzpicture}
}
\subfloat[$R_{n}R_{n+1}$]
{
\begin{tikzpicture}[scale=0.7]
\filldraw (-1,-1) circle[radius=1.5pt];
\filldraw (-0.5,-0.5) circle[radius=1.5pt];
\filldraw (0,-1) circle[radius=1.5pt];
\filldraw (.5,-.5) circle[radius=1.5pt];
\filldraw (1,0) circle[radius=1.5pt];
\filldraw (1.5,-0.5) circle[radius=1.5pt];
\node[left, outer sep=2pt, fill=white,scale=.8] at (-1,-1) {$n-1$};
\node[left, outer sep=2pt, fill=white,scale=.8] at (-0.5,-0.5) {$n$};
\node[left, outer sep=2pt, fill=white,scale=.8] at (1,0) {$n+1$};
\draw[black] (-1,-1) -- (-0.5,-0.5);
\draw[black] (-0.5,-0.5) -- (0,-1);
\draw[black] (0,-1) -- (.5,-.5);
\draw[black] (.5,-.5) -- (1,0);
\draw[black] (1,0) -- (1.5,-0.5);
\end{tikzpicture}
}
\subfloat[$R_{n}^2$]
{
\begin{tikzpicture}[scale=0.7]
\filldraw (-1,-1) circle[radius=1.5pt];
\filldraw (-0.5,-0.5) circle[radius=1.5pt];
\filldraw (0,-1) circle[radius=1.5pt];
\filldraw (.5,-.5) circle[radius=1.5pt];
\filldraw (1,-1) circle[radius=1.5pt];
\filldraw (1.5,-0.5) circle[radius=1.5pt];
\node[left, outer sep=2pt, fill=white,scale=.8] at (-1,-1) {$n-1$};
\node[left, outer sep=2pt, fill=white,scale=.8] at (-0.5,-0.5) {$n$};
\draw[black] (-1,-1) -- (-0.5,-0.5);
\draw[black] (-0.5,-0.5) -- (0,-1);
\draw[black] (0,-1) -- (.5,-.5);
\draw[black] (.5,-.5) -- (1,-1);
\draw[black] (1,-1) -- (1.5,-0.5);
\end{tikzpicture}
}
\subfloat[$R_{n-1}^2$]
{
\begin{tikzpicture}[scale=0.7]
\filldraw (-1,-1) circle[radius=1.5pt];
\filldraw (-0.5,-1.5) circle[radius=1.5pt];
\filldraw (0,-1) circle[radius=1.5pt];
\filldraw (.5,-1.5) circle[radius=1.5pt];
\filldraw (1,-1) circle[radius=1.5pt];
\filldraw (1.5,-0.5) circle[radius=1.5pt];
\node[left, outer sep=2pt, fill=white,scale=.8] at (-1,-1) {$n-1$};
\node[left, outer sep=2pt, fill=white,scale=.8] at (-0.5,-1.5) {$n-2$};
\node[left, outer sep=2pt, fill=white,scale=.8] at (1.5,-.5) {$n$};
\draw[black] (-1,-1) -- (-0.5,-1.5);
\draw[black] (-0.5,-1.5) -- (0,-1);
\draw[black] (0,-1) -- (.5,-1.5);
\draw[black] (.5,-1.5) -- (1,-1);
\draw[black] (1,-1) -- (1.5,-0.5);
\end{tikzpicture}
}
\subfloat[$R_{n+1}R_{n+2}~$]
{
\begin{tikzpicture}[scale=0.7]
\filldraw (-0.5,-1.5) circle[radius=1.5pt];
\filldraw (0,-1) circle[radius=1.5pt];
\filldraw (.5,-.5) circle[radius=1.5pt];
\filldraw (1,0) circle[radius=1.5pt];
\filldraw (1.5,-0.5) circle[radius=1.5pt];
\filldraw (2,-1) circle[radius=1.5pt];
\node[left, outer sep=2pt, fill=white,scale=.8] at (-0.5,-1.5) {$n-1$};
\node[left, outer sep=2pt, fill=white,scale=.8] at (0,-1) {$n$};
\node[left, outer sep=2pt, fill=white,scale=.8] at (.5,-.5) {$n+1$};
\node[left, outer sep=2pt, fill=white,scale=.8] at (1,0) {$n+2$};
\draw[black] (-0.5,-1.5) -- (0,-1);
\draw[black] (0,-1) -- (.5,-.5);
\draw[black] (.5,-.5) -- (1,0);
\draw[black] (1,0) -- (1.5,-0.5);
\draw[black] (1.5,-.5) -- (2,-1);
\end{tikzpicture}
}
\caption{The ten possible combinations for $V(M)= M^6$.}
\end{figure}

\section{Further examples of multicritical models}\label{m4}

In this appendix we consider further examples of multicritical matrix models and some of their properties. 

Using the method outlined in the previous appendix, we can calculate the $\binom{7}{3}=35$ contribution to $V(M)=\frac{1}{\gamma}M^8$, each consisting of seven steps multiplying three $R_{n}$: 
\begin{multline}
\frac{\gamma}{8}\int \dd \lambda V'(\lambda)P_{n-1}(\lambda)P_n(\lambda)= \Big[R_{n+1}^3+ R_{n-1}^3+R_n^3 +2 R_{n+1}^2R_{n+2}+ 2R_{n-1}^2R_{n-2}+ 2R_{n}R_{n+1}R_{n+2}\cr
+R_{n+1}R_{n+2}R_{n+3}+ 2 R_{n}R_{n-1}R_{n-2}+ R_{n-1}R_{n-2}R_{n-3}+ 4 R_{n}R_{n+1}R_{n-1}
+ R_{n+1}R_{n-1}R_{n-2}\cr
+ R_{n+1}R_{n-1}R_{n+2}+ R_{n-1}R_{n-2}^2+R_{n+1}R_{n+2}^2+R_{n-1}R_{n+1}^2
+R_{n-1}^2R_{n+1}+ 3 R_n R_{n+1}^2\cr
+3 R_{n}^2 R_{n-1}+3 R_{n}^2 R_{n+1}+3 R_{n}R_{n-1}^2\Big] h_n(\gamma)\,.
\end{multline}
Applying the technique of orthogonal polynomials we can obtain from the above expression a differential equation capturing the perturbative expansion of the $m=4$ multicritical model
\begin{equation}\label{m4model}
V(M)= \frac{1}{\gamma}\left(\frac{1}{2}M^2-\frac{1}{2}M^4+\frac{4}{15}M^6-\frac{2}{35}M^8\right)~, \quad\quad \gamma= \frac{1}{16}- \frac{1}{16}(1-a^2)^4~.
\end{equation}
The planar contribution is given by 
\begin{equation}\label{r0m4}
r_0(x,\gamma)= \frac{1}{4}+ \frac{1}{4}(1-16\gamma x)^{1/4}~.
\end{equation}
Either using orthogonal polynomials or through a saddle point approximation  we obtain the planar non-analyticity
\begin{equation}
\lim_{\gamma\rightarrow \gamma_c}\partial_{\gamma}^{(3)}\mathcal{F}_{\text{n.a.}}^{(0)}(\gamma) \sim (\gamma-\gamma_c)^{-3/4}~,
\end{equation}
where $\gamma_c= 1/16$. 
The double scaling ansatz
\begin{equation}
z= N^{8/9}\left(1-\frac{\gamma}{\gamma_c}x\right)~, \quad r(x,\gamma)= r_0(1,\gamma_c)\left(1- N^{-2/9}\delta r(z)\right)
\end{equation}
obtained by keeping $\vartheta^{-1}\equiv  (\gamma-\gamma_c)^{9/8}N$ fixed while taking $\gamma\rightarrow \gamma_c$ and $N\rightarrow \infty$, leads to 
\begin{multline}\label{eq:m4diffeq}
 -\delta r(z)^4+2\delta r(z)\delta r'(z)^2+2\delta r(z)^2\delta r''(z)-\frac{3}{5}\delta r''(z)^2-\frac{4}{5}\delta r'(z)\delta r^{(3)}(z)\cr-\frac{2}{5}\delta r(z)\delta r^{(4)}(z)
 +\frac{1}{35}\delta r^{(6)}(z)+z=0~.
\end{multline}
The perturbative expansion is given by 
\begin{equation}
\delta r_\pm(z)= z^{1/4}\left(\pm 1+ \sum_{n=1}^{\infty} c_k^\pm z^{-9n/4}\right)~,
\end{equation}
with the first few coefficients given by 
\begin{equation}
\delta r_\pm(z)=  z^{1/4}\left(\pm 1 - \frac{1}{16}\, z^{-9/4}\mp \frac{429}{5120}\, z^{-9/2} - \frac{24079}{81920}\, z^{-27/4}\mp\frac{59126331}{52428800}\,z^{-9}+\ldots \right)\,.
\end{equation}
Only $\delta r_+(z)$ connects to the double scaling limit of (\ref{r0m4}) and the coefficients of this branch are not Borel summable. At large $z$, the non-perturbative piece is determined by the equation
\begin{equation}
\frac{1}{35} \epsilon^{(6)}z-\frac{2}{5}z^{1/4}\epsilon^{(4)}z- \frac{1}{5}z^{-3/4}\epsilon'''(z)+ 2z^{1/2}\epsilon''(z)+ z^{-1/2}\epsilon'(z)- 4 z^{3/4}\epsilon(z)=0~.
\end{equation}
A WKB analysis for large $z$ yields $\epsilon(z)\sim e^{-C \, z^{9/8}}$. We find three different solutions for $C = \{2.287, 1.809\pm 0.599\}$.
\begin{center} {\it Gelfand-Dikii polynomials} \end{center}
As a final note, we remark that (\ref{eq:painlevemulticrit}) and (\ref{eq:m4diffeq}) are part of a set of equations often referred to as the string equations \cite{Gross:1989vs}. For the $m^{\mathrm{th}}$ multicritical polynomial we have
\begin{equation}
z=\frac{(-1)^{m}2^{m+1}m!}{(2m-1)!!}\,\mathcal{R}_m[\delta r(z)]~,
\end{equation}
where the $ \mathcal{R}_m[\delta r(z)]$ are the Gelfand-Dikii polynomials \cite{Gelfand:1975rn}. These polynomials also appear in Witten's conjecture \cite{Witten:1990hr, Kontsevich:1992ti} relating the intersection numbers of moduli spaces to an infinite set of partial differential equations (KdV hierarchy). 
They are obtained from the recursion formula:
\begin{equation}
\mathcal{R}_0[\delta r(z)]= \frac{1}{2}~, \quad \mathcal{R}_{m+1}'[\delta r(z)]= \frac{1}{4}\mathcal{R}_m'''[\delta r(z)]-\delta r(z)\mathcal{R}_m'[\delta r(z)]-\frac{1}{2}\delta r'(z)\mathcal{R}_m[\delta r(z)]~.
\end{equation}
We list them up to $m=4$:
\begin{align}
m=0~: ~ \mathcal{R}_0[\delta r(z)]&= \frac{1}{2}~,\cr
m=1~: ~ \mathcal{R}_1[\delta r(z)]&=-\frac{1}{4}\delta r(z)~,\cr
m=2~: ~ \mathcal{R}_2[\delta r(z)]&= \frac{1}{16}(3\delta r(z)^2-\delta r''(z))~,\cr
m=3~:~\mathcal{R}_3[\delta r(z)]&= -\frac{1}{64}\left(10\delta r(z)^3-10\delta r''(z)\delta r(z)-5\delta r'(z)^2+\delta r^{(4)}(z)\right)~,\cr
m=4~:~\mathcal{R}_4[\delta r(z)]&= \frac{1}{256}\Big( 35 \delta r(z)^4-70 \delta r(z)^2\delta r''(z)+21\delta r''(z)^2+28\delta r'(z)\delta r^{(3)}(z)\cr
&\quad -70\delta r(z)\delta r'(z)^2 +14\delta r(z)\delta r^{(4)}(z)-\delta r^{(6)}(z)\Big)~.
\end{align}
The cases $m=3$ and $m=4$ then clearly reproduce (\ref{eq:painlevemulticrit}) and (\ref{eq:m4diffeq}).
\begin{center} {\it General multicritical model} \end{center}
 
From  (\ref{eq:multcritmodel}) and (\ref{m4model}) we observe that for multicritical models $V(M)$ is a polynomial in $M$ with alternating coeffficients. This holds true for general multicritical models $V(M)= \gamma^{-1}\sum_{n=1}^{m}v_n M^{2n}$ with \cite{Ambjorn:2016lkl}
\begin{align}
v_n= \frac{(-1)^{n+1}}{4m}\binom{m}{n}B(n,1/2)~,\quad n\leq m~,
\end{align}
where $B(x,y)$ is the Euler beta function. For $m$ odd the multicritical $V(M)$ has no real maximum whereas for $m$ even we always find at least one real maximum. For the $m^{\mathrm{th}}$ multicritical model with $m$ odd the coefficients are Borel summable, while for $m$ even the coefficients are not Borel summable \cite{Ginsparg:1991ws}.
The density of eigenvalues for the $m^{\mathrm{th}}$ multicritical model with eigenvalues distributed in $[-a,a]$ is given by  \cite{Ambjorn:2016lkl}
\begin{equation}
\rho_{\mathrm{ext}}(\lambda)= \frac{(1-a^2)^m\sqrt{a^2-\lambda^2} \, _2F_1\left(1,m+\frac{1}{2},\frac{3}{2}, \frac{a^2-\lambda^2}{1-\lambda^2}\right)}{2\pi \gamma (1-\lambda^2)}~,
\end{equation}
where 
\begin{equation}
\gamma = \frac{1}{4m}- \frac{1}{4m}(1-a^2)^m~, 
\end{equation}
and $\gamma_c = 1/4m$.

\begin{figure}[H]
\begin{center}
 \includegraphics[scale=.42]{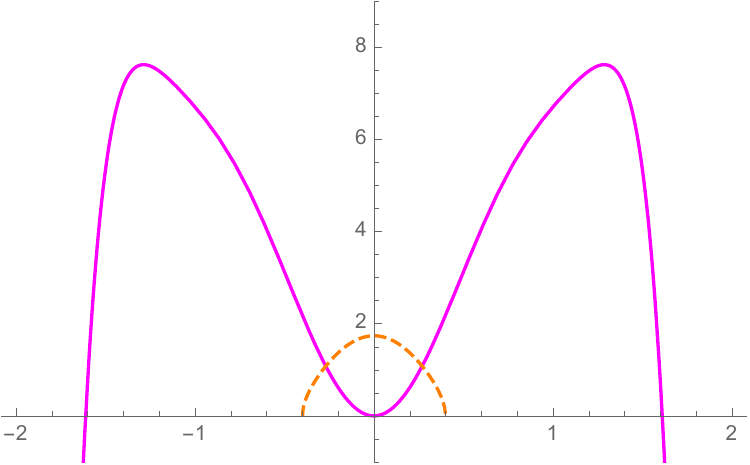} \qquad\qquad  \includegraphics[scale=.42]{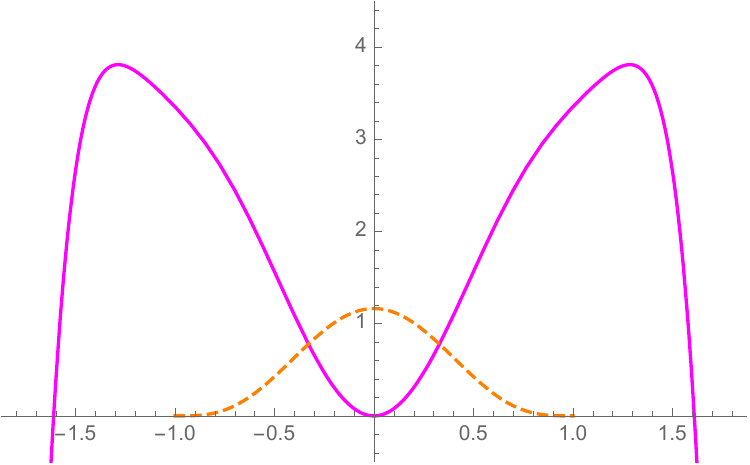}
 \caption{Multicritical polynomial $V(\lambda)$ for $m= 4$  (magenta) and corresponding eigenvalue distribution $\rho_{\mathrm{ext}}(\lambda)$ (orange, dashed) for $\gamma= 1/32$ (left) and $\gamma= \gamma_c$ (right).}
 \end{center}
\end{figure}

\section{Non-planar contributions}\label{npc}
In this appendix we will outline the calculations leading to non-planar contributions of the free energy. 

More concretely we will calculate $\mathcal{F}^{(h)}(\alpha)$ (\ref{eq:connecteddiagrams}) for $h=2$, $3$, and $4$. Solving $(\ref{eq:recrn})$ for $s=2, 3$, and $s=4$ we find
\begin{eqnarray}\label{higherR}
r_4(x,\alpha) &=& \frac{\alpha^4 r_0(x,\alpha)\left(p_1(\alpha x)+\alpha r_0(x,\alpha)\right)}{(1+48 \alpha x)^5}~,\\
r_6(x,\alpha) &=& \frac{\alpha^6 r_0(x,\alpha)\left(p_1(\alpha x)+\alpha p_1(\alpha x)r_0(x,\alpha)\right)}{(1+48\alpha x)^{15/2}}~,\\ 
r_8(x,\alpha)&=& \frac{\alpha^8r_0(x,\alpha)\left(p_2(\alpha x)+\alpha p_1(\alpha x)r_0(x,\alpha)\right)}{(1+48\alpha x)^{10}}~,
\end{eqnarray}
where $p_n(\alpha x)$ is a polynomial of degree $n$ in $\alpha x$. Introducing $\mathcal{R}_n(x,\alpha)\equiv r_{n}(\alpha,x)/r_0(x,\alpha)$ we find for $h=2$ and $h=3$ 
\begin{align}
\mathcal{F}^{(2)}(\alpha)&=-\int_0^1\dd x(1-x)\left[\mathcal{R}_4(x,\alpha)-\frac{1}{2}\mathcal{R}_2(x,\alpha)^2\right]-\frac{1}{12}\Big((1-x)\mathcal{R}_2(x,\alpha)\Big)^{(1)}\Bigg|_0^1\\ \nonumber
&+\frac{1}{720}\left((1-x)\log\frac{r_0(x,\alpha)}{x}\right)^{(3)}\Bigg|_0^1+1584\alpha^3 =\frac{\alpha  \left(84 \alpha +\left(3936 \alpha ^2-96 \alpha -1\right) r_0(1,\alpha)+1\right)}{20 (48 \alpha +1)^{5/2}}~
\end{align}
and
\allowdisplaybreaks
\begin{align}
\mathcal{F}^{(3)}(\alpha)&=-\int_0^1 (1-x)\left[\frac{1}{3}\mathcal{R}_2(x,\alpha)^3-\mathcal{R}_2(x,\alpha)\mathcal{R}_4(x,\alpha)+\mathcal{R}_6(x,\alpha)\right] \\ \nonumber
&-\frac{1}{12}\left((1-x)\left(\mathcal{R}_4(x,\alpha)-\frac{1}{2}\mathcal{R}_2(x,\alpha)^2\right)\right)^{(1)}\Bigg|_0^1+\frac{1}{720}\Big((1-x)\mathcal{R}_2(x,\alpha)\Big)^{(3)}\Bigg|_0^1\\ \nonumber
&-\frac{1}{30240}\left((1-x)\log\frac{r_0(x,\alpha)}{x}\right)^{(5)}\Bigg|_0^1+ \frac{25671168}{5}\alpha^5~\\ \nonumber
&=\frac{\alpha\, r_0(1,\alpha) \left(-954408960 \alpha ^5-61871616 \alpha ^4+1105920 \alpha ^3+23040 \alpha ^2+240 \alpha +1\right)}{84
   (48 \alpha +1)^{11/2}}\cr
  &+ \frac{\alpha  \left(110932992 \alpha ^4-889920 \alpha ^3-20448 \alpha ^2-228 \alpha -1\right)}{84 (48 \alpha +1)^{11/2}}~.
\end{align}
$\mathcal{F}^{(4)}(\alpha)$ follows similarly.  Near $\alpha_c=-1/48$  we therefore obtain  $h=2$, $3$, and $4$ the following non-analytic behaviour
\begin{eqnarray}
\lim_{\alpha\to\alpha_c} \mathcal{F}_{\mathrm{n.a.}}^{(h)}(\alpha)= f_h (\alpha- \alpha_c)^{5\chi_h/4}~, \quad\quad h \in \mathbb{N}~,
\end{eqnarray}
providing further evidence for $(\ref{eq:naF})$. The coefficients are given by
\begin{equation}\label{eq:f2f3f4}
f_2 = -\frac{7}{13271040 \sqrt{3}}~, \quad\quad f_3 =-\frac{245}{10567230160896}~, \quad \quad f_4=-\frac{259553}{23373022359076208640 \sqrt{3}}~.
\end{equation}
Additionally we have the small $\alpha$ expansion:
\begin{align}
\mathcal{F}^{(2)}(\alpha)&= 240\,\alpha^3 -32112\,\alpha^4+ \frac{14501376}{5}\,\alpha^5-220174848\,\alpha^6+15138938880\, \alpha ^7+\mathcal{O}(\alpha^8)~, \cr
\mathcal{F}^{(3)}(\alpha)&= 483840\,\alpha^5-128130048\,\alpha^6+\frac{139272044544}{7}\,\alpha^7-2366845194240\, \alpha ^8+\mathcal{O}(\alpha^9)~, \cr
\mathcal{F}^{(4)}(\alpha)&= 2767564800\,\alpha^7+ 257831155924992\,\alpha ^9-1137924495360\,\alpha ^8+\mathcal{O}(\alpha^9)~.
\end{align}

\section{Wick contractions}\label{Wickcont}

In this appendix we provide a simple example for the type of Wick contractions used to get various results. We are interested in the following 
\begin{multline}
\langle \mathrm{Fermi}| T_B\, \hat{n}(t_1,x_1) \hat{n}(t_2,x_2) |\mathrm{Fermi}\rangle =\Theta(t_1-t_2) \langle \mathrm{Fermi}| \hat{n}(t_1,x_1) \hat{n}(t_2,x_2)|\, \mathrm{Fermi}\rangle\cr
+ \Theta(t_2-t_1)\langle \mathrm{Fermi}| \hat{n}(t_2,x_2) \hat{n}(t_1,x_1) | \mathrm{Fermi}\rangle~.
\end{multline}
Using (\ref{eq:PsiPsidFS}) and the number density operator (\ref{density}) this can be expressed in Fourier space as
\begin{align}\label{eq:1stpart}
&\langle \mathrm{Fermi}| T_B\, \hat{n}(t_1,x_1) \hat{n}(t_2,x_2) |\mathrm{Fermi}\rangle \cr
&=\Big(\Theta(t_1-t_2)\int_{\mathbb{R}^4}\frac{\dd \nu_1\dd \nu_2\dd\nu_3\dd \nu_4}{(2\pi)^4}\, e^{i\nu_1t_1-i\nu_2t_1+i\nu_3t_2-i\nu_4 t_2}
\psi_{\nu_1}(x_1)\psi_{\nu_2}^*(x_1)\psi_{\nu_3}(x_2)\psi_{\nu_4}^*(x_2)\cr
&+\Theta(t_2-t_1)\int_{\mathbb{R}^4}\frac{\dd \nu_1\dd \nu_2\dd\nu_3\dd \nu_4}{(2\pi)^4}\, e^{i\nu_1t_2-i\nu_2t_2+i\nu_3t_1-i\nu_4 t_1}\psi_{\nu_1}(x_2)\psi_{\nu_2}^*(x_2)\psi_{\nu_3}(x_1)\psi_{\nu_4}^*(x_1)\Big)~\cr
&\times \langle \mathrm{Fermi}| a_{\nu_1}^\dag a_{\nu_2} a_{\nu_3}^\dag a_{\nu_4}|\mathrm{Fermi}\rangle
\end{align}
where for notational convenience we suppress the parity index.
When anti-commuting these operators we will use 
(\ref{AC}) and (\ref{eq:raisingloweringop}). We then obtain 
\begin{multline}\label{eq:anuanudag1to1}
\langle \mathrm{Fermi} | a_{\nu_1}^\dag a_{\nu_2} a_{\nu_3}^\dag a_{\nu_4} |\mathrm{Fermi}\rangle= 
(2\pi)^2\delta_{\nu_1\nu_2}\delta_{\nu_3\nu_4}\Theta(\nu_F-\nu_4)\Theta(\nu_F-\nu_2)\cr
+ (2\pi)^2\delta_{\nu_1\nu_4}\delta_{\nu_3\nu_2}\Theta(\nu_F-\nu_4)\Theta(\nu_3-\nu_F)~,
\end{multline}
where we normalised the ground state such that $\langle \mathrm{Fermi}|\mathrm{Fermi}\rangle=1$. The first term on the right hand side of the above expression corresponds to the contraction of $\Psi^\dagger(t_i,x_i)$ with $\Psi(t_i,x_i)$ and therefore does not contribute to the connected correlation function $\mathcal{Q}_c(\{t_i,x_i\})$. Dropping this term and combining (\ref{eq:1stpart}) with (\ref{eq:anuanudag1to1}) we obtain for the connected correlation function
\begin{multline}
\langle \mathrm{Fermi}| T_B\, \hat{n}(t_1,x_1) \hat{n}(t_2,x_2) |\mathrm{Fermi}\rangle_{c}=\cr
\quad\Theta(t_1-t_2)\int_{\nu_F}^{\infty}\frac{\dd \nu}{2\pi}\,e^{-i\nu(t_1-t_2)}\psi_{\nu}^*(x_1)\psi_{\nu}(x_2)\,\int_{-\infty}^{\nu_F}\frac{\dd \upsilon}{2\pi}\, e^{-i\upsilon(t_2-t_1)}\psi_{\upsilon}^*(x_2)\psi_{\upsilon}(x_1)\cr
+\Theta(t_2-t_1)\int_{\nu_F}^{\infty}\frac{\dd \nu}{2\pi}\,e^{-i\nu(t_2-t_1)}\psi_{\nu}^*(x_2)\psi_{\nu}(x_1)\, \int_{-\infty}^{\nu_F}\frac{\dd \upsilon}{2\pi}\, e^{-i\upsilon(t_1-t_2)}\psi_{\upsilon}^*(x_1)\psi_{\upsilon}(x_2)\cr
=S_F(t_1,x_1;t_2,x_2)S_F(t_2,x_2;t_1,x_1)~,
\end{multline}
where in the last line we used the definition of the Feynman propagator in $(\ref{SFexplicit})$.

\section{Perturbative expansion of the reflection coefficient} 

To evaluate the $S$-matrix in section \ref{sec6} we need the perturbative expansion of the reflection coefficient (\ref{eq:Rpurephase}). More concretely we need the perturbative expansion of  the combination $R_{\nu_F+a}R^*_{\nu_F+b}$. For the one-to-$n$ scattering the parameters $a$ and $b$ are fixed by demanding one incoming and $n\geq 1$ outgoing waves. To obtain the perturbative expansion of the scattering amplitudes we will quotient out the phase $e^{-i(a-b)\log(-\nu_F)}$ in the expansion of $R_{\nu_F+a}R^*_{\nu_F+b}$. In other words we will evaluate the $S$-matrix elements using the expansion below on the right:
\begin{multline}\label{eq:pertR}
 e^{i(a-b)\log(-\nu_F)} \, R_{\nu_F+a}R^*_{\nu_F+b}= 1- \frac{i}{2\nu_F}(a^2-b^2)\cr
-\frac{1}{24\nu_F^2}\left(3(a^2-b^2)^2-i(a(1+4a^2)-b(1+4b^2))\right)+\mathcal{O}\left(\nu_F^{-3}\right)~,
\end{multline}
with $a$ and $b$ fixed appropriately. 
We note that (\ref{eq:pertR}) can be expressed as follows
\begin{multline}\label{eq:pertRBP}
 e^{i(a-b)\log(-\nu_F)} \, R_{\nu_F+a}R^*_{\nu_F+b} =1+ \frac{i}{2\nu_F}\left(B_2(1/2-ia)- B_2(1/2+ib)\right)\cr
+\frac{1}{6\nu_F^2}\left((B_3(1/2-ia)+B_3(1/2+ib))-\frac{3}{4}\left(B_2(1/2-ia)-B_2(1/2+ib)\right)^2\right)+\mathcal{O}\left(\nu_F^{-3}\right)~,
\end{multline}
where $B_k(x)$ are the Bernoulli polynomials and we used the expansion of the Gamma function in terms of the reflection coefficient which implies
\begin{align}\label{eq:RinBp}
R_{\nu_F+a}\,e^{i\nu_F(-1+\log(-\nu_F))}=e^{\sum_{n\geq 1}\frac{(-i)^{n+2}}{n(n+1)}\frac{B_{n+1}(1/2-ia)}{\nu_F^n}}~.
\end{align}
We note that in (\ref{eq:pertRBP}) the power $n$ of $1/\nu_F$ a particular Bernoulli polynomial is multiplied with is related to its degree by deg$(B_k(x))\geq n$.
Since now the $k^{\text{th}}$ Bernoulli polynomial is a polynomial of degree $k$ we can use
\begin{equation}\label{eq:alternatingsumwk}
\sum_{\Omega\in \{\omega_1,..,\omega_{n-1}\}}(-1)^{|\Omega|}\omega(\Omega)^k=0~, \quad k<n-1~,
\end{equation}
which in combination with (\ref{eq:pertRBP}) implies the leading order power of $\nu_F$ in the perturbative expansion of the S-matrix. 
The expansion (\ref{eq:RinBp}) leads to a second nice application. The alternating sign implies that odd powers of $\nu_F$ are always multplied by differences of Bernoulli polynomials 
while for even powers of $\nu_F$ in the expansion of $R_{\nu_F+a}R^*_{\nu_F+b}$ the Bernoulli polynomials sum up. A simple change of variables now implies for $\alpha, k>0$:
\begin{align}
&\sum_{\Omega\subset\{\omega_1,..,\omega_{n-1}\}} (-1)^{|\Omega|+1}\int_{0}^{\omega(\Omega)}\dd x B_{k}(x)^\alpha= \sum_{\Omega\subset\{\omega_1,..,\omega_{n-1}\}} (-1)^{|\Omega|+1}\int_0^{\omega(\Omega)}\dd x B_{k}(\omega_n-x)^\alpha~,~n~\mathrm{odd}~,\cr
&\sum_{\Omega\subset\{\omega_1,..,\omega_{n-1}\}} (-1)^{|\Omega|+1}\int_{0}^{\omega(\Omega)}\dd x B_{k}(x)^\alpha= \sum_{\Omega\subset\{\omega_1,..,\omega_{n-1}\}} (-1)^{|\Omega|}\int_0^{\omega(\Omega)}\dd x B_{k}(\omega_n-x)^\alpha~,~ n~\mathrm{even}~. 
\end{align}
As a consequence we obtain for the case of one-to-$(n-1)$ scattering. For scattering into an even number of particles we only encounter odd powers of $\nu_F$ in the perturbative expansion of the $S$-matrix. For the case of scattering into an odd number the contrary holds true and the expansion is even in $\nu_F$.

\section{Further examples of scattering amplitudes}\label{higherscat}

In this appendix we give some additional examples of scattering amplitudes. 
\newline\newline
\textbf{One-to-four scattering.}
To obtain the one-to-four amplitude we start with 
\begin{equation}\label{eq:connected5point}
\mathcal{Q}\left( \{ t_i, x_i \}\right)  = \langle \text{Fermi} | \,T_B\, \hat{n}(t_1,x_1) \hat{n}(t_2,x_2)\hat{n}(t_3,x_3) \hat{n}(t_4,x_4)\hat{n}(t_5,x_5)\, | \text{Fermi} \rangle~.
\end{equation}
After performing Wick contractions we go to Fourier space and extract the scattering amplitude defined in (\ref{eq:amplitudegeneral}) as in the previous examples. This leads to
\begin{align}
S(\omega_1,\omega_2,\omega_3,\omega_4\,|\,\omega_5)&\equiv e^{-i\omega_5\log(-\nu_F)}\,\sum_{\Omega\, \subseteq \{\omega_1,..,\,\omega_{4}\}}\,(-1)^{|\Omega|+1}\,\int_0^{\omega(\Omega)}\dd \upsilon\, R^*_{\nu_F-\upsilon}R_{\nu_F-\upsilon-\omega_5}~,
\end{align}
where the sum ranges over all subsets $\Omega$ of $\{\omega_1,..,\omega_4\}$, $|\Omega|$ is the number of elements in $\Omega$, and $\omega(\Omega)= \sum_{\omega\in \Omega}\omega$. We note again that the additional minus sign arises because of the initial definition in (\ref{eq:amplitudegeneral}).  Inserting the expressions of the reflection coefficient (\ref{eq:Rpurephase}) and its perturbative expansion (\ref{eq:pertR}) we obtain 
\begin{align}
S(\omega_1,\omega_2,\omega_3,\omega_4\,|\,\omega_5)=&-\frac{i}{\nu_F^3}\,\omega_1\omega_2\omega_3\omega_4\omega_5(\omega_5+i)(\omega_5+2i)+\frac{i}{24\nu_F^5}\,\omega _1 \omega _2 \omega _3 \omega _4\omega_5 \cr
\times &(\omega_5 +i) (\omega_5+2 i) (\omega_5+3 i) (\omega_5+4 i) \Big(\omega _1
   \left(\omega _1-i\right)+\omega _2 \left(\omega _2-i\right)\cr
   &+\omega _3 \left(\omega _3-i\right)+\omega _4 \left(\omega _4-i\right)+1\Big)+\mathcal{O}\left(\nu_F^{-7}\right)~.
\end{align}
\textbf{One-to-five scattering.}
To obtain the one-to-five amplitude we start with 
\begin{equation}
\mathcal{Q}\left( \{ t_i, x_i \}\right)  = \langle \text{Fermi} | \,T\, \hat{n}(t_1,x_1) \hat{n}(t_2,x_2)\hat{n}(t_3,x_3) \hat{n}(t_4,x_4)\hat{n}(t_5,x_5)\hat{n}(t_6,x_6)\, | \text{Fermi} \rangle~.
\end{equation}
After performing Wick contractions we go to Fourier space and extract the scattering amplitude defined in (\ref{eq:amplitudegeneral}) along the lines explained in the last examples. This leads to
\begin{align}
S(\omega_1,\omega_2,\omega_3,\omega_4,\omega_5\,|\,\omega_6)&\equiv e^{-i\omega_6\log(-\nu_F)}\,\sum_{\Omega\, \subseteq \{\omega_1,..,\,\omega_{5}\}}\,(-1)^{|\Omega|+1}\,\int_0^{\omega(\Omega)}\dd \upsilon\, R^*_{\nu_F-\upsilon}R_{\nu_F-\upsilon-\omega_6}~.
\end{align}
Inserting the expressions of the reflection coefficient we obtain 
\begin{align}
S(\omega_1,\omega_2,\omega_3,\omega_4,\omega_5\,|\,\omega_6)=&\frac{1}{\nu_F^4}\,\omega _1 \omega _2 \omega _3 \omega _4 \omega _5\omega_6(\omega_6 +i) (\omega_6 +2 i) (\omega_6 +3 i)\cr
-&\frac{1}{24\nu_F^6}\,\omega _1 \omega _2 \omega _3 \omega _4 \omega _5\omega_6(\omega_6 +i) (\omega_6 +2 i) (\omega_6 +3 i) (\omega_6 +4 i) (\omega_6 +5 i) 
  \cr
  \times &\big(\omega _1 \left(\omega _1-i\right)+\omega _2 \left(\omega _2-i\right)+\omega _3 \left(\omega _3-i\right)+\omega _4 \left(\omega_4-i\right)\cr
  +&\omega _5 \left(\omega _5-i\right)+1\big)+\mathcal{O}\left(\nu_F^{-8}\right)~.
\end{align}
\textbf{One-to-six scattering.}
After performing Wick contractions we go to Fourier space and extract the scattering amplitude defined in (\ref{eq:amplitudegeneral}) along the lines explained in the last examples. This leads to
\begin{align}
S(\omega_1,\omega_2,\omega_3,\omega_4,\omega_5,\omega_6\,|\,\omega_7)&\equiv e^{-i\omega_7\log(-\nu_F)}\,\sum_{\Omega\, \subseteq \{\omega_1,..,\,\omega_{6}\}}\,(-1)^{|\Omega|+1}\,\int_0^{\omega(\Omega)}\dd \upsilon\, R^*_{\nu_F-\upsilon}R_{\nu_F-\upsilon-\omega_7}~.
\end{align}
Inserting the expressions of the reflection coefficient we obtain 
\begin{align}
S(\omega_1,\ldots,\omega_6\,|\,\omega_7)&=\frac{i}{\nu_F^5}\,\omega _1 \omega _2 \omega _3 \omega _4 \omega _5\omega_6\omega_7(\omega_7 +i) (\omega_7 +2 i) (\omega_7 +3 i) (\omega_7 +4 i)+  \cr
&-\frac{i}{24\nu_F^7}\,\omega _1 \omega _2 \omega _3 \omega _4 \omega _5\omega_6 \omega_7(\omega_7 +i) (\omega_7 +2 i) (\omega_7 +3 i) (\omega_7 +4 i) (\omega_7 +5 i) (\omega_7 +6 i) 
  \cr
&  \times \big(\omega _1 \left(\omega _1-i\right)+\omega _2 \left(\omega _2-i\right)+\omega _3 \left(\omega _3-i\right)+\omega _4 \left(\omega_4-i\right)+\omega _5 \left(\omega _5-i\right)\cr
  &+\omega _6 \left(\omega _6-i\right)+1\big)+\mathcal{O}\left(\nu_F^{-9}\right)~.
\end{align}
{\textbf{Three-to-two scattering.}}
We are looking for the term proportional to
\begin{align}\label{eq:2to3}
 e^{-i\omega_1u_1-i\omega_2u_2+i\omega_3u_3+i\omega_4u_4+i\omega_5 u_5}\delta(\omega_1+\omega_2+\omega_3+\omega_4+\omega_5)\,S(\omega_1,\omega_2\,|\,\omega_3,\omega_4,\omega_5)
~,
\end{align}
with $\omega_1, \omega_2$ outgoing and positive and $\omega_3, \omega_4, \omega_5$ incoming negative. Further we assume 
\begin{align}
&\omega_2>\omega_1~, \quad \omega_5>\omega_4>\omega_3~, \quad \omega_2>-\omega_3-\omega_4~,\quad \omega_1<-\omega_4-\omega_5~,
\end{align}
which implies $\omega_2$= max$\{\omega_1,\omega_2, |\omega_3|,|\omega_4|,|\omega_5|\}$. We find
\allowdisplaybreaks
\begin{align}
&S(\omega_1,\omega_2\,|\, \omega_3,\omega_4,\omega_5)\equiv e^{-i(\omega_3+\omega_4+\omega_5)\log(-\nu_F)}\, \Bigg[\int_{\omega_4}^{\omega_1+\omega_4}\dd \upsilon\, R_{\nu_F-\upsilon+\omega_4}^*R_{\nu_F-\upsilon}R_{\nu_F-\upsilon-\omega_2}^*R_{\nu_F-\upsilon+\omega_1+\omega_4}\cr
&+\int_{\omega_3}^{\omega_1+\omega_3}\dd \upsilon\, R_{\nu_F-\upsilon+\omega_3}^*R_{\nu_F-\upsilon}R^*_{\nu_F-\upsilon-\omega_2}R_{\nu_F-\upsilon+\omega_1+\omega_3}- \int_{0}^{\omega_1}\dd \upsilon\, R_{\nu_F-\upsilon+\omega_1}R^*_{\nu_F-\upsilon-\omega_2}\cr
&+\int_{\omega_5}^{\omega_1+\omega_5}\dd \upsilon\, R_{\nu_F-\upsilon+\omega_5}^*R_{\nu_F-\upsilon}R^*_{\nu_F-\upsilon-\omega_2}R_{\nu_F-\upsilon+\omega_1+\omega_5}+\int_{0}^{\omega_1}\dd \upsilon\, R_{\nu_F-\upsilon+\omega_1+\omega_2}R^*_{\nu_F-\upsilon}\cr
&-\int_{\omega_4}^{\omega_1+\omega_4}\dd \upsilon\, R_{\nu_F-\upsilon-\omega_2-\omega_5}^*R_{\nu_F-\upsilon+\omega_1+\omega_4}R^*_{\nu_F-\upsilon+\omega_4}R_{\nu_F-\upsilon-\omega_5}\cr
&-\int_{\omega_3}^{\omega_1+\omega_3}\dd \upsilon\, R_{\nu_F-\upsilon-\omega_2-\omega_4}^*R_{\nu_F-\upsilon+\omega_1+\omega_3}R_{\nu_F-\upsilon+\omega_3}^*R_{\nu_F-\upsilon-\omega_4}\cr
&-\int_{\omega_5}^{\omega_1+\omega_5}\dd \upsilon\, R_{\nu_F-\upsilon-\omega_2-\omega_3}^*R_{\nu_F-\upsilon+\omega_1+\omega_5}R_{\nu_F-\upsilon+\omega_5}^*R_{\nu_F-\upsilon-\omega_3}
\Bigg]\cr
&= -\frac{i}{\nu_F^3}\omega_1\omega_2\omega_3\omega_4\omega_5(\omega_2-i)(\omega_2-2i)+\mathcal{O}(\nu_F^{-5})~.
\end{align}

}

 \begingroup
  \section*{References}\label{references}
 \renewcommand{\section}[2]{}

\endgroup


\begin{thebibliography}{99}



\subsubsection*{A. History $\&$ Philosophy}

\bibitem{Bohr:1936zz} 
  N.~Bohr,
  ``Neutron Capture and Nuclear Constitution,''
  Nature {\bf 137}, 344 (1936).
  doi:10.1038/137344a0
  
  \bibitem{wigner} Wigner, E. Characteristic vectors of bordered matrices with infinite dimensions. Ann. of Math. v.62 (1955) no.3, 548-564; 
Wigner, E. Characteristic vectors of bordered matrices of infinite dimensions II. Ann. of Math. v.65 (1957) no.2, 203-207; 
Wigner, E. On the distribution of the roots of certain symmetric matrices. Ann. of Math. v.67 (1958) no.2, 325-326
  
  
\bibitem{wigner2}  Wigner, E. Random Matrices in Physics. SIAM Reviews v.9 (1967) No. 1, 1-23

  
  \bibitem{mehta}
M. L. Mehta, ``On the statistical properties of the level-spacings in nuclear spectra, " Nuclear Phys. 18, 395 (1960); M. L. Mehta and M. Gaudin, ``On the density of eigenvalues of a random matrix," Nuclear Phys. 18, 420 (1960); , Nucl. Phys. ; M. Gaudin, ``On the distribution of the roots of certain symmetric matrices," Nuclear Phys. 25, 447 (1961)., Nucl. Phys.

\bibitem{Regge:1961px} 
  T.~Regge,
  ``General Relativity Without Coordinates,''
  Nuovo Cim.\  {\bf 19}, 558 (1961).
  doi:10.1007/BF02733251


\bibitem{Dyson}
  F.~J.~Dyson,
  ``The Threefold Way. Algebraic Structure of Symmetry Groups and Ensembles in Quantum Mechanics,''
  J.\ Math.\ Phys.\  {\bf 3}, 6 (1962).
  doi:10.1063/1.1703863; F.~J.~Dyson,
  ``Statistical theory of the energy levels of complex systems. I,''
  J.\ Math.\ Phys.\  {\bf 3}, 140 (1962).
  doi:10.1063/1.1703773
 

\bibitem{tHooft:1973alw} 
  G.~'t Hooft,
  ``A Planar Diagram Theory for Strong Interactions,''
  Nucl.\ Phys.\ B {\bf 72}, 461 (1974).
  doi:10.1016/0550-3213(74)90154-0

\bibitem{berry} 
Berry Michael Victor, Tabor M. and Ziman John Michael ``Level clustering in the regular spectrum," 356 Proc. R. Soc. Lond. A http://doi.org/10.1098/rspa.1977.0140


\bibitem{Polyakov:1981rd} 
  A.~M.~Polyakov,
  ``Quantum Geometry of Bosonic Strings,''
  Phys.\ Lett.\  {\bf 103B}, 207 (1981).
  doi:10.1016/0370-2693(81)90743-7


\bibitem{Bohigas:1983er} 
  O.~Bohigas, M.~J.~Giannoni and C.~Schmit,
  ``Characterization of chaotic quantum spectra and universality of level fluctuation laws,''
  Phys.\ Rev.\ Lett.\  {\bf 52}, 1 (1984).
  doi:10.1103/PhysRevLett.52.1
 
\bibitem{Polyakov:1984yq} 
  A.~M.~Polyakov, A.~A.~Belavin and A.~B.~Zamolodchikov,
  ``Infinite Conformal Symmetry of Critical Fluctuations in Two-Dimensions,''
  J.\ Statist.\ Phys.\  {\bf 34}, 763 (1984).
  doi:10.1007/BF01009438

\bibitem{Friedan:1986qt} 
  D.~Friedan, Z.~A.~Qiu and S.~H.~Shenker,
  ``Conformal Invariance And Critical Exponents In Two-dimensions,''
  Preprint - FRIEDAN, D. (86,REC.AUG.) 3p
  
  
\bibitem{Callan:1985ia} 
  C.~G.~Callan, Jr., E.~J.~Martinec, M.~J.~Perry and D.~Friedan,
  ``Strings in Background Fields,''
  Nucl.\ Phys.\ B {\bf 262}, 593 (1985).
  doi:10.1016/0550-3213(85)90506-1; J.~Scherk and J.~H.~Schwarz,
``Dual Models for Nonhadrons,''
Nucl.\ Phys.\ B \textbf{81}, 118-144 (1974)
doi:10.1016/0550-3213(74)90010-8; T.~Yoneya,
``Connection of Dual Models to Electrodynamics and Gravidynamics,''
Prog.\ Theor.\ Phys.\  \textbf{51}, 1907-1920 (1974)
doi:10.1143/PTP.51.1907
  
 
  
\bibitem{Ambjorn:1985az} 
  J.~Ambj{\o}rn, B.~Durhuus and J.~Frohlich,
  ``Diseases of Triangulated Random Surface Models, and Possible Cures,''
  Nucl.\ Phys.\ B {\bf 257}, 433 (1985).
  doi:10.1016/0550-3213(85)90356-6;   J.~Ambj{\o}rn, B.~Durhuus, J.~Frohlich and P.~Orland,
  ``The Appearance of Critical Dimensions in Regulated String Theories,''
  Nucl.\ Phys.\ B {\bf 270}, 457 (1986).
  doi:10.1016/0550-3213(86)90563-8

\bibitem{David:1985nj} 
  F.~David,
  ``A Model of Random Surfaces with Nontrivial Critical Behavior,''
  Nucl.\ Phys.\ B {\bf 257}, 543 (1985).
  doi:10.1016/0550-3213(85)90363-3; A.~Billoire and F.~David,
  ``Scaling Properties of Randomly Triangulated Planar Random Surfaces: A Numerical Study,''
  Nucl.\ Phys.\ B {\bf 275}, 617 (1986).
  doi:10.1016/0550-3213(86)90577-8
  
  
\bibitem{Kazakov:1985ea} 
  V.~A.~Kazakov, A.~A.~Migdal and I.~K.~Kostov,
  ``Critical Properties of Randomly Triangulated Planar Random Surfaces,''
  Phys.\ Lett.\  {\bf 157B}, 295 (1985).
  doi:10.1016/0370-2693(85)90669-0; 
V.~Kazakov,
``Bilocal Regularization of Models of Random Surfaces,''
Phys. Lett. B \textbf{150}, 282-284 (1985)
doi:10.1016/0370-2693(85)91011-1

  
  
  
 

  
\bibitem{Boulatov:1986jd} 
  D.~V.~Boulatov, V.~A.~Kazakov, I.~K.~Kostov and A.~A.~Migdal,
  ``Analytical and Numerical Study of the Model of Dynamically Triangulated Random Surfaces,''
  Nucl.\ Phys.\ B {\bf 275}, 641 (1986).
  doi:10.1016/0550-3213(86)90578-X
  
  \bibitem{spinglass} M. Mezard, G. Parisi, and M. A. Virasoro. Spin Glass Theory and Beyond. vol. 9, Lecture Notes in Physics, (World Scientific, 1987) Google books. 9971501155
  
  
\bibitem{Witten:1990hr} 
  E.~Witten,
  ``Two-dimensional gravity and intersection theory on moduli space,''
  Surveys Diff.\ Geom.\  {\bf 1}, 243 (1991).
  doi:10.4310/SDG.1990.v1.n1.a5
  
\bibitem{Kontsevich:1992ti} 
  M.~Kontsevich,
  ``Intersection theory on the moduli space of curves and the matrix Airy function,''
  Commun.\ Math.\ Phys.\  {\bf 147}, 1 (1992).
  doi:10.1007/BF02099526
  
  
\bibitem{Maldacena:1997re}
  J.~M.~Maldacena,
  ``The Large N limit of superconformal field theories and supergravity,''
  Int.\ J.\ Theor.\ Phys.\  {\bf 38} (1999) 1113
   [Adv.\ Theor.\ Math.\ Phys.\  {\bf 2} (1998) 231]
  doi:10.1023/A:1026654312961, 10.4310/ATMP.1998.v2.n2.a1
  [hep-th/9711200].
  
  
\bibitem{Mirzakhani:2006fta} 
  M.~Mirzakhani,
  ``Simple geodesics and Weil-Petersson volumes of moduli spaces of bordered Riemann surfaces,''
  Invent.\ Math.\  {\bf 167}, no. 1, 179 (2006).
  doi:10.1007/s00222-006-0013-2
  
 
\bibitem{Eynard:2007kz} 
  B.~Eynard and N.~Orantin,
  ``Invariants of algebraic curves and topological expansion,''
  Commun.\ Num.\ Theor.\ Phys.\  {\bf 1}, 347 (2007)
  doi:10.4310/CNTP.2007.v1.n2.a4
  [math-ph/0702045].
 
 
  
  
\subsubsection*{B. Reviews on matrix models and related topics}


  

\bibitem{mehtabook}
M.L.Mehta, "Random Matrices (Revised and Enlarged Second Edition)", 
Academic Press, 1991,
isbn: 978-0-12-488051-1,
doi: 10.1016/B978-0-12-488051-1.50004-4

\bibitem{Klebanov:1991qa} 
  I.~R.~Klebanov,
  ``String theory in two-dimensions,''
  In *Trieste 1991, Proceedings, String theory and quantum gravity '91* 30-101 and Princeton Univ. - PUPT-1271 (91/07,rec.Oct.) 72 p
  [hep-th/9108019].

\bibitem{Ginsparg:1993is} 
  P.~H.~Ginsparg and G.~W.~Moore,
  ``Lectures on 2-D gravity and 2-D string theory,''
  Yale Univ. New Haven - YCTP-P23-92 (92,rec.Apr.93) 197 p. Los Alamos Nat. Lab. - LA-UR-92-3479 (92,rec.Apr.93) 197 p. e: LANL hep-th/9304011
  [hep-th/9304011].
  
\bibitem{Polchinski:1994mb} 
  J.~Polchinski,
  ``What is string theory?,''
  hep-th/9411028.

  
\bibitem{DiFrancesco:1993cyw} 
  P.~Di Francesco, P.~H.~Ginsparg and J.~Zinn-Justin,
  ``2-D Gravity and random matrices,''
  Phys.\ Rept.\  {\bf 254}, 1 (1995)
  doi:10.1016/0370-1573(94)00084-G
  [hep-th/9306153].
 
 
   \bibitem{PolchinskiBook}
 Polchinski, J. (1998). String Theory.  Vol. 1: An Introduction to the Bosonic String. (Cambridge Monographs on Mathematical Physics). 
 Cambridge: Cambridge University Press. doi:10.1017/CBO9780511816079
 
\bibitem{Taylor:2001vb} 
  W.~Taylor,
  ``M(atrix) Theory: Matrix Quantum Mechanics as a Fundamental Theory,''
  Rev.\ Mod.\ Phys.\  {\bf 73}, 419 (2001)
  doi:10.1103/RevModPhys.73.419
  [hep-th/0101126].



\bibitem{tHooft:2002ufq} 
  G.~'t Hooft,
  ``Large N,''
  doi:10.1142/97898127769140001
  hep-th/0204069.
  
  

  
\bibitem{Alexandrov:2003ut} 
  S.~Alexandrov,
  ``Matrix quantum mechanics and two-dimensional string theory in nontrivial backgrounds,''
  hep-th/0311273.
  
  
\bibitem{Nakayama:2004vk} 
  Y.~Nakayama,
  ``Liouville field theory: A Decade after the revolution,''
  Int.\ J.\ Mod.\ Phys.\ A {\bf 19}, 2771 (2004)
  doi:10.1142/S0217751X04019500
  [hep-th/0402009].
 


\bibitem{Marino:2004eq}
  M.~Marino,
  ``Les Houches lectures on matrix models and topological strings,''
  hep-th/0410165.



\bibitem{Martinec:2004td} 
  E.~J.~Martinec,
  ``Matrix models and 2D string theory,''
  hep-th/0410136.
  
    
\bibitem{Marino:2012zq} 
  M.~Marino,
  ``Lectures on non-perturbative effects in large $N$ gauge theories, matrix models and strings,''
  Fortsch.\ Phys.\  {\bf 62}, 455 (2014)
  doi:10.1002/prop.201400005
  [arXiv:1206.6272 [hep-th]].

\bibitem{Eynard:2015aea} 
  B.~Eynard, T.~Kimura and S.~Ribault,
  ``Random matrices,''
  arXiv:1510.04430 [math-ph].
  
 
 
  
  


\subsubsection*{C. Large $N$ matrices}\label{largeNmatrices}

\bibitem{tHooft:1974pnl} 
  G.~'t Hooft,
  ``A Two-Dimensional Model for Mesons,''
  Nucl.\ Phys.\ B {\bf 75}, 461 (1974).
  doi:10.1016/0550-3213(74)90088-1

\bibitem{Brezin:1977sv} 
  E.~Brezin, C.~Itzykson, G.~Parisi and J.~B.~Zuber,
  ``Planar Diagrams,''
  Commun.\ Math.\ Phys.\  {\bf 59}, 35 (1978).
  doi:10.1007/BF01614153
  
      \bibitem{Bessis:1979is} 
  D.~Bessis,
  ``A New Method in the Combinatorics of the Topological Expansion,''
  Commun.\ Math.\ Phys.\  {\bf 69}, 147 (1979).
  doi:10.1007/BF01221445

\bibitem{Bessis:1980ss} 
  D.~Bessis, C.~Itzykson and J.~B.~Zuber,
  ``Quantum field theory techniques in graphical enumeration,''
  Adv.\ Appl.\ Math.\  {\bf 1}, 109 (1980).
  doi:10.1016/0196-8858(80)90008-1
 
\bibitem{Yaffe:1981vf} 
  L.~G.~Yaffe,
  ``Large n Limits as Classical Mechanics,''
  Rev.\ Mod.\ Phys.\  {\bf 54}, 407 (1982).
  doi:10.1103/RevModPhys.54.407
  
    \bibitem{Kazakov:1989bc} 
  V.~A.~Kazakov,
  ``The Appearance of Matter Fields from Quantum Fluctuations of 2D Gravity,''
  Mod.\ Phys.\ Lett.\ A {\bf 4}, 2125 (1989).
  doi:10.1142/S0217732389002392
  
  
  
\bibitem{Staudacher:1989fy} 
  M.~Staudacher,
  ``The Yang-lee Edge Singularity on a Dynamical Planar Random Surface,''
  Nucl.\ Phys.\ B {\bf 336}, 349 (1990).
  doi:10.1016/0550-3213(90)90432-D
  
  
\bibitem{Ambjorn:2016lkl} 
  J.~Ambj{\o}rn, T.~Budd and Y.~Makeenko,
  ``Generalized multicritical one-matrix models,''
  Nucl.\ Phys.\ B {\bf 913}, 357 (2016)
  doi:10.1016/j.nuclphysb.2016.09.013
  [arXiv:1604.04522 [hep-th]].
  
  
  
  \subsubsection*{\it i) Loop equations}
  
\bibitem{Polyakov:1980ca}
A.~M.~Polyakov,
Nucl.\ Phys.\ B \textbf{164}, 171-188 (1980)
doi:10.1016/0550-3213(80)90507-6
  
\bibitem{Wadia:SD1981} 
Dyson-Schwinger equations approach to the large-$N$ limit: Model systems and string representation of Yang-Mills theory, Wadia, Spenta R.,
Phys. Rev. D, 1981
  doi: 10.1103/PhysRevD.24.970,

  
  
\bibitem{Migdal:1984gj} 
  A.~A.~Migdal,
  ``Loop Equations and 1/N Expansion,''
  Phys.\ Rept.\  {\bf 102}, 199 (1983).
  doi:10.1016/0370-1573(83)90076-5
 
\bibitem{Ambjorn:1990wg} 
  J.~Ambj{\o}rn and Y.~M.~Makeenko,
  ``Properties of Loop Equations for the Hermitean Matrix Model and for Two-dimensional Quantum Gravity,''
  Mod.\ Phys.\ Lett.\ A {\bf 5}, 1753 (1990).
  doi:10.1142/S0217732390001992
  
  
\bibitem{Kazakov:1989cq}
V.~Kazakov,
``A Simple Solvable Model of Quantum Field Theory of Open Strings,''
Phys. Lett. B \textbf{237}, 212-215 (1990)
doi:10.1016/0370-2693(90)91431-A
  
\bibitem{Kostov:1990nf} 
  I.~K.~Kostov,
  ``Exactly Solvable Field Theory of $D=0$ Closed and Open Strings,''
  Phys.\ Lett.\ B {\bf 238}, 181 (1990).
  doi:10.1016/0370-2693(90)91717-P
  

  
\bibitem{Dijkgraaf:1990rs} 
  R.~Dijkgraaf, H.~L.~Verlinde and E.~P.~Verlinde,
  ``Loop equations and Virasoro constraints in nonperturbative 2-D quantum gravity,''
  Nucl.\ Phys.\ B {\bf 348}, 435 (1991).
  doi:10.1016/0550-3213(91)90199-8
 
\bibitem{Martinec:1990qg} 
  E.~J.~Martinec,
  ``On the origin of integrability in matrix models,''
  Commun.\ Math.\ Phys.\  {\bf 138}, 437 (1991).
  doi:10.1007/BF02102036
  
\bibitem{Ambjorn:1992gw} 
  J.~Ambj{\o}rn, L.~Chekhov, C.~F.~Kristjansen and Y.~Makeenko,
  ``Matrix model calculations beyond the spherical limit,''
  Nucl.\ Phys.\ B {\bf 404}, 127 (1993)
  Erratum: [Nucl.\ Phys.\ B {\bf 449}, 681 (1995)]
  doi:10.1016/0550-3213(93)90476-6, 10.1016/0550-3213(95)00391-5
  [hep-th/9302014].
  
\bibitem{Staudacher:1993xy} 
  M.~Staudacher,
  ``Combinatorial solution of the two matrix model,''
  Phys.\ Lett.\ B {\bf 305}, 332 (1993)
  doi:10.1016/0370-2693(93)91063-S
  [hep-th/9301038].
  
\bibitem{Lin:2020mme} 
  H.~W.~Lin,
  ``Bootstraps to Strings: Solving Random Matrix Models with Positivity,''
  arXiv:2002.08387 [hep-th].

    \subsubsection*{D. Non-perturbative effects}\label{nonperturbativeref}

\bibitem{Gelfand:1975rn} 
  I.~M.~Gelfand and L.~A.~Dikii,
  ``Asymptotic behavior of the resolvent of Sturm-Liouville equations and the algebra of the Korteweg-De Vries equations,''
  Russ.\ Math.\ Surveys {\bf 30}, no. 5, 77 (1975)
  [Usp.\ Mat.\ Nauk {\bf 30}, no. 5, 67 (1975)].
  doi:10.1070/RM1975v030n05ABEH001522

\bibitem{Banks:1989df} 
  T.~Banks, M.~R.~Douglas, N.~Seiberg and S.~H.~Shenker,
  ``Microscopic and Macroscopic Loops in Nonperturbative Two-dimensional Gravity,''
  Phys.\ Lett.\ B {\bf 238}, 279 (1990).
  doi:10.1016/0370-2693(90)91736-U
 
 
\bibitem{Gross:1989aw} 
  D.~J.~Gross and A.~A.~Migdal,
  ``A Nonperturbative Treatment of Two-dimensional Quantum Gravity,''
  Nucl.\ Phys.\ B {\bf 340}, 333 (1990).
  doi:10.1016/0550-3213(90)90450-R
 
 
\bibitem{Shenker:1990uf} 
  S.~H.~Shenker, 1990
  ``The Strength of nonperturbative effects in string theory,''
  In *Brezin, E. (ed.), Wadia, S.R. (ed.): The large N expansion in quantum field theory and statistical physics* 809-819

 
\bibitem{Jevicki:1990yk} 
  A.~Jevicki and T.~Yoneya,
  ``Action Principle for Strings in Less Than One-dimension,''
  Mod.\ Phys.\ Lett.\ A {\bf 5}, 1615 (1990).
  doi:10.1142/S0217732390001840
 
\bibitem{Ginsparg:1990zc} 
  P.~H.~Ginsparg, M.~Goulian, M.~R.~Plesser and J.~Zinn-Justin,
  ``(p, q) STRING ACTIONS,''
  Nucl.\ Phys.\ B {\bf 342}, 539 (1990).
  doi:10.1016/0550-3213(90)90326-9
 
 
\bibitem{Ginsparg:1991ws} 
  P.~H.~Ginsparg and J.~Zinn-Justin,
  ``Large order behavior of nonperturbative gravity,''
  Phys.\ Lett.\ B {\bf 255}, 189 (1991).
  doi:10.1016/0370-2693(91)90234-H
 
\bibitem{Eynard:1992sg} 
  B.~Eynard and J.~Zinn-Justin,
  ``Large order behavior of 2-D gravity coupled to d $<$ 1 matter,''
  Phys.\ Lett.\ B {\bf 302}, 396 (1993)
  doi:10.1016/0370-2693(93)90416-F
  [hep-th/9301004].
 

 
  \subsubsection*{\it i) Single matrix model}
 
\bibitem{Douglas:1989ve} 
  M.~R.~Douglas and S.~H.~Shenker,
  ``Strings in Less Than One-Dimension,''
  Nucl.\ Phys.\ B {\bf 335}, 635 (1990).
  doi:10.1016/0550-3213(90)90522-F
 
\bibitem{Gross:1989vs} 
  D.~J.~Gross and A.~A.~Migdal,
  ``Nonperturbative Two-Dimensional Quantum Gravity,''
  Phys.\ Rev.\ Lett.\  {\bf 64}, 127 (1990).
  doi:10.1103/PhysRevLett.64.127
       
\bibitem{Brezin:1990rb} 
  E.~Brezin and V.~A.~Kazakov,
  ``Exactly Solvable Field Theories of Closed Strings,''
  Phys.\ Lett.\ B {\bf 236}, 144 (1990).
  doi:10.1016/0370-2693(90)90818-Q
       
\bibitem{David:1990ge} 
  F.~David,
  ``Loop Equations and Nonperturbative Effects in Two-dimensional Quantum Gravity,''
  Mod.\ Phys.\ Lett.\ A {\bf 5}, 1019 (1990).
  doi:10.1142/S0217732390001141
 
 
\bibitem{David:1992za} 
  F.~David,
  ``Nonperturbative effects in matrix models and vacua of two-dimensional gravity,''
  Phys.\ Lett.\ B {\bf 302}, 403 (1993)
  doi:10.1016/0370-2693(93)90417-G
  [hep-th/9212106].
  
 
  
 \bibitem{Kapaev_2004}
A.A. Kapaev,
  ``Quasi-linear Stokes phenomenon for the Painlevé first equation'', 2004
 
\bibitem{Bender:2015bja} 
  C.~M.~Bender and J.~Komijani,
  ``Painlevé Transcendents and PT-Symmetric Hamiltonians,''
  J.\ Phys.\ A {\bf 48}, no. 47, 475202 (2015)
  doi:10.1088/1751-8113/48/47/475202
  [arXiv:1502.04089 [math-ph]].
 
 \subsubsection*{\it ii) Two matrix model}
  
\bibitem{Gross:1989ni} 
  D.~J.~Gross and A.~A.~Migdal,
  ``Nonperturbative Solution of the Ising Model on a Random Surface,''
  Phys.\ Rev.\ Lett.\  {\bf 64}, 717 (1990).
  doi:10.1103/PhysRevLett.64.717
  
  
\bibitem{Brezin:1989db} 
  E.~Brezin, M.~R.~Douglas, V.~Kazakov and S.~H.~Shenker,
  ``The Ising Model Coupled to 2-$D$ Gravity: A Nonperturbative Analysis,''
  Phys.\ Lett.\ B {\bf 237}, 43 (1990).
  doi:10.1016/0370-2693(90)90458-I
  
\bibitem{Douglas:1989dd} 
  M.~R.~Douglas,
  ``Strings in Less Than One-dimension and the Generalized $K^- D^- V$ Hierarchies,''
  Phys.\ Lett.\ B {\bf 238}, 176 (1990).
  doi:10.1016/0370-2693(90)91716-O
  
\bibitem{Crnkovic:1989tn} 
  C.~Crnkovic, P.~H.~Ginsparg and G.~W.~Moore,
  ``The Ising Model, the Yang-Lee Edge Singularity, and 2D Quantum Gravity,''
  Phys.\ Lett.\ B {\bf 237}, 196 (1990).
  doi:10.1016/0370-2693(90)91428-E
 
 
 \subsubsection*{\it iii) Eigenvalues $\&$ instantons}
\bibitem{Hanada:2004im} 
  M.~Hanada, M.~Hayakawa, N.~Ishibashi, H.~Kawai, T.~Kuroki, Y.~Matsuo and T.~Tada,
  ``Loops versus matrices: The Nonperturbative aspects of noncritical string,''
  Prog.\ Theor.\ Phys.\  {\bf 112}, 131 (2004)
  doi:10.1143/PTP.112.131
  [hep-th/0405076].
   
\bibitem{Kawai:2004pj} 
  H.~Kawai, T.~Kuroki and Y.~Matsuo,
  ``Universality of nonperturbative effect in type 0 string theory,''
  Nucl.\ Phys.\ B {\bf 711}, 253 (2005)
  doi:10.1016/j.nuclphysb.2005.01.002
  [hep-th/0412004].
 
\bibitem{Ishibashi:2005dh} 
  N.~Ishibashi and A.~Yamaguchi,
  ``On the chemical potential of D-instantons in c=0 noncritical string theory,''
  JHEP {\bf 0506}, 082 (2005)
  doi:10.1088/1126-6708/2005/06/082
  [hep-th/0503199].
 
 
\bibitem{Sato:2004tz} 
  A.~Sato and A.~Tsuchiya,
  ``ZZ brane amplitudes from matrix models,''
  JHEP {\bf 0502}, 032 (2005)
  doi:10.1088/1126-6708/2005/02/032
  [hep-th/0412201].
  
 

  
 \subsubsection*{ E. Two large $N$ matrices}\label{twomatrixreferences}
 
 
\bibitem{Mehta:1981xt} 
  M.~L.~Mehta,
  ``A Method of Integration Over Matrix Variables,''
  Commun.\ Math.\ Phys.\  {\bf 79}, 327 (1981).
  doi:10.1007/BF01208498
 
 
\bibitem{Bershadsky:1986ws} 
  M.~A.~Bershadsky and A.~A.~Migdal,
  ``Ising Model of a Randomly Triangulated Random Surface as a Definition of Fermionic String Theory,''
  Phys.\ Lett.\ B {\bf 174}, 393 (1986).
  doi:10.1016/0370-2693(86)91023-3
 
 



\bibitem{Kazakov:1986hu} 
  V.~A.~Kazakov,
  ``Ising model on a dynamical planar random lattice: Exact solution,''
  Phys.\ Lett.\ A {\bf 119}, 140 (1986).
  doi:10.1016/0375-9601(86)90433-0
  
  
  \bibitem{Boulatov:1986sb} 
  D.~V.~Boulatov and V.~A.~Kazakov,
  ``The Ising Model on Random Planar Lattice: The Structure of Phase Transition and the Exact Critical Exponents,''
  Phys.\ Lett.\ B {\bf 186}, 379 (1987).
  doi:10.1016/0370-2693(87)90312-1
 
  


\bibitem{tada} 
  T.~Tada,
  ``(q,p) critical point from two matrix models,''
  Phys.\ Lett.\ B {\bf 259}, 442 (1991).
  doi:10.1016/0370-2693(91)91654-E
  

\bibitem{Douglas:1990pt} 
  M.~R.~Douglas and M.~Li,
  ``Free variables and the two matrix model,''
  Phys.\ Lett.\ B {\bf 348}, 360 (1995)
  doi:10.1016/0370-2693(95)00176-L
  [hep-th/9412203].
   
  
  
\bibitem{ZinnJustin:2002pk} 
  P.~Zinn-Justin and J.~B.~Zuber,
  ``On some integrals over the U(N) unitary group and their large N limit,''
  J.\ Phys.\ A {\bf 36}, 3173 (2003)
  doi:10.1088/0305-4470/36/12/318
  [math-ph/0209019].

 
  
  
\subsubsection*{F. Matrix quantum mechanics}\label{MQMreferences}

  
\bibitem{Kazakov:1988ch}
V.~Kazakov and A.~A.~Migdal,
``Recent Progress in the Theory of Noncritical Strings,''
Nucl. Phys. B \textbf{311}, 171 (1988)
doi:10.1016/0550-3213(88)90146-0

\bibitem{Brezin:1989ss} 
  E.~Brezin, V.~A.~Kazakov and A.~B.~Zamolodchikov,
  ``Scaling Violation in a Field Theory of Closed Strings in One Physical Dimension,''
  Nucl.\ Phys.\ B {\bf 338}, 673 (1990).
  doi:10.1016/0550-3213(90)90647-V

\bibitem{Parisi:1989dka}
G.~Parisi,
``On the One-dimensional Discretized String,''
Phys. Lett. B \textbf{238}, 209-212 (1990)
doi:10.1016/0370-2693(90)91722-N

\bibitem{Gross:1990ay} 
  D.~J.~Gross and N.~Miljkovic,
  ``A Nonperturbative Solution of $D=1$ String Theory,''
  Phys.\ Lett.\ B {\bf 238}, 217 (1990).
  doi:10.1016/0370-2693(90)91724-P
  
\bibitem{Ginsparg:1990as} 
  P.~H.~Ginsparg and J.~Zinn-Justin,
  ``2-d GRAVITY + 1-d MATTER,''
  Phys.\ Lett.\ B {\bf 240}, 333 (1990).
  doi:10.1016/0370-2693(90)91108-N
   

\bibitem{Das:1990kaa} 
  S.~R.~Das and A.~Jevicki,
  ``String Field Theory and Physical Interpretation of $D=1$ Strings,''
  Mod.\ Phys.\ Lett.\ A {\bf 5}, 1639 (1990).
  doi:10.1142/S0217732390001888

\bibitem{Gross:1990st} 
  D.~J.~Gross and I.~R.~Klebanov,
  ``Fermionic string field theory of c = 1 two-dimensional quantum gravity,''
  Nucl.\ Phys.\ B {\bf 352}, 671 (1991).
  doi:10.1016/0550-3213(91)90103-5
  

\bibitem{Sengupta:1990bt}
A.~M.~Sengupta and S.~R.~Wadia,
Int.\ J.\ Mod.\ Phys.\ A \textbf{6}, 1961-1984 (1991)
doi:10.1142/S0217751X91000988


\bibitem{Brezin:1990ti} 
  E.~Brezin,
  ``Two-dimensional quantum gravity,'' 1990,
  Conf.\ Proc.\ C {\bf 900802V1}
  [Conf.\ Proc.\ C {\bf 900802V1}, C900802V1:262].



\bibitem{Boulatov:1991xz} 
  D.~Boulatov and V.~Kazakov,
  ``One-dimensional string theory with vortices as the upside down matrix oscillator,''
  Int.\ J.\ Mod.\ Phys.\ A {\bf 8}, 809 (1993)
  doi:10.1142/S0217751X9300031X
  [hep-th/0012228].
  

\bibitem{Polchinski:1992ed}
  J.~Polchinski,
  ``Effective field theory and the Fermi surface,''
  In *Boulder 1992, Proceedings, Recent directions in particle theory* 235-274, and Calif. Univ. Santa Barbara - NSF-ITP-92-132 (92,rec.Nov.) 39 p. (220633) Texas Univ. Austin - UTTG-92-20 (92,rec.Nov.) 39 p
  [hep-th/9210046].
  

\bibitem{Takayanagi:2003sm} 
  T.~Takayanagi and N.~Toumbas,
  ``A Matrix model dual of type 0B string theory in two-dimensions,''
  JHEP {\bf 0307}, 064 (2003)
  doi:10.1088/1126-6708/2003/07/064
  [hep-th/0307083].
  
    
\bibitem{Douglas:2003up} 
  M.~R.~Douglas, I.~R.~Klebanov, D.~Kutasov, J.~M.~Maldacena, E.~J.~Martinec and N.~Seiberg, 2003
  ``A New hat for the c=1 matrix model,''
  In *Shifman, M. (ed.) et al.: From fields to strings, vol. 3* 1758-1827
  [hep-th/0307195].
  
\bibitem{Dijkgraaf:1992hk} 
  R.~Dijkgraaf, G.~W.~Moore and R.~Plesser,
  ``The Partition function of 2-D string theory,''
  Nucl.\ Phys.\ B {\bf 394}, 356 (1993)
  doi:10.1016/0550-3213(93)90019-L
  [hep-th/9208031].
  
\bibitem{McGreevy:2003kb} 
  J.~McGreevy and H.~L.~Verlinde,
  ``Strings from tachyons: The c=1 matrix reloaded,''
  JHEP {\bf 0312}, 054 (2003)
  doi:10.1088/1126-6708/2003/12/054
  [hep-th/0304224].


\bibitem{Maldacena:2005hi} 
  J.~M.~Maldacena,
  ``Long strings in two dimensional string theory and non-singlets in the matrix model,''
  JHEP {\bf 0509}, 078 (2005)
  [Int.\ J.\ Geom.\ Meth.\ Mod.\ Phys.\  {\bf 3}, 1 (2006)]
  doi:10.1088/1126-6708/2005/09/078, 10.1142/S0219887806001053
  [hep-th/0503112].
 
 
\bibitem{Kazhdan1978}, Hamiltonian group actions and dynamical systems of calogero type,
David Kazhdan and Bertram Kostant and Shlomo Sternberg, 1978

\bibitem{Polychronakos:2006nz} 
  A.~P.~Polychronakos,
  J.\ Phys.\ A {\bf 39}, 12793 (2006)
  doi:10.1088/0305-4470/39/41/S07
  [hep-th/0607033].
 
  
  \subsubsection*{\it i) $c=1$ scattering}
  
\bibitem{Moore:1991sf} 
  G.~W.~Moore,
  ``Double scaled field theory at c = 1,''
  Nucl.\ Phys.\ B {\bf 368}, 557 (1992).
  doi:10.1016/0550-3213(92)90214-V
  
  
\bibitem{Moore:1991zv} 
  G.~W.~Moore, M.~R.~Plesser and S.~Ramgoolam,
  ``Exact S matrix for 2-D string theory,''
  Nucl.\ Phys.\ B {\bf 377}, 143 (1992)
  doi:10.1016/0550-3213(92)90020-C
  [hep-th/9111035].

\bibitem{Mandal:1991ua} 
  G.~Mandal, A.~M.~Sengupta and S.~R.~Wadia,
  ``Interactions and scattering in d = 1 string theory,''
  Mod.\ Phys.\ Lett.\ A {\bf 6}, 1465 (1991).
  doi:10.1142/S0217732391001585
  
\bibitem{Polchinski:1991uq} 
  J.~Polchinski,
  ``Classical limit of (1+1)-dimensional string theory,''
  Nucl.\ Phys.\ B {\bf 362}, 125 (1991).
  doi:10.1016/0550-3213(91)90559-G


\subsubsection*{ G. 2d quantum gravity and Liouville theory}\label{liouvillereferences}

\bibitem{Zamolodchikov:1982vx} 
  A.~B.~Zamolodchikov,
  ``On The Entropy Of Random Surfaces,''
  Phys.\ Lett.\  {\bf 117B}, 87 (1982).
  doi:10.1016/0370-2693(82)90879-6

\bibitem{Knizhnik:1988ak} 
  V.~G.~Knizhnik, A.~M.~Polyakov and A.~B.~Zamolodchikov,
  ``Fractal Structure of 2D Quantum Gravity,''
  Mod.\ Phys.\ Lett.\ A {\bf 3}, 819 (1988).
  doi:10.1142/S0217732388000982
  
  
\bibitem{David:1988hj} 
  F.~David,
  ``Conformal Field Theories Coupled to 2D Gravity in the Conformal Gauge,''
  Mod.\ Phys.\ Lett.\ A {\bf 3}, 1651 (1988).
  doi:10.1142/S0217732388001975
  
\bibitem{Distler:1988jt} 
  J.~Distler and H.~Kawai,
  ``Conformal Field Theory and 2D Quantum Gravity,''
  Nucl.\ Phys.\ B {\bf 321}, 509 (1989).
  doi:10.1016/0550-3213(89)90354-4

  
\bibitem{Polchinski:1990mf} 
  J.~Polchinski,
  ``Critical Behavior of Random Surfaces in One-dimension,''
  Nucl.\ Phys.\ B {\bf 346}, 253 (1990).
  doi:10.1016/0550-3213(90)90280-Q

\bibitem{Seiberg:1990eb} 
  N.~Seiberg,
  ``Notes on quantum Liouville theory and quantum gravity,''
  Prog.\ Theor.\ Phys.\ Suppl.\  {\bf 102}, 319 (1990).
  doi:10.1143/PTPS.102.319
  
\bibitem{Krasnov:2000zq} 
  K.~Krasnov,
  ``Holography and Riemann surfaces,''
  Adv.\ Theor.\ Math.\ Phys.\  {\bf 4}, 929 (2000)
  doi:10.4310/ATMP.2000.v4.n4.a5
  [hep-th/0005106].
   
   
   
\bibitem{Zamolodchikov:2001dz} 
  A.~Zamolodchikov,
  ``Scaling Lee-Yang model on a sphere. 1. Partition function,''
  JHEP {\bf 0207}, 029 (2002)
  doi:10.1088/1126-6708/2002/07/029
  [hep-th/0109078].
  

  
\bibitem{Zamolodchikov:2005jb} 
  A.~B.~Zamolodchikov,
  ``Perturbed conformal field theory on fluctuating sphere,'' 2005,
  hep-th/0508044.

\bibitem{Harlow:2011ny} 
  D.~Harlow, J.~Maltz and E.~Witten,
  ``Analytic Continuation of Liouville Theory,''
  JHEP {\bf 1112}, 071 (2011)
  doi:10.1007/JHEP12(2011)071
  [arXiv:1108.4417 [hep-th]].

  
  \subsubsection*{\it i) Correlation functions in Liouville theory}
  
\bibitem{Liu:1987nz} 
  J.~Liu and J.~Polchinski,
  ``Renormalization of the Mobius Volume,''
  Phys.\ Lett.\ B {\bf 203}, 39 (1988).
  doi:10.1016/0370-2693(88)91566-3
  
   
  
\bibitem{Dorn:1994xn} 
  H.~Dorn and H.~J.~Otto,
  ``Two and three point functions in Liouville theory,''
  Nucl.\ Phys.\ B {\bf 429}, 375 (1994)
  doi:10.1016/0550-3213(94)00352-1
  [hep-th/9403141].
 
\bibitem{Teschner:2001rv} 
  J.~Teschner,
  ``Liouville theory revisited,''
  Class.\ Quant.\ Grav.\  {\bf 18}, R153 (2001)
  doi:10.1088/0264-9381/18/23/201
  [hep-th/0104158].
  
 
\bibitem{Zamolodchikov:1995aa} 
  A.~B.~Zamolodchikov and A.~B.~Zamolodchikov,
  ``Structure constants and conformal bootstrap in Liouville field theory,''
  Nucl.\ Phys.\ B {\bf 477}, 577 (1996)
  doi:10.1016/0550-3213(96)00351-3
  [hep-th/9506136].
  
 
\bibitem{Zamolodchikov:2005fy} 
  A.~B.~Zamolodchikov,
  ``Three-point function in the minimal Liouville gravity,''
  Theor.\ Math.\ Phys.\  {\bf 142}, 183 (2005)
  doi:10.1007/s11232-005-0003-3
  [hep-th/0505063].

\subsubsection*{\it ii) Minimal Liouville Theory}  


\bibitem{Seiberg:2004at} 
  N.~Seiberg and D.~Shih,
  ``Minimal string theory,''
  Comptes Rendus Physique {\bf 6}, 165 (2005)
  doi:10.1016/j.crhy.2004.12.007
  [hep-th/0409306].
 
  
  
\bibitem{Belavin:2005jy} 
  A.~Belavin and A.~Zamolodchikov,
  ``Polyakov's string: Twenty five years after. Proceedings,'' 2005
  hep-th/0510214.
  
\bibitem{Belavin:2006ex} 
  A.~A.~Belavin and A.~B.~Zamolodchikov,
  ``Integrals over moduli spaces, ground ring, and four-point function in minimal Liouville gravity,''
  Theor.\ Math.\ Phys.\  {\bf 147}, 729 (2006)
  [Teor.\ Mat.\ Fiz.\  {\bf 147}, 339 (2006)].
  doi:10.1007/s11232-006-0075-8
  
  
  
\bibitem{Belavin:2008kv} 
  A.~A.~Belavin and A.~B.~Zamolodchikov,
  ``On Correlation Numbers in 2D Minimal Gravity and Matrix Models,''
  J.\ Phys.\ A {\bf 42}, 304004 (2009)
  doi:10.1088/1751-8113/42/30/304004
  [arXiv:0811.0450 [hep-th]].
  
\bibitem{Belavin:2010ba} 
  A.~Belavin and C.~Rim,
  ``Bulk one-point function on disk in one-matrix model,''
  Phys.\ Lett.\ B {\bf 687}, 264 (2010)
  doi:10.1016/j.physletb.2010.03.020
  [arXiv:1001.4356 [hep-th]].
  
\bibitem{Belavin:2010sr} 
  V.~Belavin,
  ``Torus Amplitudes in Minimal Liouville Gravity and Matrix Models,''
  Phys.\ Lett.\ B {\bf 698}, 86 (2011)
  doi:10.1016/j.physletb.2011.03.003
  [arXiv:1010.5508 [hep-th]].
 
\bibitem{Belavin:2015ffa} 
  V.~Belavin and Y.~Rud,
  J.\ Phys.\ A {\bf 48}, no. 18, 18FT01 (2015)
  doi:10.1088/1751-8113/48/18/18FT01
  [arXiv:1502.05575 [hep-th]].
  
      \subsubsection*{\it iii) Scattering from the continuum}
 
\bibitem{Gross:1991qp} 
  D.~J.~Gross and I.~R.~Klebanov,
  ``S = 1 for c = 1,''
  Nucl.\ Phys.\ B {\bf 359}, 3 (1991).
  doi:10.1016/0550-3213(91)90291-5
  
\bibitem{deBoer:2003hd} 
  J.~de Boer, A.~Sinkovics, E.~P.~Verlinde and J.~T.~Yee,
  ``String interactions in c = 1 matrix model,''
  JHEP {\bf 0403}, 023 (2004)
  doi:10.1088/1126-6708/2004/03/023
  [hep-th/0312135].
    
\bibitem{Balthazar:2017mxh} 
  B.~Balthazar, V.~A.~Rodriguez and X.~Yin,
  ``The $c$ = 1 string theory S-matrix revisited,''
  JHEP {\bf 1904}, 145 (2019)
  doi:10.1007/JHEP04(2019)145
  [arXiv:1705.07151 [hep-th]].
  
\bibitem{Balthazar:2018qdv} 
  B.~Balthazar, V.~A.~Rodriguez and X.~Yin,
  ``Long String Scattering in c $=$ 1 String Theory,''
  JHEP {\bf 1901}, 173 (2019)
  doi:10.1007/JHEP01(2019)173
  [arXiv:1810.07233 [hep-th]].
  
  
\bibitem{Erbin:2019uiz} 
  H.~Erbin, J.~Maldacena and D.~Skliros,
  ``Two-Point String Amplitudes,''
  JHEP {\bf 1907}, 139 (2019)
  doi:10.1007/JHEP07(2019)139
  [arXiv:1906.06051 [hep-th]].
  
  \subsubsection*{H. Non-perturbative effects in Liouville}\label{NonpertLiouville}
 
\bibitem{Witten:1991yr} 
  E.~Witten,
  ``On string theory and black holes,''
  Phys.\ Rev.\ D {\bf 44}, 314 (1991).
  doi:10.1103/PhysRevD.44.314
  
\bibitem{Mandal:1991tz} 
  G.~Mandal, A.~M.~Sengupta and S.~R.~Wadia,
  ``Classical solutions of two-dimensional string theory,''
  Mod.\ Phys.\ Lett.\ A {\bf 6}, 1685 (1991).
  doi:10.1142/S0217732391001822
  
\bibitem{Fukuma:1999tj} 
  M.~Fukuma and S.~Yahikozawa,
  ``Comments on D instantons in $c < 1$ strings,''
  Phys.\ Lett.\ B {\bf 460}, 71 (1999)
  doi:10.1016/S0370-2693(99)00744-3
  [hep-th/9902169].
  
\bibitem{Fateev:2000ik} 
  V.~Fateev, A.~B.~Zamolodchikov and A.~B.~Zamolodchikov,
  ``Boundary Liouville field theory. 1. Boundary state and boundary two point function,'' 2000,
  hep-th/0001012.
  
\bibitem{Teschner:2000md} 
  J.~Teschner,
  ``Remarks on Liouville theory with boundary,''
  PoS tmr {\bf 2000}, 041 (2000)
  doi:10.22323/1.006.0041
  [hep-th/0009138].

\bibitem{Zamolodchikov:2001ah} 
  A.~B.~Zamolodchikov and A.~B.~Zamolodchikov, (2001) ,
  ``Liouville field theory on a pseudosphere,''
  hep-th/0101152.
  
\bibitem{Alexandrov:2003nn} 
  S.~Y.~Alexandrov, V.~A.~Kazakov and D.~Kutasov,
  ``Nonperturbative effects in matrix models and D-branes,''
  JHEP {\bf 0309}, 057 (2003)
  doi:10.1088/1126-6708/2003/09/057
  [hep-th/0306177].
  
\bibitem{Klebanov:2003km} 
  I.~R.~Klebanov, J.~M.~Maldacena and N.~Seiberg,
  ``D-brane decay in two-dimensional string theory,''
  JHEP {\bf 0307}, 045 (2003)
  doi:10.1088/1126-6708/2003/07/045
  [hep-th/0305159].
  
  
\bibitem{Martinec:2003ka} 
  E.~J.~Martinec,
  ``The Annular report on noncritical string theory,'' 2003,
  hep-th/0305148.
  


  
\bibitem{Kazakov:2000pm} 
  V.~Kazakov, I.~K.~Kostov and D.~Kutasov,
  ``A Matrix model for the two-dimensional black hole,''
  Nucl.\ Phys.\ B {\bf 622}, 141 (2002)
  doi:10.1016/S0550-3213(01)00606-X
  [hep-th/0101011].
  
\bibitem{Sen:2002nu}
A.~Sen,
``Rolling tachyon,''
JHEP \textbf{04}, 048 (2002)
doi:10.1088/1126-6708/2002/04/048
[arXiv:hep-th/0203211 [hep-th]].
 
\bibitem{Seiberg:2003nm} 
  N.~Seiberg and D.~Shih,
  ``Branes, rings and matrix models in minimal (super)string theory,''
  JHEP {\bf 0402}, 021 (2004)
  doi:10.1088/1126-6708/2004/02/021
  [hep-th/0312170].
 
\bibitem{Kutasov:2004fg} 
  D.~Kutasov, K.~Okuyama, J.~w.~Park, N.~Seiberg and D.~Shih,
  ``Annulus amplitudes and ZZ branes in minimal string theory,''
  JHEP {\bf 0408}, 026 (2004)
  doi:10.1088/1126-6708/2004/08/026
  [hep-th/0406030].
 
\bibitem{Sato:2004tz} 
  A.~Sato and A.~Tsuchiya,
  ``ZZ brane amplitudes from matrix models,''
  JHEP {\bf 0502}, 032 (2005)
  doi:10.1088/1126-6708/2005/02/032
  [hep-th/0412201].

  
\bibitem{Zamolodchikov:2006xs} 
  A.~B.~Zamolodchikov and A.~B.~Zamolodchikov,
  ``Decay of Metastable Vacuum in Liouville Gravity,''
  Conf.\ Proc.\ C {\bf 060726}, 1223 (2006)
  [hep-th/0608196].
 
\bibitem{Gaiotto:2005gd} 
  D.~Gaiotto,
  ``Long strings condensation and FZZT branes,''
  hep-th/0503215.



\bibitem{Betzios:2017yms} 
  P.~Betzios and O.~Papadoulaki,
  ``FZZT branes and non-singlets of Matrix Quantum Mechanics,''
  arXiv:1711.04369 [hep-th].
  
\bibitem{Balthazar:2019rnh} 
  B.~Balthazar, V.~A.~Rodriguez and X.~Yin,
  ``ZZ Instantons and the Non-Perturbative Dual of c = 1 String Theory,''
  arXiv:1907.07688 [hep-th].
  
\bibitem{Balthazar:2019ypi} 
  B.~Balthazar, V.~A.~Rodriguez and X.~Yin,
  ``Multi-Instanton Calculus in c = 1 String Theory,''
  arXiv:1912.07170 [hep-th].
  
\bibitem{Sen:2020oqr} 
  A.~Sen,
  ``Divergent to Complex Amplitudes in Two Dimensional String Theory,''
  arXiv:2003.12076 [hep-th].


\subsubsection*{ I. Supersymmetric matrix models}\label{susymmreferences}

\bibitem{Baake:1984ie} 
  M.~Baake, M.~Reinicke and V.~Rittenberg,
  ``Fierz Identities for Real Clifford Algebras and the Number of Supercharges,''
  J.\ Math.\ Phys.\  {\bf 26}, 1070 (1985).
  doi:10.1063/1.526539

\bibitem{Marinari:1990jc} 
  E.~Marinari and G.~Parisi,
  ``The Supersymmetric One-dimensional String,''
  Phys.\ Lett.\ B {\bf 240}, 375 (1990).
  doi:10.1016/0370-2693(90)91115-R

\bibitem{Cooper:1994eh} 
  F.~Cooper, A.~Khare and U.~Sukhatme,
  ``Supersymmetry and quantum mechanics,''
  Phys.\ Rept.\  {\bf 251}, 267 (1995)
  doi:10.1016/0370-1573(94)00080-M
  [hep-th/9405029].

\bibitem{Banks:1996vh} 
  T.~Banks, W.~Fischler, S.~H.~Shenker and L.~Susskind,
  ``M theory as a matrix model: A Conjecture,''
  Phys.\ Rev.\ D {\bf 55}, 5112 (1997)
  doi:10.1103/PhysRevD.55.5112
  [hep-th/9610043].
  
\bibitem{Ishibashi:1996xs}
  N.~Ishibashi, H.~Kawai, Y.~Kitazawa and A.~Tsuchiya,
  ``A Large N reduced model as superstring,''
  Nucl.\ Phys.\ B {\bf 498} (1997) 467
  doi:10.1016/S0550-3213(97)00290-3
  [hep-th/9612115].
  
\bibitem{Dijkgraaf:1997vv} 
  R.~Dijkgraaf, E.~P.~Verlinde and H.~L.~Verlinde,
  ``Matrix string theory,''
  Nucl.\ Phys.\ B {\bf 500}, 43 (1997)
  doi:10.1016/S0550-3213(97)00326-X
  [hep-th/9703030].
  
\bibitem{Itzhaki:1998dd} 
  N.~Itzhaki, J.~M.~Maldacena, J.~Sonnenschein and S.~Yankielowicz,
  ``Supergravity and the large N limit of theories with sixteen supercharges,''
  Phys.\ Rev.\ D {\bf 58}, 046004 (1998)
  doi:10.1103/PhysRevD.58.046004
  [hep-th/9802042].
  
\bibitem{Frohlich:1999zf} 
  J.~Frohlich, G.~M.~Graf, D.~Hasler, J.~Hoppe and S.~T.~Yau,
  ``Asymptotic form of zero energy wave functions in supersymmetric matrix models,''
  Nucl.\ Phys.\ B {\bf 567}, 231 (2000)
  doi:10.1016/S0550-3213(99)00649-5
  [hep-th/9904182].
 
\bibitem{Berenstein:2002jq} 
  D.~E.~Berenstein, J.~M.~Maldacena and H.~S.~Nastase,
  ``Strings in flat space and pp waves from N=4 superYang-Mills,''
  JHEP {\bf 0204}, 013 (2002)
  doi:10.1088/1126-6708/2002/04/013
  [hep-th/0202021].
 
\bibitem{Berenstein:2004kk} 
  D.~Berenstein,
  ``A Toy model for the AdS / CFT correspondence,''
  JHEP {\bf 0407}, 018 (2004)
  doi:10.1088/1126-6708/2004/07/018
  [hep-th/0403110].
 
\bibitem{Anagnostopoulos:2007fw} 
  K.~N.~Anagnostopoulos, M.~Hanada, J.~Nishimura and S.~Takeuchi,
  ``Monte Carlo studies of supersymmetric matrix quantum mechanics with sixteen supercharges at finite temperature,''
  Phys.\ Rev.\ Lett.\  {\bf 100}, 021601 (2008)
  doi:10.1103/PhysRevLett.100.021601
  [arXiv:0707.4454 [hep-th]].
  
  
\bibitem{Wiseman:2013cda} 
  T.~Wiseman,
  ``On black hole thermodynamics from super Yang-Mills,''
  JHEP {\bf 1307}, 101 (2013)
  doi:10.1007/JHEP07(2013)101
  [arXiv:1304.3938 [hep-th]].
  
\bibitem{Lin:2014wka} 
  Y.~H.~Lin and X.~Yin,
  ``On the Ground State Wave Function of Matrix Theory,''
  JHEP {\bf 1511}, 027 (2015)
  doi:10.1007/JHEP11(2015)027
  [arXiv:1402.0055 [hep-th]].
  
  
\bibitem{Filev:2015hia} 
  V.~G.~Filev and D.~O'Connor,
  ``The BFSS model on the lattice,''
  JHEP {\bf 1605}, 167 (2016)
  doi:10.1007/JHEP05(2016)167
  [arXiv:1506.01366 [hep-th]].
  
\bibitem{Anous:2017mwr} 
  T.~Anous and C.~Cogburn,
  ``Mini-BFSS matrix model in silico,''
  Phys.\ Rev.\ D {\bf 100}, no. 6, 066023 (2019)
  doi:10.1103/PhysRevD.100.066023
  [arXiv:1701.07511 [hep-th]].
 
  
  
\subsubsection*{ J. Topological String $\&$ 2d Yang-Mills}\label{YMreferences}
  
\bibitem{Migdal:1975zg} 
  A.~A.~Migdal,
  ``Recursion Equations in Gauge Theories,''
  Sov.\ Phys.\ JETP {\bf 42}, 413 (1975)
  [Zh.\ Eksp.\ Teor.\ Fiz.\  {\bf 69}, 810 (1975)].
  
  
\bibitem{Makeenko:1979pb} 
  Y.~M.~Makeenko and A.~A.~Migdal,
  ``Exact Equation for the Loop Average in Multicolor QCD,''
  Phys.\ Lett.\  {\bf 88B}, 135 (1979)
  Erratum: [Phys.\ Lett.\  {\bf 89B}, 437 (1980)].
  doi:10.1016/0370-2693(79)90131-X
 
 
  
\bibitem{Rusakov:1990rs} 
  B.~E.~Rusakov,
  ``Loop averages and partition functions in U(N) gauge theory on two-dimensional manifolds,''
  Mod.\ Phys.\ Lett.\ A {\bf 5}, 693 (1990).
  doi:10.1142/S0217732390000780
     
\bibitem{Kazakov:1983fn} 
 V.~Kazakov and I.~Kostov,
``Computation of the Wilson Loop Functional in Two-dimensional U(infinite) Lattice Gauge Theory,''
Phys. Lett. B \textbf{105}, 453-456 (1981)
doi:10.1016/0370-2693(81)91203-X;  V.~A.~Kazakov,
  ``U(infinity) Lattice Gauge Theory As A Free Lattice String Theory,''
  Phys.\ Lett.\  {\bf 128B}, 316 (1983).
  doi:10.1016/0370-2693(83)90267-8
     
\bibitem{Witten:1992xu} 
  E.~Witten,
  ``Two-dimensional gauge theories revisited,''
  J.\ Geom.\ Phys.\  {\bf 9}, 303 (1992)
  doi:10.1016/0393-0440(92)90034-X
  [hep-th/9204083].
  
\bibitem{Gross:1992tu} 
  D.~J.~Gross,
  ``Two-dimensional QCD as a string theory,''
  Nucl.\ Phys.\ B {\bf 400}, 161 (1993)
  doi:10.1016/0550-3213(93)90402-B
  [hep-th/9212149];
  D.~J.~Gross and W.~Taylor,
  ``Two-dimensional QCD and strings,''
  In *Berkeley 1993, Proceedings, Strings '93* 214-225, and MIT Cambridge - CTP-2250 (93/10,rec.Nov.) 12 p. Princeton U. - PUPT-1431 (93/10,rec.Nov.) 12 p
  [hep-th/9311072];
  D.~J.~Gross and W.~Taylor,
  ``Two-dimensional QCD is a string theory,''
  Nucl.\ Phys.\ B {\bf 400}, 181 (1993)
  doi:10.1016/0550-3213(93)90403-C
  [hep-th/9301068].
  
  
\bibitem{Douglas:1993iia}
M.~R.~Douglas and V.~A.~Kazakov,
``Large N phase transition in continuum QCD in two-dimensions,''
Phys. Lett. B \textbf{319}, 219-230 (1993)
doi:10.1016/0370-2693(93)90806-S
[arXiv:hep-th/9305047 [hep-th]].
  
  
\bibitem{Minahan:1993np} 
  J.~A.~Minahan and A.~P.~Polychronakos,
  ``Equivalence of two-dimensional QCD and the C = 1 matrix model,''
  Phys.\ Lett.\ B {\bf 312}, 155 (1993)
  doi:10.1016/0370-2693(93)90504-B
  [hep-th/9303153].
  
\bibitem{Douglas:1993xv} 
  M.~R.~Douglas,
  ``Conformal field theory techniques for large N group theory,'' (1993),
  hep-th/9303159.

\bibitem{Periwal:1993yu} 
  V.~Periwal,
  ``Topological closed string interpretation of Chern-Simons theory,''
  Phys.\ Rev.\ Lett.\  {\bf 71}, 1295 (1993)
  doi:10.1103/PhysRevLett.71.1295
  [hep-th/9305115].
  

  
\bibitem{Douglas:1994pq} 
  M.~R.~Douglas, K.~Li and M.~Staudacher,
  ``Generalized two-dimensional QCD,''
  Nucl.\ Phys.\ B {\bf 420}, 118 (1994)
  doi:10.1016/0550-3213(94)90377-8
  [hep-th/9401062].
  

  
  
\bibitem{Gopakumar:1998ki} 
  R.~Gopakumar and C.~Vafa,
  ``On the gauge theory / geometry correspondence,''
  Adv.\ Theor.\ Math.\ Phys.\  {\bf 3}, 1415 (1999)
  [AMS/IP Stud.\ Adv.\ Math.\  {\bf 23}, 45 (2001)]
  doi:10.4310/ATMP.1999.v3.n5.a5
  [hep-th/9811131].
  
\bibitem{Ooguri:2002gx} 
  H.~Ooguri and C.~Vafa,
  ``World sheet derivation of a large N duality,''
  Nucl.\ Phys.\ B {\bf 641}, 3 (2002)
  doi:10.1016/S0550-3213(02)00620-X
  [hep-th/0205297].
  
  
\bibitem{Marino:2002fk} 
  M.~Marino,
  ``Chern-Simons theory, matrix integrals, and perturbative three manifold invariants,''
  Commun.\ Math.\ Phys.\  {\bf 253}, 25 (2004)
  doi:10.1007/s00220-004-1194-4
  [hep-th/0207096].
  
\bibitem{Tierz:2002jj} 
  M.~Tierz,
  ``Soft matrix models and Chern-Simons partition functions,''
  Mod.\ Phys.\ Lett.\ A {\bf 19}, 1365 (2004)
  doi:10.1142/S0217732304014100
  [hep-th/0212128].
  




  \subsubsection*{ K. Fermionic matrices}\label{FMMreferences}



\bibitem{Semenoff:1996vm} 
  G.~W.~Semenoff and R.~J.~Szabo,
  ``Fermionic matrix models,''
  Int.\ J.\ Mod.\ Phys.\ A {\bf 12}, 2135 (1997)
  doi:10.1142/S0217751X97001328
  [hep-th/9605140].
  
\bibitem{Berenstein:2004hw} 
  D.~Berenstein,
  ``A Matrix model for a quantum Hall droplet with manifest particle-hole symmetry,''
  Phys.\ Rev.\ D {\bf 71}, 085001 (2005)
  doi:10.1103/PhysRevD.71.085001
  [hep-th/0409115].
  

  
\bibitem{Anninos:2015eji} 
  D.~Anninos, F.~Denef and R.~Monten,
  ``Grassmann Matrix Quantum Mechanics,''
  JHEP {\bf 1604}, 138 (2016)
  doi:10.1007/JHEP04(2016)138
  [arXiv:1512.03803 [hep-th]].


\bibitem{Anninos:2016klf} 
  D.~Anninos and G.~A.~Silva,
  ``Solvable Quantum Grassmann Matrices,''
  J.\ Stat.\ Mech.\  {\bf 1704}, no. 4, 043102 (2017)
  doi:10.1088/1742-5468/aa668f
  [arXiv:1612.03795 [hep-th]].
  
\bibitem{Tierz:2017nvl} 
  M.~Tierz,
  ``Polynomial solution of quantum Grassmann matrices,''
  J.\ Stat.\ Mech.\  {\bf 1705}, no. 5, 053203 (2017)
  doi:10.1088/1742-5468/aa6c84
  [arXiv:1703.02454 [hep-th]].
  
\bibitem{Azeyanagi:2017drg}
  T.~Azeyanagi, F.~Ferrari and F.~I.~Schaposnik Massolo,
  ``Phase Diagram of Planar Matrix Quantum Mechanics, Tensor, and Sachdev-Ye-Kitaev Models,''
  Phys.\ Rev.\ Lett.\  {\bf 120} (2018) no.6,  061602
  doi:10.1103/PhysRevLett.120.061602
  [arXiv:1707.03431 [hep-th]].
  
  

\bibitem{Klebanov:2018nfp} 
  I.~R.~Klebanov, A.~Milekhin, F.~Popov and G.~Tarnopolsky,
  ``Spectra of eigenstates in fermionic tensor quantum mechanics,''
  Phys.\ Rev.\ D {\bf 97}, no. 10, 106023 (2018)
  doi:10.1103/PhysRevD.97.106023
  [arXiv:1802.10263 [hep-th]].


\bibitem{Berenstein:2019esh} 
  D.~Berenstein and R.~de Mello Koch,
  ``Gauged fermionic matrix quantum mechanics,''
  JHEP {\bf 1903}, 185 (2019)
  doi:10.1007/JHEP03(2019)185
  [arXiv:1903.01628 [hep-th]].
  
\bibitem{Gaitan:2020zbm} 
  G.~Gaitan, I.~R.~Klebanov, K.~Pakrouski, P.~N.~Pallegar and F.~K.~Popov, (2020)
  ``Hagedorn Temperature in Large $N$ Majorana Quantum Mechanics,''
  arXiv:2002.02066 [hep-th].
 

  \subsubsection*{ L. Entanglement entropy $\&$ matrix models}\label{entanglementreferences}

   


\bibitem{Holzhey:1994we} 
  C.~Holzhey, F.~Larsen and F.~Wilczek,
  ``Geometric and renormalized entropy in conformal field theory,''
  Nucl.\ Phys.\ B {\bf 424}, 443 (1994)
  doi:10.1016/0550-3213(94)90402-2
  [hep-th/9403108].
  
\bibitem{Calabrese:2004eu} 
  P.~Calabrese and J.~L.~Cardy,
  ``Entanglement entropy and quantum field theory,''
  J.\ Stat.\ Mech.\  {\bf 0406}, P06002 (2004)
  doi:10.1088/1742-5468/2004/06/P06002
  [hep-th/0405152].

\bibitem{Das:1995vj}
  S.~R.~Das,
  ``Geometric entropy of nonrelativistic fermions and two-dimensional strings,''
  Phys.\ Rev.\ D {\bf 51} (1995) 6901
  doi:10.1103/PhysRevD.51.6901
  [hep-th/9501090].

\bibitem{Jacobson:1995ab} 
  T.~Jacobson,
  ``Thermodynamics of space-time: The Einstein equation of state,''
  Phys.\ Rev.\ Lett.\  {\bf 75}, 1260 (1995)
  doi:10.1103/PhysRevLett.75.1260
  [gr-qc/9504004].

\bibitem{Ryu:2006bv} 
  S.~Ryu and T.~Takayanagi,
  ``Holographic derivation of entanglement entropy from AdS/CFT,''
  Phys.\ Rev.\ Lett.\  {\bf 96}, 181602 (2006)
  doi:10.1103/PhysRevLett.96.181602
  [hep-th/0603001].


    
\bibitem{Casini:2011kv} 
  H.~Casini, M.~Huerta and R.~C.~Myers,
  ``Towards a derivation of holographic entanglement entropy,''
  JHEP {\bf 1105}, 036 (2011)
  doi:10.1007/JHEP05(2011)036
  [arXiv:1102.0440 [hep-th]].

\bibitem{Lashkari:2013koa} 
  N.~Lashkari, M.~B.~McDermott and M.~Van Raamsdonk,
  ``Gravitational dynamics from entanglement 'thermodynamics',''
  JHEP {\bf 1404}, 195 (2014)
  doi:10.1007/JHEP04(2014)195
  [arXiv:1308.3716 [hep-th]].
  
\bibitem{Anninos:2014ffa} 
  D.~Anninos, S.~A.~Hartnoll, L.~Huijse and V.~L.~Martin,
  ``Large N matrices from a nonlocal spin system,''
  Class.\ Quant.\ Grav.\  {\bf 32}, no. 19, 195009 (2015)
  doi:10.1088/0264-9381/32/19/195009
  [arXiv:1412.1092 [hep-th]].
  
\bibitem{Hartnoll:2015fca} 
  S.~A.~Hartnoll and E.~Mazenc,
  ``Entanglement entropy in two dimensional string theory,''
  Phys.\ Rev.\ Lett.\  {\bf 115}, no. 12, 121602 (2015)
  doi:10.1103/PhysRevLett.115.121602
  [arXiv:1504.07985 [hep-th]].
  
\bibitem{Hartnoll:2019pwe} 
  S.~A.~Hartnoll, E.~A.~Mazenc and Z.~D.~Shi,
  ``Topological order in matrix Ising models,''
  SciPost Phys.\  {\bf 7}, no. 6, 081 (2019)
  doi:10.21468/SciPostPhys.7.6.081
  [arXiv:1908.07058 [hep-th]].
  
\bibitem{Donnelly:2019zde} 
  W.~Donnelly, S.~Timmerman and N.~Valdés-Meller,
  ``Entanglement entropy and the large $N$ expansion of two-dimensional Yang-Mills theory,''
  arXiv:1911.09302 [hep-th].

 
\bibitem{Anous:2019rqb} 
  T.~Anous, J.~L.~Karczmarek, E.~Mintun, M.~Van Raamsdonk and B.~Way,
  ``Areas and entropies in BFSS/gravity duality,''
  arXiv:1911.11145 [hep-th].
  
\bibitem{1789261}
S.~R.~Das, A.~Kaushal, G.~Mandal and S.~P.~Trivedi,
``Bulk Entanglement Entropy and Matrices,''
[arXiv:2004.00613 [hep-th]].

  
  
  \subsubsection*{ M. Wigner's black holes $\&$ SYK}
  
\bibitem{Gurau:2010ba} 
  R.~Gurau,
  ``The 1/N expansion of colored tensor models,''
  Annales Henri Poincare {\bf 12}, 829 (2011)
  doi:10.1007/s00023-011-0101-8
  [arXiv:1011.2726 [gr-qc]].
  
\bibitem{Bonzom:2011zz} 
  V.~Bonzom, R.~Gurau, A.~Riello and V.~Rivasseau,
  ``Critical behavior of colored tensor models in the large N limit,''
  Nucl.\ Phys.\ B {\bf 853}, 174 (2011)
  doi:10.1016/j.nuclphysb.2011.07.022
  [arXiv:1105.3122 [hep-th]].
  
 
  
\bibitem{Anninos:2013nra} 
  D.~Anninos, T.~Anous, P.~de Lange and G.~Konstantinidis,
  ``Conformal quivers and melting molecules,''
  JHEP {\bf 1503}, 066 (2015)
  doi:10.1007/JHEP03(2015)066
  [arXiv:1310.7929 [hep-th]].
  
  
\bibitem{Benini:2015eyy}
  F.~Benini, K.~Hristov and A.~Zaffaroni,
  ``Black hole microstates in AdS$_{4}$ from supersymmetric localization,''
  JHEP {\bf 1605} (2016) 054
  doi:10.1007/JHEP05(2016)054
  [arXiv:1511.04085 [hep-th]].
  
\bibitem{Cabo-Bizet:2019eaf} 
  A.~Cabo-Bizet and S.~Murthy,
  ``Supersymmetric phases of 4d $N=4$ SYM at large $N$,''
  arXiv:1909.09597 [hep-th].
  
\bibitem{Sachdev:2015efa} 
  S.~Sachdev,
  ``Bekenstein-Hawking Entropy and Strange Metals,''
  Phys.\ Rev.\ X {\bf 5}, no. 4, 041025 (2015)
  doi:10.1103/PhysRevX.5.041025
  [arXiv:1506.05111 [hep-th]].
  
  
\bibitem{Witten:2016iux} 
  E.~Witten,
  ``An SYK-Like Model Without Disorder,''
  J.\ Phys.\ A {\bf 52}, no. 47, 474002 (2016)
  doi:10.1088/1751-8121/ab3752
  [arXiv:1610.09758 [hep-th]].
  
  
\bibitem{Maldacena:2016hyu} 
  J.~Maldacena and D.~Stanford,
  ``Remarks on the Sachdev-Ye-Kitaev model,''
  Phys.\ Rev.\ D {\bf 94}, no. 10, 106002 (2016)
  doi:10.1103/PhysRevD.94.106002
  [arXiv:1604.07818 [hep-th]].
  
\bibitem{Polchinski:2016xgd} 
  J.~Polchinski and V.~Rosenhaus,
  ``The Spectrum in the Sachdev-Ye-Kitaev Model,''
  JHEP {\bf 1604}, 001 (2016)
  doi:10.1007/JHEP04(2016)001
  [arXiv:1601.06768 [hep-th]].
 

\bibitem{Maldacena:2016upp} 
  J.~Maldacena, D.~Stanford and Z.~Yang,
  ``Conformal symmetry and its breaking in two dimensional Nearly Anti-de-Sitter space,''
  PTEP {\bf 2016}, no. 12, 12C104 (2016)
  doi:10.1093/ptep/ptw124
  [arXiv:1606.01857 [hep-th]].
  
\bibitem{Klebanov:2016xxf} 
  I.~R.~Klebanov and G.~Tarnopolsky,
  ``Uncolored random tensors, melon diagrams, and the Sachdev-Ye-Kitaev models,''
  Phys.\ Rev.\ D {\bf 95}, no. 4, 046004 (2017)
  doi:10.1103/PhysRevD.95.046004
  [arXiv:1611.08915 [hep-th]].
  
  
\bibitem{Kitaev:2017awl} 
  A.~Kitaev and S.~J.~Suh,
  ``The soft mode in the Sachdev-Ye-Kitaev model and its gravity dual,''
  JHEP {\bf 1805}, 183 (2018)
  doi:10.1007/JHEP05(2018)183
  [arXiv:1711.08467 [hep-th]].
  
\bibitem{Anninos:2016szt} 
  D.~Anninos, T.~Anous and F.~Denef,
  ``Disordered Quivers and Cold Horizons,''
  JHEP {\bf 1612}, 071 (2016)
  doi:10.1007/JHEP12(2016)071
  [arXiv:1603.00453 [hep-th]].
  

\bibitem{Cotler:2016fpe} 
  J.~S.~Cotler {\it et al.},
  ``Black Holes and Random Matrices,''
  JHEP {\bf 1705}, 118 (2017)
  Erratum: [JHEP {\bf 1809}, 002 (2018)]
  doi:10.1007/JHEP09(2018)002, 10.1007/JHEP05(2017)118
  [arXiv:1611.04650 [hep-th]].
  
\bibitem{Saad:2019lba} 
  P.~Saad, S.~H.~Shenker and D.~Stanford, (2019)
  ``JT gravity as a matrix integral,''
  arXiv:1903.11115 [hep-th].
  
  
  
 
  \subsubsection*{ N. Finiteness \& Cosmology?}\label{deSitterreferences}

\bibitem{Gibbons:1977mu} 
  G.~W.~Gibbons and S.~W.~Hawking,
  ``Cosmological Event Horizons, Thermodynamics, and Particle Creation,''
  Phys.\ Rev.\ D {\bf 15}, 2738 (1977).
  doi:10.1103/PhysRevD.15.2738

\bibitem{Hartle:1983ai} 
  J.~B.~Hartle and S.~W.~Hawking,
  ``Wave Function of the Universe,''
  Phys.\ Rev.\ D {\bf 28}, 2960 (1983)
  [Adv.\ Ser.\ Astrophys.\ Cosmol.\  {\bf 3}, 174 (1987)].
  doi:10.1103/PhysRevD.28.2960
  
\bibitem{Coleman:1988tj} 
  S.~R.~Coleman,
  ``Why There Is Nothing Rather Than Something: A Theory of the Cosmological Constant,''
  Nucl.\ Phys.\ B {\bf 310}, 643 (1988).
  doi:10.1016/0550-3213(88)90097-1
  
\bibitem{Giddings:1988wv} 
  S.~B.~Giddings and A.~Strominger,
  ``Baby Universes, Third Quantization and the Cosmological Constant,''
  Nucl.\ Phys.\ B {\bf 321}, 481 (1989).
  doi:10.1016/0550-3213(89)90353-2

\bibitem{Li:2001ky} 
  M.~Li,
  ``Matrix model for de Sitter,''
  JHEP {\bf 0204}, 005 (2002)
  [AIP Conf.\ Proc.\  {\bf 607}, no. 1, 146 (2002)]
  doi:10.1063/1.1454368, 10.1088/1126-6708/2002/04/005
  [hep-th/0106184].
  
\bibitem{Silverstein:2001xn} 
  E.~Silverstein,
  ``(A)dS backgrounds from asymmetric orientifolds,''
  Clay Mat.\ Proc.\  {\bf 1}, 179 (2002)
  [hep-th/0106209].
  
    
\bibitem{Strominger:2001gp} 
  A.~Strominger,
  ``Inflation and the dS / CFT correspondence,''
  JHEP {\bf 0111}, 049 (2001)
  doi:10.1088/1126-6708/2001/11/049
  [hep-th/0110087].
  
\bibitem{Kachru:2003aw} 
  S.~Kachru, R.~Kallosh, A.~D.~Linde and S.~P.~Trivedi,
  ``De Sitter vacua in string theory,''
  Phys.\ Rev.\ D {\bf 68}, 046005 (2003)
  doi:10.1103/PhysRevD.68.046005
  [hep-th/0301240].
  
\bibitem{Goheer:2002vf}
N.~Goheer, M.~Kleban and L.~Susskind,
``The Trouble with de Sitter space,''
JHEP \textbf{07}, 056 (2003)
doi:10.1088/1126-6708/2003/07/056
[arXiv:hep-th/0212209 [hep-th]]; 
  M.~K.~Parikh and E.~P.~Verlinde,
  ``De Sitter holography with a finite number of states,''
  JHEP {\bf 0501} (2005) 054
  doi:10.1088/1126-6708/2005/01/054
  [hep-th/0410227];
  T.~Banks, B.~Fiol and A.~Morisse,
  ``Towards a quantum theory of de Sitter space,''
  JHEP {\bf 0612} (2006) 004
  doi:10.1088/1126-6708/2006/12/004
  [hep-th/0609062];
  X.~Dong, B.~Horn, E.~Silverstein and G.~Torroba,
  ``Micromanaging de Sitter holography,''
  Class.\ Quant.\ Grav.\  {\bf 27} (2010) 245020
  doi:10.1088/0264-9381/27/24/245020
  [arXiv:1005.5403 [hep-th]];
  D.~Anninos, S.~A.~Hartnoll and D.~M.~Hofman,
  ``Static Patch Solipsism: Conformal Symmetry of the de Sitter Worldline,''
  Class.\ Quant.\ Grav.\  {\bf 29}, 075002 (2012)
  doi:10.1088/0264-9381/29/7/075002
  [arXiv:1109.4942 [hep-th]];
S.~Leuven, E.~Verlinde and M.~Visser,
``Towards non-AdS Holography via the Long String Phenomenon,''
JHEP \textbf{06}, 097 (2018)
doi:10.1007/JHEP06(2018)097
[arXiv:1801.02589 [hep-th]];
H.~Geng, S.~Grieninger and A.~Karch,
``Entropy, Entanglement and Swampland Bounds in DS/dS,''
JHEP \textbf{06}, 105 (2019)
doi:10.1007/JHEP06(2019)105
[arXiv:1904.02170 [hep-th]];
A.~Lewkowycz, J.~Liu, E.~Silverstein and G.~Torroba,
``$T \bar T$ and EE, with implications for (A)dS subregion encodings,''
[arXiv:1909.13808 [hep-th]].

  
\bibitem{Polyakov:2007mm} 
  A.~M.~Polyakov,
  ``De Sitter space and eternity,''
  Nucl.\ Phys.\ B {\bf 797}, 199 (2008)
  doi:10.1016/j.nuclphysb.2008.01.002
  [arXiv:0709.2899 [hep-th]].

\bibitem{Anninos:2012qw} 
  M.~Spradlin, A.~Strominger and A.~Volovich,
  ``Les Houches lectures on de Sitter space,''
  hep-th/0110007;
  D.~Anninos,
  ``De Sitter Musings,''
  Int.\ J.\ Mod.\ Phys.\ A {\bf 27}, 1230013 (2012)
  doi:10.1142/S0217751X1230013X
  [arXiv:1205.3855 [hep-th]].


\bibitem{Bautista:2015wqy} 
  T.~Bautista and A.~Dabholkar,
  ``Quantum Cosmology Near Two Dimensions,''
  Phys.\ Rev.\ D {\bf 94}, no. 4, 044017 (2016)
  doi:10.1103/PhysRevD.94.044017
  [arXiv:1511.07450 [hep-th]].
  
\bibitem{Betzios:2016lne} 
  P.~Betzios, U.~Gürsoy and O.~Papadoulaki,
  ``Matrix Quantum Mechanics on $S^{1}/{\mathbb Z}_{2}$,''
  Nucl.\ Phys.\ B {\bf 928}, 356 (2018)
  doi:10.1016/j.nuclphysb.2018.01.019
  [arXiv:1612.04792 [hep-th]].
  
\bibitem{1789241}
P.~Betzios and O.~Papadoulaki,
``Liouville theory and Matrix models A Wheeler DeWitt perspective,''
[arXiv:2004.00002 [hep-th]].
  
\bibitem{Denef:2011ee} 
  F.~Denef,
  ``TASI lectures on complex structures,''
  doi:10.1142/9789814350525 0007
  arXiv:1104.0254 [hep-th].
  
\bibitem{Anninos:2011kh} 
  D.~Anninos and F.~Denef,
  ``Cosmic Clustering,''
  JHEP {\bf 1606}, 181 (2016)
  doi:10.1007/JHEP06(2016)181
  [arXiv:1111.6061 [hep-th]].
  
 


\bibitem{Anninos:2017eib} 
  D.~Anninos, F.~Denef, R.~Monten and Z.~Sun,
  ``Higher Spin de Sitter Hilbert Space,''
  JHEP {\bf 1910}, 071 (2019)
  doi:10.1007/JHEP10(2019)071
  [arXiv:1711.10037 [hep-th]].




\bibitem{Anninos:2017hhn} 
  D.~Anninos and D.~M.~Hofman,
  ``Infrared Realization of dS$_2$ in AdS$_2$,''
  Class.\ Quant.\ Grav.\  {\bf 35}, no. 8, 085003 (2018)
  doi:10.1088/1361-6382/aab143
  [arXiv:1703.04622 [hep-th]].


\bibitem{Anninos:2018svg} 
  D.~Anninos, D.~A.~Galante and D.~M.~Hofman,
  ``De Sitter Horizons \& Holographic Liquids,''
  JHEP {\bf 1907}, 038 (2019)
  doi:10.1007/JHEP07(2019)038
  [arXiv:1811.08153 [hep-th]].
  
  

\bibitem{Gross:2019ach}
  D.~J.~Gross, J.~Kruthoff, A.~Rolph and E.~Shaghoulian,
  ``$T\overline{T}$ in AdS$_2$ and Quantum Mechanics,''
  arXiv:1907.04873 [hep-th].
  
\bibitem{Gross:2019uxi} 
  D.~J.~Gross, J.~Kruthoff, A.~Rolph and E.~Shaghoulian,
  ``Hamiltonian deformations in quantum mechanics, $T\bar T$, and SYK,''
  arXiv:1912.06132 [hep-th].

 
  
  
\bibitem{Maldacena:2019cbz} 
  J.~Maldacena, G.~J.~Turiaci and Z.~Yang,
  ``Two dimensional Nearly de Sitter gravity,''
  arXiv:1904.01911 [hep-th].
  
\bibitem{Cotler:2019nbi} 
  J.~Cotler, K.~Jensen and A.~Maloney,
  ``Low-dimensional de Sitter quantum gravity,''
  arXiv:1905.03780 [hep-th].

  
\bibitem{Cotler:2019dcj} 
  J.~Cotler and K.~Jensen,
  ``Emergent unitarity in de Sitter from matrix integrals,''
  arXiv:1911.12358 [hep-th].
  
  


  
  
  
  

\end{thebibliography}
\end{document}